\documentclass[]{jkpaper}
\usepackage{tikz}
\usepackage{tikz-cd}
\usepackage{bigints}
\usepackage{biblatex}
\usepackage{aas_macros}
\usepackage{soul}

\usetikzlibrary{decorations.pathreplacing}
\usetikzlibrary{arrows.meta}

 \newcommand{\pah}[1]{#1}
 \newcommand{\red}[1]{#1}
 \newcommand{\jjvk}[1]{#1}

\newcommand{\OIST}{\raisebox{-0.08em}{\includegraphics[height=0.8em]{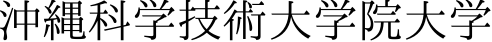}}}

\title{Diffeomorphism-invariant observables and dynamical frames in gravity: reconciling bulk locality with general covariance}
\author{Christophe Goeller,\texorpdfstring{\textsuperscript{1\,*}}{} Philipp A.\ H\"ohn,\texorpdfstring{\textsuperscript{2,3\,$\dagger$}}{} and Josh Kirklin\texorpdfstring{\textsuperscript{2\,$\ddagger$}}{}}
\institution{\texorpdfstring{\textsuperscript{1}}{}Arnold Sommerfeld Center for Theoretical Physics,\\ Ludwig-Maximilians-Universit\"at M\"unchen,\\
Theresienstrasse 37, 80333 M\"unchen, Germany\texorpdfstring{\\\vspace*{1.5em}}{ } \texorpdfstring{\textsuperscript{2}}{}Qubits and Spacetime Unit,\texorpdfstring{\\}{ } Okinawa Institute of Science and Technology \emph{(}\OIST\emph{)},\texorpdfstring{\\}{ } 1919-1 Tancha, Onna-son, Kunigami-gun, Okinawa, Japan 904-0495
\texorpdfstring{\\\vspace*{1.5em}}{ } \texorpdfstring{\textsuperscript{3}}{}Department of Physics and Astronomy,\\ University College London,\\ Gower Street, London, WC1E 6BT, United Kingdom}

\email{\textsuperscript{*}\emaillink{C.Goeller@physik.uni-muenchen.de}\\ \textsuperscript{$\dagger$}\emaillink{philipp.hoehn@oist.jp}\\ \textsuperscript{$\ddagger$}\emaillink{joshua.kirklin@oist.jp}}

\abstr{
    We describe a completely general and fully non-perturbative framework for constructing dynamical reference frames in generally covariant theories, and for understanding the gauge-invariant observables that they yield. Our approach makes use of a `universal dressing space', which contains as a subset every possible dynamical frame. We describe examples of such frames, including matter frames, a popular construction based on boundary-anchored geodesics and one using minimal surfaces -- but our formalism does not depend on the existence of a boundary. The class of observables we construct generalises and unifies the dressed and relational approaches to constructing  gravitational observables, including single-integral and canonical power-series constructions. All these (possibly gravitationally charged) relational observables describe physics in a precise sense  relative to the dynamical frame and respect a notion of `relational' locality based on the relationships between fields. By using `relational atlases', i.e.\ collections of dynamical frames glued together by field-dependent maps (which are relational observables too), we can construct relationally local observables throughout spacetime. This further establishes a framework for dynamical frame covariance that permits us to change between arbitrary relational frame perspectives. Relational locality obeys many desirable properties: we prove that it satisfies microcausality in the bulk (in tension with previous work done mainly in a perturbative setting which we comment on), and show that it permits a relational version of local bulk dynamics. Relational locality is therefore arguably more physically meaningful than the ordinary notion of locality. Thus, our formalism -- which we argue to be an updated, gauge-invariant version of general covariance -- refutes the commonly claimed non-existence of local gravitational bulk physics. 
}

\bibliography{refs}

\begin{document}
\maketitleandtoc

\section{Introduction}
\label{Section: Introduction}

One of the central pillars underlying gravitational physics is the principle of \emph{general covariance}, which (loosely speaking) says that the laws of physics are the same in every frame of reference. Precise interpretations of this statement can vary \pah{(we provide one below)}, but usually the `laws of physics' are tensor equations on spacetime, and a `frame of reference' is a set of \pah{non-dynamical} spacetime coordinates. One maps between the different frames of reference using standard coordinate transformations, and general covariance says that the form of the tensor equations stays the same, no matter which coordinates we use.

The general covariance that plays a role in gravity is background-independent, and a well-known implication of this is that acting on the fields with a broad class of diffeomorphisms (which includes all those which are trivial near \pah{any possible} boundaries of spacetime), leaves the physical state that they represent unchanged -- so these diffeomorphisms are gauge symmetries~\cite{rovelliQuantumGravity2004,Norton_1993}. This adds a layer of conceptual complexity to gravitational systems that is absent in ordinary field theories, owing to the fact that such diffeomorphisms move spacetime points around.

Observables which are invariant under such transformations are not too difficult to come across. For example, we could take any top form covariantly constructed from the \jjvk{local} fields, and integrate that top form over spacetime. Under the action of a diffeomorphism on the fields (i.e.\ an `active' diffeomorphism), such an integral is invariant. Similarly, we could take some bilocal function that is covariantly constructed from the fields, and integrate it twice over spacetime with respect to the volume form from the metric. Again, such a double integral will be invariant under active diffeomorphisms, and so correspond to a physical observable. Unfortunately, observables such as these usually do not capture much of the interesting physics in the theory. Indeed, since they are global over spacetime, they will be oblivious to any concrete local bulk dynamics\pah{, unless we can sufficiently and covariantly restrict the support of the integrand \cite{DeWitt:1962cg,Marolf:1994wh,Giddings:2005id,Marolf:2015jha,Donnelly:2016rvo}}.

On the other hand, if we try to construct a physical observable that depends non-trivially on the fields at a single fixed point in the bulk, we will clearly fail, because diffeomorphisms can move that point around arbitrarily. \pah{In line with this, it has been shown that non-perturbative gauge-invariant observables in general relativity (under certain conditions) involve infinitely many derivatives \cite{Torre:1993fq}, and that gauge-invariant observables describing perturbations around Minkowski or AdS spacetimes necessarily have a non-local support and violate microcausality as defined relative to the background~\cite{Donnelly:2015hta,Donnelly:2016rvo,Giddings:2018umg} (see also~\cite{Dittrich:2006ee,Brunetti:2013maa,Brunetti:2016hgw,Rejzner:2016yuy,Khavkine:2015fwa}). This seems to lead to a tension between general covariance and the cherished principle of locality; objects that are local relative to (fixed) spacetime points are typically not gauge-invariant and, conversely, observables that are gauge-invariant are not local to fixed events. Related to this is the fact that the bulk dynamics in generally covariant theories is generated by a Hamiltonian constraint which appears at first to render time evolution indistinguishable from a gauge transformation (this is the so-called `problem of time'~\cite{Kuchar:1991qf,Isham:1992ms,Hoehn:2019fsy}); a physical notion of time evolution can thus not be defined with respect to a fixed time coordinate.

Such observations} have sometimes led to the claim that there are no physical local bulk degrees of freedom in gravity. But this claim is misleading, and in this paper we will advocate for a different perspective: that we have to rethink what we mean by locality, \pah{subsystems and time evolution} \jjvk{in the gravitational context}. Indeed, the standard notion of bulk locality loses much of its physical meaning in the presence of diffeomorphism invariance, and so is not appropriate for gravitational physics.

%This fact and various other arguments have sometimes led to the claim that there are no physical local bulk degrees of freedom in gravity. But this claim is misleading, and in this paper we will advocate for a different perspective: that we have to rethink what we mean by locality and subsystems in gravity. Indeed, the standard notion of bulk locality loses much of its physical meaning in the presence of diffeomorphism invariance, and so is not appropriate for gravitational physics (see~\cite{Torre:1993fq,Donnelly:2015hta,Donnelly:2016rvo,Giddings:2018umg,Dittrich:2006ee,Brunetti:2013maa,Brunetti:2016hgw,Rejzner:2016yuy,Khavkine:2015fwa,Giddings:2022jda} for related discussion).

Our aim will be to present a better notion of locality -- one which maintains its physical significance despite the generally covariant nature of the theory. What properties should such a physical notion of locality have? The following wishlist seems like a good starting point:
\begin{enumerate}
    \item \textbf{Gauge-invariant local observables.} It should be possible to construct non-trivial observables that are local, in the sense that they are associated with points in spacetime.
    \item \textbf{Non-trivial local dynamics.} Most of these observables should evolve non-trivially in time.
    \item \textbf{Mutually compatible reference frames.} There should be a large variety of reference frames which respect the notion of locality. Here by `reference frames' we roughly mean systems for organising the local observables and the spacetime points to which they are associated. Additionally, these systems should be compatible with one another, in the sense that one can freely map between them.
    \item \textbf{Microcausality.} The local observables should respect the causal structure of spacetime. In particular, they should Poisson commute when spacelike separated.
\end{enumerate}
The non-existence of gauge-invariant observables that depend on the fields at a single fixed point in the bulk means the standard notion of locality clearly doesn't fulfil the first requirement in this wishlist. Similarly, it cannot satisfy the requirement of local dynamics, \pah{due to the `problem of time' alluded to above}. If we didn't insist on gauge-invariance for our local observables, then the requirement of microcausality would be difficult to address, because the Poisson bracket is only unambiguously defined for gauge-invariant observables. But the standard notion of locality does fulfil the third requirement in the wishlist. The reference frames are  \pah{(non-dynamical)} coordinate systems, and general covariance guarantees their mutual compatibility.

It is not particularly radical to assert that there is a sensible and physically meaningful kind of locality that plays a role in gravitational theories. Intuitively speaking this is obviously true. Each of this paper's readers will know that they are a `local' observer, and as local observers we have access to our local neighbourhoods, which contain local degrees of freedom, from which we can make physical local observations -- despite existing in a gravitational universe. A fundamental observation is that this kind of locality depends on the \emph{relationship} between the observer and the rest of the universe. For this reason, we will call it `relational locality'. We will show that relational locality satisfies all the requirements in the above wishlist (in the bulk).

The gauge-invariant observables which respect relational locality, and therefore which fulfil the first requirement in our wishlist, are called `relational observables'. \pah{These are intimately tied to a choice of dynamical frame and the} basic idea behind \pah{them} is \pah{to} describe the \pah{remaining degrees of freedom, which by themselves might otherwise be gauge-dependent, relative to this frame in a gauge-invariant fashion. Relational observables have appeared in various guises in the literature}. In gravity, some of the first attempts at constructing such observables were in~\cite{Bergmann:1960wb,Bergmann:1961wa,Bergmann:1961zz} and~\cite{DeWitt:1962cg}. They were then made more explicit in~\cite{Rovelli:1989jn,Rovelli:1990ph,Rovelli:1990pi,Rovelli:2001bz,rovelliQuantumGravity2004,Dittrich:2004cb,Thiemann:2004wk,Gambini:2000ht}, before being generalised for classical gravity in~\cite{Dittrich:2005kc,Dittrich:2006ee,Dittrich:2007jx,Thiemann:2007pyv,Pons:2009cz,Giesel:2007wi,Dapor:2013hca,Khavkine:2015fwa}; a review of these developments may be found in~\cite{Tambornino:2011vg}. They have also appeared further in various approaches to quantum gravity, such as in loop quantum gravity / loop quantum cosmology~\cite{rovelliQuantumGravity2004,Thiemann:2007pyv,Giesel:2012rb,Husain:2011tk,Domagala:2010bm,Rovelli:1993bm,Kaminski:2008td,Kaminski:2009qb,Bojowald:2011zzb,Ashtekar:2011ni}, effective quantum gravity~\cite{Giddings:2005id,Giddings2006,Marolf:2015jha}, the perturbative algebraic QFT approach to quantum gravity~\cite{Brunetti:2013maa,Brunetti:2016hgw,Rejzner:2016yuy}, and asymptotic safety~\cite{Baldazzi:2021fye}. They have been used in cosmology~\cite{Giesel:2017roz,Giesel:2018tcw,Giesel:2020bht}, and in quantum cosmology more generally~\cite{Marolf:1994nz,Marolf:1994wh,Marolf:2009wp,Hohn:2018iwn,Chataignier:2020fap}, including effective approaches \cite{Bojowald:2010xp,Bojowald:2010qw,Hohn:2011us,}. Besides gravity, they can also be used in gauge theories~\cite{Carrozza:2020bnj} and play a prominent role in the literature on quantum reference frames \cite{delaHamette:2021oex,Hoehn:2021wet,Hoehn:2019fsy,Hoehn:2020epv,Hoehn:2021flk,Loveridge:2017pcv,loveridge2017relativity,miyadera2016approximating,Vanrietvelde:2018pgb,Vanrietvelde:2018dit,Krumm:2020fws,Bartlett:2006tzx}. They are notoriously difficult to quantize; for some recent progress in mechanical/cosmological models see~\cite{Hoehn:2019fsy,Hoehn:2020epv,delaHamette:2021oex,Chataignier:2019kof,Chataignier:2020fys,Dittrich:2016hvj}.

Another set of gauge-invariant observables that have been considered separately in the literature are `dressed observables'. These are constructed by starting with a so-called `bare' gauge-dependent observable, and using some other degrees of freedom to modify it in such a way that it becomes gauge-invariant. The prescription for carrying out this modification is sometimes called a `dressing procedure'. Dressed observables have seen much use in various approaches to quantum gravity such as holography~\cite{Shenker1,Shenker2,Shenker3,Maldacena,Heemskerk,Kabat,QEC1,Blommaert:2019hjr}, perturbative quantum gravity~\cite{Donnelly:2015hta,Donnelly:2016rvo,Donnelly:2017jcd,Giddings:2018umg,Giddings:2019hjc}, and in Jackiw-Teitelboim gravity~\cite{Blommaert:2019hjr,Harlow:2021dfp}, among many others. 

 \pah{Note that dressed observables are based on the relationship between the bare degrees of freedom, and the auxiliary degrees of freedom that are used to dress them. One observation we will make in this work is that the dressing degrees of freedom are nothing but a dynamical frame in the same sense as above.} Thus, dressed observables are relational observables\pah{, describing the bare field relative to the `dressing frame'; within the framework that we shall describe, they can be constructed in the same manner}.\footnote{\pah{Relational observables have usually been constructed relative to (spacetime) local frames, while the frames of dressed observables are typically boundary-anchored and non-local. However, as we will see, they transform in the same way under diffeomorphisms and thereby lead to the same construction and interpretation of relational observables}.} \jjvk{We will mainly just use the name `relational observables' to refer to both types.}

To give some intuition for how relational observables work, let us temporarily depart from gravity. Key to understanding any physical system is the identification of its physical degrees of freedom. In the simplest models, these degrees of freedom are independent of the observer who wishes to access them, and their analysis is relatively simple. However, in more physically accurate scenarios things are complicated by the fact that the observer itself is a physical object whose state is inextricably linked with that of the system under observation.\footnote{Nowhere is this more evident than in gravity, where local observers are modelled as entities living at points in a spacetime whose dynamics they wish to measure.}

One way to get around this link between observer and system is to use a `gauged' description of the physics, i.e.\ to introduce a set of so-called kinematical states, and to employ a correspondence between kinematical states and physical states. In general, there are more kinematical degrees of freedom than physical ones, so multiple kinematical states can correspond to the same physical state -- this is known as `gauge redundancy'. Moreover, some kinematical states do not correspond to physical states at all -- restricting to those which do is known as `imposing constraints'.\footnote{\pah{This refers strictly to the classical case, where we refer to states in the constraint surface still as kinematical states and to their gauge equivalence classes as physical states. In the quantum theory, imposing the constraints directly produces physical states.}} At the price of these complications, by gauging one can establish a kinematical picture within which the observer and system are no longer linked, and so associate with the system a set of observer-independent kinematical degrees of freedom.

Let us consider a down to earth example: an electrician testing an electronic circuit. The electrician may describe different parts of the circuit as having different voltages. But this is a gauged, kinematic description, and by themselves the voltages of individual components have no true significance. Rather, it is the differences between the voltages of separate components (defined in terms of the integral of the electric field along a path between them) that carry physical meaning. Still, the electrician considers certain components to be special, and says that they are at `zero volts'. Really, this is a relational statement -- it is shorthand for saying that the electrician has the same voltage as the component. The electrician can touch such components without receiving a shock.

Another example is that of Yang-Mills theory. In this case, the gauged description of the physics involves a Lie algebra valued gauge potential $A$, as well as various charged local objects. Because they carry charge, the states of the objects are gauge-dependent, so they can not be observed directly. However, one can obtain gauge-independent information by considering two charged objects and a Wilson line connecting them (i.e.\ the path-ordered exponential of $A$ in some representation along a curve between the two objects), e.g.\ see \cite{Giddings:2015lla,Donnelly:2015hta}. This is relevant to the present story because, just as the electrician has a voltage, a general observer carries Yang-Mills charge. Such an observer can measure other charged objects by attaching to them with Wilson lines. The resulting observables depend on the relationship between the charged state of the observer, and the charged state of the object. Thus, they are relational observables \cite{Carrozza:2020bnj}.

In both of these examples, the gauged version of the physics used local kinematical degrees of freedom. However, the relational observables that were produced were inherently non-local. This reflects the presence of non-local degrees of freedom in the underlying physical theory. Such non-local degrees of freedom can be hard to deal with conceptually and computationally, and these are the main reasons we tend to use the gauged description. Note that even though there are non-local physical degrees of freedom in theories such as Yang-Mills theory, there are also local ones.

Let us now return to gravity, whose ordinary kinematical description involves a set of fields on a spacetime manifold. This picture involves a certain kind of locality which can be understood in terms of a fixed \pah{(i.e.\ field-independent)} labelling of the points in spacetime. This could involve some sets of coordinates, or it could take a more abstract form. Within the kinematical picture, each such point has its own special role to play, and each of them carries a certain number of local kinematical degrees of freedom, which taken altogether make up the fields.

Just as electricians have a voltage, and Yang-Mills observers carry charge, in gravity local observers have their own kinematical degrees of freedom describing how they relate to the rest of the kinematical fields, and in particular how they are embedded in spacetime: their position, their momentum, and so on. Within the kinematical picture these degrees of freedom are distinct and special. But consistency with general covariance requires diffeomorphisms to act on the observer's degrees of freedom as well as the fields. As a consequence, the observer's position, momentum, etc.\ lose meaning in the physical picture, when taken alone. However, by combining the observer's kinematical degrees of freedom with those of the fields in a relational manner, we can obtain observables which are gauge-independent. One such observable might be the value of the Ricci scalar at the location of the observer; another might be the components of the Ricci curvature along the momentum of the observer. Clearly these observables are local \emph{in relation to} the observer. This is `relational locality'.

Relational locality is unglued from the fixed labelling of spacetime points that went into the kinematical construction\pah{, as it depends on the dynamical fields}. This is fine, because the kinematical locality was only ever a crutch that was used when specifying the theory. By promoting our viewpoint to a relational one, we are discarding this crutch, gaining in exchange a much more powerful and physically compelling perspective into the dynamical nature of locality in gravity.

Having defined relational locality, we can then address the second item in the wishlist: local dynamics. Because relational observables are gauge-invariant, they do not evolve with respect to diffeomorphisms that are compactly supported in the bulk. This would appear to be at odds with a notion of local time evolution. However, as we will show, \pah{gravitationally charged} relationally local observables \emph{do} evolve with respect to global time translations, which correspond to diffeomorphisms which act non-trivially on the boundary. Moreover, we will show that there is a well-defined gauge-covariant function in the bulk which acts as a local clock for the global time evolution \pah{and can be considered a bulk-extended boundary time}. Thus, the relational notion of locality satisfies the requirement of non-trivial local dynamics. In the case of a spacetime without boundary, there is no global time translation. However, we will show that \pah{one can nevertheless construct (under certain conditions) a relational bulk time evolution for relational observables with trivial gravitational charge. This encompasses the notion of relational dynamics that has appeared primarily in the canonical literature, e.g.\ see~\cite{Rovelli:1989jn,Rovelli:1990ph,rovelliQuantumGravity2004,Thiemann:2007pyv,Tambornino:2011vg,Dittrich:2005kc,Dittrich:2007jx} and includes the notion of relational dynamics underlying the Page-Wootters formalism~\cite{Page:1983uc,Giovannetti:2015qha,Smith:2017pwx} which has been shown to be equivalent to the relational observables framework~\cite{Hoehn:2019fsy,Hoehn:2020epv}.}

Let us now move on to the third item in the wishlist. Each local observer in a gravitational theory can carry with it some set of coordinates for its local neighbourhood. These coordinates are the observer's reference frame. However, the observer is a dynamical entity, so (unlike an ordinary set of coordinates) the reference frame carried by the observer must also be dynamical -- in other words it must depend on and evolve with the rest of the degrees of freedom in the theory. The necessity of such dynamical frames is far from a novel insight, and this perspective has informally played a role in gravitational physics essentially since the very beginnings of general relativity. Indeed, Einstein himself recognised their importance~\cite{einstein1970autobiographical}:
    \begin{quote}
        ``The theory.... introduces two kinds of physical
things, i.e., (1) measuring rods and clocks, (2) all other things, e.g., the electro-magnetic field,
the material point, etc. This, in a certain sense, is inconsistent; strictly speaking measuring rods
and clocks would have to be represented as solutions of the basic equations..., not, as it were,
as theoretically self-sufficient entities...''
    \end{quote}
In our opinion, what has been lacking up until now is a satisfactory unified framework within which to formalise and develop the mathematical structures underlying such dynamical frames, and the relational and dressed observables that are based upon them. In this paper, we will describe such a framework. To our knowledge it is the most general relational formalism that has so far been presented in the literature, and we believe that it is capable of encompassing as special cases essentially all the previous examples of gravitational relational observables. \pah{It applies to spacetimes with and without boundaries.} We are hopeful that by exposing in a relatively rigorous manner the basic structures of the ideas discussed above, we will lay some groundwork for a better understanding of the foundational components of gravity -- one that does not rely upon the kinematical crutch.

\pah{In order to extend the notion of general covariance to dynamical frames, we will need a general space of frames.}
We will describe a general dressing procedure, and explain how it gives rise to a general notion of dynamical coordinates for spacetime. \pah{This leads to what we call the `universal dressing space', which \jjvk{contains as subsets all possible} gauge-covariant dynamical frames for spacetime, which in particular need not be globally defined.} We will show that different choices of dressings correspond to different dynamical frames, and in particular explain how this implies that dressed observables are therefore nothing but relational observables relative to certain choices of frame\jjvk{, as alluded to above}.

\pah{Besides establishing \jjvk{in this way} an equivalence between dressed and relational observables, our framework will also permit us to unify and generalise previous approaches to gravitational relational observables. We will show that our covariant relational observables are equivalent to the `single-integral' representation of relational observables studied in ~\cite{DeWitt:1962cg,Marolf:1994wh,Giddings:2005id,Giddings2006,Marolf:2015jha,Donnelly:2016rvo} and (under certain restrictions) to the canonical power series representation of relational observables developed in~\cite{Dittrich:2004cb,Dittrich:2005kc,Dittrich:2006ee,Dittrich:2007jx,Thiemann:2004wk}. In fact, our construction will yield a generalisation of the single-integral and power-series representations to encompass relational observables that are gravitationally charged. Furthermore, it will expand the single-integral representation to dynamical frames that are not globally defined in spacetime. Together with the notion of a relational atlas (see below), this will permit us to address (at the classical level) the issue, e.g.\ raised in \cite{Giddings:2005id,Donnelly:2016rvo}, that the relational locality implied by these observables is `state-dependent'.}

%Dressed observables have typically been more popular in covariant treatments of gravity, while relational observables have typically been more popular in canonical treatments. We thus hope that linking these two types of gravitational observables will connect the covariant and canonical settings.

Our formalism will give rise to a completely general and \emph{gauge-invariant} notion of dynamical frame covariance that allows us to map between different dynamical frames and their associated relational observables. These maps come in the form of field-dependent changes of coordinates, which leads to a dynamical form of general covariance, fulfilling the third requirement of the wishlist. \pah{In particular, we establish the notion of `relational atlases' to cover all of spacetime with dynamical coordinates.} This dynamical general covariance is a consequence of the usual notion of general covariance. However, the usual general covariance involves fictional coordinate systems (like the rods and clocks in the quote above), and the components of any spacetime objects in these coordinate systems, as well as the maps between their components in different coordinate systems, are gauge-dependent. By contrast, the dynamical general covariance that we describe involves coordinate systems built out of the physical degrees of freedom already present in the theory, with respect to which the components of spacetime objects are gauge-invariant, and so are the maps between the different coordinate systems \pah{(\jjvk{indeed, these maps} are \jjvk{themselves} relational observables)}. Thus, our dynamical formulation of general covariance is arguably more physically relevant than the original picture\pah{, \jjvk{since it} account\jjvk{s} for the fact that in practice every reference frame  is a physical system itself (e.g.\ a lab). As we shall see, this construction will support the colloquial phrase that ``all the laws of physics are the same in every \emph{dynamical} frame''.}

It should be noted that edge modes~\cite{Donnelly:2016auv,Speranza:2017gxd,Freidel:2020xyx,Freidel:2021dxw,Geiller:2019bti} can be identified as dynamical frames~\cite{Carrozza:2020bnj,CEH} (see also \cite{Kabel:2022efn}). More specifically, edge modes can be viewed as embedding fields, which are a type of reference frame. Thus, our formalism can be applied in particular to finite subregions, for which relational observables will contain gauge-invariant information about how the subregion relates to its complement, via the frame~\cite{CEH}.

\pah{
The notion of dynamical frames has recently attracted interest in the study of quantum reference frames~\cite{delaHamette:2021oex,Hoehn:2021wet,Hoehn:2019fsy,Hoehn:2021flk,Castro-Ruiz:2021vnq,delaHamette:2020dyi,Giacomini:2017zju,Giacomini:2018gxh,Vanrietvelde:2018pgb,Vanrietvelde:2018dit,Hoehn:2020epv,Hohn:2018iwn,Castro-Ruiz:2019nnl,Ballesteros:2020lgl,Krumm:2020fws,delaHamette:2021iwx}, where the key question is of how the different descriptions of quantum physics that arise from using different quantum reference frames\pah{, that may be in relative superposition,} are related to one another (this is called `quantum frame covariance'). \pah{Such frames are always associated with a (finite-dimensional) gauge group.} The results in this paper may be viewed as a classical gravitational version of quantum frame covariance, \pah{i.e.\ as an extension to dynamical frames associated with the infinite-dimensional diffeomorphism group. For dynamical frame covariance in non-gravitational gauge field theories, see \cite{Carrozza:2020bnj}.}

}

The final property on our wishlist is microcausality. This property says that the notion of locality we are defining is relativistically meaningful. In particular, it implies that measurements of different local observables cannot affect each other if they are spacelike separated, and is intuitively equivalent to the statement that the information carried by these observables cannot propagate faster than light.

The relational locality that we describe in this paper obeys \emph{bulk} microcausality. We show this by computing the Poisson bracket of two relationally local bulk observables using the Peierls bracket (which is equivalent to the Poisson bracket)~\cite{Peierls:1952cb,Gieres:2021ekc}, showing that it vanishes at spacelike separation. Our \pah{non-perturbative} argument is similar to, but much more general than, the one suggested in~\cite{Marolf:2015jha} \pah{(see also \cite{Kabat})}. It should be noted that another (unpublished) argument~\cite{bianca} shows that certain canonical relational observables~\cite{Dittrich:2005kc} obey microcausality. Again, our derivation is much more general, and applies to all the frames and relational observables that can be constructed using our formalism. \pah{We will also comment on possible relations to the findings in~\cite{Donnelly:2015hta,Donnelly:2016rvo,Giddings:2018koz} that microcausality is violated in perturbative gravity.}

It should also be noted that our argument does not work for observables that are relationally local to the boundary -- so relationally local bulk observables \pah{that are gravitationally charged} should not be expected to commute with relationally local boundary observables. This is consistent with the fact that the Hamiltonian (which is a boundary observable) generates a non-trivial time evolution for \pah{such} relational observables. Notably, although relationally local observables may be constructed in any theory, (bulk) relational microcausality is only generally obeyed in gravitational ones. \pah{This failure of gravitationally charged bulk observables to commute with observables on the boundary even at spacelike separation is related to observations on boundary unitarity in quantum gravity \cite{Marolf:2008mf,Jacobson:2019gnm} and quantum error correction in holography \cite{QEC1}.}

Thus, relational locality satisfies the first three items in our wishlist for a physical notion of locality, and \emph{in the bulk} it satisfies all four. The purpose of this paper is to prove these claims and discuss their consequences. For illustrative purposes, and to motivate the various ingredients that go into our general formalism, we will describe various examples of its application. These include a popular gravitational dressing procedure involving geodesics~\cite{Donnelly:2015hta,Donnelly:2016rvo,Giddings:2018umg,Shenker1,Shenker2,Shenker3,Maldacena,Heemskerk,Kabat,QEC1,Blommaert:2019hjr,Harlow:2021dfp,CEH}, the case of parametrised field theory~\cite{Isham:1984sb,Kuchar:1989bk,Kuchar:1989wz,Torre:1992rg,Lee:1990nz,Torre:1992bg,Andrade:2010hx}, the dust fields described by Brown and Kucha\v{r}~\cite{Brown:1994py}, as well as a more exotic construction involving minimal surfaces, the main purpose of which is to demonstrate the versatility of our approach, which may be useful in holography~\cite{MinimalSurfaces1,MinimalSurfaces2,MinimalSurfaces3,MinimalSurfaces4,MinimalSurfaces5}.

Our approach to relational observables and dynamical covariance will make extensive use of the covariant phase space formalism, which is a very efficient framework for understanding the Hamiltonian structure of generally covariant theories~\cite{Lee:1990nz,Iyer:1994ys,Harlow:2021dfp}. We will also employ the notion of equivariant bundles to define local objects; we will introduce their basic properties as we need them, but a separate review may be found in~\cite{steenrod1999topology,ZouBundle}.

The rest of the paper proceeds as follows. The rest of the Introduction contains some remarks on our terminology and conventions. In Section~\ref{Section: geodesic example}, we describe a motivating example in which reference frames and relational observables are constructed using geodesics propagating in from the boundary of spacetime. This example allows us to demonstrate in a familiar setting many of the ingredients that will go into the general formalism, which we then describe in Section~\ref{Section: formalism} in some detail. Afterwards, to demonstrate the versatility of our approach, we discuss some more examples in Section~\ref{Section: more examples}. We conclude with some discussion in Section~\ref{Section: conclusion}.

We also include some appendices. Appendix~\ref{Appendix: small diffeos} contains a brief discussion of small diffeomorphisms in evolving spacetimes, Appendix~\ref{Appendix: noether} contains a derivation of some common formulae arising in the covariant phase space formalism, and Appendix~\ref{Appendix: Glossary} contains a glossary of the concepts we define.

\subsection{\pah{Definition of general covariance and some} conventions}
\label{Section: Introduction / conventions}

The term `general covariance' has different meanings to different people (see~\cite{Norton_1993,Giulini:2006yg,Freidel:2021bmc,Anderson} for some examples), owing perhaps to the fact that it was never precisely defined in the original formulations of general relativity. Before continuing, we would like to clarify exactly what we mean when we use it and a few other related terms. We are not necessarily claiming that our definitions are the ``correct'' ones, but they do seem to be the most useful for our purposes.

First, let us remind the reader of the distinction between `dynamical fields' and `background fields'. The former, which we denote $\phi$, represent physical degrees of freedom, and thus take on different values in different physical states. On the other hand, the latter, which we denote $\gamma$, are the same for every physical state, and so can just be viewed as some fixed parameters that go into the definition of the theory. Whether a field is a dynamical field or background field depends on the theory under consideration. For example, the metric is dynamical in gravity, but it is a background field in special relativity.

We'll assume that both the dynamical fields and background fields have a well-defined action of the spacetime diffeomorphism group. We could consider a joint transformation
\begin{equation}
    \phi\to f_*\phi, \qquad \gamma \to f_*\gamma,
    \label{Equation: terminology all field diffeomorphism}
\end{equation}
where $f$ is a spacetime diffeomorphism, and $f_*\phi$, $f_*\gamma$ represent the fields after being acted on by $f$. However, since the background fields are fixed within any given theory, $f$ would have to satisfy $\gamma=f_*\gamma$, in which case we say it is a symmetry of the background fields. For example, if the metric is a background field, then $f$ would have to be an isometry. An alternative is to just act on the dynamical fields:
\begin{equation}
    \phi\to f_*\phi, \qquad \gamma \to \gamma,
    \label{Equation: terminology active diffeomorphism}
\end{equation}
which is an `active' diffeomorphism. In~\eqref{Equation: terminology active diffeomorphism}, $f$ doesn't have to be a symmetry of the background fields.

It should be noted that gravitational theories are `background independent', meaning that there are no background fields, at least in the bulk. Thus, (in the bulk), in gravitational theories~\eqref{Equation: terminology all field diffeomorphism} and~\eqref{Equation: terminology active diffeomorphism} are essentially the same thing.

Another thing a spacetime diffeomorphism can do is simply act on all the points in spacetime, moving them around like $x\to f(x)$. This is a `passive' diffeomorphism. It can be viewed as equivalent to leaving the spacetime points fixed $x\to x$, and instead acting with the diffeomorphism on objects which are defined on spacetime.

Covariance is a property that relates active and passive diffeomorphisms. Suppose we have some quantity $A[\phi]$ that depends on the dynamical fields, and on spacetime. Under an active diffeomorphism, we have $A[\phi]\to A[f_*\phi]$, whereas under a passive diffeomorphism, we have $A[\phi] \to f_*\big(A[\phi]\big)$. The results of these two transformations do not have to be the same. But if they are, i.e.\ if
\begin{equation}
    A[f_*\phi] = f_*\big(A[\phi]\big),
\end{equation}
then we say that $A$ is `covariant'. It may be that $A$ is covariant only for some $f$.

For a quick contrived example, suppose the dynamical fields include a scalar field $\varphi_d$, and the background fields include a scalar field $\varphi_b$. Then we can construct three observables:
\begin{equation}
    O_1 = \varphi_d, \qquad O_2 = \varphi_b, \qquad O_3 = \varphi_d\varphi_b.
\end{equation}
Note that each of these observables are functions on spacetime. Therefore, the diffeomorphism group can act on these observables via pushforward:
\begin{equation}
    O\to f_* O = O\circ f^{-1}.
\end{equation}
This is a passive diffeomorphism. It is important to recognise that it is a transformation at the level of the \emph{observables}. It is a change from one observable $O$ to a different one $f_*O$. This change happens completely independently of the underlying fields $\varphi_d,\varphi_b$. Under an active diffeomorphism $\varphi_d\to f_*\varphi_d$, $\varphi_b\to\varphi_b$, we have
\begin{align}
    O_1 &\to f_*\varphi_d = f_*O_1, \\
    O_2 &\to \varphi_b = O_2, \\
    O_3 &\to f_*\varphi_d\,\varphi_b \ne f_* O_3.
\end{align}
For $O_1$, we can see that an active diffeomorphism and a passive diffeomorphism lead to the exact same change. Thus, $O_1$ is covariant. For $O_2$, we see that an active diffeomorphism leads to no change -- we say that $O_2$ is `diffeomorphism-invariant'. Lastly, we see that $O_3$ is neither covariant nor diffeomorphism-invariant.

Now consider the equations of motion, which typically take the form of some set of spacetime equations $E[\phi,\gamma]=0$ that depends on both the dynamical fields and background fields. If $E[\phi,\gamma]$ obeys
\begin{equation}
    E[f_*\phi,\gamma] = f_*\big(E[\phi,\gamma]\big)
    \label{Equation: eom covariance}
\end{equation}
for all $f$, i.e.\ if it is \emph{covariant} for \emph{general} diffeomorphisms, then we call the theory \emph{generally covariant}.\footnote{\pah{In~\cite{Anderson}, this was called `general invariance' (see also \cite{Norton_1993,Giulini:2006yg} for some discussion on this).}} \pah{With this definition, one can also read the second part of the title as ``reconciling bulk locality with active diffeomorphism invariance''.}

Some other formulations of general covariance allow the background fields to transform as well, replacing $\gamma$ with $f_*\gamma$ in the left-hand side of~\eqref{Equation: eom covariance}. However, in any theory there is some ambiguity in what one considers to be a background field, and what one considers to be a separate fundamental structure. For example, in special relativity one could consider the Minkowski metric to be a background field, or one could consider it to be just some numbers in the equations of motion. Thus, such a definition would inherit this ambiguity. Additionally, by defining the background fields appropriately one could make `generally covariant' many theories which have nothing to do with gravity (depending on your point of view this may or may not be a problem). In any case, as we have already commented, gravity is background independent, so from now on we will leave any dependence on $\gamma$ implicit. Our definitions allow us to do this in an unambiguous way, because $\gamma$ is always fixed.

This definition of general covariance \pah{implies} the statement that the equations of motion are the same in all coordinate systems, \pah{which is a way of saying that ``all the laws of physics are the same in every frame''}. To make this precise, suppose $\sigma$ is a \pah{(non-dynamical)} coordinate system in some open subset of spacetime, i.e.\ a sufficiently smooth map to an open subset of $\RR^D$, where $D$ is the dimension of spacetime. The `components' of the fields $\phi$ in this coordinate system are essentially given by the pushforward $\bar\phi=\sigma_*\phi$. If $E[\phi]=0$ are the equations of motion, then the components of $\phi$ clearly obey $E[\sigma^*\bar\phi] = 0$. The components of the equations of motion are given by again pushing forward through $\sigma$:
\begin{equation}
    \bar{E}[\bar\phi] = 0, \qq{where} \bar{E}[\bar\phi] = \sigma_*\big(E[\sigma^*\bar\phi]\big).
\end{equation}
Now suppose $\sigma_1, \sigma_2$ are two different coordinate systems, and define $\bar{\phi}_1,\bar{\phi}_2$ and $\bar{E}_1,\bar{E}_2$ as above. If $\bar\phi_1=\bar\phi$ is a solution to $\bar{E}_1[\bar\phi_1] = 0$, then (in the region where both coordinate systems are defined)
\begin{multline}
    \bar{E}_2[\bar\phi] = \sigma_{2\,*} \big(E[\sigma_2^*\bar\phi]\big) = \sigma_{2\,*}\big(E[(\sigma_2^{-1}\circ\sigma_1)_*\sigma_1^*\bar\phi]\big) \\
    = \sigma_{2\,*} \Big((\sigma_2^{-1}\circ\sigma_1)_* \big(E[\sigma_1^*\bar\phi]\big)\Big) = \sigma_{1\,*} \big(E[\sigma_1^*\bar\phi]\big) = \bar{E}_1[\bar\phi]=0.
\end{multline}
Here the second line follows from general covariance. Thus, $\bar\phi_2=\bar\phi$ is also a solution to $\bar{E}_2[\bar\phi_2] = 0$ -- so the equations of motion are the same in all coordinate systems.

\pah{The above is usually restricted to non-dynamical frames and coordinates. In this paper, we generalise this notion of general covariance to dynamical frames and so field-dependent coordinate maps $\sigma[\phi]$. The frame dressed fields $\bar\phi=(\sigma[\phi])_*\phi$ are then gauge-invariant relational observables and we will see that the field-dependent transition functions $(\sigma_2[\phi])^{-1}\circ\sigma_1[\phi]$ will themselves be gauge-invariant relational observables. In this manner, we will obtain a dynamical frame covariance, i.e.\ a gauge-invariant framework of frame-dependent physics. This supports the statement that
\begin{quote}
``all the laws of physics are the same in every \emph{dynamical} frame'',
\end{quote}
while retaining in gauge-invariant fashion that different frames give different (internal) descriptions of the world. We will argue that this constitutes a more physical update of the notion of general covariance that takes into account that the world is relational; physical systems are described relative to one another, rather than to fictituous or external reference structures.

}

\section{Warm-up: geodesic dressing example}
\label{Section: geodesic example}

To illustrate the main principles underlying our approach to relational locality, it is helpful to first demonstrate them in a basic example. To that end, in this section we will show how one can use geodesic dressings to construct dynamical reference frames. Geodesic dressings have been invoked frequently in various perturbative or non-perturbative scenarios to construct gauge-invariant observables, e.g.\ see~\cite{Donnelly:2015hta,Donnelly:2016rvo,Giddings:2018umg,Shenker1,Shenker2,Shenker3,Maldacena,Heemskerk,Kabat,QEC1,Blommaert:2019hjr,Harlow:2021dfp,CEH}. We will generalise the construction and make its relation to the dynamical reference frame program explicit, in particular showing that dressed observables are nothing but relational observables relative to a geodesic frame. While the authors of~\cite{Donnelly:2015hta,Donnelly:2016rvo,Giddings:2018umg} observed non-local properties of the dressed observables relative to the background spacetime in perturbative constructions, we will see later that the dynamical frame gives rise to a (non-perturbative) \emph{relational} notion of locality. The reader mainly interested in the formulation of the general formalism can safely skip this section and move to Section~\ref{Section: formalism}.

Let us consider general relativity coupled to a scalar field $\psi$ on a $D$-dimensional manifold $\mathcal{M}$. In this paper, we will make extensive use of the language of bundles over spacetime. In general, a field configuration is a section of some bundle $\Phi$ over $\mathcal{M}$, which we will refer to as the `field bundle'. The dynamical field of general relativity is the metric $g$, which is a section of the vector bundle $S^2(\mathrm{T}^*\mathcal{M})$ of rank $(0,2)$ symmetric tensors over $\mathcal{M}$. The scalar field $\psi$ is a function on $\mathcal{M}$, but may also be viewed as a section of the product bundle $\RR\times\mathcal{M}$. Since we are considering a theory of both the metric and scalar field, the total field bundle $\Phi$ is given by the fibre product of these two bundles:
\begin{equation}
    \Phi = S^2(\mathrm{T}^*\mathcal{M}) \times_{\mathcal{M}} (\RR\times\mathcal{M}) = S^2(\mathrm{T}^*\mathcal{M})\times \RR.
\end{equation}
We will use the notation $\phi$ to refer to the total field configuration, which is a section of $\Phi$. In this case, the total field configuration is made up of the metric configuration and the scalar field configuration, and so may be written $\phi:x\mapsto(g(x),\psi(x))$.

The action of this theory is a functional of $\psi$ and $g$, and is given by
\begin{equation}
    S = \int_{\mathcal{M}} \qty(\frac1{16\pi G} R - \frac12\nabla_a\psi\nabla^a\psi)\sqrt{\abs{g}}\dd[D]{x} + \frac1{8\pi G}\int_{\partial\mathcal{M}} K \sqrt{\abs{h}} \dd[D-1]{x}.
\end{equation}
Here, $R$ is the Ricci scalar of $g$, $h_{ab}$ is the induced metric on $\partial\mathcal{M}$ (which for simplicity we will assume is everywhere timelike), and $K=h_{ab} K^{ab}$ is the trace of the extrinsic curvature $K^{ab}$ of $\partial\mathcal{M}$. As is well-known, under a linearised variation $g_{ab}\to g_{ab}+\delta g_{ab}$, $\psi\to\psi+\delta\psi$ of the dynamical fields, the action changes by
\begin{multline}
   \delta S = \int_{\mathcal{M}}\qty(\frac1{16\pi G} \qty(8\pi G\,T^{ab}-R^{ab}+\frac12 R g^{ab})\delta g_{ab} + (\nabla^a\nabla_a\psi)\delta\psi)\sqrt{\abs{g}}\dd[D]{x} \\
    - \frac12 \int_{\partial\mathcal{M}} \qty(\frac1{8\pi G}(K^{ab}-K h^{ab})\delta h_{ab} + \hat{n}^a\partial_a\psi\,\delta\psi)\sqrt{\abs{h}}\dd[D-1]{x},
\end{multline}
where $R_{ab}$ is the Ricci curvature, $T_{ab} = \nabla_a\psi\nabla_b\psi - \frac12 g_{ab}\nabla_c\psi\nabla^c\psi$ is the energy-momentum tensor, and $\hat{n}$ is the outward-pointing unit normal vector of $\partial\mathcal{M}$. Thus, if we set some boundary conditions such that the boundary term above vanishes, then the extremisation of the action implies the equations of motion
\begin{equation}\label{eq:eom}
    R^{ab}-\frac12 R g^{ab} = 8\pi G\, T^{ab}, \qquad \nabla^a\nabla_a\psi = 0.
\end{equation}
In other words, the variational principle is consistent. Let us employ the standard Dirichlet boundary conditions $\delta h_{ab}=0$ and $\delta\psi=0$, i.e.\ we fix the induced metric and $\psi$ on $\partial\mathcal{M}$.

This theory is diffeomorphism invariant, meaning that if the equations of motion are satisfied, then under the action of any diffeomorphism $f:\mathcal{M}\to\mathcal{M}$ on the fields
\begin{equation}
    g_{ab}\to f_* g_{ab}, \qquad \psi \to f_*\psi,
\end{equation}
the equations of motion will continue to be satisfied. This property holds for any diffeomorphism $f$, but we must restrict to diffeomorphisms which preserve the boundary conditions at $\partial\mathcal{M}$. The boundary conditions are part of the definition of the theory, so diffeomorphisms without this property would map field configurations $g_{ab},\psi$ which are valid in a certain theory, to ones which are invalid in that theory (or alternatively to configurations which are only valid in a modified theory). Thus, they cannot play a consistent role within the fixed theory we are considering here. Let $G\subset\operatorname{Diff}(\mathcal{M})$ be the group of all boundary condition preserving diffeomorphisms.

Some diffeomorphisms in $G$ are gauge transformations, and these form a subgroup $H\subset G$. Roughly speaking, this subgroup can be identified by the way in which it acts on the fields near the boundary of any Cauchy surface. For the example that we are discussing, it consists of all diffeomorphisms which act trivially on $\partial\mathcal{M}$, i.e.\ $f\in H$ if $f(x)=x$ for all $x\in\partial\mathcal{M}$. The reason for this is not so important for the present discussion, but for more details see Appendix~\ref{Appendix: small diffeos}. We will refer to diffeomorphisms in $H$ as `small', and we will refer to any diffeomorphisms in $G$ but not in $H$ as `large'.\footnote{This use of `small' and `large' is commonly used by physicists. However, many mathematicians use a different definition: to them, `small' diffeomorphisms of compact manifolds are those within the identity component of $\operatorname{Diff}(\mathcal{M})$, and `large' diffeomorphisms are those in other components. For now, we will just use the physicist's definition, but it is possible that by defining an appropriate topology on $\operatorname{Diff}(\mathcal{M})$ (one which takes into account the privileged role played by $\partial\mathcal{M}$), the two definitions can be made consistent with each other.}

In this way, we have determined a hierarchy of groups of diffeomorphisms:
\begin{equation}
    H \subset G \subset \operatorname{Diff}(\mathcal{M}).
\end{equation}
Here $\operatorname{Diff}(\mathcal{M})$ is the group of all diffeomorphisms of $\mathcal{M}$, $G$ is the group of diffeomorphisms which preserve the boundary conditions at $\partial\mathcal{M}$, and $H$ is the group of small diffeomorphisms.

\subsection{Dressed observables}

Since small diffeomorphisms are gauge symmetries, they must leave all physical observables invariant. Suppose we want to construct an observable $\psi(x)$ that gives the value of the scalar field $\psi$ at some fixed point $x\in\mathcal{M}$. If we act on the fields with a small diffeomorphism $f$, so that $\psi\to f_*\psi = \psi\circ f^{-1}$, then under this change we have
\begin{equation}
    \psi(x) \to (\psi\circ f^{-1})(x) = \psi(f^{-1}(x)).
\end{equation}
If $x\in\partial\mathcal{M}$, then since $f$ is small we have $f^{-1}(x)=x$ and so $\psi(x)$ is gauge-invariant and hence physical. On the other hand, if $x\not\in\partial\mathcal{M}$, then for any $y\not\in\partial\mathcal{M}$ there exists a small diffeomorphism satisfying $f^{-1}(x)=y$. The only way for $\psi(x)$ to be gauge-invariant would be for $\psi$ to be constant over such $y$. This is too physically restrictive, since $\psi$ will not in general be constant. Thus, $\psi(x)$ is only a physical observable for fixed $x$ if $x\in \partial\mathcal{M}$.

Na\"ively this seems to indicate that we can only observe the value of $\psi$ on the boundary. Clearly, however, this is not true, and one way to circumvent this conclusion is to allow $x\in\mathcal{M}$ to be determined in terms of the fields $g_{ab},\psi$. If we could do this in such a way that
\begin{equation}
    g_{ab}\to f_*g_{ab},\, \psi\to f_*\psi \implies x \to f(x) \text{ for all }f\in H,
\end{equation}
then we would have
\begin{equation}
    \psi(x) \to (\psi\circ f^{-1})(f(x)) = \psi(f^{-1}(f(x))) = \psi(x)\text{ for all }f\in H,
\end{equation}
so $\psi(x)$ would be gauge-invariant and physical. It is worth stressing the conceptual difference between the $x$ that appears in this paragraph, and the one that appears in the preceding paragraph. Before, $x$ was a kinematically fixed point in $\mathcal{M}$, whereas now $x$ is \emph{dynamically} determined via its dependence on the fields, $x=x[\phi]$.\footnote{In this paper, we will use square brackets whenever a quantity depends on the fields.}  Conceptually, this amounts to defining a dynamical coordinate system on spacetime. The aim of this paper is to introduce a general way of constructing such a dynamical coordinate system.

\begin{figure}
    \centering
    \begin{tikzpicture}[scale=1.5]
        \begin{scope}[shift={(1.2,0.7)}]
            \fill[blue!10] (0,0) -- (3,0) -- (6,1) -- (6,4) -- (3,4) -- (0,0);
            \draw[blue!50,fill=blue!15,line width=0.8pt] (0,0) -- (3,1) -- (3,4) -- (0,3) -- (0,0);
        \end{scope}
        \draw[-{Latex},line width=0.8pt] (1.5,2) -- (1,2) node[left] {\large$\hat{n}$};
        \draw[-{Latex},line width=0.8pt,blue] (1.5,2) -- (1.6,2.9) node[above] {\large$W$};
        \draw[-{Latex},line width=1.2pt,red!70!black] (1.5,2) -- (2.1,2.9) node[above right] {\large$W-\hat{n}$};
        \draw[red!60!black,line width=2.8pt] (1.5,2) .. controls (2.1,2.9) and (2.8,2.7) .. (6,3.4);
        \draw[red!60,line width=1.4pt] (1.5,2) .. controls (2.1,2.9) and (2.8,2.7) .. (6,3.4);
        \draw[black,line width=1pt,-{Latex},shift={(0.1,-0.15)},dashed] (1.5,2) .. controls (2.1,2.9) and (2.8,2.7) .. (6,3.4) node[pos=0.9,below] {$\tau$};
        \fill (1.5,2) circle (0.04) node[below] {\large$z$};
        \fill[red!40!black] (6,3.4) circle (0.06) node[right] {\large$x[g]$};
        \node at (3.7,3.7) {\Large$\partial\mathcal{M}$};
        \node at (5.7,2.2) {\Large$\mathcal{M}$};
    \end{tikzpicture}
    \caption{Shooting a geodesic in from the boundary $\partial\mathcal{M}$ to construct a dressing in $\mathcal{M}$.}
    \label{Figure: geodesic shot in}
\end{figure}
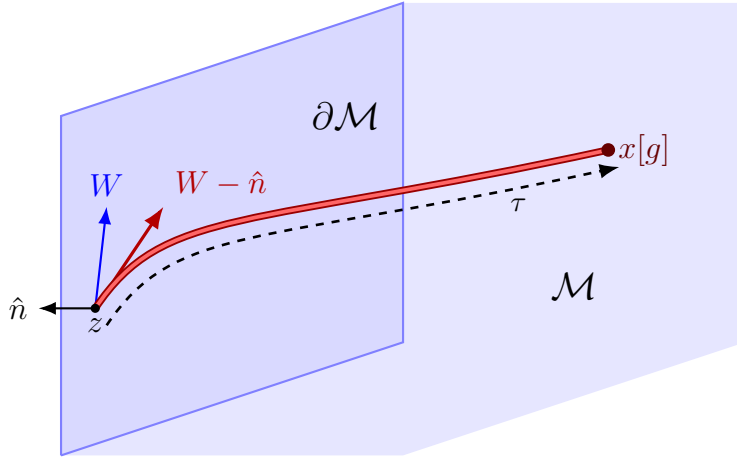

One way to obtain such an $x$ involves `shooting in' a geodesic from $\partial\mathcal{M}$.  In particular, suppose we fix a point $z \in \partial\mathcal{M}$, a vector $W\in\mathrm{T}_z\partial\mathcal{M}$, and a number $\tau$. Then, let $x$ be the point at a distance $\tau$ along the geodesic which starts at $z$ with tangent vector $W-\hat{n}$, as shown in Figure~\ref{Figure: geodesic shot in}. For simplicity, let us assume that $\abs{W}^2\ne -1$ so that this tangent vector is not null. Then by `distance' we mean either the proper time or proper length, depending on whether the geodesic is timelike or spacelike.

The point $x$ defined in this way transforms as $x\to f(x)$ when the fields transform according to the action of a small diffeomorphism $f$. In other words, $x[f_*\phi] = f(x[\phi])$. Let us see why this is true.

First, let us compute how $\hat{n}$ changes. Let $b:\mathcal{M} \to \mathbb{R}$ be some fixed (field-independent) function which is constant on $\partial\mathcal{M}$ and whose derivative is non-zero and outward-pointing on $\partial\mathcal{M}$. Then $n=\dd{b}$ is a field-independent non-vanishing outward-pointing normal 1-form to $\partial\mathcal{M}$. The normalised vector $\hat{n}$ is then defined in terms of the metric via
\begin{equation}
    \hat{n}^a = \hat{n}^a [g] = \frac{g^{ab}n_b}{\sqrt{g^{cd}n_cn_d}}.
\end{equation}
So for some diffeomorphism $f\in\operatorname{Diff}(\mathcal{M})$ we have
\begin{equation}
    \hat{n}^a [f_*g] = \frac{(f_* g)^{ab} n_b}{\sqrt{(f_*g)^{cd}n_cn_d}}.
    \label{Equation: unit normal transformation}
\end{equation}
Since pushforwards commute with exterior derivatives, we have $f_*n = f_*\dd{b} = \dd(f_*b)$. But $f_*b$ is a function which is constant on $\partial\mathcal{M}$ and whose derivative is non-zero and outward-pointing on $\partial\mathcal{M}$, meaning $\dd(f_*b)$ must be a non-vanishing outward-pointing normal 1-form to $\partial\mathcal{M}$. Therefore at $\partial\mathcal{M}$ we can write $n = \alpha f_*n$ for some positive function $\alpha$, and the factors of $\alpha$ cancel in~\eqref{Equation: unit normal transformation}, so we have
\begin{equation}
    \hat{n}^a[f_*g] = \frac{(f_* g)^{ab}(f_*n)_b}{\sqrt{(f_*g)^{cd}(f_*n)_c(f_*n)_d}} = f_*\hat{n}^a [g].
\end{equation}
Thus, the unit normal $\hat{n}$ transforms covariantly by pushforward.

Let $\gamma=\gamma[g]\subset \mathcal{M}$ be the geodesic according to the metric $g$ of length $\tau$ starting from $z\in\partial\mathcal{M}$ with tangent vector $W-\hat{n}[g]$. Its endpoint defines $x$. By the tensorial nature of the geodesic equation, $f(\gamma[g])$ is a geodesic according to the transformed metric $f_* g$. Moreover, it starts at $f(z)$ with tangent vector $f_*(W-\hat{n}[g])$, and has length $\tau$ (as measured by $f_*g)$. But note that, since small diffeomorphisms reduce to the identity at $\partial\mathcal{M}$, for $f\in H$ we have $f(z)=z$, and moreover $f_*W=W$\footnote{This is independent of how we choose to extend the definition of $W$ to a neighbourhood of $z$.} which implies
\begin{equation}
    f_*(W-\hat{n}[g]) = W-\hat{n}[f_*g].
\end{equation}
So $f(\gamma(g))$ is the geodesic according to the metric $f_*g$ of length $\tau$ starting from $z$ with tangent vector $W-\hat{n}[f_*g]$. By definition this implies
\begin{equation}
    f(\gamma[g]) = \gamma[f_*g].
\end{equation}
Thus, if $x$ is the endpoint of $\gamma(g)$, then $f(x)$ is the endpoint of $\gamma(f_*g)$. So $x\to f(x)$ under $g_{ab}\to f_* g_{ab}$ for $f\in H$, which is exactly what we set out to show.

Now that we have constructed the point $x$, we can evaluate the scalar field $\psi$ there, and since $x\to f(x)$ under the action of a small diffeomorphism on the fields, $\psi(x)$ will be gauge-invariant in the way discussed above. In this way, we have constructed a gauge-invariant observable that returns the value of the scalar field a certain distance along a geodesic. It is an observable which measures a property of the scalar field relative to a property of the metric; the metric degrees of freedom constitute a `dressing' of the bare scalar field $\psi$, and observables constructed in this way are accordingly sometimes known as `dressed' observables~\cite{Donnelly:2015hta,Donnelly:2016rvo,Giddings:2018umg,Harlow:2021dfp,Carrozza:2020bnj,CEH}.

One way to think of the dressed observable we have described is by imagining that there is an observer who has travelled along the geodesic we constructed. After their journey, the observer accesses the observable by simply examining the value of $\psi$ at their location. But there are many more observables that such an observer would be able to measure. For example, they would be able to measure the rate of change of $\psi$ with respect to their proper time $\tau$. They could also, in principle, measure how the scalar field at their location would differ if they had started at a different boundary point $z$, heading in a different direction $W-\hat n$. To formalise these different observables, it is useful to associate the dressing procedure with the notion of reference frames and frame orientations.

\subsection{Dynamical reference frames}\label{ssec_dynframe}

The set of all possible geodesic dressings constitutes a field-dependent spacetime reference frame, in the sense that it provides a dynamical coordinate system for some region of spacetime. This coordinate system may not be invertible. However, as we shall see, different restrictions of it give rise to different choices of `dynamical frame fields' on spacetime, which permit us to identify dressed observables as relational observables relative to these frame fields (in the sense of~\cite{Rovelli:1989jn,Rovelli:1990ph,Rovelli:1990pi,rovelliQuantumGravity2004,Dittrich:2005kc,Dittrich:2004cb,Tambornino:2011vg,Thiemann:2007pyv,Carrozza:2020bnj,delaHamette:2021oex,Brunetti:2013maa,Brunetti:2016hgw,Rejzner:2016yuy,Baldazzi:2021fye}).

We begin by formalising the spacetime frame associated with the dressing. We will refer to each possible configuration of the collection of parameters $\tau,z,W$ as a `local orientation' of the frame, and we denote the set of all local orientations by $\mathscr{O}$. Each possible local orientation corresponds to a different dressing, in the form of the corresponding geodesic.

Let us explicitly write $x=x_{\tau,z,W}[g]$ to denote the dependence of $x$ on the metric $g$ and the local orientation $\tau,z,W$. It is helpful to be a bit more precise about the nature of the parameters $\tau,z,W$. First, $z\in\partial\mathcal{M}$ and $W\in\mathrm{T}_y\partial\mathcal{\mathcal{M}}$ can be viewed as a point $(z,W)$ in $\mathrm{T}\partial\mathcal{M}$, the tangent bundle of $\partial\mathcal{M}$. Note that we also have to restrict to $\abs{W}^2\ne-1$, since we are excluding null geodesics.\footnote{Note that this condition only depends on the fields through the induced metric on the boundary. But our boundary conditions fix the induced metric, so really this condition doesn't have any real dependence on the fields.} The other parameter $\tau$ is just a real non-negative number. Thus, the complete space of all possible local orientations (for our geodesic dressing prescription) is
\begin{equation}
    \mathscr{O}= \RR_{\ge0}\times \mathrm{T}\partial\mathcal{M}_{\abs{W}^2\ne-1},
\end{equation}
where
\begin{equation}
    \mathrm{T}\partial\mathcal{M}_{\abs{W}^2\ne-1} = \{(z,W)\in\mathrm{T}\partial\mathcal{M}\mid1+\abs{W}^2\ne0\}.
\end{equation}
is the tangent bundle $\mathrm{T}\partial\mathcal{M}$ with vectors $W$ for which $\abs{W}^2=-1$ is removed. We can then recast the parameterised dressing as a field-dependent map from $\mathscr{O}$ to $\mathcal{M}$:
\begin{equation}\label{stframe}
    R[g]: \quad \mathscr{O} \to \mathcal{M}, \quad (\tau,(z, W)) \mapsto x_{\tau,z,W}[g].
\end{equation}
As we have seen, under the action of a small diffeomorphism on the metric $g\to f_*g$, this map changes as
\begin{equation}
    R[g]\to R[f_*g] = f\circ R[g],
    \label{Equation: reference frame covariance}
\end{equation}
This is the fundamental gauge-covariance property that defines physical dynamical reference frames in generally covariant theories.

The reference frame $R[g]$ is not injective, since there is a continuous set of (non-null) geodesics connecting a given spacetime point with the boundary $\partial\mathcal{M}$, and so continuous subsets of local orientations that correspond to the same spacetime point. The labels $(\tau,z,W)$ thus provide us with an overcomplete coordinate system on spacetime. In this sense, $R[g]$ constitutes an overcomplete spacetime frame. There is, however, also a sense in which it is incomplete: since $R[g]$ will typically also fail to be surjective, it will only provide a coordinate system for a (generally field-dependent) subregion of spacetime.

Let us now consider the notion of a dynamical frame field. This is a field on (a subregion of) spacetime that constitutes a coordinate system. To construct one from the reference frame $R[g]$, we have to restrict to a $D$-dimensional subset of $\mathscr{O}$. If $R[g]$ is an injective map on this subset, then it will be invertible on its image in $\mathcal{M}$. The inverse is the frame field.

\begin{figure}
    \centering
    \begin{tikzpicture}[scale=1.5]
        \begin{scope}[shift={(1.2,0.7)}]
            \fill[blue!10] (0,0) -- (3,0) -- (6,1) -- (6,4) -- (3,4) -- (0,0);
            \draw[blue!50,fill=blue!15,line width=0.8pt] (0,0) -- (3,1) -- (3,4) -- (0,3) -- (0,0);
            \draw[red!60,fill=red!10,rounded corners=10pt,dashed] (0.5,0.8) -- (3,1.1) -- (4.8,1.8) -- (4.15,3.1) .. controls (3,2.9) .. (2.6,3.2) -- (0.5,2.6) -- cycle;
            \begin{scope}
                \clip (0.5,0.8) -- (3,1.1) -- (4.8,1.8) -- (4.15,3.1) .. controls (3,2.9) .. (2.6,3.2) -- (0.5,2.6) -- cycle;
                \draw[blue!20,line width=0.8pt] (0,0) -- (3,1) -- (3,4) -- (0,3) -- (0,0);
            \end{scope}
            \draw[blue!50,fill=purple!30,line width=0.8pt,rounded corners=10pt] (0.5,0.8) -- (2.5,1.4) -- (2.6,3.2) -- (0.5,2.6) -- cycle;
        \end{scope}
        \foreach \ia in {1,2,3,4} {
            \foreach \ib in {1,2,3} {
                \begin{scope}[shift={(\ia/2,\ib/2+\ia/6)}]
                    \draw[-{Latex},line width=0.8pt,blue] (1.4-0.03*\ia,1+0.05*\ia) -- (1.6,1.5-0.02*\ia);
                \end{scope}
            }
        }
        \foreach \ib in {1,2,3} {
            \foreach \ia in {4,3,2,1} {
                \begin{scope}[shift={(\ia/2,\ib/2+\ia/6)}]
                    \draw[red!60!black,line width=2.6pt] (1.4-0.03*\ia,1+0.05*\ia) .. controls (1.6,1.22-0.02*\ia) and (2.8,1.1) .. (4-\ia*\ib*0.07,1.6-\ib*0.1);
                    \draw[red!60,line width=1.4pt] (1.4-0.03*\ia,1+0.05*\ia) .. controls (1.6,1.22-0.02*\ia) and (2.8,1.1) .. (4-\ia*\ib*0.07,1.6-\ib*0.1);
                    \fill (1.4-0.03*\ia,1+0.05*\ia) circle (0.04);
                \end{scope}
            }
        }
        \foreach \ia in {1,2,3,4} {
            \foreach \ib in {1,2,3} {
                \begin{scope}[shift={(\ia/2,\ib/2+\ia/6)}]
                    \fill[red!40!black] (4-\ia*\ib*0.07,1.6-\ib*0.1) circle (0.06);
                \end{scope}
            }
        }
        \node[blue!60!black] at (2.2,3.7) {\large$\mathcal{U}_1$};
        \node[red!60!black] at (6.1,3.3) {\large$\mathcal{N}_1[g]$};
    \end{tikzpicture}
    \caption{Different choices of sets of values taken by the parameters $\tau$, $z$ and $W$ lead to different sets of geodesic dressings. Each such set may be viewed as a different frame. For example, here we choose a maximum length $\tau_1$, and a vector field $W_1$ in a boundary subregion $\mathcal{U}_1$. Then we vary over all $z\in\mathcal{U}_1$ and $\tau\in[0,\tau_1)$, and set $W=W_1(z)$ at $z$. The region of spacetime $\mathcal{N}_1[g]$ accessible by the frame is called its image.}
    \label{Figure: different geodesic frames}
\end{figure}
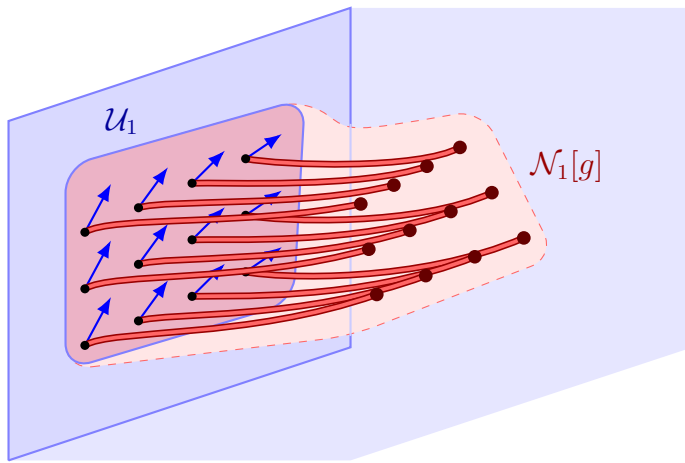

Clearly, there are many ways to achieve this. For example, we can fix a boundary vector field $W_1$ and consider
\begin{equation}
    \mathscr{O}_1=\{(\tau,z,W)\in\mathscr{O}\mid \tau\in[0,\tau_1),\,z\in\mathcal{U}_1,\,\text{and } W(z)=W_1(z) \text{ for all $z$}\}\subset\mathscr{O}
\end{equation}
where $\tau_1\in\RR_{\geq0}$ and $\mathcal{U}_1\subset\partial\mathcal{M}$ are chosen such that
\begin{equation}
    R_1[g]=R[g]\big|_{\mathscr{O}_1}:\mathscr{O}_1\rightarrow\mathcal{M}
\end{equation}
is injective (this choice may be valid only locally in field space).\footnote{Since a congruence of geodesics fired in from the boundary may develop caustics or orbits, or the spacetime may feature singularities, we generally have to restrict to a subset of $(\tau,z,W_1)$.} This is illustrated in Figure~\ref{Figure: different geodesic frames}. The dynamical frame field corresponding to the boundary vector field $W_1$ is then given by the inverse
\begin{equation}\label{eq:geoframe1}
    (R_1[g])^{-1}:\quad \mathcal{N}_1[g]\rightarrow\mathscr{O}_1,\quad x\mapsto(T_1(x),Z_1(x)),
\end{equation}
where $\mathcal{N}_1[g]=R_1[g](\mathscr{O}_1)\subset\mathcal{M}$ is the frame's image subregion in spacetime and
\begin{equation}
    T_1(x_{\tau,z,W_1}[g])=\tau\in[0,\tau_1),\qquad Z_1(x_{\tau,z,W_1}[g])=z\in\mathcal{U}_1.
\end{equation}
Since the values this map assigns to spacetime points depend on the metric configuration, we have that $T_1[g],Z_1[g]$ are metric-dependent dynamical fields in spacetime. In particular, we can choose a coordinate system $\varphi$ on $\mathcal{U}_1$ and define $Z^k=\varphi^k\circ Z:\mathcal{N}_1[g]\to\mathbb{R}$, $k=1,\dots,D-1$, and note that, owing to \eqref{Equation: reference frame covariance}, we have for any small diffeomorphism $f\in H$
\begin{equation}\label{eq:geoframe2}
    f_* T_1 = T_1\circ f^{-1},\qquad f_*Z_1^k=Z_1^k\circ f^{-1}.
\end{equation}
In other words, $(T_1,Z_1^k)$ constitute $D$ (metric-dependent) scalar fields on spacetime. The construction establishes the coordinates $(\tau=T_1(x),z^k=Z_1^k(x))$ both on spacetime and the space $\mathscr{O}_1$ of local frame orientations. This is very similar to the Brown-Kucha\v{r} dust models~\cite{Brown:1994py}, which we shall discuss in Section~\ref{Section: more examples / dust} and in which the comoving dust coordinates furnish $D$ scalar fields on spacetime.

Choosing a different boundary vector field $W_2$ would lead to a different dynamical frame field $(R_2[g])^{-1}$ (and hence distinct metric-dependent scalar fields $T_2,Z_2^k$) with a similarly defined local orientation space $\mathscr{O}_2$ and spacetime domain $\mathcal{N}_2[g]\subset\mathcal{M}$. Alternatively, one could fix a boundary anchor point $z\in\partial\mathcal{M}$ and keep $\tau,W$ free to vary, thus firing in a fan of geodesics in different directions from a single location. Clearly, there exist many other possibilities. This will become useful below when discussing changes of dynamical frame field. 

For the purpose of our general formalism in Section~\ref{Section: formalism}, we note that all possible dynamical frame fields associated with the spacetime frame $R[g]$ constitute (local) sections $s[g]:\mathcal{M}\to\mathscr{O}$ of the space $\mathscr{O}$, which may be viewed as a bundle over spacetime $\mathcal{M}$, with projection given by $R[g]$ in~\eqref{stframe} itself. These frame fields constitute dynamical reference frames in the same sense in which they have appeared in the context of relational observables in gravity~\cite{Rovelli:1990ph,Rovelli:1990pi,rovelliQuantumGravity2004,Dittrich:2005kc,Tambornino:2011vg,Thiemann:2007pyv,Brunetti:2013maa,Brunetti:2016hgw,Rejzner:2016yuy,Baldazzi:2021fye}, as edge modes in gauge theories~\cite{Carrozza:2020bnj} and gravity~\cite{CEH}, and in the recent literature on quantum reference frames~\cite{delaHamette:2021oex,Hoehn:2021wet,Hoehn:2019fsy,Hoehn:2021flk,Castro-Ruiz:2021vnq,delaHamette:2020dyi,Giacomini:2017zju,Giacomini:2018gxh,Vanrietvelde:2018pgb,Vanrietvelde:2018dit,Hoehn:2020epv,Hohn:2018iwn,Castro-Ruiz:2019nnl,Ballesteros:2020lgl}.

Let us say a few words about terminology. In the previous literature on dynamical reference frames and relational observables, the frames would have been identified with the frame fields on spacetime themselves. However, in view of the general formalism of Section~\ref{Section: formalism}, we will find it convenient to relax the notion of dynamical frame and apply it instead to (possibly non-invertible) field-dependent coordinate maps, such as $R[g]$, from some local orientation space into spacetime. We will reserve the more specific term `frame field' for the case that the frame map is invertible. In other words, each dynamical frame field is associated with a canonical frame (its inverse), whereas a dynamical frame does not come with a preferred choice of frame field when it is not injective.

\subsection{Dressed observables are relational observables}\label{ssec_georelobs}

Dynamical reference frames and local orientation spaces permit us to consider a variety of different types of gauge-invariant dressed observables. Let us begin with the non-injective spacetime frame $R[g]$ given in \eqref{stframe}. We can extend the dressed scalar
\begin{equation}
    O_{\psi,R}(\tau,z,W)=\big((R[g])^*\psi\big)(\tau,z,W)=\psi\left(R[g](\tau,z,W)\right)
\end{equation}
to a function on the local orientation space $\mathscr{O}$. It measures the value of the scalar field at the end point of the boundary anchored geodesic parametrised by $(\tau,z,W)$.

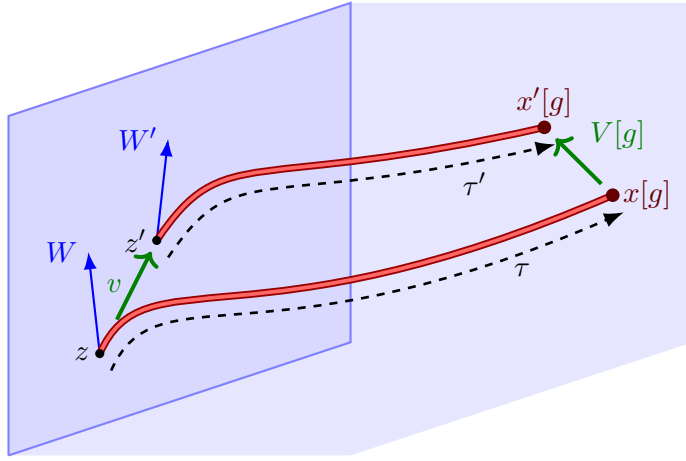
\begin{figure}
    \centering
    \begin{tikzpicture}[scale=1.5]
        \begin{scope}[shift={(1.2,0.7)}]
            \fill[blue!10] (0,0) -- (3,0) -- (6,1) -- (6,4) -- (3,4) -- (0,0);
            \draw[blue!50,fill=blue!15,line width=0.8pt] (0,0) -- (3,1) -- (3,4) -- (0,3) -- (0,0);
        \end{scope}
        \begin{scope}[shift={(0.5,-0.4)}]
            \draw[-{Latex},line width=0.8pt,blue] (1.5,2) -- (1.4,2.9) node[left] {$W$};
            \draw[red!60!black,line width=2.8pt] (1.5,2) .. controls (1.9,2.9) and (2.8,2) .. (6,3.4);
            \draw[red!60,line width=1.4pt] (1.5,2) .. controls (1.9,2.9) and (2.8,2) .. (6,3.4);
            \draw[black,line width=1pt,-{Latex},shift={(0.1,-0.15)},dashed] (1.5,2) .. controls (1.9,2.9) and (2.8,2) .. (6,3.4) node[pos=0.9,below] {$\tau$};
            \fill (1.5,2) circle (0.04) node[left] {$z$};
            \fill[red!40!black] (6,3.4) circle (0.06) node[right] {$x[g]$};
        \end{scope}
        \begin{scope}[shift={(1,0.6)}]
            \draw[-{Latex},line width=0.8pt,blue] (1.5,2) -- (1.6,2.9) node[left] {$W'$};
            \draw[red!60!black,line width=2.8pt] (1.5,2) .. controls (2.1,2.9) and (2.3,2.4) .. (4.9,3);
            \draw[red!60,line width=1.4pt] (1.5,2) .. controls (2.1,2.9) and (2.3,2.4) .. (4.9,3);
            \draw[black,line width=1pt,-{Latex},shift={(0.1,-0.15)},dashed] (1.5,2) .. controls (2.1,2.9) and (2.3,2.4) .. (4.9,3) node[pos=0.9,below] {$\tau'$};
            \fill (1.5,2) circle (0.04) node[left] {$z'$};
            \fill[red!40!black] (4.9,3) circle (0.06) node[above] {$x'[g]$};
        \end{scope}
        \draw[green!50!black,->,line width=1.5pt] (2.15,1.9) -- (2.45,2.5) node[midway, left] {$v$};
        \draw[green!50!black,->,line width=1.5pt] (6.4,3.1) -- (6,3.5) node[midway, above right] {$V[g]$};
    \end{tikzpicture}
    \caption{A vector $v$ on $\mathscr{O}$ is an infinitesimal change in the parameters of a dressing, $(\tau,z,W)\to(\tau',z',W')$. Pushing $v$ forward to spacetime with the frame yields a spacetime vector $V[g]$ that gives the corresponding infinitesimal change in the end of the geodesic.}
    \label{Figure: geodesic vector pushforward}
\end{figure}

This point of view allows us to measure more complicated relational observables. For example, if one varies the parameters $\tau,z,W$, the point $x_{\tau,z,W}[g]$ will also change, and we can compute the corresponding change of the value of $\psi$ at that point. More precisely, a change in local orientation is a vector $v\in \mathrm{T}\mathscr{O}$. Pushing this vector forward through the frame map $R[g]$, we get a vector $V[g]=(R[g])_*v$, which points from the endpoint of the original geodesic to the endpoint of the one resulting from the change in parameters (see Figure~\ref{Figure: geodesic vector pushforward}). Then we can measure $\partial_a\psi$ in the direction of $V$:
\begin{equation}
    V[g]^a\partial_a\psi = V[g](\psi) = v(O_{\psi,R}).
\end{equation}
This is gauge-invariant because $O_{\psi,R}$ is gauge-invariant and $v$ is a fixed vector.

Next, suppose $A$ is some type $(0,k)$ tensor quantity on spacetime $\mathcal{M}$. Provided $R[\phi]$ is sufficiently smooth, we can use it to pull back such quantities to $\mathscr{O}$. In particular, the dressed object
\begin{equation}
    O_{A,R}(\tau,z,W)=\big((R[g])^*A\big)(\tau,z,W),
\end{equation}
is a gauge-invariant type $(0,k)$ tensor on local orientation space $\mathscr{O}$, because for $f\in H$
\begin{equation}
   O_{A,R}\mapsto (R[f_*g])^*f_*A=(R[g])^*f^*f_*A=O_{A,R}.
\end{equation}
The observable $O_{A,R}(\tau,z,W)$ measures the components of the tensor $A$ at the end point of the geodesic specified by $(\tau,z,W)$, along directions corresponding to changes of geodesic arising from variations in these parameters. For instance, we can choose $A=g$ to be the metric itself. Then we get the dressed metric $O_{g,R}=(R[g])^*g$, which is a rank $2$ symmetric tensor field on $\mathscr{O}$ that measures the metric components in a diffeomorphism-invariant fashion.

Let us now turn to dynamical frame fields and relational observables. For example, consider the dynamical frame field $(R_1[g])^{-1}$ as defined in the previous subsection. Since this furnishes a diffeomorphism from $\mathcal{N}_1[g]$ to $\mathscr{O}_1$, we can pushforward any tensor quantity $B$ on spacetime $\mathcal{M}$ to a corresponding gauge-invariant tensor field on the local orientation space $\mathscr{O}_1$:
\begin{equation}\label{eq:br1}
    O_{B,R_1}[\phi](\tau,z^k)=\big((R_1[g])^{-1}_* B\big)(\tau,z^k)=\big((R_1[g])^*B\big)(\tau,z^k).
\end{equation}
This constitutes a relational observable that measures the components of the tensor $B$ relative to the frame field $(R_1[g])^{-1}$. For example, for $B=\psi$, the observable $O_{\psi,R_1}(\tau,z^k)=\psi(x_{\tau,z^k,W_1}[g])$ encodes the question ``what is the value of the scalar field $\psi$ at the event in spacetime $\mathcal{M}$ where the frame field $(R_1[g])^{-1}$ assumes the local orientation $(\tau,z^k)$?'' Similarly, for the metric $B=g$, we have the gauge-invariant metric components
\begin{equation}\label{eq:dressedg}
    g_{\mu\nu}(y)=\big(O_{g,R_1}\big)_{\mu\nu}(y)=\partial_\mu R^a_1(y)\,\partial_\nu R_1^b(y)\, g_{ab}(R_1(y)),
\end{equation}
where, to lighten the notation, we have written $y^\mu$ for the coordinate system $(\tau,z^k)$ and $R^a_1[g]$ is the composition of the frame mapping $R_1[g]$ with an arbitrary coordinate system on the neighbourhood $\mathcal{N}_1$ in spacetime. $g_{\mu\nu}$ measures the components of the metric at the event in spacetime, where $(R_1[g])^{-1}$ is in local orientation $(\tau,z^k)$, along directions $\partial_\mu R^a_1$ encoding changes of geodesic as the parameters $(\tau,z^k)$ are varied.

Let us note a subtlety concerning the counting of independent degrees of freedom. While at first sight it may now seem that \emph{all} the metric components can be turned into gauge-invariant observables, recall that the relational observables are describing the metric relative to a frame field that through the geodesic dressing is itself constructed from the metric. More precisely, $g_{\mu\nu}$ describes the metric relative to other metric degrees of freedom. As a consequence, there is some redundancy in the components of the dressed metric. For example, the $\tau\tau$ component of the dressed metric will always be $g_{\tau\tau}=\pm1$, since $\tau$ measures the proper time or proper length along a curve. Another important fact is that the dressing procedure is non-local. If we restrict our attention to a subregion that doesn't touch the boundary, so that the boundary anchored geodesics may have support in the causal complement of the subregion, one can construct the metric components as independent gauge-invariant observables. These then describe the metric relative to a spacetime non-local frame in the complement~\cite{CEH}. By contrast, for subregions that touch the boundary and contain the complete dressing geodesics, the dressed metric components, while gauge-invariant, will not all be independent.

The quantities $O_{B,R_1}[\phi]$ are versions in the \emph{covariant} setting of the relational observables usually considered in the canonical formulation~\cite{Dittrich:2004cb,Dittrich:2005kc,rovelliQuantumGravity2004,Tambornino:2011vg,Thiemann:2007pyv,delaHamette:2021oex,Hoehn:2019fsy}, and have appeared in similar form before in~\cite{Brunetti:2013maa,Brunetti:2016hgw,Rejzner:2016yuy,Baldazzi:2021fye} (however, not for geodesic dressings). In Section~\ref{Section: formalism}, we shall substantially generalise these constructions of dynamical frames and relational observables for generally covariant theories, and clarify the link with their canonical incarnation in Section~\ref{Section: general formalism / covariant to canonical}. Frame field dressed observables thus coincide with relational observables, an observation that was previously made for non-generally covariant gauge theories in~\cite{Carrozza:2020bnj}.

In~\cite{Donnelly:2015hta,Donnelly:2016rvo,Giddings:2018umg}, it has been shown that geodesically dressed matter observables feature non-local properties (e.g.\ they violate relativistic microcausality) \emph{relative to the background spacetime} in perturbative expansions. This led to the question of how locality and subsystems should be meaningfully defined in the diffeomorphism-invariant context of gravity. The present identification of dressed observables with relational observables in our construction will permit us to propose a possible answer in this work. In Section~\ref{Section: general formalism / relational phase space / microcausality}, we will see that non-perturbative relational observables will satisfy locality properties \emph{relative to the local orientation space $\mathscr{O}$} on which the dressed metric $g_{\mu\nu}(y)$ defines a causal structure too. In particular, they will satisfy microcausality: if local orientations $(\tau,z^k)$ and $(\tau',z'^k)$ correspond to spacelike separated points in spacetime $\mathcal{M}$, then the Peierls bracket (which is equivalent to the covariant phase space Poisson bracket~\cite{Kirklin:2019xug,Gieres:2021ekc}) of relational observables $O_{A,R_1}(\tau,z^k)$ and $O_{B,R_1}(\tau',z'^k)$  (or smeared versions thereof) will vanish.\footnote{\jjvk{We will discuss the conflict between this result and~\cite{Donnelly:2015hta,Donnelly:2016rvo,Giddings:2018umg} in the conclusion.}} This suggests that a gauge-invariant notion of locality in gravity should be defined in a relational manner. In the context of perturbative expansions of relational observables, the tension between relational and spacetime background locality has previously been observed in~\cite{Dittrich:2006ee,Brunetti:2013maa,Brunetti:2016hgw,Rejzner:2016yuy}.

\subsection{Changes of dynamical frame}\label{ssec_geoframechange}

Having various frame fields at our disposal, we can now consider what it means to change between the gauge-invariant relational descriptions that correspond to different dynamical frames. This will amount to a gauge-invariant field-dependent changes of coordinates. For the gauge theory version of this, see~\cite{Carrozza:2020bnj}.

Let us consider two dynamical frame fields $(R_1[g])^{-1}:\mathcal{N}_1[g]\to\mathscr{O}_1$ and $(R_2[g])^{-1}:\mathcal{N}_2[g]\to\mathscr{O}_2$ and let us assume that their spacetime domains overlap non-trivially $\mathcal{N}_1\cap\mathcal{N}_2\neq\emptyset$. The relational observable
\begin{equation}\label{eq:fcm}
    R_{1\to2}[g]=O_{R_2^{-1},R_1}[g]=(R_1[g])^*(R_2[g])^{-1} = (R_2[g])^{-1}\circ R_1[g]
\end{equation}
describes the second frame field relative to the first in a gauge-invariant manner,\footnote{Under $f\in H$, it transforms as
$ R_{1\rightarrow2}[f_* g] = (R_2[f_* g])^{-1} \circ R_1[f_* g] = (R_2[g])^{-1} \circ f^{-1} \circ f \circ R_1[g] = R_{1\rightarrow2}[g]$.
} and is a map from (a subset of) $\mathscr{O}_1$ to (a subset of) $\mathscr{O}_2$.
We can use it to map a relational observable $O_{B,R_1}[\phi]$, describing some dynamical spacetime tensor quantity $B$ relative to the first frame, into the corresponding relational observable $O_{B,R_2}[\phi]$, describing $B$ relative to the second frame. Indeed,
\begin{equation}\label{eq:relobschange}
    O_{B,R_2}[\phi] = (R_{1\to 2}[g])_* O_{B,R_1}[\phi] = (R_2[g])^*(R_1[g])_*(R_1[g])^*B[\phi]=(R_2[g])^*B[\phi],
\end{equation}
which has the form of a change of coordinates. The quantum analog of such a change of relational observables (for mechanical systems) can be found in~\cite{delaHamette:2021oex}. 

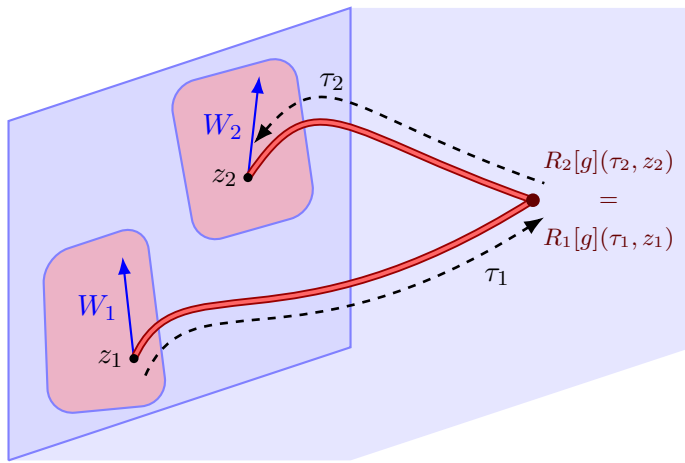
\begin{figure}
    \centering
    \begin{tikzpicture}[scale=1.5]
        \begin{scope}[shift={(0.9,0.7)}]
            \fill[blue!10] (0,0) -- (3,0) -- (6,1) -- (6,4) -- (3,4) -- (0,0);
            \draw[blue!50,fill=blue!15,line width=0.8pt] (0,0) -- (3,1) -- (3,4) -- (0,3) -- (0,0);
            \draw[blue!50,fill=purple!30,line width=0.8pt,rounded corners=10pt,shift={(0.2,0)}] (0.15,0.4) -- (1.2,0.5) -- (1,2.1) -- (0.1,1.8) -- cycle;
            \draw[blue!50,fill=purple!30,line width=0.8pt,rounded corners=10pt,shift={(1.5,1.5)}] (0.15,0.4) -- (1.2,0.7) -- (1,2.1) -- (-0.1,1.8) -- cycle;
        \end{scope}
        \begin{scope}[shift={(0.5,-0.4)}]
            \draw[-{Latex},line width=0.8pt,blue] (1.5,2) -- (1.4,2.9) node[left,midway] {$W_1$};
            \draw[red!60!black,line width=2.8pt] (1.5,2) .. controls (1.9,2.9) and (2.8,2) .. (5,3.4);
            \draw[red!60,line width=1.4pt] (1.5,2) .. controls (1.9,2.9) and (2.8,2) .. (5,3.4);
            \draw[black,line width=1pt,-{Latex},shift={(0.1,-0.15)},dashed] (1.5,2) .. controls (1.9,2.9) and (2.8,2) .. (5,3.4) node[pos=0.9,below right] {$\tau_1$};
            \fill (1.5,2) circle (0.04) node[left] {$z_1$};
        \end{scope}
        \begin{scope}[shift={(1.5,1.2)}]
            \draw[-{Latex},line width=0.8pt,blue] (1.5,2) -- (1.6,2.9) node[left,midway] {$W_2$};
            \draw[red!60!black,line width=2.8pt] (1.5,2) .. controls (2.1,2.9) and (2.3,2.4) .. (4,1.8);
            \draw[red!60,line width=1.4pt] (1.5,2) .. controls (2.1,2.9) and (2.3,2.4) .. (4,1.8);
            \draw[black,line width=1pt,{Latex}-,shift={(0.1,0.15)},dashed] (1.45,2.15) .. controls (2.1,2.9) and (2.3,2.4) .. (4,1.8) node[pos=0.4,above] {$\tau_2$};
            \fill (1.5,2) circle (0.04) node[left] {$z_2$};
            \fill[red!40!black] (4,1.8) circle (0.06) node[right,align=center] {\footnotesize  $R_2[g](\tau_2,z_2)$\\\footnotesize  $=$\\\footnotesize $R_1[g](\tau_1,z_1)$};
        \end{scope}
    \end{tikzpicture}
    \caption{Given two frames of geodesics specified by vector fields $W_1,W_2$ in boundary subregions $\mathcal{U}_1,\mathcal{U}_2$, the map $R_{1\to 2}[g]$ implementing the change of frames involves going along one geodesic tangent to $W_1-\hat{n}$ from $\mathcal{U}_1$ into the bulk, and then along a second geodesic to $\mathcal{U}_2$ where it is tangent to $W_2-\hat{n}$. This gives a map $\tau_1,z_1\mapsto \tau_2,z_2$.}
    \label{Figure: geodesic frame change}
\end{figure}

It is instructive to see this more explicitly in an example. Suppose we have constructed two frames corresponding to the two geodesic congruences specified by the fixed choices $W_1$ and $W_2$ of the boundary vector field, as discussed in Section~\ref{ssec_dynframe}. One can change between the parameters for these two frames by moving into spacetime along a geodesic in one of the congruences, and then back out to the boundary along a geodesic in the other congruence, as depicted in Figure~\ref{Figure: geodesic frame change}. The two corresponding frame fields $(R_1[g])^{-1},(R_2[g])^{-1}$ are given by the sets of fields $Y_i^\mu[g]=(T_i[g],Z^k_i[g])$ with associated coordinates $y^\mu_i=(\tau_i,z^k_i)$ on the local orientation space $\mathscr{O}_i$, $i=1,2$ and $\mu=0,1,\ldots,D-1$. The frame change map in \eqref{eq:fcm} then reads more concretely (with a slight abuse of notation)
\begin{equation}\label{eq:gframectrans}
    Y_2^\mu(y_1)=R^\mu_{1\to2}[g](\tau_1,z_1^k)=Y^\mu_2(x_{\tau_1,z_1^k,W_1}[g]).
\end{equation}
For a fixed metric configuration $g$, this defines the coordinate change $y_2^\mu(y_1)$, encoding the question ``what is the value of the scalar field $Y^\mu_2$ in metric configuration $g$ at the event in spacetime $\mathcal{M}$ where the frame field is in local orientation $y_1$?''. For the change of relational observable in \eqref{eq:relobschange} let us choose the metric, i.e.\ $B=g$, and recall \eqref{eq:dressedg}. For a fixed metric configuration $g$, the change of frame map \eqref{eq:relobschange} becomes a mere coordinate transformation
\begin{equation}
    g_{\mu_2\nu_2}(y_2)=\frac{\partial y^{\mu_1}_1}{\partial y^{\mu_2}_2}\,\frac{\partial y_1^{\nu_1}}{\partial y_2^{\nu_2}}\,g_{\mu_1\nu_1}(y_1(y_2)).
\end{equation}
There are, however, two crucial differences to standard coordinate changes on a spacetime manifold: (i) the coordinate transformation is \emph{field-dependent} and invariant under small spacetime diffeomorphisms, and (ii) $g_{\mu_i\nu_i}$ are the \emph{gauge-invariant} components of the dressed metric on the local orientation space $\mathscr{O}_i$.

Using the geodesic dressings to define dynamical frame fields and relational observables thereby gives rise to a gauge-invariant, yet frame-dependent description of the local physics in spacetime. The generalization of this will later permit us to formalise a dynamical and gauge-invariant (and so arguably more physical) notion of general covariance.

\subsection{Frame reorientations}

Before turning to the general formalism for generally covariant theories, let us briefly note that there exists another type of dynamical frame transformation. While the changes of frame just described amount to a change of description of a given physical situation (the field configuration is unchanged), we can also consider \emph{frame reorientations}, which correspond to a change of field configuration that reconfigures the frame field, but not the remaining degrees of freedom. In our present example, where we use a geodesic frame, this means essentially changing the metric configuration, but leaving the scalar $\psi$ untouched. Reorientations have also been discussed in the context of edge modes in gauge theory and gravity~\cite{Carrozza:2020bnj,CEH}, and quantum reference frames~\cite{delaHamette:2021oex}.

For concreteness, let us consider once more the frame field $(R_1[g])^{-1}$ associated with the fixed boundary vector field $W_1$. Suppose two field configurations $\phi=(g,\psi)$ and $\bar\phi=(\bar g,\psi)$ have the same configuration $\psi$ for the scalar, but different configurations $g$ and $\bar{g}$ for the metric. Then we can define a diffeomorphism on the local orientation space $\mathscr{O}_1$
\begin{equation}
  F[g,\bar{g}] = (R_1[g])^{-1} \circ R_1[\bar{g}]
\end{equation}
which depends on both $g$ and $\bar{g}$. This diffeomorphism can be thought of as implementing the frame reorientation associated with the change of metric $g\to \bar{g}$. It obeys
\begin{equation}
    x_{\tau,z^k,W_1}[\bar g]=(F[g,\bar g])^*x_{\tau,z^k,W_1}[g]=x_{\tau,z^k,W_1}[g]\circ F[g,\bar g]. 
\end{equation}
Thus, while gauge transformations $f\in H$ act on the frame $R_1[g]$ from the left (see \eqref{Equation: reference frame covariance}), frame reorientations act from the right.
Applying this to the scalar field relational observable with $B=\psi$ in \eqref{eq:br1} yields simply $(F[g,\bar g])^* O_{\psi,R_1}[\phi] = O_{\psi,R_1}[\bar\phi]$, or
\begin{equation}
    (F[g,\bar g])^*\psi(x_{\tau,z^k,W_1}[g])=\psi(x_{\tau,z^k,W_1}[\bar g]).
\end{equation}
This amounts to evaluating the scalar field at two distinct events in spacetime for the same scalar field configuration and hence will in general return distinct values.

Here, we have made use of the kinematical splitting between the metric $g$ and scalar field $\psi$. Since these are coupled by the equations of motion \eqref{eq:eom}, $\bar\phi$ will typically not be a solution  if $\phi$ is one. Thus, such a frame reorientation will only in special cases be tangential to the space of solutions and accordingly be a symmetry of the theory.\footnote{However, when studying subregions using a geodesic dressing frame that originates in the spacetime complement, such reorientations are compatible with the solution space \cite{CEH}.} There is a second issue with this kinematical notion of frame reorientations: the underlying kinematical splitting between the degrees of freedom of the frame and the degrees of freedom of everything else is not gauge-invariant. Essentially, this reflects the fact that there is no intrinsic manner in which one can reorient a given frame in a relational world. In Section~\ref{Section: formalism}, we will describe this in more detail, and remedy both issues by directly defining frame reorientations on-shell, and in a relational manner. That is, we will reorient different frames relative to one another.

\section{The general formalism}
\label{Section: formalism}

In this section we will describe in detail the general formalism that underlies our approach to relational locality and reference frames in gravity. First, we will establish some preliminaries: in Section~\ref{Section: general formalism / locality}, we will give a general definition of what it means for an object to be local to some space, and in Section~\ref{Section: general formalism / kinematical picture}, we describe the usual kinematical picture of generally covariant theories. Then, in Section~\ref{Section: general formalism / relational picture}, we will describe the relational picture. In it, we explore the general structures of local dressings and dynamical reference frames, and how they can be used to construct dressed and relational observables. We will explain how the choice of a dynamical reference frame leads to a notion of relational locality. We will discuss how different frames are related to each other, how a single frame can be physically reoriented, and how multiple frames can be patched together into a `relational atlas'. Afterwards, in Section~\ref{Section: general formalism / relational phase space}, we will discuss the phase space and algebraic properties of relational observables. In particular, we obtain an expression for the presymplectic form in terms of the so-called `relational field configuration', and we prove that relationally local observables obey microcausality \emph{in the bulk}, i.e.\ they Poisson commute when they are spacelike separated. We also show how large diffeomorphisms cause relationally local observables to transform non-trivially, and demonstrate that these transformations are generated by appropriate boundary charges. In Section~\ref{Section: general formalism / further generalisations}, we describe some further possible generalisations of our formalism. Finally, in Section~\ref{Section: general formalism / covariant to canonical} we explain how our relational observables are related to those previously obtained in the canonical setting~\cite{Dittrich:2005kc}.

In the course of this section we will make many definitions, so for the convenience of the reader a glossary of these may be found in Appendix~\ref{Appendix: Glossary}.
\subsection{Locality}
\label{Section: general formalism / locality}

Our first task is to nail down what is meant by locality. The concepts we discuss here are probably familiar to most readers, although perhaps in a less mathematical form.

\subsubsection{Local quantities with equivariant bundles}

Let us suppose that a local observer is living in some space $\mathcal{A}$. In our construction, $\mathcal{A}$ will mostly be either spacetime $\mathcal{M}$ or the local orientation space $\mathscr{O}$ of the frame, but at this point we keep it general. What exactly does this mean? It is clear that the observer should have a location $a\in\mathcal{A}$. The observer also might have some other properties that describe their relationship with $\mathcal{A}$, such as a velocity $v\in\mathrm{T}_a\mathcal{A}$ and a spin. Each of these are what we shall call `local quantities', because they are associated with the location of the observer $a\in\mathcal{A}$. This is clearly true for the location $a$. But it is also true for the velocity of the observer since it is a member of the tangent space at $a$. The spin is an element of a spin bundle over $\mathcal{A}$, but more specifically it is contained within the fibre of that bundle over $a$.

More generally, suppose $b$ is an element of some space $\mathcal{B}$. If we want to call $b$ a `local quantity' on $\mathcal{A}$, we need to be able to associate it with a specific location $a\in\mathcal{A}$. We do this by assuming that $\mathcal{B}$ is equipped with a map $\pi:\mathcal{B}\to\mathcal{A}$ for which $a=\pi(b)$. For example, in the case of the velocity $v$, the space $\mathcal{B}$ is the tangent bundle $\mathrm{T}\mathcal{A}$, and $\pi$ is the usual projection onto $\mathcal{A}$. In fact, in general the triple $(\mathcal{B},\pi,\mathcal{A})$ is what defines a \emph{bundle}. Thus, we get a very general mathematical definition of local quantities -- they are elements of bundles over $\mathcal{A}$. The structure of the bundle (in particular the function $\pi$) determines at which point in $\mathcal{A}$ each local quantity exists. Note that this definition includes the trivial case where $\mathcal{B}=\mathcal{A}$ and $\pi$ is the identity. This then just gives the statement that each point $a\in\mathcal{A}$ is local to $\pi(a)=a\in\mathcal{A}$.

The fundamental transformations in gravitational theories are diffeomorphisms. The diffeomorphism group $\operatorname{Diff}(\mathcal{A})$ of a differential manifold $\mathcal{A}$ clearly acts on $\mathcal{A}$. Thus, it will move around the location of the observer:
\begin{equation}
    a \to f(a), \quad f\in \operatorname{Diff}(\mathcal{A}).
\end{equation}
But in many cases of interest it also acts on local quantities defined on $\mathcal{A}$. Moreover, it acts \emph{equivariantly} on those quantities, meaning it moves in the appropriate way the point in $\mathcal{A}$ to which they are local.

To be more precise, consider a bundle $\mathcal{B}$ over $\mathcal{A}$ with projection map $\pi:\mathcal{B}\to\mathcal{A}$, and let $\operatorname{Diff}(\mathcal{A})$ act equivariantly on $\mathcal{B}$. This means that for each $f\in\operatorname{Diff}(\mathcal{A})$ there is a map $X_f:\mathcal{B}\to\mathcal{B}$, and these maps obey the group action axioms
\begin{equation}
    X_{\operatorname{Id}_{\mathcal{A}}} = \operatorname{Id}_{\mathcal{B}},\quad X_{f\circ g} = X_f\circ X_g,
\end{equation}
as well as the equivariance property
\begin{equation}
    f \circ \pi = \pi \circ X_f.
\end{equation}
Intuitively speaking, this property means that if $b\in\mathcal{B}$ is local to $a=\pi(b)\in\mathcal{A}$, then $X_f(b)\in\mathcal{B}$ is local to $\pi(X_f(b)) = f(a)\in \mathcal{A}$. So diffeomorphisms of $\mathcal{A}$ move local quantities on $\mathcal{A}$ appropriately. We call $X_f$ the `transformation law' for local quantities in $\mathcal{B}$.\footnote{Often $X_f$ is referred to as the `pushforward' map. However, in order to emphasise the distinction between $X_f$ and the diffeomorphism action $f_*$ on local maps and fields (as described in the next section), we choose here to reserve that name for the latter. \pah{More precisely, $f_*$ defines the pushforward for local maps and fields, while $X_f$ defines it for local quantities.}}

We call such a bundle, equipped with an equivariant action of the diffeomorphism group of its base space, an equivariant bundle. We will assume unless stated otherwise that the local quantities we deal with are elements of equivariant bundles.

As an example, for the tangent bundle we have $X_f=\dd{f}$, the differential of $f$. Thus, the velocity of the observer changes like
\begin{equation}
    v \to (\dd{f})(v), \quad f\in \operatorname{Diff}(\mathcal{A}).
\end{equation}
Note that the equivariance property means that if $v\in \mathrm{T}_a\mathcal{A}$, then $(\dd{f})(v)\in\mathrm{T}_{f(a)}\mathcal{A}$, so the velocity of the observer remains local to the observer after the transformation.

\subsubsection{Local maps and fields}

Often, we will want to consider an entire `field' of local quantities, which is the specification of a local quantity at each point in a space. More precisely, a field in a bundle $\mathcal{B}$ over $\mathcal{A}$ is the choice of an element $s(a)\in\mathcal{B}$ for each $a\in\mathcal{A}$ such that $s(a)$ is local to $a$, i.e.\ $\pi(s(a))=a$. More concisely, we can write
\begin{equation}
    \operatorname{Id}_{\mathcal{A}} = \pi \circ s.
\end{equation}
In other words, a field is just a section of a bundle over $\mathcal{A}$. We use $\Gamma(\mathcal{B})$ to denote the set of all fields / sections of $\mathcal{B}$.

Local quantities can be manipulated in various ways to form other local quantities. For example, an observer's spin and momentum may be combined to compute their helicity. Each of these are local quantities, and moreover they are local to the same point $a\in\mathcal{A}$. A map between local quantities with this property is called a `local map'.

To be more precise, suppose $\sigma:\mathcal{B}_1\to\mathcal{B}_2$ is a function between local quantities, for two bundles $\pi_1:\mathcal{B}_1\to\mathcal{A}$, $\pi_2:\mathcal{B}_2\to\mathcal{A}$. Then $\sigma$ maps local quantities to local quantities. However, in general such a $\sigma$ does not act locally, by which we mean it does not map quantities that are local to a certain point to quantities that are local to the same point. Indeed, if $b_1\in\mathcal{B}_1$ is local to $a_1=\pi_1(b_1)\in\mathcal{A}$, then $b_2=\sigma(b_1)\in\mathcal{B}_2$ will be local to $a_2=\pi_2(b_2)\in\mathcal{A}$, and in general there is no reason why $a_1=\pi_1(b_1)$ should equal $a_2=\pi_2(\sigma(b_1))$. This motivates the definition of a `local map' as a function between local quantities that \emph{does} act locally, i.e.\ for which $a_1=a_2$. This can be concisely written as
\begin{equation}
    \pi_1 = \pi_2 \circ \sigma.
\end{equation}
In mathematics, such functions are known as bundle maps. Note that the composition of two local maps is also a local map. We will use the notation $\Gamma(\mathcal{B}_1,\mathcal{B}_2)$ to denote the set of all local maps between $\mathcal{B}_1$ and $\mathcal{B}_2$.

Note that fields are special cases of local maps that we get by setting the first bundle to $\mathcal{B}_1=\mathcal{A}$ and $\pi_1=\operatorname{Id}_{\mathcal{A}}$.\footnote{Any space is trivially a bundle over itself in this way, with the projection just being the identity on that space.} Also, functions $F:\mathcal{A}\to \mathcal{F}$ may be viewed as sections of the bundle $\mathcal{B}=\mathcal{A}\times\mathcal{F}$ whose map $\pi:\mathcal{B}\to\mathcal{A}$ just projects onto the first factor.\footnote{Technically, a section of the bundle $\pi:\mathcal{A}\times\mathcal{F}\to\mathcal{A}$ is a map of the form $a \mapsto (a,F(a))$, where $F(a)$ is a function $F:\mathcal{A}\to\mathcal{F}$. Such sections are in one-to-one correspondence with the functions $F$, and we will often use (or abuse) this correspondence to just identify the sections with the functions.} In particular, this means that functions are also local maps.

If $\mathcal{B}_1,\mathcal{B}_2$ have equivariant actions $X^1_f,X^2_f$ respectively for $f\in\operatorname{Diff}(\mathcal{A})$, then there is an induced action of $\operatorname{Diff}(\mathcal{A})$ on local maps in $\Gamma(\mathcal{B}_1,\mathcal{B}_2)$. This action is called the pushforward, and is defined via
\begin{equation}
    f_*\sigma = X^2_f \circ \sigma \circ X^1_{f^{-1}}.
\end{equation}
Note that
\begin{equation}
    \pi_2\circ f_*\sigma = \pi_2\circ X^2_f \circ \sigma \circ X^1_{f^{-1}} = f \circ \pi_2 \circ \sigma \circ X^1_{f^{-1}} = f\circ \pi_1 \circ X^1_{f^{-1}} = f\circ f^{-1}\circ\pi_1 = \pi_1,
\end{equation}
so $f_*\sigma$ is a local map. We also have $f_*(g_*\sigma) = (f\circ g)_*\sigma$. Thus, $f_*$ gives a group action of $\operatorname{Diff}(\mathcal{A})$ on $\Gamma(\mathcal{B}_1,\mathcal{B}_2)$.

The pullback $f^*\sigma$ of any local map $\sigma$ is defined by $(f^{-1})_*\sigma$. In other words, the pullback is the inverse of the pushforward, or equivalently the pushforward of the inverse.

In the case where $\mathcal{B}_1=\mathcal{A}$, $\mathcal{B}_2=\mathcal{B}$, $\pi_1=\operatorname{id}_{\mathcal{A}}$ (so that $\sigma$ is a field / section of $\mathcal{B}$), we have $X^1_f = f$, $X^2_f=X_f$, and
\begin{equation}
    f_*\sigma = X_f\circ \sigma \circ f^{-1}.
\end{equation}
So $f_*$ also gives a group action of $\operatorname{Diff}(\mathcal{A})$ on $\Gamma(\mathcal{B})$.

The properties of bundles $\pi:\mathcal{B}\to\mathcal{A}$, bundle maps $\sigma$, equivariant actions $X_f$, and pushforwards $f_*$ are summarised by the following commutative diagram:
\begin{equation}
    \begin{tikzcd}
        \mathcal{B}_1 \arrow[rd,"\pi_1"] \arrow[rrr,"X_f^{1}"] \arrow[dd,"\sigma"] &  & & \mathcal{B}_1 \arrow [ld,"\pi_1"] \arrow[dd,"f_*\sigma"] \\
                                                   & \mathcal{A} \arrow[r,"f"] & \mathcal{A} \\
        \mathcal{B}_2 \arrow[ru,"\pi_2"] \arrow[rrr,"X_f^{2}"] &  & & \mathcal{B}_2 \arrow [lu,"\pi_2"]
    \end{tikzcd}
\end{equation}

The basic example of this structure is provided by the case where $\mathcal{B}=\mathrm{T}\mathcal{A}$. Then we have $X_f=\dd{f}$, and the pushforward defined above coincides with the usual pushforward of vector fields. This extends in the usual way to other types of tensor bundles.

\subsection{The kinematical picture}
\label{Section: general formalism / kinematical picture}
The covariant kinematical description of gravitational theories is based on local objects in a $D$-dimensional spacetime $\mathcal{M}$. Thus, in keeping with the above notion of local quantities, there is a certain equivariant bundle $\Phi$ over $\mathcal{M}$, which we call the kinematical field bundle. Elements of this bundle comprise the set of local values taken by the kinematical fields at points in spacetime. For example, if the only kinematical field is a metric then $\Phi$ would be the bundle of rank 2 symmetric tensors on $\mathcal{M}$. If there are also kinematical matter fields, then elements of $\Phi$ would also contain their values.

A kinematical field configuration is a choice of value for each of the kinematical fields at each point in spacetime. In other words, it is a section $\phi\in\Gamma(\Phi)$ of the kinematical field bundle $\Phi$.

However, not every section of $\Phi$ is valid as a configuration of the kinematical fields. Indeed, depending on the particular theory under consideration, $\phi\in\Gamma(\Phi)$ should have a certain degree of differentiability or smoothness, it should obey certain boundary conditions at $\partial\mathcal{M}$ or asymptotic falloffs in some region near infinity, and it should obey certain constraints or equations of motion. Altogether, these conditions restrict the allowed kinematical field configurations to fall within some subspace $\mathcal{S}\subset\Gamma(\Phi)$ of the full space of sections of $\Phi$. We call $\mathcal{S}$ the `space of solutions', and we refer to kinematical field configurations in $\mathcal{S}$ as `on-shell'. If a kinematical field configuration $\phi$ does not lie within $\mathcal{S}$, then it is simply not valid as a description of a physical state within the theory under consideration. (This does not preclude it from being a valid description of a physical state within some \emph{other} theory.)

Within the space of solutions, there is some gauge redundancy. After eliminating this gauge redundancy (by a procedure we will describe shortly), we get a certain space $\mathcal{P}$ whose elements are in one-to-one correspondence with the possible physical states of the theory.

To summarise, the space of all kinematical field configurations is $\Gamma(\Phi)$. After imposing the constraints, boundary conditions, equations of motion, etc., we get the space of solutions $\mathcal{S}\subset\Gamma(\Phi)$. Then, after gauge reducing, we get the space of physical states $\mathcal{P}$.
\begin{equation}
    \Gamma(\Phi) \supset \mathcal{S} \longrightarrow \mathcal{P}.
\end{equation}
We will refer to everything that depends on structures that come before reaching $\mathcal{P}$, i.e.\ before gauge reducing, as `kinematical'. We will refer to objects that depend only on structures coming after reaching $\mathcal{S}$ as `dynamical'. Note that objects can be both kinematical and dynamical under these definitions, if they can be defined on $\mathcal{S}$ but not on $\mathcal{P}$.\footnote{It should be noted that sometimes a subtly different version of `kinematical' is used, under which objects that depend only on the structure of $\mathcal{S}$ are not `kinematical'. However, it is convenient for us to think of $\mathcal{S}$ as kinematical, because it still involves gauge redundancy.} On the other hand `physical' objects are those which can be defined using the structure of $\mathcal{P}$ alone.

To motivate this perspective, it may be useful to view as `kinematical' anything which is not fully deterministic. Indeed, the purpose of a set of kinematical variables is to allow us to describe in some way a physical system. The deterministic nature of such a physical system requires us to restrict to the space of solutions $\mathcal{S}$. Additionally, gauge-equivalent kinematical descriptions can never be deterministically distinguished. Thus, to get a fully deterministic description we have to gauge reduce to $\mathcal{P}$. Anything before this gauge reduction is not fully deterministic, and thus (at least to some extent) kinematical.

\subsubsection{Local observables}
\label{Section: general formalism / kinematical picture / local observables}
An observable $A$ is any object that has a non-trivial dependence on the kinematical field configuration $\phi$.\footnote{We do not at this stage require `observables' to be gauge-invariant -- although of course only observables which are gauge-invariant will be physically accessible.} We will use square brackets to denote the value of an observable for a certain field configuration, so for example $A[\phi]$ is the value of the observable $A$ when the field configuration is $\phi$. We will keep this definition completely general, so that $A[\phi]$ can be any type of object whatsoever -- in particular, it need not be local on spacetime, or any other space.

On the other hand, if an observable $A$ is such that $A[\phi]$ \emph{is} a local quantity or local map, then we will call it a `local observable'. Note that we will not require in this definition that the space on which $A[\phi]$ is local must be spacetime $\mathcal{M}$. Rather, in general $A[\phi]$ can be local on any space $\mathcal{A}$ (even though the kinematical fields have to be local on spacetime). In particular, later we will also consider locality in the local orientation space $\mathscr{O}$ of the frame.

When $A[\phi]$ is a local quantity on a space $\mathcal{A}$, we can view the local observable $A$ as a map
\begin{equation}
    A: \mathcal{S}\to\mathcal{B},\text{ where $\mathcal{B}$ is a bundle over $\mathcal{A}$.}
\end{equation}
When $A[\phi]$ is a local map on a space $\mathcal{A}$, we can view the local observable $A$ as a map
\begin{equation}
    A: \mathcal{S}\to\Gamma(\mathcal{B}_1,\mathcal{B}_2),\text{ where $\mathcal{B}_1,\mathcal{B}_2$ are bundles over $\mathcal{A}$.}
\end{equation}

\subsubsection{Active and passive diffeomorphisms}
Since the kinematical field bundle $\Phi$ is equivariant, pushforwards $f_*$ give a group action of the spacetime diffeomorphism group $\operatorname{Diff}(\mathcal{M})$ on kinematical field configurations in $\Gamma(\Phi)$. These are known as \emph{active} diffeomorphisms, because they \emph{actively} change the kinematical degrees of freedom.

It should be noted that not all active diffeomorphisms can actually be carried out. Indeed, some active diffeomorphisms may map configurations in the space of solutions $\mathcal{S}$ to configurations outside the space of solutions, and so would not be valid within any fixed theoretical setup. Let $G[\phi]\subset\operatorname{Diff}(\mathcal{M})$ be the set of active diffeomorphisms which can be carried out on a field configuration $\phi\in\mathcal{S}$:
\begin{equation}
    G[\phi] = \{f\in\operatorname{Diff}(\mathcal{M}) \mid f_*\phi \in \mathcal{S}\}.
\end{equation}
Often, $G[\phi]$ will be a group, but this is not strictly required in the present formalism.

Under an active diffeomorphism $\phi\to f_*\phi$, the value taken by an observable $A$ changes like $A[\phi]\to A[f_*\phi]$. Thus, viewing the observable as a function on $\mathcal{S}$, the effect of an active diffeomorphism is to change all observables as $A\to A\circ f_*$. Active diffeomorphisms do not have any effect on non-observables, i.e.\ objects which do not depend on $\phi\in\mathcal{S}$, such as background fields.

Besides active diffeomorphisms, we also need to consider \emph{passive} diffeomorphisms. The name \emph{passive} reflects the fact that these diffeomorphisms do not `actively' produce any change in the kinematical field configuration $\phi$. Rather they act directly on the points that make up spacetime $\mathcal{M}$, or any other space $\mathcal{A}$ in which we are considering local objects. Suppose we have some function $F$ on such a space $\mathcal{A}$, and we act on $\mathcal{A}$ with a passive diffeomorphism $f\in\operatorname{Diff}(\mathcal{A})$. Then the value of $F$ at some point $a\in\mathcal{A}$ changes like $F(a)\to F(f^{-1}(a))$, so the effect of a passive diffeomorphism is to modify $F\to F\circ f^{-1} = f_*F$. More generally, a passive diffeomorphism changes a local map via the pushforward, $\sigma \to f_*\sigma$, and it changes a local quantity via the map $X_f$ that defines the equivariant action of $\operatorname{Diff}(\mathcal{A})$ on that quantity.

To briefly summarise, active diffeomorphisms act on the kinematical field configuration $\phi$, and have an effect on all observables, whereas passive diffeomorphisms act on the space on which local objects are defined, and affect all such objects, even those which do not depend on the kinematical fields, such as background fields.\footnote{In a covariant phase space approach, this means that active diffeomorphisms are generated by field space Lie derivatives, while the passive ones are generated by spacetime Lie derivatives.}

The only type of object that both passive and active diffeomorphisms have an effect on are local observables. Under an active diffeomorphism $f\in\operatorname{Diff}(\mathcal{M})$, such an observable changes like $A\to A\circ f_*$, while under a passive diffeomorphism $g\in\operatorname{Diff}(\mathcal{A})$, it changes like
\begin{equation}
    A[\phi] \to X_g (A[\phi]),\qq{ i.e.\ } A \to X_g\circ A
\end{equation}
if $A[\phi]$ is a local quantity, whereas it changes like
\begin{equation}
    A[\phi] \to g_* (A[\phi]), \qq{ i.e.\ } A \to g_* \circ A
\end{equation}
if it is a local map.

If we think of local observables as functions on field space, then active diffeomorphisms act on such functions from the right, while passive diffeomorphisms act from the left.

\subsubsection{Covariance}
In the special case where $\mathcal{A}=\mathcal{M}$, we can use the same $f\in\operatorname{Diff}(\mathcal{M})$ as both an active diffeomorphism and a passive diffeomorphism on the same local observable $A$. If the effects of the two transformations are the same, so if
\begin{equation}
    A[f_*\phi] = X_f(A[\phi]), \qq{ i.e.\ } A\circ f_* = X_f \circ A
    \label{Equation: local quantity covariance}
\end{equation}
for local quantities and
\begin{equation}
    A[f_*\phi] = f_*(A[\phi]), \qq{ i.e.\ } A\circ f_* = f_* \circ A
    \label{Equation: local map covariance}
\end{equation}
for local maps, then we say that the local observable $A$ is `covariant'.

The most basic observable is just the kinematical field configuration $\phi$ itself. This is a local observable which clearly obeys~\eqref{Equation: local map covariance} (since $f_*\phi=f_*\phi$), so it is covariant. We can obtain other covariant observables by doing standard local operations on $\phi$, such as taking covariant derivatives and performing tensorial summations. For example, if the kinematical fields contain a metric $g_{ab}$, then we can construct the volume form or the Riemann tensor in this way.

\subsubsection{Kinematical locality}
In the case where $A[\phi]$ is a local observable on spacetime, we will call $A[\phi]$ `kinematically local' if
\begin{equation}
    A[\phi] \text{ depends on $\phi$ only through its value and derivatives at $\pi(A[\phi])$}
\end{equation}
when $A[\phi]\in\mathcal{B}$ is a local quantity in a bundle $\pi:\mathcal{B}\to\mathcal{M}$, and
\begin{equation}
    A[\phi](b) \text{ depends on $\phi$ only through its values and derivatives at $\pi(b)$, for all $b\in\mathcal{B}$},
\end{equation}
when $A[\phi]\in\Gamma(\mathcal{B},\mathcal{B}')$ is a local map between bundles $\pi:\mathcal{B}\to\mathcal{M}$ and $\pi':\mathcal{B}'\to\mathcal{M}$.
Intuitively speaking, kinematical locality ties the locality of the observable $A$ to the locality of the kinematical field configuration $\phi$.

A consequence of this definition is that under a field variation $\phi\to\phi+\delta\phi$, the change in any kinematically local observable $A$ can be written in the form
\begin{equation}
    \delta(A[\phi]) = A'[\phi]\cdot \delta\phi|_{\pi(A[\phi])} + A''[\phi] \cdot \partial(\delta\phi)|_{\pi(A[\phi])} + A'''[\phi]\cdot\partial^2(\delta\phi)|_{\pi(A[\phi])} + \dots
    \label{Equation: kinematically local quantity}
\end{equation}
when $A[\phi]$ is a local quantity, and
\begin{equation}
    \delta(A[\phi](b)) = A'[\phi](b)\cdot \delta\phi|_{\pi(b)} + A''[\phi](b) \cdot \partial(\delta\phi)|_{\pi(b)} + A'''[\phi](b)\cdot\partial^2(\delta\phi)|_{\pi(b)} + \dots
    \label{Equation: kinematically local map}
\end{equation}
when $A[\phi]$ is a local map. Here, $A',A'',\dots$ are some observables (they are local quantity valued and local map valued for~\eqref{Equation: kinematically local quantity} and~\eqref{Equation: kinematically local map} respectively), $\partial^n\phi$ collectively denotes all the possible $n^{\text{th}}$ derivatives of $\phi$, and the $\cdot$ denotes a summation over the various components of the fields and their derivatives that are involved. Note that the variation of a kinematically local observable depends on $\delta\phi$ in a kinematically local way.

Kinematical locality is a strictly stronger property than just locality. For theories where the version of locality that plays a role in the kinematical description is preserved at the physical level, kinematically local observables are physically useful. However, in gravitational theories the kinematical locality is lost at the physical level, so requiring the observables we consider to respect it would be too restrictive. Indeed, the observables we construct in this paper are highly kinematically non-local. Despite this, they will still be local in the sense of the definitions in Section~\ref{Section: general formalism / kinematical picture / local observables}. Moreover, they will be \emph{relationally} local, which is a property that we will define soon.

\subsubsection{Gauge invariance}
Up to now, what we have described can be applied to any covariant field theory. What sets gravitational theories apart is the relationship between active diffeomorphisms and gauge symmetries.

In any theory, the presence of gauge symmetry can be understood as a redundancy in the kinematical description of the state. In other words, two distinct kinematical configurations can correspond to the same physical state. In the present discussion, this redundancy can be concisely described using an equivalence relation $\sim$ on $\mathcal{S}$. For two kinematical field configurations $\phi,\phi'\in\mathcal{S}$, the equivalence $\phi\sim\phi'$ is the statement that $\phi$ and $\phi'$ both correspond to the same physical state.

We can then define the physical space of states as the quotient space
\begin{equation}
    \mathcal{P} = \mathcal{S}/\!\sim,
\end{equation}
each element of which is an equivalence class of field configurations modulo gauge redundancy. The map
\begin{equation}
    P:\mathcal{S}\to\mathcal{P}
\end{equation}
from each field configuration to its physical equivalence class makes $\mathcal{S}$ into a bundle over $\mathcal{P}$.

Sections of this bundle are a choice of field configuration within each equivalence class, i.e.\ they are gauge choices. Given such a section $z:\mathcal{P}\to \mathcal{S}$, we get a function $\bar\phi:\mathcal{S}\to\mathcal{S}$ via $\bar\phi = z\circ P$. For each $\phi\in\mathcal{S}$, $\bar{\phi}[\phi]$ is the `gauge-fixed' version of $\phi$. Given any observable $A$ and a gauge choice, we can construct the `gauge-fixed' observable $O = A\circ \bar\phi$.\footnote{This is the covariant phase space version of gauge-invariant extensions of gauge-fixed quantities~\cite{Henneaux:1992ig,Dittrich:2005kc,Dittrich:2004cb,Hoehn:2019fsy,Chataignier:2019kof}.} This is an example of a `gauge-invariant' observable, because it obeys
\begin{equation}
    \phi\sim\phi' \implies O[\phi] = O[\phi'].
\end{equation}
Note that there is a 1-to-1 correspondence between gauge-invariant observables $O$ and functions $O_{\text{phys.}}$ of the physical state $\phi_{\text{phys.}}\in\mathcal{P}$, provided by the equation $O_{\text{phys.}}\circ P = O$.

Gauge redundancies can take many forms. However, since the aim of the present paper is to focus on gravitational effects, we will restrict to gravitational gauge symmetries, which are spacetime diffeomorphisms. Thus, we will assume that the physical equivalence relation obeys
\begin{equation}
    \phi\sim\phi'\implies \phi' = f_*\phi \text{ for some } f\in G[\phi]\subset\operatorname{Diff}(\mathcal{M}).
\end{equation}
In other words, two kinematical field configurations are physically equivalent only if they are related by an active diffeomorphism. However, in general the opposite is not true -- not all active diffeomorphisms are gauge transformations, e.g.\ those that change boundary data are not. Thus, we shall define
\begin{equation}
    H[\phi] = \{f\in G[\phi] \mid \phi\sim f_*\phi\}.
\end{equation}
This is the set of active diffeomorphisms which are also gauge transformations. Just like $G[\phi]$, it need not be a group in general. Given a kinematical field configuration $\phi$, we will call elements of $H[\phi]$ `small' diffeomorphisms, and elements of $G[\phi]$ but not $H[\phi]$ `large' diffeomorphisms. Large diffeomorphisms are the only allowed diffeomorphisms with physical consequences.

Diffeomorphisms whose action is trivial in a neighbourhood of the boundary of spacetime are always gauge transformations in generally covariant theories. For some intuition on this, see Appendix~\ref{Appendix: small diffeos} or refer to Section~\ref{Section: geodesic example}. Such diffeomorphisms do form a group which is independent of $\phi$. However, the full set of small diffeomorphisms can be larger than just these, in a way that depends on the theory under consideration. 

Suppose we make a gauge choice $z:\mathcal{P}\to \mathcal{S}$ and construct the gauge-fixing function $\bar{\phi} = z\circ P$ as above. Then we have $\bar{\phi}[\phi] \sim \phi$ for all $\phi$, which implies that there exists some field-dependent diffeomorphism $U[\phi]$ obeying
\begin{equation}
    \bar{\phi}[\phi] = (U[\phi])^*\phi, \qquad (U[\phi])^{-1}\in G[\phi].
\end{equation}
Since $\bar{\phi}$ is gauge-invariant, $U[\phi]$ must obey
\begin{equation}
    U[f_*\phi] = f \circ U[\phi] \text{ for all } f\in H[\phi].
    \label{Equation: gauge-fixing diffeomorphism}
\end{equation}
We call a field-dependent diffeomorphism with this property a `gauge-fixing diffeomorphism'.

If $A$ is covariant, then we may write the gauge-fixed observable $O=A\circ\bar\phi$ as
\begin{equation}
    O[\phi] = A\big[\bar\phi[\phi]\big] = A[(U[\phi])^*\phi] = (U[\phi])^* A[\phi].
    \label{Equation: U dressed O}
\end{equation}
As we will see later, $O$ can be viewed as a relational observable with respect to a dynamical frame provided by $U$.

It is useful to define a restricted kind of covariance that only applies to small diffeomorphisms. In particular, we will call a local observable `gauge-covariant' if it obeys~\eqref{Equation: local quantity covariance} or~\eqref{Equation: local map covariance} for all $f\in H[\phi]$. Note that if an observable is covariant then it is clearly also gauge-covariant.

\subsection{The relational picture}
\label{Section: general formalism / relational picture}
The kinematical field configuration $\phi$ is a covariant observable. It is also kinematically local. However, it is clearly generically not gauge-invariant, except in the most trivial cases.

In fact, requiring a local observable to be both covariant and gauge-invariant is  highly restrictive. Gauge-invariance is a statement about invariance under active diffeomorphisms, and covariance is a statement relating active diffeomorphisms and passive diffeomorphisms. In combination, we get a statement about invariance under passive diffeomorphisms. To be more precise, suppose $A$ is such an observable. If it is local map valued, then we have
\begin{equation}
    A[\phi]
    \underset{\hspace{0.5em}\mathclap{\substack{\\\text{gauge}\\\text{invariance}}}\hspace{0.5em}}{=}
    A[f_*\phi]
    \underset{\hspace{0.5em}\mathclap{\substack{\\\text{covariance}}}\hspace{0.5em}}{=}
    f_*(A[\phi])
    \qq{ for all }
    f\in H[\phi].
\end{equation}
Since usually small diffeomorphisms can act arbitrarily away from the boundary of spacetime, this equation is saying that the value taken by $A$ for any fixed $\phi$ must be constant throughout the bulk. On the other hand if $A$ is local quantity valued then for similar reasons we have $A[\phi] = X_f(A[\phi])$ for all small diffeomorphisms $f$. Again assuming small diffeomorphisms act arbitrarily away from the boundary of spacetime, this equation says that $A[\phi]$ must be local to the boundary.

Similarly, kinematically local observables cannot be gauge-invariant and non-trivial in the bulk. Indeed, suppose $A$ is a kinematically local observable. Then, if it is local map valued, we have for any open subset $\mathcal{U}$ of the bulk of spacetime and small diffeomorphism $f$
\begin{align}
    A[\phi]|_{\mathcal{U}} &\text{ depends only on } \phi|_{\mathcal{U}},\footnotemark
                             \label{Equation: kinematical local trivial in bulk 1} \\
    \text{and }A[f_*\phi]_{\mathcal{U}} &\text{ depends only on } (f_*\phi)|_{\mathcal{U}} \nonumber\\
                                        &\text{ which depends only on } \phi|_{f^{-1}(\mathcal{U})}.
                             \label{Equation: kinematical local trivial in bulk 2}
\end{align}
\footnotetext{\jjvk{Note that a dependence on $\phi$ in the open set $\mathcal{U}$ implies also a dependence on all derivatives of $\phi$ in that set.}}
If $A$ is additionally gauge-invariant, then $A[\phi]=A[f_*\phi]$. Thus, $A[\phi]|_{\mathcal{U}}$ only depends on $\phi|_{\mathcal{U}}$, but it also only depends on $\phi|_{f^{-1}(\mathcal{U})}$. By choosing $f$ appropriately we can always choose for $\mathcal{U}\cap f^{-1}(\mathcal{U})=\emptyset$, so that $A[\phi]|_{\mathcal{U}}$ can have no dependence on $\phi$ (since by~\eqref{Equation: kinematical local trivial in bulk 1} and~\eqref{Equation: kinematical local trivial in bulk 2} it only depends on $\phi|_{\mathcal{U}\cap f^{-1}(\mathcal{U})}$). Thus, $A[\phi]$ must be trivial in the bulk. A similar argument applies to local quantity valued observables.

So requiring covariance or kinematical locality will not allow us to construct any non-trivial gauge-invariant bulk observables. Clearly, if we wish to do so, we must drop some of these requirements. We are not allowed to drop gauge-invariance, because we want the observables we obtain to be physical. Thus, we must abandon covariance and kinematical locality.

Unfortunately, the most basic methods of constructing observables out of the kinematical fields $\phi$ (for example taking covariant derivatives, summing over tensor indices, and so on) preserve covariance and kinematical locality. So we need to do something more involved to construct gauge-invariant observables.

\subsubsection{Local dressings and dressed observables}
Perhaps the most elementary thing we could do is use \emph{dressings}, which are objects which allow us to convert covariant, gauge-dependent observables into non-covariant, gauge-invariant observables.

In particular, we will focus on `local dressings', which are gauge-covariant spacetime points. Thus, a local dressing is a map $x:\mathcal{S}\to \mathcal{M}$ obeying~\eqref{Equation: local quantity covariance} for small diffeomorphisms. Since on $\mathcal{M}$ we have $X_f=f$, this property may be written
\begin{equation}
    x[f_*\phi] = f(x[\phi]) \text{ for all } f\in H[\phi].
\end{equation}
The endpoint of a boundary anchored geodesic is a local dressing, as was described in Section~\ref{Section: geodesic example}.

Suppose we have some spacetime function observable $F[\phi]:\mathcal{M}\to \mathcal{F}$. Let us assume it is covariant, so that under an active diffeomorphism it transforms via the pushforward
\begin{equation}
    F[f_*\phi] = f_* (F[\phi]) = F[\phi] \circ f^{-1}.
\end{equation}
As discussed above, such a function can only be gauge-invariant if it is constant in spacetime. But let us not insist on this.

We can combine $F$ with a local dressing to form a gauge-invariant observable. We simply evaluate $F[\phi]$ at $x[\phi]$:
\begin{equation}
    O_{F,x}[\phi] = F[\phi](x[\phi]).
\end{equation}
Under an active small diffeomorphism $f\in G[\phi]$, we have
\begin{equation}
    O_{F,x}[f_*\phi] = F[f_*\phi] (x[f_*\phi]) = (F[\phi]\circ f^{-1}) (f(x[\phi])) = F[\phi](x[\phi]) = O_{F,x}[\phi],
\end{equation}
so it is indeed gauge-invariant. We call $O_{F,x}$ a `dressed observable'; in Section~\ref{Section: geodesic example} we saw some examples of geodesically dressed observables.

$O_{F,x}$ is a local observable. Indeed, $O_{F,x}[\phi]$ is a local quantity on spacetime $\mathcal{M}$, local to $x[\phi]$.\footnote{This makes sense only if we view $F$ as a section of a bundle $\mathcal{F}\times\mathcal{M}\to\mathcal{M}$.} Notably, even though $O_{F,x}$ is gauge-invariant, the point $x$ to which it is local is not gauge-invariant. This is consistent with the fact that $x$ gives a relationship between the na\"ive kinematical locality, and a more physical type of locality. Since the kinematical locality is non-physical, we should not expect this relationship to be gauge-invariant.

In this way, we have used a local dressing to convert a covariant observable into a gauge-invariant observable, as promised. However, the types of dressed observables we can construct using a single local dressing are quite limited. Indeed, if we only use a single local dressing $x[\phi]$, we can only ever dress function observables, and we can only ever get dressed observables which are local to the single point $x[\phi]$. There are clearly many more physical bulk observables that we would like to access. This is the motivation for the construction in the next section.

\subsubsection{Dynamical reference frames}
A gravitational `dynamical reference frame', or just `frame', is a set of local dressings. It is useful to formulate this in terms of what we will call the `universal dressing space' $\mathscr{D}$, which is defined as the set of all possible local dressings:
\begin{equation}
    \mathscr{D} = \{x:\mathcal{S}\to\mathcal{M} \mid x[f_*\phi] = f(x[\phi]) \text{ for all } f\in H[\phi]\}.
\end{equation}
Thus, a `frame' is a subset $\mathscr{R}\subset\mathscr{D}$ of the universal dressing space.

Note that, for each fixed kinematical field configuration $\phi\in\mathcal{S}$, the universal dressing space has the structure of a bundle over $\mathcal{M}$ whose projection is just given by the evaluation of each dressing:
\begin{equation}
    \pi[\phi]:\quad \mathscr{D}\to\mathcal{M},\quad x\to x[\phi].
\end{equation}
This structure is clearly field dependent, and it has the gauge-covariance property
\begin{equation}
    \pi[f_*\phi] = f\circ \pi[\phi] \text{ for all } f\in H[\phi].
\end{equation}

The `image' of a frame $\mathscr{R}$ is defined as the region of spacetime which its dressings cover, i.e.\ it is the set $\mathcal{N}[\phi] = \pi[\phi](\mathscr{R})$. The image of a frame is basically the region in which we can use the frame to construct local observables. If the image is all of spacetime $\mathcal{N}[\phi] = \mathcal{M}$, then we will call the frame surjective, in which case, the frame can be used to construct local observables anywhere in spacetime. If the map $\pi[\phi]$ is injective when restricted to $\mathscr{R}$, then we call the frame injective. To construct certain observables, often we will need the frame to be injective. Note that an injective frame may be viewed as the image of a partial section of the bundle $\pi[\phi]:\mathscr{D}\to\mathcal{M}$. When the frame is both surjective and injective, we call it bijective, and in this case it is the image of a complete section. In general the surjectivity and injectivity of a frame will depend on the kinematical field configuration $\phi$.

Usually, we will want to pair frames with parametrisations of their constituent dressings. To be more precise, given a frame $\mathscr{R}\subset\mathscr{D}$, a parametrisation of that frame is the choice of some space of parameters $\mathscr{O}$, as well as a bijection $R:\mathscr{O}\to\mathscr{R}$. In fact, it is efficient to extend this definition in the following way. A `parametrised frame' is the choice of a parameter space $\mathscr{O}$, and an \emph{injective} map $R:\mathscr{O}\to\mathscr{D}$. The image of this map $\mathscr{R}=R(\mathscr{O})$ is the frame itself (without the parametrisation), and the restriction of $R$ to a function $\mathscr{O}\to\mathscr{R}$ is the parametrisation of that frame.

In the example with geodesics described in Section~\ref{Section: geodesic example}, we initially constructed a parametrised frame with local orientation space $\mathscr{O}=\RR_{\ge0}\times \mathrm{T}\partial\mathcal{M}_{\abs{W}^2\ne-1}$. But this space has higher dimension than spacetime itself, which prevented the frame from being injective. To get around this, we restricted to subspaces of $\mathscr{O}$, by e.g.\ fixing a boundary vector field $W_1$.

\begin{figure}
    \centering
    \begin{tikzpicture}
        \begin{scope}[shift={(-1.5,0.5)}]
            \draw[rounded corners=15pt, thick, fill=green!15] (0,0) rectangle (3,3);
            \node[above] at (1.5,0.6) {\Large$\mathscr{O}$};
            \node[below] at (1.5,-0.1) {\footnotesize\parbox{3cm}{\centering local orientation space}};
        \end{scope}
        \draw[yellow!40!black,fill=yellow!15] (4.5,-1) -- (8,-1) -- (8,3) -- (6,5) -- (3,5) -- (3,1) -- cycle;
        \draw[yellow!40!black,dotted] (6,5) -- (6,1);
        \draw[yellow!40!black,dotted] (3,1) -- (6,1);
        \draw[yellow!40!black,dotted] (8,-1) -- (6,1);
        \draw[rounded corners=10pt, thick, fill=green!15] (5,1) -- (6.2,0) -- (6.2,3) -- (5,4) -- cycle;
        \draw[yellow!80!black] (3,5) -- (4.5,3);
        \draw[yellow!80!black] (4.5,-1) -- (4.5,3);
        \draw[yellow!80!black] (8,3) -- (4.5,3);
        \node at (5.6,1.8) {\Large$\mathscr{R}$};
        \node at (5.6,1.3) {\footnotesize\parbox{3cm}{\centering frame}};
        \node[above] at (6,-0.85) {\Large$\mathscr{D}$};
        \node[below] at (6,-1.1) {\footnotesize\parbox{3cm}{\centering universal dressing space}};
        \begin{scope}[shift={(9.5,0)},scale=0.9]
            \draw[blue!40!black,fill=blue!15] (0,0)
                .. controls (1,-1) and (2,-1) .. (3,-1)
                .. controls (4,1) and (4.5,4) .. (4,5)
                .. controls (3,5) and (2,5) .. (1,6)
                .. controls (1.2,4) and (1.2,3) .. (0,0);
        \end{scope}
        \begin{scope}[shift={(6.5,0.3)},scale=0.9]
            \draw[rounded corners=10pt, thick, fill=green!15] (4.4,0.5) -- (6,0)
                .. controls (6.5,2) and (6.5,3) .. (6.5,3.5) -- (5,4)
                .. controls (5,3.5) and (5,2.5) .. (4.7,1.5) -- cycle;
        \end{scope}
        \node at (11.4,1.4) {\large$\mathcal{N}[\phi]$};
        \node at (11.4,0.9) {\footnotesize\parbox{3cm}{\centering image}};
        \node[above] at (11,-0.55) {\large$\mathcal{M}$};
        \node[below] at (11,-1.05) {\footnotesize\parbox{3cm}{\centering spacetime}};

        \draw[gray,line width=1pt,-{latex}] (0.1,2) .. controls (2,2.5) and (4,2.5) .. (5.4,2.5) node[pos=0.375,above,black] {\Large$R$};
        \draw[gray,line width=1pt,-{latex}] (5.6,2.5) .. controls (6,2.5) and (10,2.5) .. (11.5,2) node[pos=0.64,above,black] {\Large$\pi[\phi]$};
        \draw[gray,line width=1pt,-{latex}] (0.1,2.6) .. controls (3,8) and (8,7) .. (11.5,2.1) node[midway,above,black] {\Large$R[\phi]$};
        %\draw[gray,line width=1pt,{latex}-] (1,1.4) .. controls (4,-4) and (8,-4) .. (10.9,1) node[midway,below,black] {\Large$(R[\phi])^{-1}$};
    \end{tikzpicture}
    \caption{A frame $\mathscr{R}$ is a subset of the universal dressing space $\mathscr{D}$. A parametrisation of the frame is a bijection $R$ from a space of local orientations $\mathscr{O}$ to $\mathscr{R}$. For each kinematical field configuration $\phi$, the image of the frame is the subset $\mathcal{N}[\phi]=\pi[\phi](\mathscr{R})$ of spacetime $\mathcal{M}$ obtained by evaluating all the dressings in the frame. If the frame is injective, then we can invert $R[\phi]$ on its image to obtain $(R[\phi])^{-1}$, the dynamical frame field on the region of spacetime $\mathcal{N}[\phi]$.}
    \label{Figure: parametrised frame}
\end{figure}
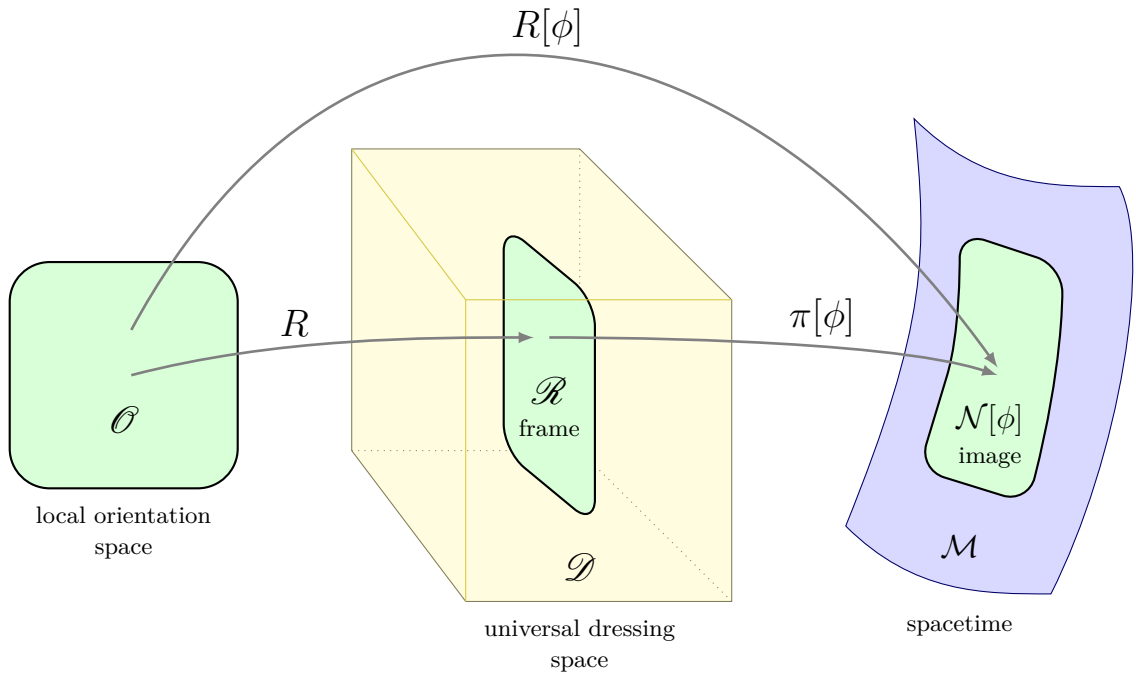

Intuitively speaking, the frame $\mathscr{R}\subset\mathscr{D}$ provides a relationship between the kinematical notion of locality and a physical version of locality. The parametrisation provides us with a means to use this relationship. Sometimes we will refer to the elements of $\mathscr{O}$, i.e.\ the possible values taken by the parameters, as `local orientations' of the frame. The setup is depicted in Figure~\ref{Figure: parametrised frame}.

Let $R:\mathscr{O}\to\mathscr{D}$ be a parametrised frame. For each fixed kinematical field configuration $\phi\in\mathcal{S}$, this frame provides us with a map $R[\phi]$ from the space of local orientations to spacetime:
\begin{equation}
    R[\phi] = \pi[\phi] \circ R : \mathscr{O}\to\mathcal{M}.
\end{equation}
Under a small diffeomorphism we have
\begin{equation}
    R[f_*\phi] = \pi[f_*\phi]\circ R = f\circ\pi[\phi]\circ R = f\circ R[\phi] \text{ for all } f\in H[\phi].
\end{equation}
This is the fundamental gauge-covariance property of dynamical frames in generally covariant theories that we already encountered in \eqref{Equation: reference frame covariance} in the geodesic dressing example. The image of $R[\phi]$ is the image of the frame $\mathcal{N}[\phi]$.

A `frame field' is an observable $s$ which for each kinematical field configuration $\phi$ is a section of $R[\phi]$, i.e.\ it is a map $s[\phi]:\mathcal{N}[\phi]\to\mathscr{O}$ obeying
\begin{equation}
    R[\phi]\circ s[\phi] = \operatorname{Id}_{\mathcal{N}[\phi]}.
\end{equation}
The frame field constitutes a dynamical field on spacetime, whose local values are the local frame orientations.  Let
\begin{equation}
    \mathcal{O} = \{s[\phi] \mid \phi \in \mathcal{S}\}
\end{equation}
be the space of possible configurations of the frame field. Assuming we have chosen a particular frame field, we shall call each element of $\mathcal{O}$ a `global orientation' of the frame.

For a small diffeomorphism $f\in H[\phi]$, we have
\begin{multline}
    R[f_*\phi] \circ s[\phi]\circ f^{-1} = f\circ R[\phi]\circ s[\phi] \circ f^{-1} = f\circ\operatorname{Id}_{\mathcal{N}[\phi]}\circ f^{-1} = \operatorname{Id}_{f(\mathcal{N}[\phi])} = \operatorname{Id}_{\mathcal{N}[f_*\phi]},
\end{multline}
i.e.\ $s[\phi]\circ f^{-1}$ is a section of $R[f_*\phi]$. Thus, it would be consistent, but not strictly necessary, for the frame field to transform like a gauge-covariant scalar, i.e.\ $s[f_*\phi]= s[\phi]\circ f^{-1}$ for all $f\in H[\phi]$. 

In the case of an injective frame, $R[\phi]$ is also injective and thus has a unique section, which is given by its inverse $(R[\phi])^{-1}:\mathcal{N}[\phi] \to \mathscr{O}$ on its image. Thus, this is the only possible choice of frame field. In this case we \emph{do} have that $(R[\phi])^{-1}$ is a gauge-covariant scalar:
\begin{equation}\label{eq:framefield}
    (R[\phi])^{-1} = (R[f_*\phi])^{-1} = (f\circ R[\phi])^{-1} = (R[\phi])^{-1}\circ f^{-1}
\end{equation}
for all $f\in H[\phi]$. For example, in (\ref{eq:geoframe1}--\ref{eq:geoframe2}) we encountered the frame scalar fields associated with geodesic dressings. Note that the frame field is not fully covariant, as the above equations need not hold for a more general diffeomorphism $f$. 

From now on, we will assume for simplicity that, unless stated otherwise, we are using an injective frame, so that the frame field is always just the inverse of $R[\phi]$.

\subsubsection{Relational observables}

Having chosen a parametrised frame $R$, we can use it to construct a gauge-invariant observable from a covariant one. In particular, suppose $A$ is a covariant local map valued observable on spacetime $\mathcal{M}$.\footnote{Strictly speaking, we do not need $A$ to be fully covariant. It can just be gauge-covariant, and then the relational observable $O_{A,R}$ will be gauge-invariant. But in practical terms it is usually simpler to use a fully covariant $A$.} Then we can construct a new observable $O_{A,R}$ by taking the pullback of $A[\phi]$ through $R[\phi]$:
\begin{equation}
    O_{A,R}[\phi] = (R[\phi])^*A[\phi].
\end{equation}
Of course, the viability of this construction depends on whether this pullback exists and is well-defined. Let us for the moment assume that it is, and that it obeys the standard compatibility condition of the pullback, $(f\circ g)^*=g^*f^*$. Then $O_{A,R}$ is gauge-invariant, since
\begin{equation}
    O_{A,R}[f_*\phi] = (R[f_*\phi])^*A[f_*\phi] = (f\circ R[\phi])^*f_*A[\phi] = (R[\phi])^*f^* f_*A[\phi] = (R[\phi])^* A[\phi] = O_{A,R}[\phi].
\end{equation}
Moreover (as we will explain shortly), the pullback $(R[\phi])^*$ here converts local maps on spacetime into local maps on the local orientation space $\mathscr{O}$. Therefore, $O_{A,R}$ is a gauge-invariant local observable on $\mathscr{O}$. We call it a relational observable, because it measures $A$ relative to the frame $R$. To be precise $O_{A,R}$ is the observable that answers the question: ``What is the value of $A$ at the event in spacetime $\mathcal{M}$ where the local orientation of the frame field is $o\in\mathscr{O}$, given that the global orientation of the frame field is $(R[\phi])^{-1}\in\mathcal{O}$?'' Various examples of such relational observables were provided for geodesic dressings in Section~\ref{ssec_georelobs}.

Let us now address the meaning and existence of the pullback. Recall that equivariant bundles $\mathcal{B}\to\mathcal{A}$ come equipped with an equivariant action $X_f$ of $f\in \operatorname{Diff}(\mathcal{A})$ on $\mathcal{B}$. We wish to extend the definition of $X_f$ to maps $f:\mathcal{A}\to\mathcal{A}'$ between different spaces $\mathcal{A}$, $\mathcal{A}'$, with associated bundles $\pi:\mathcal{B}\to\mathcal{A}$, $\pi':\mathcal{B}'\to\mathcal{A}'$. Thus, for such an $f$ we require a map $X_f:\mathcal{B}\to\mathcal{B}'$ with the generalised equivariance property
\begin{equation}
    \pi'\circ X_f = f\circ \pi.
\end{equation}
Moreover, $X_f$ must obey
\begin{equation}
    X_{g'fg} = X_{g'} \circ X_f \circ X_g \text{ for all }g\in \operatorname{Diff}(\mathcal{A}),\,g'\in\operatorname{Diff}(\mathcal{A}').
\end{equation}
Such a generalisation of $X_f$ exists only for certain pairs $(\mathcal{B},\mathcal{B}')$, and moreover its existence depends highly on properties of $f$. The main example that demonstrates this is the case when $\mathcal{B}=\mathrm{T}\mathcal{A}$ is the tangent bundle of $\mathcal{A}$. Then we must have that $\mathcal{B}'=\mathrm{T}\mathcal{A}'$ is the tangent bundle of $\mathcal{A}'$, and the map $X_f$ is just given by the differential $\dd{f}$ -- which only exists if $f$ is differentiable. Another example is the case when $\mathcal{B}=\mathrm{T}_*\mathcal{A}$ is the cotangent bundle. Then $\mathcal{B}'=\mathrm{T}_*\mathcal{A}'$ must also be the cotangent bundle, and the map $X_f$ is given by the inverse of the dual of the differential -- which additionally requires $f$ to be invertible.

Assuming that such a generalisation of $X_f$ exists, we can define the pushforward of a local map $\sigma\in\Gamma(\mathcal{B}_1,\mathcal{B}_2)$ on a space $\mathcal{A}$ through a map $f:\mathcal{A}\to\mathcal{A}'$ via the same definition as previously:
\begin{equation}
    f_*\sigma = X^2_f\circ \sigma \circ (X^1_f)^{-1}.
\end{equation}
Here $f_*\sigma\in\Gamma(\mathcal{B}'_1,\mathcal{B}'_2)$ can be shown to be a local map on $\mathcal{A}'$. Given a local map $\sigma'\in\Gamma(\mathcal{B}'_1,\mathcal{B}'_2)$ on $\mathcal{A}'$, we can similarly define the pullback via
\begin{equation}
    f^*\sigma' = (X^2_f)^{-1} \circ \sigma \circ X^1_f,
\end{equation}
which is a local map on $\mathcal{A}$. Clearly, both of these definitions require $X^1_f$ or $X^2_f$ to be invertible, which usually reduces to the condition that $f$ be invertible. When it is, one can show that $f^* = (f^{-1})_*$. However, we can weaken this to the requirement that $X^1_f,X^2_f$ are injective, which usually reduces to requiring $f$ to be injective. Then these maps are invertible on their images, and the objects $f_*\sigma, f^*\sigma'$ will be well-defined in certain restricted domains.

The structure of the pushforward for $f:\mathcal{A}\to\mathcal{A}'$ is summarised by the following commutative diagram:
\begin{equation}
    \begin{tikzcd}
        \mathcal{B}_1 \arrow[rd,"\pi_1"] \arrow[rrr,"X_f^{1}"] \arrow[dd,"\sigma"] &  & & \mathcal{B}'_1 \arrow [ld,"\pi'_1"] \arrow[dd,"f_*\sigma"] \\
                                                   & \mathcal{A} \arrow[r,"f"] & \mathcal{A}' \\
        \mathcal{B}_2 \arrow[ru,"\pi_2"] \arrow[rrr,"X_f^{2}"] &  & & \mathcal{B}'_2 \arrow [lu,"\pi'_2"]
    \end{tikzcd}
\end{equation}

In this paper, we'll always assume unless stated otherwise that the map $X_f$ exists for sufficiently well-behaved $f$. This assumption implies certain things about the spaces $f$ maps between. For example, if $X_f$ acts on tensor bundles, then the spaces must be differential manifolds, or if $X_f$ acts on spinor bundles, then the spaces must admit spin structures.

Let us return to the relational observable $O_{A,R}$. Spacetime $\mathcal{M}$ is a fixed space, so depending on $A$, the existence of $O_{A,R}$ will require the local orientation space $\mathscr{O}$ to have certain properties. For example, if $A$ is a tensor field, then $\mathscr{O}$ needs to be a differential manifold so that it admits tensor bundles.

The existence of $O_{A,R}$ also puts requirements on the frame itself, in order for the pullback $(R[\phi])^*$ to exist. Often it will require $R[\phi]$ to be sufficiently differentiable. Unless stated otherwise, in what follows we will assume that $R[\phi]$ is smooth, so that it is always sufficiently differentiable.

All-in-all, the only assumption we have left to make is that $R[\phi]$ is injective. If we make this assumption then we are guaranteed that the relational observable $O_{A,R}$ exists (so long as each of the other assumptions we have described are satisfied).

Note that we have been formulating $O_{A,R}$ as the pullback of $A$ through $R[\phi]$. We can also formulate it as the pushforward of $A$ through the frame field $(R[\phi])^{-1}$. Mathematically, these two perspectives are equivalent. However, conceptually they are subtly different. The former emphasises the fact that the frame is constructed as a set of local dressings, whereas the latter emphasises that the relational observable measures some dynamical fields (in this case $A$) relative to other dynamical fields (in this case the frame field). The latter point of view is closer in spirit to previous approaches to relational observables~\cite{Rovelli:1989jn,Rovelli:1990ph,Rovelli:1990pi,Rovelli:2001bz,rovelliQuantumGravity2004,Dittrich:2004cb,Dittrich:2005kc,rovelliQuantumGravity2004,Tambornino:2011vg,Thiemann:2007pyv,delaHamette:2021oex,Hoehn:2019fsy,Hoehn:2020epv,Dittrich:2007jx,Thiemann:2004wk,Giddings:2005id,Brunetti:2013maa,Brunetti:2016hgw,Rejzner:2016yuy,Marolf:1994nz,Marolf:1994wh,Chataignier:2019kof,Chataignier:2020fys,Pons:2009cz,Giesel:2007wi,Dapor:2013hca}.

$O_{A,R}$ is the relational observable that we get from using the parametrised frame $R$ and the local map observable $A$. We can also construct relational observables from parametrised frames and local quantity observables $B$, by using the map $X_f$:
\begin{equation}
    O_{B,R}[\phi] = (X_{R[\phi]})^{-1} (B[\phi]).
\end{equation}
$O_{B,R}$ is a gauge-invariant observable that is a local quantity on $\mathscr{O}$. Again, the existence of $O_{B,R}$ requires the frame to be injective. Moreover, we need $B[\phi]$ to be local to a point in the image of $R[\phi]$.

Let us take a step back to consider what this construction means physically. We started with covariant observables, usually constructed locally from the kinematical fields, and a frame, usually (although not necessarily -- see for example the case of parametrised field theory in Section~\ref{Section: more examples / pft}, or the case of Brown-Kucha\v{r} dust in Section~\ref{Section: more examples / dust}) constructed non-locally from the kinematical fields. Neither of these alone were gauge-invariant. However, by combining them we obtained a gauge-invariant relational observable. If the frame is kinematically non-local, the relational observable is also kinematically non-local. However, the relational observable is a local object on the local orientation space $\mathscr{O}$. Unlike the kinematical version of locality, the locality that the relational observables respect is physically significant -- simply because we can construct gauge-invariant observables which respect it.

    \subsubsection{Single integral representation and smearing}
    \label{Section: general formalism / relational picture / single integrals smearing}

    In many works~\cite{DeWitt:1962cg,Marolf:1994wh,Giddings:2005id,Giddings2006,Marolf:2015jha,Donnelly:2016rvo}, gravitational observables are written using the so-called `single integral' representation:
    \begin{equation}
        A[\phi] = \int_{\mathcal{M}}a[\phi].
    \end{equation}
    where $a[\phi]$ is a top form on spacetime. This form is usually taken to be \pah{fully} covariant; thus, for any active diffeomorphism $f\in G[\phi]$, the observable changes to
    \begin{equation}
        A[f_*\phi] = \int_{\mathcal{M}} a[f_*\phi] = \int_{\mathcal{M}}f_*(a[\phi]) = \int_{f^{-1}(\mathcal{M})}a[\phi] = \int_{\mathcal{M}}a[\phi] = A[\phi].
        \label{Equation: single integral invariance}
    \end{equation}
    Thus, $A[\phi]$ is invariant under \emph{all} diffeomorphisms. The form $a[\phi]$ is also usually taken to be kinematically local. 

    A particularly relevant example of such an observable is as follows. Let us assume that $\mathcal{M}$ permits a global coordinate system $x^\mu$, $\mu=0,\dots,D-1$, and the kinematical fields $\phi$ contain a set of $D$ scalar fields $Z^a$, $i=0,\dots,D-1$, and that $B[\phi]$ is a covariant function on spacetime. Then we define the observable
    \begin{equation}
        A[\phi](Z_0) = \int_{\mathcal{M}}\dd[D]{x}\delta^{(D)}(Z^a_0-Z^a(x))\abs{\det\pdv{Z^a}{x^\mu}}\, B[\phi],
    \end{equation}
    where $Z^a_0$, $a=0,\dots,D-1$ is a set of $D$ real numbers. This is clearly a single integral observable, and the integrand is covariant. On the other hand, we can simply evaluate the delta function to obtain 
    \begin{multline}
        A[\phi](Z_0) = \int_{\mathcal{M}}\dd[D]{x}\delta^{(D)}(x-Z^{-1}(Z^a_0))\, B[\phi]=B[\phi]\big(Z^{-1}(Z_0)\big),\\
        \qq{or} A[\phi] = B[\phi]\circ Z^{-1} = Z_* (B[\phi]),
    \end{multline}
    where we are assuming that the function $Z:\mathcal{M}\to \RR^D, x \mapsto (Z^0(x),\dots,Z^{D-1}(x))$ is invertible. If we define a parametrised frame $R$ by setting $R[\phi]=Z^{-1}$, then it is clear that $A= O_{B,R}$ is a relational observable of the kind we have defined.

    The point of this section is to demonstrate that this extends to the general relational observables we have defined, To that end, consider now a general frame $R:\mathscr{O}\to\mathscr{D}$, and a covariant local spacetime observable $B[\phi]$. For simplicity, we consider the case where $B[\phi]$ is a scalar, but the below discussion would extend straightforwardly to more general fields. We then have for all local orientations $o\in\mathscr{O}$
    \begin{nalign}
        O_{B,R}(o) = \big((R[\phi])^*B[\phi]\big)(o) &= B[\phi](R[\phi](o)) \\
                                                     &= \int_{\mathcal{M}}\dd[D]{x}\delta^{(D)}(x-R[\phi](o))\,  B[\phi],
        \label{Equation: relational observable single integral 1}
    \end{nalign}
    We can rewrite the delta function in this integral as
    \begin{equation}
        \delta^{(D)}(x-R[\phi](o)) =\delta^{(D)}\Big(o - (R[\phi])^{-1}(x)\Big)\, \abs{\det\pdv{(R[\phi])^{-1}}{x}}.
    \end{equation}
    The right-hand side is only defined where the frame field $(R[\phi])^{-1}$ exists, i.e.\ in the image $\mathcal{N}[\phi]$ of the frame. But the delta function on the left-hand side only has support in $\mathcal{N}[\phi]$, so we can write the relational observable as
    \begin{nalign}
        O_{B,R}(o) &= \int_{\mathcal{N}[\phi]}\dd[D]{x}\delta^{(D)}\Big(o - (R[\phi])^{-1}(x)\Big)\, \abs{\det\pdv{(R[\phi])^{-1}}{x}}\, B[\phi] \\
                   &= \int_{\mathcal{M}}\dd[D]{x}\delta^{(D)}\Big(o - (R[\phi])^{-1}(x)\Big)\, \abs{\det\pdv{(R[\phi])^{-1}}{x}}\,1_{\mathcal{N}[\phi]}\, B[\phi].
        \label{Equation: relational observable single integral 2}
    \end{nalign}
    In the second line we have extended the integral to the whole of spacetime by including a characteristic function $1_{\mathcal{N}[\phi]}$, defined to be equal to one within $\mathcal{N}[\phi]$, and zero elsewhere.

    It is clear that in either form~\eqref{Equation: relational observable single integral 1} or~\eqref{Equation: relational observable single integral 2}, the relational observable $O_{B,R}$ resembles the type of single integral observable described above. However, there are some key differences with the ordinary type of single integral observable. One difference is that the top form which is being integrated is here in general not fully covariant, but rather just gauge-covariant, due to the fact that the same is true of the frame $R[\phi]$. This of course still defines a valid gravitational observable, since~\eqref{Equation: single integral invariance} holds for all small diffeomorphisms $f\in H[\phi]$, so $A[\phi]$ is gauge-invariant. But~\eqref{Equation: single integral invariance} will not hold for large diffeomorphisms, so single integral observables formed from integrands which are only gauge-covariant can have non-trivial large diffeomorphism charges. The other key difference is that even if $B[\phi]$ is kinematically local, the frame itself will often not be, and so the top form integrand will in general not be kinematically local. As will be explained in Section~\ref{Section: large diffeos}, kinematically non-local frames are essential for producing relational observables which are non-trivially charged under large diffeomorphisms.

    To summarise, \pah{generalising} to gauge-covariant and non-kinematically local integrands, \pah{the above implies that all our} relational observables can be written as single integral observables. Later, in Section~\ref{Section: general formalism / covariant to canonical} we will demonstrate how (after appropriately restricting to certain types of frame), the relational observables defined in this paper can be understood using a canonical point of view involving a power series of Poisson brackets with constraints~\cite{Dittrich:2004cb,Dittrich:2005kc,Dittrich:2006ee,Dittrich:2007jx}. As a consequence, this paper demonstrates the mutual compatibility of these three approaches to relational observables.

    Let us also point out that by generalising the integrals above, we can produce \emph{smeared} relational observables. Indeed, by picking a top form $w$ on local orientation space $\mathscr{O}$, we can define a smeared observable $O_{B,R;w}$ via
    \begin{equation}
        O_{B,R;w}[\phi] = \int_{\mathscr{O}} w \, O_{B,R}[\phi].
        \label{Equation: smeared observable local orientations}
    \end{equation}
    The form $w$ plays the role of a weight for the smearing; indeed, by choosing a set of coordinates $o^a$, $a=1,\dots n$ on $\mathscr{O}$ we may write $w=W \dd[n]{o}$, where $W$ is a weight function. Here, we are taking $w$ to be independent of $\phi$, but more generally $w$ could depend on $\phi$ in a gauge-invariant way. This would guarantee that $O_{B,R;w}$ is gauge-invariant. The integral in~\eqref{Equation: smeared observable local orientations} implements a smearing over local orientations, but we can reinterpret this as a smearing in spacetime by pushing everything forward through the frame. Indeed, we have
    \begin{equation}
        O_{B,R;w}[\phi] = \int_{\mathscr{O}} w\, (R[\phi])^*B[\phi] = \int_{\mathcal{N}[\phi]} w_R[\phi]\, B[\phi],
    \end{equation}
    where $w_R[\phi] = (R[\phi])_* w$. Thus $O_{B,R;w}$ may be understood as the covariant observable smeared over the image of the frame by the weight form $w_R[\phi]$. Note that even if $w$ doesn't depend on $\phi$, its pushforward through the frame $w_R[\phi]$ must. Indeed it must be gauge-covariant:
    \begin{equation}
        w_R[f_*\phi] = (R[f_*\phi])_* w = (f\circ R[\phi])_* w = f_* (R[\phi])_* w = f_* w_R[\phi], \text{ for all } f\in H[\phi].
    \end{equation}
    Note that we may extend the above integral to one over all of spacetime by again using a characteristic function:
    \begin{equation}
        O_{B,R;w}[\phi] = \int_{\mathcal{M}} w_R[\phi]\,1_{\mathcal{N}[\phi]}\, B[\phi].
        \label{Equation: smeared relational observable spacetime}
    \end{equation}
    So $O_{B,R;w}$ may be understood as being smeared over spacetime by the weight form $w_R[\phi]\,1_{\mathcal{N}[\phi]}$, which is again gauge-covariant. Such observables encompass all smearings previously considered in the single integral approach, as well as any others\pah{, including relative to frame fields that are not globally defined on spacetime}.

    The other direction works too. One may construct gauge-invariant smeared observables by picking any gauge-covariant spacetime weight form $\tilde w[\phi]$ against which we integrate $B[\phi]$:
    \begin{equation}
        O[\phi] = \int_{\mathcal{M}} \tilde w[\phi] B[\phi].
    \end{equation}
    If $\tilde{w}[\phi]$ has support in $\mathcal{N}[\phi]$, then we can restrict the integral to $\mathcal{N}[\phi]$ and pull everything back to $\mathscr{O}$:
    \begin{equation}
        O[\phi] = \int_{\mathscr{O}} w[\phi] O_{B,R}[\phi], \qq{where} w[\phi] = (R[\phi])^*\tilde w[\phi].
    \end{equation}

    Note that, so long as the frame map $R[\phi]$ is continuous, if the spacetime weight form is compactly supported, then the corresponding orientation space weight form will also be compactly supported, and vice versa.\footnote{This may be important when one wishes to quantise relational observables.}

\subsubsection{Relational locality}
\label{subsection:relational_locality}
Suppose we have a parametrised frame $R:\mathscr{O}\to\mathscr{D}$. We will call a local observable $B$ on $\mathscr{O}$ `relationally local' with respect to $R$ if $R[\phi]_*B[\phi]$ is kinematically local (in the case $B[\phi]$ is a local map), or if $X_{R[\phi]}(B[\phi])$ is kinematically local (in the case $B[\phi]$ is a local quantity). Just like kinematical locality, relational locality provides a link between the locality of the kinematical fields and the locality of the local observable. However, unlike kinematical locality, the link itself depends on the kinematical fields, in a way that is determined by the frame.

Furthermore, unlike kinematical locality, there is no obstruction to the existence of relationally local gauge-invariant observables that are non-trivial in the bulk. The easiest way to get gauge-invariant relationally local observables with respect to a parametrised injective frame $R$ is to take a kinematically local covariant observable $A$, and construct the relational observable $O_{A,R}$ of $A$ with respect to $R$. Such a relational observable $O_{A,R}$ is relationally local by definition.

Consider a variation of the fields $\phi\to\phi+\delta\phi$. Recall that the linearised variations of kinematically local observables can be written as sums~\eqref{Equation: kinematically local quantity} and~\eqref{Equation: kinematically local map} over the components of $\delta\phi$ (and its derivatives). In particular, these sums only depend on $\delta\phi$ (and its derivatives) at the point in $\mathcal{M}$ to which those observables are local.

Suppose $B[\phi]$ is a gauge-invariant relationally local observable. We would like to have a similar kind of expression for the variation of $B[\phi]$\jjvk{, as it will assist us later in Section~\ref{Section: general formalism / relational phase space / microcausality} when we come to discuss relational microcausality}. The possible non-local dependence of $B[\phi]$ on the kinematical fields would seem to make this difficult to achieve \jjvk{in a controlled manner}. However, there is in fact a similar expression (at least in the bulk), as we will now demonstrate.

Let us first address the case where $B[\phi]$ is a local quantity at a local orientation $o = \pi(B[\phi])$. The frame $R[\phi]$ associates $o$ with the point $R[\phi](o)\in\mathcal{M}$. We will assume that this point is in the bulk, i.e.\ $R[\phi](o)\not \in \partial\mathcal{M}$. Then we can pick a diffeomorphism $F[\phi,\delta\phi]$ of $\mathcal{M}$ that agrees with $R[\phi+\delta\phi]\circ (R[\phi])^{-1}$ in an open neighbourhood of $R[\phi](o)$, but which vanishes in an open neighbourhood of $\partial\mathcal{M}$. This implies that $F[\phi,\delta\phi]$ is a small diffeomorphism, so by the gauge invariance of $B[\phi]$ we may write
\begin{equation}
    B[\phi+\delta\phi] = B[(F[\phi,\delta\phi])^*(\phi+\delta\phi)].
\end{equation}
Also,
\begin{equation}
    R[(F[\phi,\delta\phi])^*(\phi+\delta\phi)] = (F[\phi,\delta\phi])^{-1} \circ R[\phi+\delta\phi]
    = R[\phi]\circ (R[\phi+\delta\phi])^{-1} \circ R[\phi+\delta\phi] = R[\phi]
\end{equation}
where the second equality only holds in an open neighbourhood of $o$. By expanding out $F[\phi,\delta\phi]$ to linear order in $\delta\phi$, we can write
\begin{equation}
    (F[\phi,\delta\phi])^*(\phi+\delta\phi) = \phi + \delta_R\phi,
    \label{Equation: F delta R phi}
\end{equation}
where $\delta_R\phi$ is some field variation that depends on $\delta\phi$ and the frame (we will give a more explicit expression for $\delta_R\phi$ shortly). Since $X_{R[\phi]}(B[\phi])$ is kinematically local, we have
\begin{nalign}
    B[\phi+\delta\phi] &= X_{(R[\phi])^{-1}}\Big(X_{R[(F[\phi,\delta\phi])^*(\phi+\delta\phi)]}\big(B[(F[\phi,\delta\phi])^*(\phi+\delta\phi)]\big)\Big) \\
                       &= X_{(R[\phi])^{-1}}\Big(X_{R[\phi+\delta_R\phi]}\big(B[\phi+\delta_R\phi]\big)\Big) \\
                       &= X_{(R[\phi])^{-1}}\Big(\tilde B[\phi] + \tilde{B}'[\phi]\cdot \delta_R\phi|_{R[\phi](o)} + \tilde{B}''[\phi]\cdot \partial(\delta_R\phi)|_{R[\phi](o)} + \dots\Big) \\
                       &= B[\phi] + B'[\phi] \cdot \delta_R\phi|_{R[\phi](o)} + B''[\phi] \cdot \partial(\delta_R\phi)|_{R[\phi](o)} + \dots,
\end{nalign}
where we have set $\tilde B[\phi]:=X_{R[\phi]}\big(B[\phi]\big)$. Here, the third line follows from an application of~\eqref{Equation: kinematically local quantity}, the fourth line follows from the chain rule, while $\tilde{B}'$, $\tilde{B}''$, $\dots$ are some observables that depend on $B$ and $R$, and we have set $B'[\phi] = X_{(R[\phi])^{-1}}(\tilde{B}'[\phi])$, $B''[\phi] = X_{(R[\phi])^{-1}}(\tilde{B}''[\phi])$, $\dots$. Therefore, the variation of $B[\phi]$ takes the form
\begin{equation}
    \delta(B[\phi]) = B'[\phi] \cdot \delta_R\phi|_{R[\phi](\pi(B[\phi]))} + B''[\phi] \cdot \partial(\delta_R\phi)|_{R[\phi](\pi(B[\phi]))} + \dots
    \label{Equation: relationally local quantity}
\end{equation}
In the case where $B[\phi]$ is a local map, similar arguments (replacing $X_f$ by $f_*$ for all the diffeomorphisms $f$ that appear) yield
\begin{equation}
    \delta(B[\phi](b)) = B'[\phi](b) \cdot \delta_R\phi|_{R[\phi](\pi(b)} + B''[\phi](b) \cdot \partial(\delta_R\phi)|_{R[\phi](\pi(b))} + \dots
    \label{Equation: relationally local map}
\end{equation}
for some local map observables $B',B'',\dots$. In other words, the variation of a relationally local gauge-invariant observable depends on $\delta_R\phi$ in a relationally local way. To reiterate, these expressions in general only apply in the bulk, i.e.\ when $R[\phi](\pi(B[\phi]))\not\in\partial\mathcal{M}$ and $R[\phi](b)\not\in\partial\mathcal{M}$ for~\eqref{Equation: relationally local quantity} and~\eqref{Equation: relationally local map} respectively.\footnote{If we can pick the small diffeomorphism $F[\phi,\delta\phi]$ such that it agrees with $R[\phi+\delta\phi]\circ (R[\phi])^{-1}$ on the boundary, then these expressions will apply on the boundary too. This may require relaxation of the requirement that $F[\phi,\delta\phi]$ vanishes in a neighbourhood of the boundary, while still remaining small. This may be possible for some frames, but for general frames it is not possible.}

Thus, although the possible kinematical non-locality of the frame may prevent us from writing the variations of gauge-invariant relationally local observables as sums over local expressions involving $\delta\phi$, in the bulk we still can write them as sums over local expressions involving $\delta_R\phi$. There is no contradiction here, because $\delta_R\phi$ is itself formed non-locally from $\delta\phi$. Note that, in an open neighbourhood of each of the points at which $\delta_R\phi$ is used in~\eqref{Equation: relationally local quantity} and~\eqref{Equation: relationally local map}, $F[\phi,\delta\phi]$ agrees with $R[\phi+\delta\phi]\circ (R[\phi])^{-1}$. Thus, we can replace the definition~\eqref{Equation: F delta R phi} by
\begin{equation}
    \big(R[\phi+\delta\phi]\circ (R[\phi])^{-1}\big)^*(\phi+\delta\phi) = \phi + \delta_R\phi
    \label{Equation: R delta R phi}
\end{equation}
Also, we can expand $R[\phi+\delta\phi]\circ (R[\phi])^{-1}$ to linear order in $\delta\phi$ and obtain
\begin{equation}
    R[\phi+\delta\phi]\circ (R[\phi])^{-1} = \operatorname{Id}_{\mathcal{N}[\phi]} + V_R[\phi,\delta\phi],
\end{equation}
where $V_R[\phi,\delta\phi]$ is a vector field in the image $\mathcal{N}[\phi]$ of the frame with a generically highly kinematically non-local dependence on $\phi$ and $\delta\phi$. Then we have
\begin{equation}
    \delta_R\phi = \delta\phi - \lie_{V_R[\phi,\delta\phi]}\phi,
\end{equation}
where $\lie$ denotes the Lie derivative. Thus, $\delta\phi$ and $\delta_R\phi$ are just related by a diffeomorphism of $\phi$. In Section~\ref{Section: general formalism / further generalisations}, we will see how $\delta_R$ can be interpreted as a kind of field space covariant derivative, and also how $V_R$ may be thought of as a generalisation of the Maurer-Cartan form on field space~\cite{Gomes:2016mwl,Donnelly:2016auv,Speranza:2017gxd,Speranza:2022lxr,Freidel:2020xyx,Freidel:2021dxw,CEH}.

The reader may be concerned that the relationally local observables we have constructed have somehow `forgotten' about spacetime, because relational locality is based on the local orientation space $\mathscr{O}$. This is not the case. Indeed, if we choose for the local orientation space to be a subset of spacetime, then relational locality is indeed a form of spacetime locality. Moreover, given an injective parametrised frame $R:\mathscr{O}\to\mathscr{D}$, we can always construct a new parametrisation for the frame which has this property. Indeed, since the frame is injective, $R[\phi]$ is invertible on its image $\mathcal{N}[\phi]$. Thus, fixing a $\phi_0\in\mathcal{S}$, we can construct
\begin{equation}
    R_0 = R\circ (R[\phi_0])^{-1}.
\end{equation}
This is a parameterised frame whose local orientation space is $\mathcal{N}[\phi_0]$. Moreover, the image of $R_0$ is the same as the image of $R$, so if we ignore the parametrisations, the two frames are the same, because they are made up of the same underlying set of dressings in $\mathscr{D}$. It is also worth noting that we have already met a parametrised frame of this kind -- the gauge-fixing diffeomorphisms $U[\phi]$ defined in~\eqref{Equation: gauge-fixing diffeomorphism}. Indeed, for such a $U[\phi]$, $\phi\mapsto U[\phi](x)$ is a local dressing for all $x\in\mathcal{M}$. Thus, the map taking $x\in\mathcal{M}$ to this local dressing constitutes a parametrised frame whose local orientation space is all of spacetime. The observable $O$ appearing in~\eqref{Equation: U dressed O} is an example of a relational observable using this frame.

\subsubsection{Changes of frame}
Suppose we have two frames $\mathscr{R}_1,\mathscr{R}_2 \subset \mathscr{D}$. The images of these frames $\mathcal{N}_1[\phi],\mathcal{N}_2[\phi]$ may overlap for certain kinematical field configurations $\phi$. This can happen even if the frames themselves are completely disjoint, i.e.\ $\mathscr{R}_1\cap\mathscr{R}_2=\emptyset$.

If $x_1\in\mathscr{R}_1$, and $x_2\in\mathscr{R}_2$ are dressings in the first and second frames respectively such that for a fixed $\phi$
\begin{equation}
    x_1[\phi] = x_2[\phi],
\end{equation}
then the dressings are essentially equivalent for that $\phi$. For example, in the case of two frames constructed from geodesics in the way described in Section~\ref{Section: geodesic example}, this equation says that the endpoints $x_1[\phi]$ and $x_2[\phi]$ of two distinct geodesics from the different frames coincide. Of course, this property is highly dependent on the metric configuration, and so will only hold for certain $\phi$.

If $x_1[\phi] \in \mathcal{N}_1[\phi]\cap\mathcal{N}_2[\phi]$, then there is always an $x_2$ satisfying this equation. This $x_2$ is unique if the frames are injective. Indeed, let us assume the frames are injective, and define
\begin{nalign}
    \mathscr{R}_{1|2}[\phi] &= \{x_1\in \mathscr{R}_1 \mid x_1[\phi]\in \mathcal{N}_2[\phi]\}, \\
    \mathscr{R}_{2|1}[\phi] &= \{x_2\in \mathscr{R}_2 \mid x_2[\phi]\in \mathcal{N}_1[\phi]\}.
\end{nalign}
Then for each $\phi$ there is a unique bijection
\begin{equation}
    \mathscr{R}_{1\to 2}[\phi]: \mathscr{R}_{1|2}[\phi]\to \mathscr{R}_{2|1}[\phi]
\end{equation}
satisfying
\begin{equation}
    x_1[\phi] = x_2[\phi], \qq{ where } x_2 = \mathscr{R}_{1\to 2}[\phi](x_1).
\end{equation}
This is the `change of frames' map. It is a local map on spacetime, with respect to the bundle structure given by $\pi[\phi]:\mathscr{D}\to\mathcal{M}$. Indeed, one has
\begin{equation}
    \pi[\phi]\circ \mathscr{R}_{1\to 2}[\phi] = \pi[\phi].
\end{equation}
In this way, we can think of a change of frames as a local procedure in spacetime.

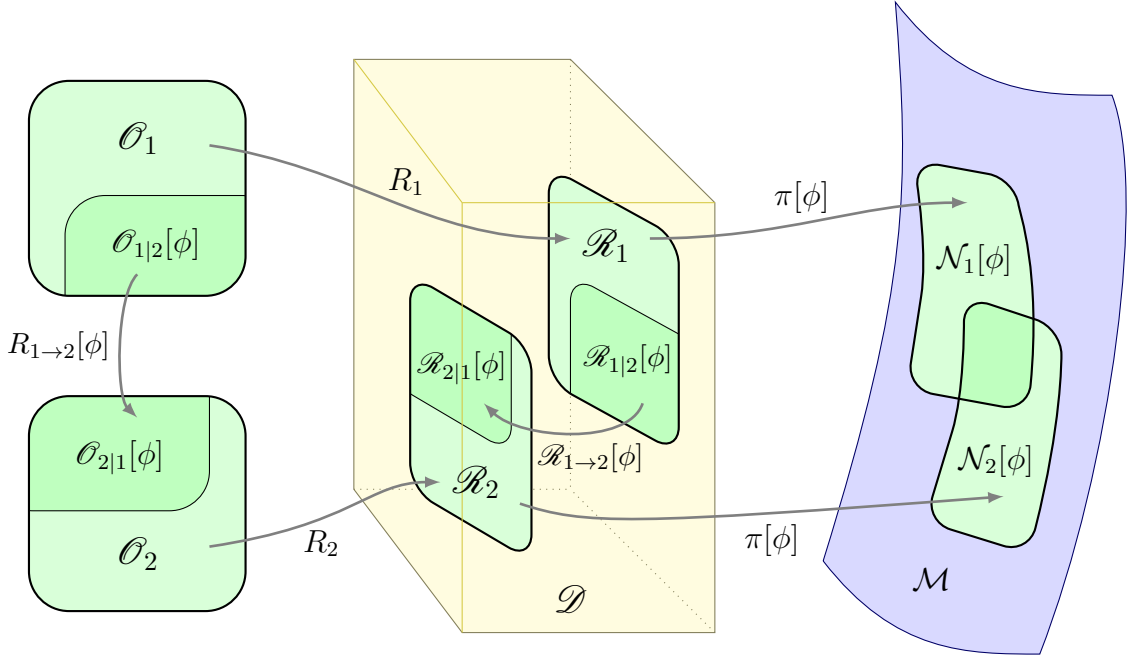
\begin{figure}
    \centering
    \begin{tikzpicture}[scale=0.95]
        \begin{scope}[shift={(-1.5,3.7)}]
            \fill[rounded corners=15pt, green!15] (0,0) rectangle (3,3);
            \begin{scope}
                \clip[rounded corners=15pt] (0,0) rectangle (3,3);
                \draw[rounded corners=15pt, fill=green!25, shift={(0.5,-1.6)}] (0,0) rectangle (3,3);
            \end{scope}
            \draw[rounded corners=15pt, thick] (0,0) rectangle (3,3);
            \node at (1.5,2.2) {\Large$\mathscr{O}_1$};
            \node at (1.75,0.7) {\large$\mathscr{O}_{1|2}[\phi]$};
        \end{scope}
        \begin{scope}[shift={(-1.5,-0.7)}]
            \fill[rounded corners=15pt, green!15] (0,0) rectangle (3,3);
            \begin{scope}
                \clip[rounded corners=15pt] (0,0) rectangle (3,3);
                \draw[rounded corners=15pt, fill=green!25, shift={(-0.5,1.4)}] (0,0) rectangle (3,3);
            \end{scope}
            \draw[rounded corners=15pt, thick] (0,0) rectangle (3,3);
            \node at (1.5,0.8) {\Large$\mathscr{O}_2$};
            \node at (1.25,2.15) {\large$\mathscr{O}_{2|1}[\phi]$};
        \end{scope}

        \draw[yellow!40!black,fill=yellow!15] (4.5,-1) -- (8,-1) -- (8,5) -- (6,7) -- (3,7) -- (3,1) -- cycle;
        \draw[yellow!40!black,dotted] (6,7) -- (6,1);
        \draw[yellow!40!black,dotted] (3,1) -- (6,1);
        \draw[yellow!40!black,dotted] (8,-1) -- (6,1);

        \begin{scope}[xscale=1.5,shift={(-1,0)}]
            \fill[rounded corners=10pt, green!15, shift={(-0.2,1.5)}] (5,1) -- (6.2,0) -- (6.2,3) -- (5,4) -- cycle;
            \begin{scope}
                \clip[rounded corners=10pt, shift={(-0.2,1.5)}] (5,1) -- (6.2,0) -- (6.2,3) -- (5,4) -- cycle;
                \draw[rounded corners=10pt, fill=green!25] (5,1) -- (6.2,0) -- (6.2,3) -- (5,4) -- cycle;
            \end{scope}
            \draw[rounded corners=10pt, thick, shift={(-0.2,1.5)}] (5,1) -- (6.2,0) -- (6.2,3) -- (5,4) -- cycle;
        \end{scope}

        \begin{scope}[xscale=1.4,shift={(-2.3,0)}]
            \fill[rounded corners=10pt, green!15] (5,1) -- (6.2,0) -- (6.2,3) -- (5,4) -- cycle;
            \begin{scope}
                \clip[rounded corners=10pt] (5,1) -- (6.2,0) -- (6.2,3) -- (5,4) -- cycle;
                \draw[rounded corners=10pt, fill=green!25, shift={(-0.2,1.5)}] (5,1) -- (6.2,0) -- (6.2,3) -- (5,4) -- cycle;
            \end{scope}
            \draw[rounded corners=10pt, thick] (5,1) -- (6.2,0) -- (6.2,3) -- (5,4) -- cycle;
        \end{scope}

        \draw[yellow!80!black] (3,7) -- (4.5,5);
        \draw[yellow!80!black] (4.5,-1) -- (4.5,5);
        \draw[yellow!80!black] (8,5) -- (4.5,5);

        \node at (6.5,4.4) {\Large$\mathscr{R}_1$};
        \node at (6.8,2.8) {$\mathscr{R}_{1|2}[\phi]$};
        \node at (4.7,1.1) {\Large$\mathscr{R}_2$};
        \node at (4.5,2.7) {$\mathscr{R}_{2|1}[\phi]$};

        \node[above] at (6,-0.85) {\Large$\mathscr{D}$};
        \begin{scope}[shift={(9.5,0)},yscale=1.3]
            \draw[blue!40!black,fill=blue!15] (0,0)
                .. controls (1,-1) and (2,-1) .. (3,-1)
                .. controls (4,1) and (4.5,4) .. (4,5)
                .. controls (3,5) and (2,5) .. (1,6)
                .. controls (1.2,4) and (1.2,3) .. (0,0);
        \end{scope}

        \begin{scope}[shift={(6.8,1.4)},scale=0.9,rotate=8]
            \fill[rounded corners=10pt, green!15] (4.4,0.5) -- (6.2,-0.1)
                .. controls (6.7,2) and (6.5,3) .. (6.5,3.5) -- (5,4)
                .. controls (5,3.5) and (5,2.5) .. (4.7,1.5) -- cycle;
        \end{scope}
        \begin{scope}[shift={(6.95,0.1)},scale=0.9]
            \fill[rounded corners=10pt, green!15] (4.4,0.5) -- (6,0)
                .. controls (6.5,2) and (6.5,3) .. (6.5,3.5) -- (5,4)
                .. controls (5,3.5) and (5,2.5) .. (4.7,1.5) -- cycle;
        \end{scope}
        \begin{scope}
            \clip[rounded corners=10pt,shift={(6.95,0.1)},scale=0.9] (4.4,0.5) -- (6,0)
                .. controls (6.5,2) and (6.5,3) .. (6.5,3.5) -- (5,4)
                .. controls (5,3.5) and (5,2.5) .. (4.7,1.5) -- cycle;
            \fill[rounded corners=10pt, green!25,shift={(6.8,1.4)},scale=0.9,rotate=8] (4.4,0.5) -- (6.2,-0.1)
                .. controls (6.7,2) and (6.5,3) .. (6.5,3.5) -- (5,4)
                .. controls (5,3.5) and (5,2.5) .. (4.7,1.5) -- cycle;
        \end{scope}
        \begin{scope}[shift={(6.8,1.4)},scale=0.9,rotate=8]
            \draw[rounded corners=10pt, thick] (4.4,0.5) -- (6.2,-0.1)
                .. controls (6.7,2) and (6.5,3) .. (6.5,3.5) -- (5,4)
                .. controls (5,3.5) and (5,2.5) .. (4.7,1.5) -- cycle;
        \end{scope}
        \begin{scope}[shift={(6.95,0.1)},scale=0.9]
            \draw[rounded corners=10pt, thick] (4.4,0.5) -- (6,0)
                .. controls (6.5,2) and (6.5,3) .. (6.5,3.5) -- (5,4)
                .. controls (5,3.5) and (5,2.5) .. (4.7,1.5) -- cycle;
        \end{scope}

        \node at (11.6,4.2) {\large$\mathcal{N}_1[\phi]$};
        \node at (11.9,1.4) {\large$\mathcal{N}_2[\phi]$};
        \node[above] at (11,-0.55) {\large$\mathcal{M}$};

        \draw[gray,line width=1pt,-{latex}] (1,5.8) .. controls (3,5.5) and (4,4.5) .. (6,4.5) node[pos=0.55,above,black] {\large$R_1$};
        \draw[gray,line width=1pt,-{latex}] (1,0.2) .. controls (3,0.5) and (3.2,1.1) .. (4.2,1.1) node[pos=0.35,below,black] {\large$R_2$};
        \draw[gray,line width=1pt,-{latex}] (7.1,4.5) .. controls (9,4.5) and (10,5) .. (11.5,5) node[pos=0.44,above,black] {\large$\pi[\phi]$};
        \draw[gray,line width=1pt,-{latex}] (5.3,0.8) .. controls (6.2,0.5) and (6.8,0.5) .. (12,0.9) node[pos=0.74,below,black] {\large$\pi[\phi]$};
        \draw[gray,line width=1pt,-{latex}] (0,4) .. controls (-0.3,3.7) and (-0.3,2.3) .. (0,2) node[midway,left,black] {$R_{1\to 2}[\phi]$};
        \draw[gray,line width=1pt,-{latex}] (7,2.2) .. controls (6.8,1.6) and (5.3,1.7) .. (4.8,2.2) node[pos=0.4,below,black] {$\mathscr{R}_{1\to 2}[\phi]$};
    \end{tikzpicture}
    \caption{For certain kinematical field configurations $\phi$, the images of two frames $\mathscr{R}_1,\mathscr{R}_2$ may overlap. This means there are dressings in the first frame which give the same spacetime points as dressings in the second frame. When the frames are injective, this can be understood in terms of a map $\mathscr{R}_{1\to 2}[\phi]$ between subsets of dressings $\mathscr{R}_{1|2}[\phi]\subset\mathscr{R}_1$, $\mathscr{R}_{2|1}[\phi]\subset\mathscr{R}_2$. Similarly, when the frames also have orientations, we get a gauge-invariant map $R_{1\to 2}[\phi]$ between subsets of their local orientation spaces $\mathscr{O}_{1|2}[\phi]\subset\mathscr{O}_1$, $\mathscr{O}_{2|1}[\phi]\subset\mathscr{O}_2$.}
    \label{Figure: change of frame}
\end{figure}

When we have two injective parametrised frames $R_1:\mathscr{O}_1\to\mathscr{D}$, $R_2:\mathscr{O}_2\to\mathscr{D}$, the change of frames map relates the local orientations of the two frames. Indeed, it gives for each field configuration $\phi$ a map from a subset
\begin{equation}
    \mathscr{O}_{1|2}[\phi] = R_1^{-1}(\mathscr{R}_{1|2}[\phi])
\end{equation}
of $\mathscr{O}_1$ to a subset
\begin{equation}
    \mathscr{O}_{2|1}[\phi] = R_2^{-1}(\mathscr{R}_{2|1}[\phi])
\end{equation}
of $\mathscr{O}_2$:
\begin{equation}
    R_{1\to 2}[\phi] = R_2^{-1}\circ \mathscr{R}_{1\to 2}[\phi] \circ R_1 = (R_2[\phi])^{-1}\circ R_1[\phi].
\end{equation}
This map is defined such that if $R_{1\to 2}[\phi](o_1)=o_2$, then the local orientations $o_1$ and $o_2$ yield the same spacetime point, i.e.\ $R_1[\phi](o_1)=R_2[\phi](o_2)$. Note that we may view $R_{1\to 2}[\phi]$ as a relational observable -- it is the second frame field relative to the first frame:
\begin{equation}
    R_{1\to 2}[\phi] = (R_1[\phi])^* (R_2[\phi])^{-1}.
\end{equation}
In particular, the change of frame map is gauge-invariant. This again emphasises the fact that the change of frames is a relational procedure. The whole setup is depicted in Figure~\ref{Figure: change of frame}.

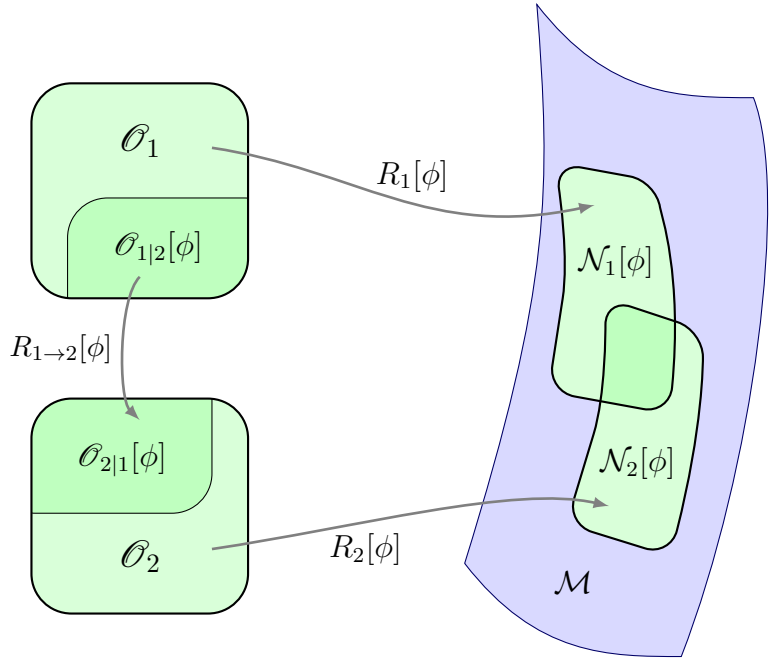
\begin{figure}
    \centering
    \begin{tikzpicture}[scale=0.95]
        \begin{scope}[shift={(-1.5,3.7)}]
            \fill[rounded corners=15pt, green!15] (0,0) rectangle (3,3);
            \begin{scope}
                \clip[rounded corners=15pt] (0,0) rectangle (3,3);
                \draw[rounded corners=15pt, fill=green!25, shift={(0.5,-1.6)}] (0,0) rectangle (3,3);
            \end{scope}
            \draw[rounded corners=15pt, thick] (0,0) rectangle (3,3);
            \node at (1.5,2.2) {\Large$\mathscr{O}_1$};
            \node at (1.75,0.7) {\large$\mathscr{O}_{1|2}[\phi]$};
        \end{scope}
        \begin{scope}[shift={(-1.5,-0.7)}]
            \fill[rounded corners=15pt, green!15] (0,0) rectangle (3,3);
            \begin{scope}
                \clip[rounded corners=15pt] (0,0) rectangle (3,3);
                \draw[rounded corners=15pt, fill=green!25, shift={(-0.5,1.4)}] (0,0) rectangle (3,3);
            \end{scope}
            \draw[rounded corners=15pt, thick] (0,0) rectangle (3,3);
            \node at (1.5,0.8) {\Large$\mathscr{O}_2$};
            \node at (1.25,2.15) {\large$\mathscr{O}_{2|1}[\phi]$};
        \end{scope}

        \begin{scope}[shift={(-5,0)}]
            \begin{scope}[shift={(9.5,0)},yscale=1.3]
                \draw[blue!40!black,fill=blue!15] (0,0)
                    .. controls (1,-1) and (2,-1) .. (3,-1)
                    .. controls (4,1) and (4.5,4) .. (4,5)
                    .. controls (3,5) and (2,5) .. (1,6)
                    .. controls (1.2,4) and (1.2,3) .. (0,0);
            \end{scope}

            \begin{scope}[shift={(6.8,1.4)},scale=0.9,rotate=8]
                \fill[rounded corners=10pt, green!15] (4.4,0.5) -- (6.2,-0.1)
                    .. controls (6.7,2) and (6.5,3) .. (6.5,3.5) -- (5,4)
                    .. controls (5,3.5) and (5,2.5) .. (4.7,1.5) -- cycle;
            \end{scope}
            \begin{scope}[shift={(6.95,0.1)},scale=0.9]
                \fill[rounded corners=10pt, green!15] (4.4,0.5) -- (6,0)
                    .. controls (6.5,2) and (6.5,3) .. (6.5,3.5) -- (5,4)
                    .. controls (5,3.5) and (5,2.5) .. (4.7,1.5) -- cycle;
            \end{scope}
            \begin{scope}
                \clip[rounded corners=10pt,shift={(6.95,0.1)},scale=0.9] (4.4,0.5) -- (6,0)
                    .. controls (6.5,2) and (6.5,3) .. (6.5,3.5) -- (5,4)
                    .. controls (5,3.5) and (5,2.5) .. (4.7,1.5) -- cycle;
                \fill[rounded corners=10pt, green!25,shift={(6.8,1.4)},scale=0.9,rotate=8] (4.4,0.5) -- (6.2,-0.1)
                    .. controls (6.7,2) and (6.5,3) .. (6.5,3.5) -- (5,4)
                    .. controls (5,3.5) and (5,2.5) .. (4.7,1.5) -- cycle;
            \end{scope}
            \begin{scope}[shift={(6.8,1.4)},scale=0.9,rotate=8]
                \draw[rounded corners=10pt, thick] (4.4,0.5) -- (6.2,-0.1)
                    .. controls (6.7,2) and (6.5,3) .. (6.5,3.5) -- (5,4)
                    .. controls (5,3.5) and (5,2.5) .. (4.7,1.5) -- cycle;
            \end{scope}
            \begin{scope}[shift={(6.95,0.1)},scale=0.9]
                \draw[rounded corners=10pt, thick] (4.4,0.5) -- (6,0)
                    .. controls (6.5,2) and (6.5,3) .. (6.5,3.5) -- (5,4)
                    .. controls (5,3.5) and (5,2.5) .. (4.7,1.5) -- cycle;
            \end{scope}

            \node at (11.6,4.2) {\large$\mathcal{N}_1[\phi]$};
            \node at (11.9,1.4) {\large$\mathcal{N}_2[\phi]$};
            \node[above] at (11,-0.55) {\large$\mathcal{M}$};
        \end{scope}

        \draw[gray,line width=1pt,-{latex}] (1,5.8) .. controls (3,5.5) and (4,4.5) .. (6.3,5) node[pos=0.55,above,black] {\large$R_1[\phi]$};
        \draw[gray,line width=1pt,-{latex}] (1,0.2) .. controls (3,0.5) and (5.2,1.1) .. (6.5,0.8) node[pos=0.35,below,black] {\large$R_2[\phi]$};
        \draw[gray,line width=1pt,-{latex}] (0,4) .. controls (-0.3,3.7) and (-0.3,2.3) .. (0,2) node[midway,left,black] {$R_{1\to 2}[\phi]$};
    \end{tikzpicture}
    \caption{At a fixed kinematical field configuration $\phi$, a change of frames is essentially the same as a change of coordinates. Thus, varying over all $\phi$, we can view a change of frames as a field-dependent (i.e.\ dynamical) change of coordinates. Note that the change of frames map is gauge-invariant.}
    \label{Figure: change of frame fixed phi}
\end{figure}

Note that for any fixed kinematical field configuration $\phi$, the structure of the universal dressing space becomes irrelevant, and we are just left with the map $R_{1\to2}[\phi]$, which essentially just describes the change of coordinates\footnote{Or more generally whichever abstract parameters make up each local orientation.} entailed by switching between the two parametrised frames $R_1$ and $R_2$. This is shown in Figure~\ref{Figure: change of frame fixed phi}, which is basically just Figure~\ref{Figure: change of frame} with the universal dressing space removed. Notably, we get a different change of coordinates for each $\phi$. So a change of frames can really be understood as a field-dependent, dynamical change of coordinates. We saw an example of this when changing between different geodesic dressing frames in Section~\ref{ssec_geoframechange}.

The map $R_{1\to 2}[\phi]$ also tells us how relational observables change when we change the frame. Indeed, we have
\begin{equation}
    O_{A,R_2}[\phi] = (R_{1\to 2}[\phi])_*O_{A,R_1}[\phi].
\end{equation}

By specifying a field-dependent map between local orientation spaces, and a parametrised frame, we can construct a second frame. Indeed, suppose we have a parametrised frame $R_1:\mathscr{O}_1\to\mathscr{D}$, and let $F$ be an observable which for each $\phi$ is a map $F[\phi]:\mathscr{O}_2\to\mathscr{O}_1$. Let us define
\begin{equation}
    R_2[\phi] = R_1[\phi]\circ F[\phi] = \pi[\phi] \circ R_1 \circ F[\phi].
\end{equation}
$R_2[\phi]$ is a map from $\mathscr{O}_2$ to $\mathcal{M}$. It can be viewed as a field-dependent reparametrisation of $R_1$. Suppose that we can write
\begin{equation}
    R_2[\phi] = \pi[\phi]\circ R_2,
    \label{Equation: reparametrised frame}
\end{equation}
for some other parametrised frame $R_2:\mathscr{O}_2\to\mathscr{D}$. If $R_1$ is injective, then a necessary and sufficient condition for $R_2$ to exist is that $F$ is gauge-invariant, i.e.\ $F[\phi] = F[f_*\phi]$ for all $f\in H[\phi]$. To see this, first note that for~\eqref{Equation: reparametrised frame} to hold we must have, for all $f\in H[\phi]$,
\begin{equation}
    f \circ R_1[\phi] \circ F[\phi] = f \circ R_2[\phi] = R_2[f_*\phi] = R_1[f_*\phi] \circ F[f_*\phi] = f \circ R_1[\phi] \circ F[f_*\phi].
\end{equation}
Injectivity of $R_1[\phi]$ implies that $F[\phi]=F[f_*\phi]$, as required. For the other way round, suppose that $F$ is gauge-invariant. Then we can explicitly construct the function $R_2$ by defining $R_2(o)$ for $o\in\mathscr{O}$ via
\begin{equation}
    R_2(o)[\phi] = R_2[\phi](o) = R_1[\phi](F[\phi](o)).
\end{equation}
It can be confirmed that $R_2(o)$ defined in this way is indeed a local dressing, provided $F$ is gauge-invariant. Thus, $R_2$ is indeed a parametrised frame, and moreover we have~\eqref{Equation: reparametrised frame} by construction. The map between the parametrised frames is just $R_{1\to 2}[\phi] = F[\phi]$.

It is possible that $R_1$ and $R_2$ are just different parametrisations of the same frame. Indeed, let us strip away the parametrisations of $R_1$ and $R_2$. Recall that the frames without their parametrisations are just the images of the parametrised frames:
\begin{equation}
    \mathscr{R}_1 = R_1(\mathscr{O}_1),\qquad \mathscr{R}_2 = R_2(\mathscr{O}_2).
\end{equation}
If there is no true change of frames, then by definition $\mathscr{R}_1=\mathscr{R}_2$. For this to be the case, for every $o_1\in\mathscr{O}_1$, there must be an $o_2\in\mathscr{O}_2$ such that $R_1(o_1)=R_2(o_2)$, and vice versa. Thus, there must be some bijection $p:o_2\mapsto o_1$ obeying $R_1\circ p = R_2$, which is equivalent to
\begin{equation}
    R[\phi]\circ p = \tilde{R}[\phi] = R[\phi]\circ F[\phi].
\end{equation}
Thus, when $R_1$ is injective, for there to be no change of frames we must have that $F[\phi]$ is a bijection and completely independent of $\phi$. Otherwise, the non-trivial dependence of $F[\phi]$ on $\phi$ must be absorbed by a change of frames.

Consider a local observable $A$ on a local orientation space $\mathscr{O}$, and let $R_1,R_2:\mathscr{O}\to\mathscr{D}$ be two parametrised frames. The observable $A$ can be relationally local with respect to one of these frames, but not the other. Indeed, if $A$ is relationally local with respect to $R_1$, then by definition $(R_1[\phi])_*A[\phi]$ is kinematically local. But it is only relationally local with respect to $R_2$ if
\begin{equation}
    (R_2[\phi])_*A[\phi] = (R_2[\phi]\circ(R_1[\phi])^{-1})_* (R_1[\phi])_* A[\phi]
\end{equation}
is kinematically local. This would require the map $R_2[\phi]\circ(R_1[\phi])^{-1}$ to be kinematically local, which is not guaranteed. On the other hand, if $A$ is relationally local to $R_1$, then the pushforward of $A$ through the change of frames map will be relationally local to $R_2$, since
\begin{equation}
    (R_2[\phi])_*(R_{1\to 2}[\phi])_*A[\phi] = (R_2[\phi])_*\qty((R_2[\phi])^{-1}\circ R_1[\phi])_* A[\phi] = (R_1[\phi])_* A[\phi]
\end{equation}
is kinematically local. In this way, the change of frames map `glues together' the relational localities associated with the different frames.

It should be noted that a change of frames does not correspond to a change in the physical state. It is merely a change in the way we describe that state. This is not to say that there are no physical questions we can ask about the change of frames. For example, an observer using one frame might ask questions about the relationship between the relational observables that they measure, and those of another observer using a different frame. Such questions are essentially answered by the change of frames map, and since the change of frames map is gauge-invariant, it is clear that such questions are physically meaningful.

\subsubsection{Frame reorientations}\label{sssec_reorient}

Let $R[\phi]$ be an injective parametrised frame whose image is $\mathcal{N}[\phi]\subset\mathcal{M}$. As previously noted, the inverse $(R[\phi])^{-1}:\mathcal{N}[\phi]\to \mathscr{O}$ may be viewed as a local orientation valued field on spacetime (or more precisely on part $\mathcal{N}[\phi]$ of spacetime), which we call the frame field. Under small diffeomorphisms $\phi\to f_*\phi$, it transforms like a scalar, $(R[\phi])^{-1}\to (R[\phi])^{-1}\circ f^{-1}$. For each value of $\phi$, the configuration of the frame field is known as its global orientation, and is an element of $\mathcal{O}$.

In some physical situations, there is a clear separation between the degrees of freedom that make up the frame field, and the degrees of freedom that make up all the `other' fields (we will see two examples of this in Sections~\ref{Section: more examples / pft} and~\ref{Section: more examples / dust}). The frame is then used to observe the other fields. Below, we shall explain various ways to define such a separation in fairly general scenarios. For the moment, however, let us assume that we have such a separation, and that the configuration of the `other' fields take values in some space $\mathcal{S}_{\text{other}}$. Given a kinematical field configuration $\phi\in\mathcal{S}$, there is an associated $\phi_{\text{other}}[\phi]\in\mathcal{S}_{\text{other}}$. This can be understood as an observable
\begin{equation}
    \mathcal{S}\to\mathcal{S}_{\text{other}},\qquad \phi\mapsto\phi_{\text{other}}[\phi].
\end{equation}
Together, the frame field and other degrees of freedom yield a map 
\begin{equation}
    \mathcal{S}\to \mathcal{O}\times\mathcal{S}_{\text{other}},\qquad \phi\mapsto \big((R[\phi])^{-1},\phi_{\text{other}}[\phi]\big).
    \label{Equation: splitting map}
\end{equation}
If we have chosen $\phi_{\text{other}}$ so that this map is injective, then together the configurations of the frame field and other fields suffice to fully determine the state in $\mathcal{S}$. We will assume this is the case. On the other hand, in general, the frame and other fields cannot be independently specified, so~\eqref{Equation: splitting map} is not surjective.

The existence of $\phi_{\text{other}}$ singles out certain transformations $\phi\to\tilde{\phi}$ of the kinematical field configuration as special. Indeed, suppose that such a transformation obeys
\begin{equation}
    \phi_{\text{other}}[\phi] = \phi_{\text{other}}[\tilde\phi],
    \label{Equation: other field invariance}
\end{equation}
i.e.\ it leaves the `other' fields invariant. On the other hand, unless it is trivial, it will modify the frame field:
\begin{equation}
    (R[\phi])^{-1} \ne (R[\tilde\phi])^{-1}
\end{equation}
Indeed, if it didn't modify the frame field, then the injectivity of~\eqref{Equation: splitting map} would guarantee that $\phi=\tilde\phi$. We call transformations of this kind `frame reorientations'.

Suppose we have some map $\mathscr{F}$ on the space of solutions such that $\phi\to\mathscr{F}[\phi]$ is a frame reorientation. The transformation of the frame field can be understood in terms of a field-dependent reparametrisation of the frame
\begin{equation}
    R[\phi] \to R[\phi]\circ F[\phi], \text{ where } F[\phi] = (R[\phi])^{-1}\circ R[\mathscr{F}[\phi]].
\end{equation}
As discussed previously, if we want the right-hand side to be of the form $\pi[\phi]\circ \tilde{R}$ for some other parametrised frame $\tilde{R}$, then $F[\phi]$ should be gauge-invariant, i.e.
\begin{equation}
    (R[\phi])^{-1}\circ R[\mathscr{F}[\phi]] = (R[f_*\phi])^{-1}\circ R[\mathscr{F}[f_*\phi]] = (R[\phi])^{-1} \circ f^{-1} \circ R[\mathscr{F}[f_*\phi]] \text{ for all } f\in H[\phi].
\end{equation}
One way to guarantee this is to only consider gauge-covariant maps $\mathscr{F}$, i.e.\ those which obey
\begin{equation}
    \mathscr{F}[f_*\phi] = f_*\mathscr{F}[\phi] \text{ for all } f\in H[\phi].
\end{equation}

It is important to point out at this stage that there is a difference between two concepts that we have described: \emph{changes of frame} and \emph{frame reorientations}. Despite the fact that a frame reorientation is, in a sense, a `change' of the frame configuration, it is not a \emph{change of frame} in the sense that we have described. Indeed, a change of frames is just a different choice of dynamical degrees of freedom that we use to describe gauge-invariant physics. A change of frames should not be viewed as having any physical significance; it amounts to a change of description of the physical situation, but not a change of physical situation. On the other hand, a frame reorientation is a certain dynamical transformation that modifies the configuration of a fixed frame. We can view it as a physical symmetry, with genuine physical consequences. Reorientations as symmetries have also been discussed in the context of edge modes in gauge theories~\cite{Carrozza:2020bnj} and gravity~\cite{CEH}, and in the study of quantum reference frames~\cite{delaHamette:2021oex}.

To be precise, it causes a change in the values of the physical observables that depend on the frame, such as the relational observables. In particular, under a frame reorientation $\phi\to\tilde\phi$ we have
\begin{equation}
    O_{A,R}[\phi] = (R[\phi])^* A[\phi] \to (R[\tilde\phi])^* A[\phi] = \big((R[\tilde\phi])^{-1}\circ R[\phi]\big)_* O_{A,R}[\phi].
\end{equation}
Here we are assuming that $A$ doesn't depend on the frame, i.e.\ it only depends on $\phi$ through the other fields $\phi_{\text{other}}[\phi]$, so that $A[\tilde\phi] = A[\phi]$. Thus, under a frame reorientation, such relational observables transform according to a diffeomorphism $(R[\tilde\phi])^{-1}\circ R[\phi]$ of the local orientation space $\mathscr{O}$ of the frame.

Note that this transformation can be reinterpreted as a transformation that leaves the frame invariant but modifies $A$. Indeed,
\begin{equation}
    R[\phi]\to R[\phi], \quad A[\phi]\to \big(R[\phi]\circ (R[\tilde\phi])^{-1}\big)_*A[\phi] = A\qty[\big(R[\phi]\circ (R[\tilde\phi])^{-1}\big)_*\phi]
\end{equation}
yields the same result. This is essentially due to the relational nature of these observables.

Let us now address the issue of exactly how to carry out the split between frame field and `other fields'. To do this, we need to pick the function $\phi_{\text{other}}$. For some frames, there is an obvious choice,\footnote{Parametrised field theory provides a good example -- see Section~\ref{Section: more examples / pft}.} but in general it is not so simple. Indeed, there is no unique choice of such a function, and in the absence of some physical principles to guide the decision, it is not a priori clear which one we should choose. It is important to recognise that different choices of $\phi_{\text{other}}$ will lead to different sets of field transformations $\phi\to\tilde\phi$ which obey~\eqref{Equation: other field invariance}, and therefore different meanings of `frame reorientation'. With this in mind, our aim now is to specify some possible canonical ways in which we could choose $\phi_{\text{other}}$.

Let us invoke the covariant phase space formalism and assume that $\mathcal{S}$ comes equipped with a presymplectic form $\Omega$. This form associates gauge-invariant observables $\alpha:\mathcal{S}\to \RR$ with vector fields $V\in\mathrm{T}\mathcal{S}$ via
\begin{equation}
    \iota_V \Omega = - \delta\alpha.
\end{equation}
Since $\Omega$ is only a presymplectic form, it has degenerate directions, and the above equation only determines $V$ up to these degenerate directions. In other words, it only determines the field variation $\delta\phi$ associated with $V$ up to (possibly field-dependent) gauge transformations, a.k.a.\ small diffeomorphisms. The observable $\alpha$ is said to `generate' the class of gauge equivalent field variations $\delta\phi$. In this way, $\Omega$ yields a pairing between physical degrees of freedom which generate transformations of each other.

The condition we would like to impose is that the frame fields and the other fields are not paired by $\Omega$. To be precise, we would like to impose that a gauge-invariant observable $\alpha$ only depends on the other fields $\phi_{\text{other}}[\phi]$ if and only if the transformation $\delta\phi$ that it generates does not affect the frame, so $\delta\qty((R[\phi])^{-1})=0$. Frame reorientations could then be identified as those field variations which leave such $\alpha$ invariant. In other words, frame reorientations should be \emph{presymplectically orthogonal} to field variations that don't affect the frame.

Unfortunately, there are a number of problems with this condition. First, the variation $\delta\phi$ generated by $\alpha$ is only defined up to a gauge transformation, and under a gauge transformation the frame field transforms like a scalar. Thus, $\delta\qty((R[\phi])^{-1})=0$ cannot be true for all gauge-equivalent $\delta\phi$. To get around this, we could try weakening the equation to allow for $\delta\qty((R[\phi])^{-1}) = \lie_\xi \big((R[\phi])^{-1}\big)$ for some infinitesimal small diffeomorphism $\xi$ (which inevitably would have to depend on $\phi$ and $\delta\phi$). In other words, $\alpha$ would be allowed to generate field variations which look like gauge transformations when acting on the frame field. However, another problem now arises when considering frames whose images $\mathcal{N}[\phi]$ do not touch the boundary $\partial\mathcal{M}$.\footnote{Meaning $\mathcal{N}[\phi]\cap \mathcal{K}=\emptyset$ for some open neighbourhood $\mathcal{K}$ of $\partial\mathcal{M}$.} In this case, any field variation $\phi\to\phi+\delta\phi$ changes the frame via
\begin{equation}
    R[\phi]\to R[\phi+\delta\phi] = \big(R[\phi+\delta\phi]\circ (R[\phi])^{-1}\big) \circ R[\phi]
\end{equation}
Assuming the frame is smooth, the map $R[\phi+\delta\phi]\circ (R[\phi])^{-1}$ is a diffeomorphism from $\mathcal{N}[\phi]$ to $\mathcal{N}[\phi+\delta\phi]$. But since neither of these spacetime subsets touch the boundary $\partial\mathcal{M}$, we can arbitrarily extend $R[\phi+\delta\phi]\circ (R[\phi])^{-1}$ to a diffeomorphism $f\in\operatorname{Diff}(\mathcal{M})$ which acts trivially in an open neighbourhood of $\partial\mathcal{M}$. This is a small diffeomorphism, so under the transformation $\phi\to\phi+\delta\phi$ the frame only changes by a small diffeomorphism. But this is true for \emph{all} $\delta\phi$, so it must be true that \emph{any} gauge-invariant $\alpha$ generates this kind of transformation. Thus, every field is an `other field', so there would appear to be no frame reorientations, other than the trivial $\phi\to \phi$. But this is too restrictive -- we would like to have a condition that allows for non-trivial reorientations of frames of this kind.\footnote{As an alternative to the following, we could just be satisfied with this result and leave the definition as it is. Then $\mathcal{N}[\phi]$ would have to touch the boundary for there to be non-trivial reorientations. This could be related to gravitational edge modes.}

The na\"ive condition given above is thus clearly not good enough, essentially because the frame field $(R[\phi])^{-1}$ is not a gauge-invariant object. There are basically two ways we can avoid this problem:
\begin{itemize}
    \item \jjvk{The first way is to use the structure provided to us by the kinematical description of the theory. The kinematical field configuration typically decomposes into a set of individual `fundamental'\footnote{N.B.\ these fields are only `fundamental' from a kinematical perspective.} fields $\phi_1,\phi_2,\phi_3,\dots,\phi_N$, where $\phi_1$ might be the metric, $\phi_2$ might be a particular matter field, and so on. It may then be the case that the frame can be written as a function of only some of these fields:
            \begin{equation}
                R[\phi]= R[\phi_1,\phi_2,\dots,\phi_K], \quad K<N.
            \end{equation}
            If this is true, then we can simply choose to view the rest of the fundamental fields as the `other' fields:
            \begin{equation}
                \phi_{\text{other}}[\phi] = (\phi_{K+1},\phi_{K+2},\dots,\phi_N)
            \end{equation}

            For example, in the geodesic example described in Section~\ref{Section: geodesic example}, the fundamental fields are the metric $g$ and the scalar field $\psi$. But the reference frame may be written only in terms of the metric, so $R[\phi] = R[g]$. Thus it is natural to choose $\phi_{\text{other}}=\psi$.
            
            In this way, frame reorientations are those field transformations which change $\phi_1,\phi_2,\dots,\phi_K$, but leave $\phi_{K+1},\phi_{K+2},\dots,\phi_N$ invariant. Whem some of $\phi_1,\phi_2,\dots,\phi_K$ are coupled to some of $\phi_{K+1},\phi_{K+2},\dots,\phi_N$ by the equations of motion, a general such transformation will not remain on-shell. However, in certain cases, some frame rorientations defined in this way will be on-shell, and can thus be viewed as symmetries. We will see examples of this in Sections~\ref{Section: more examples / pft} and~\ref{Section: more examples / dust}. \pah{Another example appears in the context of finite subregions: understanding edge modes as immaterial non-local frames originating in the complement of the subregion (e.g.\ as geodesic dressing frames connecting the finite with a possibly distant boundary) renders them into degrees of freedom on the boundary that are new from the subregion perspective, but not from the global one. Since they do not couple to the subregion fields in the action, they can be varied independently (e.g., by varying the field configuration in the causal complement of the subregion). One can argue that such reorientations of the edge mode frame constitute symmetries of the solution space quite generically \cite{CEH}.}

        }
    \item An alternative possibility is to stay within the physical space of solutions $\mathcal{S}$, but to modify what we mean by `leaving the frame invariant'. In particular, we want to replace the gauge-dependent frame field $(R[\phi])^{-1}$ by a different observable that measures its gauge-invariant properties. Thus, we would get rid of the problems discussed above.

        Within the relational framework, a very natural way to do this is to introduce a second frame $\tilde{R}[\phi]$. The change of frames map from $R[\phi]$ to $\tilde{R}[\phi]$:
        \begin{equation}
            (\tilde{R}[\phi])^{-1}\circ R[\phi]
        \end{equation}
        is a gauge-invariant observable that measures the first frame relative to the second. We can substitute this observable for the frame field in the above. To be explicit, we can impose that a gauge-invariant observable $\alpha$ only depends on the other fields if and only if the transformation that it generates obeys $\delta\qty((\tilde{R}[\phi])^{-1}\circ R[\phi])=0$. Frame reorientations are field variations that leave such $\alpha$ invariant.

        In the interest of precision, we could call these reorientations of $R[\phi]$ relative to $\tilde{R}[\phi]$. It should be clear that reorientations of $R[\phi]$ relative to $\tilde{R}[\phi]$ are also reorientations of $\tilde{R}[\phi]$ relative to $R[\phi]$. Such reorientations are presymplectically orthogonal to field variations that leave the change of frames map invariant.

        This definition of frame reorientation does not suffer from the ambiguities of the approach involving the kinematical phase space $\mathcal{P}_{\text{kin.}}$. On the other hand, one might be troubled by the fact that the set of field variations which are reorientations are not an intrinsic property of any one frame. Rather, they depend on a choice of \emph{two} frames. In a relational world, however, this seems to be quite a natural property.
\end{itemize}

\subsubsection{Relational atlases}\label{sssec_relatlas}

It would be great if we were able to use frames that are both injective and surjective. Without injectivity, we cannot construct many relational observables, because the pullback $(R[\phi])^*$ does not exist. Without surjectivity, we cannot use the frame to reach all of spacetime $\mathcal{M}$, and thus cannot use it to access all the local degrees of freedom encoded in the kinematical field configuration. In principle, frames with both of these properties exist, but in practice it is often difficult to construct them. In this section, we'll address one way to get around this issue.

We still want to be able to construct relational observables, so let us keep the requirement of injectivity. On the other hand, let us drop the requirement of surjectivity. This means we can't use the frame to construct relational observables over all of spacetime.

But this is only really a problem if we insist on using a single frame. An alternative is to use a collection of injective frames with the property that their images cover spacetime. We call such a collection a `relational atlas'.

More precisely, a relational atlas is a set $\mathscr{A}$ of injective reference frames obeying
\begin{equation}
    \bigcup_{\mathclap{\mathscr{R}\in\mathscr{A}}} \pi[\phi](\mathscr{R}) = \mathcal{M}.
\end{equation}
The name `relational \emph{atlas}' is very much inspired by the differential atlases of manifolds. Indeed, the frames that make up a relational atlas are analogous to the charts that make up a differential atlas. Similarly, the transition functions between the charts of a differential atlas are analogous to the change of frame maps between the frames of a relational atlas.

Note that we can view a relational atlas as a single frame which is just the union of all the constituent frames:
\begin{equation}
    \mathscr{R}_{\mathscr{A}} = \bigcup_{\mathclap{\mathscr{R}\in\mathscr{A}}}\mathscr{R}.
\end{equation}
This frame is surjective by definition. However, in general it is not injective, so we can't use it to construct relational observables. Thus, the relational atlas carries some extra information, namely how to decompose $\mathscr{R}_{\mathscr{A}}$ into a collection of injective frames, and thus construct relational observables.

A \emph{parametrised} relational atlas is a relational atlas equipped with a parametrisation for each frame. If we wanted to even more closely align with the structure of a differential atlas, we could require the local orientation space for each frame to be an open subset of $\RR^D$. However, we will keep the definition more general than this, allowing the local orientation spaces to be arbitrary.

Given a covariant observable $B$, and a parametrised relational atlas made up of parametrised injective frames $R_i:\mathscr{O}_i\to \mathscr{D}$, we may construct the relational observable in each frame
\begin{equation}
    O_{B,R_i}[\phi] = (R_i[\phi])^{*}B[\phi].
\end{equation}
These relational observables are automatically consistent with the change of frames between the frames making up the atlas:
\begin{equation}
    O_{B,R_i}[\phi] = (R_{j\to i}[\phi])_*O_{B,R_j}[\phi].
\end{equation}
We will sometimes refer to the entire collection of maps $O_{B,R_i}$ as the relational observable of $B$ relative to the relational atlas $\mathscr{A}$.

More generally, suppose $A_i$ is a collection of local observables, one on each of the local orientation spaces $\mathscr{O}_i$ of the parametrised atlas. We can view this set of local observables as a single consistent object if they are consistent with the change of frames maps, i.e.
\begin{equation}
    A_i[\phi] = (R_{j\to i}[\phi])_*A_j[\phi].
    \label{Equation: atlas observable gluing}
\end{equation}
We will call such a collection of local observables relationally local with respect to the atlas if $A_i$ is relationally local with respect to $R_i$. From the discussion on relational locality in the change of frames section, it is clear that this is consistent with~\eqref{Equation: atlas observable gluing}.

Another practical point is that it may be difficult to construct a set of frames that is a relational atlas for all possible kinematical field configurations $\phi\in\mathcal{S}$. One way to get around this is to use different constructions for different sets of field configurations. That is, we can cover $\mathcal{S}$ by a collection of sets of field configurations $\mathcal{U}_I\subset\mathcal{S}$, and use a different relational atlas $\mathscr{A}_I$ when the kinematical field configuration is in each different subset, $\phi\in\mathcal{U}_I$. When $\phi\in\mathcal{U}_I\cap\mathcal{U}_J$, we can use either $\mathscr{A}_I$ or $\mathscr{A}_J$, and the change of frame maps give a relationship between these atlases. Thus, we get a kind of field-space `meta-atlas'.

    \subsubsection{Dynamical frame covariance: a gauge-invariant update of general covariance}
\label{sssec_dyncov}

\pah{
The relational picture we have advocated in this section establishes a (classical) dynamical frame covariance for generally covariant theories. In particular, our dynamical frame change maps and relational atlases give rise to a gauge-invariant, yet frame-dependent description of spacetime physics. While clearly rooted in the usual notion of general covariance (cf.\ Section~\ref{Section: Introduction / conventions}), let us clarify in which sense it can be considered a more physical, truly relational update of it.}

The standard \pah{(coordinate)} version of general covariance employs \emph{fixed} reference frames \pah{in the form of} systems of coordinates. These reference frames are field-independent, and so in this sense they are obviously gauge-invariant \pah{under active diffeomorphisms}. However, the observables constructed \pah{relative to} them are \emph{not} gauge-invariant. \pah{Indeed, non-dynamical coordinates correspond to a passive choice of gauge, which is why non-perturbative gauge-invariant observables do not depend on them.\footnote{Perturbative diffeomorphism-invariant observables may depend on background coordinates, see Section~\ref{ssec_DGcomparison} and~\cite{Donnelly:2015hta,Donnelly:2016rvo,Giddings:2018umg}.} This can be seen from the fact that they are objects on some local orientation space $\mathscr{O}$, rather than spacetime; it is especially evident in their single-integral representation, discussed in Section~\ref{Section: general formalism / relational picture / single integrals smearing}, which clearly is independent of the choice of non-dynamical bulk coordinates. \emph{A gauge-invariant description is thus independent of non-dynamical frames} and thus trivially invariant under changes of them. Nevertheless, the colloquial statement `all the laws of physics are the same in every frame' is usually not interpreted in terms of gauge-invariant observables, but in terms of coordinate descriptions of tensors. They capture a notion of invariance and covariance at the same time: the overall tensorial form of the laws is unaffected by a change of non-dynamical frame, but not their explicit coordinate expression. This is the usual manner in which one encodes that different frames yield different accounts of the same physics. But it is \emph{a frame-dependent  description of physics that is not gauge-invariant}.}  This is what we mean when we say that the standard notion of general covariance is not gauge-invariant.

    The relational picture we are proposing involves dynamical reference frames, i.e.\ \pah{ones that} depend on the field configuration. In this way, they address the problem noted by Einstein in the quote in the introduction -- namely that the clocks and rods that make up a fixed reference frame are not part of the set of physical degrees of freedom, but gravitational physics requires them to be. Moreover, the dynamical reference frames we have defined are required to be gauge-covariant. Although this means the frames themselves are not gauge-\emph{invariant}, the relational observables constructed from them \emph{are} gauge-invariant \pah{and describe the remaining degrees of freedom relative to them. In particular, different choices of dynamical frame lead to different gauge-invariant descriptions of the same physics; dressing a scalar field with a geodesic frame or another matter frame results in two \emph{distinct} relational observables. Through the universal dressing space $\mathscr{D}$ we encompass all possible choices of gauge-covariant dynamical frames and through the frame change maps and the relational atlases we have all the tools to transform from any one (injective) dynamical frame perspective into any other. In particular, the frame change maps are relational observables themselves and thus both gauge-invariant and dynamical. This yields \emph{a gauge-invariant description of physics that depends on the choice of dynamical frame}, but is independent of non-dynamical coordinate frames. 
    
    It permits us to revisit the statement `all the laws of physics are the same in every frame' and to equip it with a new meaning. Using (sufficiently differentiable) frame fields, we can push forward all the usual tensorial objects making up the laws of physics to the respective local frame orientation spaces $\mathscr{O}_i$ in order to rewrite them (and thereby the laws themselves) as gauge-invariant tensors on $\mathscr{O}_i$, i.e.\ as relational observables. This still captures an invariance and covariance at the same time, but now in a gauge-invariant manner: the overall tensorial form of the laws is unaffected by a change of \emph{dynamical} frame, but not their explicit relational observable expression. This dynamical frame covariance is a gauge-invariant and dynamical update of general covariance and can thereby be viewed as an arguably more physical version of it. It fully captures that physics is relational: a physical account of the world amounts to describing systems relative to one another, rather than to fictitious or external reference structures, and to interlink the many different and non-privileged ways in doing so.
    
  Our construction extends the dynamical frame covariance for gauge field theories~\cite{Carrozza:2020bnj} to the generally covariant setting, and prepares the ground for ultimately expanding  the notion of quantum frame covariance~\cite{Giacomini:2017zju,Hoehn:2019fsy,Hoehn:2021flk,Castro-Ruiz:2021vnq,delaHamette:2020dyi,Giacomini:2018gxh,Vanrietvelde:2018pgb,Vanrietvelde:2018dit,Hoehn:2020epv,Hohn:2018iwn,Castro-Ruiz:2019nnl,Ballesteros:2020lgl,Krumm:2020fws, delaHamette:2021iwx}, see especially~\cite{delaHamette:2021oex, Hoehn:2021wet} in this context, to the field theory setting. Such dynamical frames are always associated with a gauge group and usually taken to transform covariantly under it; here, we extend this to the infinite-dimensional diffeomorphism group (see also~\cite{CEH}).

    }

%    In both cases, the change of frames maps are gauge-invariant. But for dynamical reference frames, this is only guaranteed by the gauge-covariance of the frames.

\subsection{Properties of relational degrees of freedom}
\label{Section: general formalism / relational phase space}

Having defined relational observables in the previous subsection, in this subsection we will explore some physical properties of the relational degrees of freedom that they represent. 

\subsubsection{Relational field variations}

Recall that, for any field variation $\delta\phi$, we have defined
\begin{equation}
    \delta_R\phi = \delta\phi - \lie_{V_R[\phi,\delta\phi]}\phi,
    \label{Equation: delta_R definition}
\end{equation}
where $V_R[\phi,\delta\phi]$ is a vector field in $\mathcal{N}[\phi]$ given by the linear order in $\delta\phi$ expansion
\begin{equation}
    R[\phi+\delta\phi]\circ(R[\phi])^{-1} = \operatorname{Id}_{\mathcal{N}[\phi]} + V_R[\phi,\delta\phi].
    \label{Equation: V_R definition}
\end{equation}
We call $\delta_R\phi$ a `relational field variation'.

Note that $\delta_R\phi$ might not be on-shell even if $\delta\phi$ is. Put another way, even if we require $\phi\to\phi+\delta\phi$ to be a transformation that stays within the space of solutions $\mathcal{S}$, the transformation $\phi\to\phi+\delta_R\phi$, may not necessarily do so. This is because the diffeomorphism corresponding to $V_R[\phi,\delta\phi]$ may not preserve the boundary conditions of the theory.

If $\delta\phi$ is an infinitesimal small diffeomorphism, then $\delta_R\phi$ vanishes. To see this, let $\phi+\delta\phi = f_*\phi$ for some small diffeomorphism $f\in H[\phi]$. Assuming $f$ is close to the identity, we can assume that it is generated by some spacetime vector field $\xi$, so at linear order we may write $\delta\phi=\lie_\xi\phi$. Then by the gauge-covariance of the frame we have
\begin{equation}
    R[\phi+\delta\phi]\circ(R[\phi])^{-1} = f\circ R[\phi] \circ (R[\phi])^{-1} = f,
\end{equation}
so comparing with~\eqref{Equation: V_R definition}, we see that
\begin{equation}
    V_R[\phi,\lie_\xi\phi] = \xi.
    \label{Equation: V_R small}
\end{equation}
Thus, from~\eqref{Equation: delta_R definition} we have
\begin{equation}\label{eq:phirvanish}
    \delta_R\phi = \lie_\xi\phi - \lie_\xi\phi = 0,
\end{equation}
as claimed. In this way, we can view $\delta_R\phi$ as a kind of gauge-fixed field variation, with the gauge-fixing determined by the frame.

Suppose we have a parametrised injective frame $R:\mathscr{O}\to\mathscr{D}$. The `relational field configuration' $\phi_R[\phi]$ is the pullback of the kinematical field configuration through this frame:
\begin{equation}
    \phi_R[\phi] = (R[\phi])^*\phi.
\end{equation}
Since $\phi$ is a covariant observable, $\phi_R[\phi]$ is a relational observable. It is a section of some bundle $\Phi_{\mathscr{O}}$ over the local orientation space $\mathscr{O}$, which we call the `relational field bundle'. Only certain sections of $\Phi_{\mathscr{O}}$ arise from kinematical field configurations in the space of solutions $\mathcal{S}$. Those that do are elements of
\begin{equation}
    \mathcal{S}_R = \phi_R[\mathcal{S}]\subset\Gamma(\Phi_{\mathscr{O}}),
\end{equation}
i.e.\ the image of $\phi_R$. Note that $\Phi_{\mathscr{O}}$ doesn't depend on the frame, only on the space of local orientations to be used. On the other hand $\mathcal{S}_R$ depends explicitly on the frame used.

The relational field configuration $\phi_R$ and the relational field variation $\delta_R\phi$ are connected by the frame via the formula $\delta_R\phi = (R[\phi])_*\delta(\phi_R[\phi])$. Indeed, we have
\begin{nalign}
    (R[\phi])_*\delta(\phi_R[\phi]) &= (R[\phi])_*\big(\phi_R[\phi+\delta\phi]-\phi_R[\phi]\big) \\
                                    &= (R[\phi])_*\Big((R[\phi+\delta\phi])^*(\phi+\delta\phi) - (R[\phi])^*\phi\Big) \\
                                    &= \big((R[\phi+\delta\phi]\circ (R[\phi])^{-1}\big)^*(\phi+\delta\phi) - \phi \\
                                    &= \phi+\delta\phi - \lie_{V_R[\phi,\delta\phi]}(\phi+\delta\phi) - \phi \\
                                    &= \delta\phi - \lie_{V_R[\phi,\delta\phi]}\phi = \delta_R\phi,
                                    \label{Equation: delta phi_R vs delta_R phi}
\end{nalign}
where in the fourth line we used~\eqref{Equation: V_R definition}, and everything is to linear order in $\delta\phi$. This is a generalisation of a formula appearing previously in the edge mode literature~\cite{Donnelly:2016auv,Speranza:2017gxd,Speranza:2022lxr,Freidel:2020xyx,Freidel:2021dxw,CEH} that describes the commutation between the field variation $\delta$ and the frame dressing $(R[\phi])_*$.

It is worth commenting that~\eqref{Equation: V_R small} means that $V_R$ has some of the structure of a connection on the bundle $\mathcal{S}\to\mathcal{S}_R$, and that the relational field variation $\delta_R$ may therefore be viewed as a kind of field space covariant derivative. However, this is complicated by the fact that $\delta_R\phi$ can go off-shell. This may be connected to the proposals in~\cite{Gomes:2016mwl}. \jjvk{It should also be noted that $V_R$ is a generalisation of the Maurer-Cartan form on field space that appears for example in~\cite{Donnelly:2016auv,Speranza:2017gxd,Speranza:2022lxr,Freidel:2021dxw,CEH}.}\footnote{\jjvk{In those contexts, one has a field-dependent diffeomorphism $U$, and one defines the Maurer-Cartan form as a spacetime-vector-field-valued field space 1-form $\chi= U^{-1}\circ\delta U$, where $\delta$ is an exterior derivative on field space. In the present paper, $U$ may be identified with the frame map $R[\phi]$. If one contracts into $\chi$ the field space vector $\xi$ corresponding to a certain field variation $\phi\to\phi+\delta\phi$, one finds $I_{\xi}\chi = -V_{R[\phi,\delta\phi]}$, where $I$ denotes the field space interior product. The Maurer-Cartan form is a connection and so allows one to define a covariant field space exterior derivative $\delta+\chi$, which in the present context could sensibly be denoted $\delta_R$.}} We will leave a more full exploration of this to future work.

\subsubsection{The presymplectic form}

In the image $\mathcal{N}[\phi]$ of the frame, the field configuration $\phi$ can be determined in terms of $(R[\phi])^{-1}$ and $\phi_R[\phi]$ via $\phi|_{\mathcal{N}[\phi]} = (R[\phi])_*\phi_R[\phi]$. Thus, instead of treating general observables as functions of $\phi$, we can treat them as functions of the relational field configuration (an element of $\mathcal{S}_R$), the global orientation of the frame (an element of $\mathcal{O}$), and $\phi|_{\mathcal{M}\setminus\mathcal{N}[\phi]}$ (i.e.\ the kinematical field configuration outside the image of the frame).

Our objective in this section is to write the presymplectic form $\Omega$ on the space of solutions $\mathcal{S}$ in terms of these variables. \pah{The following is not strictly necessary for the subsequent discussion and may be skipped on a first reading.} Our starting point will be the covariant phase space formalism~\cite{Iyer:1994ys,Lee:1990nz,Harlow:2019yfa}, which (assuming the theory has a local Lagrangian formulation) yields a natural presymplectic form:
\begin{equation}
    \Omega[\phi,\delta_1\phi,\delta_2\phi] = \int_\Sigma \omega[\phi,\delta_1\phi,\delta_2\phi].
\end{equation}
Here, $\Sigma$ is any Cauchy surface, $\delta_1\phi,\delta_2\phi\in\mathrm{T}\mathcal{S}$ are vectors on the space of solutions, a.k.a.\ on-shell field variations, and $\omega$ is a spacetime $(D-1)$-form called the `presymplectic current' which is locally and covariantly constructed from $\phi,\delta_1\phi,\delta_2\phi$ and their derivatives and is linear and antisymmetric in $\delta_1\phi,\delta_2\phi$. $\Omega$ is closed, and has the property that it is independent of the choice of Cauchy surface $\Sigma$. Moreover, its degenerate directions correspond exactly with the gauge transformations of the theory, which for us are the small diffeomorphisms. Historically, there have been some ambiguities in the boundary contributions to $\Omega$ at $\partial\Sigma$, but these ambiguities can be resolved with a careful treatment of their relationship with the boundary conditions that define the theory~\cite{Harlow:2019yfa,Carrozza:2020bnj,CEH,Chandrasekaran:2020wwn,Speranza:2022lxr}.

Let us assume that the frame is smooth enough that its image $\mathcal{N}[\phi]$ is a submanifold of $\mathcal{M}$, and let us define $\Sigma[\phi] = \Sigma\cap\mathcal{N}[\phi]$ and $\overline{\Sigma}[\phi] = \Sigma\setminus\Sigma[\phi]$, so that we can write
\begin{equation}
    \Omega[\phi,\delta_1\phi,\delta_2\phi] = \int_{\Sigma[\phi]} \omega[\phi,\delta_1\phi,\delta_2\phi] + \int_{\overline{\Sigma}[\phi]} \omega[\phi,\delta_1\phi,\delta_2\phi].
    \label{Equation: presymplectic image split}
\end{equation}
Inside $\Sigma[\phi]$, we can write $\phi$ in terms of $\phi_R[\phi]$ and $R[\phi]$ via $\phi = R[\phi]_*\phi_R[\phi]$. Using~\eqref{Equation: delta phi_R vs delta_R phi}, we can also write field variations in terms of these objects via
\begin{equation}
    \delta\phi = (R[\phi])_*\delta(\phi_R[\phi]) + \lie_{V_R[\phi,\delta\phi]} \phi.
\end{equation}
One finds
\begin{nalign}
    \int_{\Sigma[\phi]}\omega[\phi,\delta_1\phi,\delta_2\phi]
    &= \int_{\Sigma[\phi]}\omega[\phi,(R[\phi])_*\delta_1(\phi_R[\phi]),(R[\phi])_*\delta_2(\phi_R[\phi])] \\
    &+ \int_{\Sigma[\phi]}\omega[\phi, \lie_{V_R[\phi,\delta_1\phi]}\phi,\delta_2\phi] + \int_{\Sigma[\phi]}\omega[\phi, \delta_1\phi, \lie_{V_R[\phi,\delta_2\phi]}\phi] \\
    & - \int_{\Sigma[\phi]}\omega[\phi, \lie_{V_R[\phi,\delta_1\phi]}\phi, \lie_{V_R[\phi,\delta_2\phi]}\phi]
    \label{Equation: presymplectic four terms}
\end{nalign}
We can use the fact that $\omega$ is covariantly constructed from $\phi,\delta_1\phi,\delta_2\phi$ to rewrite the first term on the right-hand side as 
\begin{nalign}
    \MoveEqLeft \int_{\Sigma[\phi]}\omega[\phi,(R[\phi])_*\delta_1(\phi_R[\phi]),(R[\phi])_*\delta_2(\phi_R[\phi])] \\
    &= \int_{\Sigma[\phi]}\omega[(R[\phi])_*\phi_R[\phi],(R[\phi])_*\delta_1(\phi_R[\phi]),(R[\phi])_*\delta_2(\phi_R[\phi])] \\
    &= \int_{\Sigma[\phi]}(R[\phi])_* \Big(\omega_R[\phi_R[\phi],\delta_1(\phi_R[\phi]),\delta_2(\phi_R[\phi])]\Big) \\
    &= \int_{(R[\phi])^{-1}(\Sigma[\phi])}\omega_R[\phi_R[\phi],\delta_1(\phi_R[\phi]),\delta_2(\phi_R[\phi])],
\end{nalign}
where $\omega_R$ is a $(D-1)$-form on $\mathscr{O}$. For the latter three terms, let us use the covariant phase space result\footnote{This formula~\eqref{Equation: omega noether}, and the later formula~\eqref{Equation: omega noether 2}, are standard results in the covariant phase space formalism. For the convenience of the reader we have given a quick derivation of them in Appendix~\ref{Appendix: noether}.}
\begin{equation}
    \omega[\phi,\delta\phi,\lie_V\phi] = \dd{\big(\delta (Q_V[\phi]) - Q_{\delta V}[\phi] - \iota_V \theta[\phi,\delta\phi]\big)}
    \label{Equation: omega noether}
\end{equation}
for a certain form $\theta$ known as the `presymplectic potential density', and another form $Q_V$, known as the `Noether charge', that depends linearly on $V$. This holds for any $V$, and the two are related by
\begin{equation}
    \theta[\phi,\lie_V\phi] = \iota_VL[\phi] + \dd(Q_V[\phi]),
\end{equation}
where $L[\phi]$ is the Lagrangian density. $\theta$ itself is related to $\omega$ by
\begin{equation}
    \omega[\phi,\delta_1\phi,\delta_2\phi] = \delta_1\big(\theta[\phi,\delta_2\phi]\big) - \delta_2\big(\theta[\phi,\delta_1\phi]\big) - \theta[\phi,\delta_{12}\phi],
\end{equation}
where $\delta_{12}\phi$ is the Lie bracket of $\delta_1\phi$ and $\delta_2\phi$ viewed as vector fields on $\mathcal{S}$ (in other words $\omega$ is the field space exterior derivative of $\theta$).
We also have
\begin{equation}
    \omega[\phi,\lie_V\phi,\lie_W\phi] = \dd{\big(\iota_V\dd(Q_W[\phi]) - \iota_W\dd(Q_V[\phi]) - Q_{[V,W]}[\phi] + \iota_V\iota_W L[\phi]\big)}.
    \label{Equation: omega noether 2}
\end{equation}
Thus, we find that the latter three terms in~\eqref{Equation: presymplectic four terms} reduce to an integral over $\partial(\Sigma[\phi])$. In total, the symplectic form may be written as
\begin{multline}
    \Omega[\phi,\delta_1\phi,\delta_2\phi] = \int_{\Sigma_R[\phi]} \omega_R[\phi_R,\delta_1\phi_R,\delta_2\phi_R] + \int_{\overline{\Sigma}[\phi]} \omega[\phi,\delta_1\phi,\delta_2\phi] \\
    + \int_{\partial(\Sigma[\phi])} \Big(\delta_1(Q_{V_2}[\phi]) - \delta_2 (Q_{V_1}[\phi]) - Q_{\delta_1 V_2 - \delta_2 V_1 + [V_1,V_2]}[\phi] \\
    + \iota_{V_1}\big(\theta[\phi,\delta_2\phi]-\dd(Q_{V_2}[\phi])\big) - \iota_{V_2}\big(\theta[\phi,\delta_1\phi]-\dd(Q_{V_1}[\phi])\big) - \iota_{V_1}\iota_{V_2}L[\phi]\Big),
    \label{Equation: presymplectic relational variables}
\end{multline}
where $\Sigma_R[\phi] = (R[\phi])^{-1}(\Sigma[\phi])$, $V_1 = V_R[\phi,\delta_1\phi]$, $V_2 = V_R[\phi,\delta_2\phi]$, and for notational simplicity we are omitting the dependence of $\phi_R$ and $V_R$ on the kinematical fields $\phi$. \jjvk{This formula is an alternative derivation and slight generalisation of the one found in~\cite{Speranza:2017gxd}, in that here the frame need not be surjective over all of $\Sigma$.}

The first term in~\eqref{Equation: presymplectic relational variables} governs the relational fields $\phi_R$. The second term governs the degrees of freedom in $\overline{\Sigma}[\phi]$, i.e.\ those which cannot be accessed by the frame in a relationally local way. The third term (i.e.\ the integral over $\partial(\Sigma[\phi])$) governs the way in which the frame field $R[\phi]$ is paired with the rest of the degrees of freedom in the theory. Indeed, if $\delta_1(R[\phi])=0$ or $\delta_2(R[\phi])=0$, then $V_1=0$ or $V_2=0$, and so the third term would vanish.

Note that, when the symplectic form is written in this way, the only contribution of the frame is at $\partial(\Sigma[\phi])$ through $V_R$. Thus, the frame may be viewed as an edge mode. For more discussion on this, see~\cite{Donnelly:2016auv,Speranza:2017gxd,CEH,Freidel:2020xyx,Freidel:2021bmc}. 

\subsubsection{Relational microcausality}
\label{Section: general formalism / relational phase space / microcausality}
Consider the Poisson bracket of gauge-invariant observables $A,B$. This can be constructed in the usual way using the covariant phase space presymplectic form $\Omega$. Generally, $A,B$ generate field transformations $\delta_A\phi,\delta_B\phi$ through the presymplectic structure:
\begin{equation}
    \delta(A[\phi]) = \Omega[\phi,\delta\phi,\delta_A\phi[\phi]],\qquad
    \delta(B[\phi]) = \Omega[\phi,\delta\phi,\delta_B\phi[\phi]].
    \label{Equation: A B generate}
\end{equation}
The Poisson bracket of $A$ and $B$ is then the observable $\pb{A}{B}$ defined by
\begin{equation}
    \pb{A}{B}[\phi] = \Omega[\phi,\delta_A\phi[\phi],\delta_B\phi[\phi]].
\end{equation}
Note that~\eqref{Equation: A B generate} only defines $\delta_A\phi[\phi],\delta_B\phi[\phi]$ up to gauge transformations, but $\pb{A}{B}$ is uniquely defined.

We could try to substitute into this the expression~\eqref{Equation: presymplectic relational variables} for the presymplectic form obtained in the previous section. However, it turns out that the Poisson bracket just defined is equivalent to another object known as the Peierls bracket (provided we use the $\Omega$ given to us by the covariant phase space formalism, with a proper treatment of boundary conditions)~\cite{Kirklin:2019xug,Harlow:2019yfa}. The Peierls bracket gives much more direct access to the properties of the dynamical fields, in particular providing an explicit link between the equations of motion and the Poisson bracket. In this section, we will exploit this link to prove that relationally local observables obey microcausality, i.e.\ if two such observables are spacelike separated then their bracket vanishes. \jjvk{Importantly, this result only holds if both observables are relationally local to regions in the \emph{bulk} of spacetime, i.e.\ away from $\partial\mathcal{M}$. Our \pah{argument is} similar to, but significantly generalise\pah{s} and expand\pah{s} on, \pah{the one briefly} described in~\cite{Marolf:2015jha}.} \pah{We comment further on this below.}

The Peierls bracket of two observables $A$ and $B$ is defined in the following way. One first deforms the action
\begin{equation}
    S_\lambda [\phi] = S[\phi] - \lambda A[\phi],
    \label{Equation: Peierls deformed action}
\end{equation}
where $S[\phi]$ is the ordinary action of the theory, and $\lambda$ is some small parameter. This deformed action accordingly has deformed equations of motion, to which we now obtain two solutions $\phi_r,\phi_a$ which at small $\lambda$ we can write as:
\begin{align}
    \phi_r = \phi_0 + \lambda\, \delta\phi_r + \order{\lambda^2}, \qquad \phi_a=\phi_0 + \lambda\, \delta\phi_a + \order{\lambda^2}.
\end{align}
Here $\phi_0$ is a fixed solution of the original equations of motion of the undeformed action, and $\delta\phi_r,\delta\phi_a$ are linearised variations away from $\phi_0$. The `retarded' solution $\phi_r$ should be such that $\delta\phi_r$ vanishes on a Cauchy surface sufficiently far in the past, while the `advanced' solution $\phi_a$ should be such that $\delta\phi_a$ vanishes on a Cauchy surface sufficiently far in the future. Note that
\begin{equation}
    \phi_\lambda = \phi_0+\lambda(\delta\phi_r-\delta\phi_a)
\end{equation}
obeys the original undeformed equations of motion to linear order in $\lambda$.\footnote{
    To see this, we can expand order by order in $\lambda$. Writing $\delta_r(\dots),\delta_a(\dots)$ as the variation of any quantity with respect to the retarded and advanced field variations $\delta\phi_r,\delta\phi_a$, we have
    \begin{equation}
        0 = E[\phi_{r,a}] - \lambda \epsilon\fdv{A}{\phi}[\phi_{r,a}] = E[\phi_0] + \lambda\qty(\delta_{r,a}E[\phi_0] - \epsilon\fdv{A}{\phi}[\phi_0]) + \order{\lambda^2},
        \label{Equation: delta r a phi eom}
    \end{equation}
    where $\epsilon$ and $\fdv{A}{\phi}$ are defined in~\eqref{Equation: functional derivative}.
    Since $E[\phi_0]=0$, we have that~\eqref{Equation: delta r a phi eom} implies $\delta_{r,a}E[\phi_0] = \epsilon\fdv{A}{\phi}[\phi_0] + \order{\lambda}$. On the other hand,
    \begin{nalign}
        E[\phi_\lambda] = E[\phi_0 + \lambda(\delta\phi_r-\delta\phi_a)] &= \underbrace{E[\phi_0]}_{=0} + \lambda \delta_r E[\phi_0]-\lambda\delta_a E[\phi_0]+\order{\lambda^2} \\
                                                                         &= \lambda\qty(\epsilon\fdv{A}{\phi}[\phi_0]-\epsilon\fdv{A}{\phi}[\phi_0]) + \order{\lambda^2} =\order{\lambda^2},
    \end{nalign}
    as required.
} We also require that $\phi_r$ and $\phi_a$ are chosen such that $\phi_\lambda$ obeys the boundary conditions of the theory. Thus, $\delta\phi_r-\delta\phi_a$ is an on-shell field variation, and indeed \jjvk{in the limit $\lambda\to0$} we can identify it with $\delta_A\phi[\phi_0]$, where $\delta_A\phi$ was defined in~\eqref{Equation: A B generate}. Similarly, we can obtain $\delta_B\phi$ in this way by deforming the action with $B[\phi]$ instead of $A[\phi]$. The Peierls bracket is defined as
\begin{equation}
    \pb{A}{B}[\phi] = \delta_AB[\phi] = \dv{\lambda}\left.\qty(B\big[\phi+\lambda\delta_A\phi[\phi]\big])\right|_{\lambda=0},
\end{equation}
which turns out to match the definition of the Poisson bracket from the presymplectic form given above.

As a warm-up to the Peierls bracket of relationally local observables in a theory with diffeomorphism gauge symmetry, we instead first consider the Peierls bracket of kinematically local observables in a theory without diffeomorphism gauge symmetry. To be more precise, we will let $A:\mathcal{S}\to\RR$ be an observable constructed from a collection of kinematically local observables $A_i$, $i=1,2,\dots$. \jjvk{For example, $A$ could be a kinematical field smeared over some spacetime region.} Using the chain rule with~\eqref{Equation: kinematically local quantity} and~\eqref{Equation: kinematically local map}, we may write the variation of $A$ as
\begin{equation}
    \delta(A[\phi]) = \int_{\mathcal{M}} A^{(0)}[\phi]\cdot \delta\phi + A^{(1)}[\phi]\cdot \partial(\delta\phi) + A^{(2)}[\phi] \cdot \partial^2(\delta\phi) + \dots,
\end{equation}
for some observables $A^{(0)},A^{(1)},\dots$. We define the `support' of $A$ as
\begin{equation}
    \operatorname{supp}(A)[\phi] = \bigcup_{k} \,\operatorname{supp}\big(A^{(k)}[\phi]\big).
\end{equation}
Thus, it is the subset of spacetime over which $\delta(A[\phi])$ has a dependence on $\delta\phi$. We will assume that $A$ is a bulk observable, meaning that $\operatorname{supp}(A)$ doesn't touch the boundary $\partial\mathcal{M}$ (i.e.\ there is some open neighbourhood of the boundary which doesn't intersect with $\operatorname{supp}(A)$). This implies that $A^{(k)}[\phi]$ vanishes in an open neighbourhood of the boundary for all $k$, and thus we can integrate by parts to write
\begin{equation}
    \delta(A[\phi]) = \int_{\mathcal{M}}\epsilon\, \fdv{A}{\phi}[\phi]\cdot \delta\phi,
    \label{Equation: functional derivative}
\end{equation}
where $\epsilon=\sqrt{\abs{\det g}}\dd[D]{x}$ is the spacetime volume form, and the observable $\fdv{A}{\phi}$ is known as the functional derivative of $A$. The support of $A$ coincides with the support of $\fdv{A}{\phi}$, so $\fdv{A}{\phi}$ vanishes in an open neighbourhood of $\partial\mathcal{M}$. We will make similar assumptions about the second observable $B:\mathcal{S}\to\RR$, so that its variation may be written
\begin{equation}
    \delta(B[\phi]) = \int_{\mathcal{M}}\epsilon\, \fdv{B}{\phi}[\phi]\cdot \delta\phi,
\end{equation}
where its functional derivative $\fdv{B}{\phi}$ also vanishes in a neighbourhood of $\partial\mathcal{M}$.

The undeformed action may be written as the integral of the bulk Lagrangian plus a boundary term:
\begin{equation}
    S[\phi] = \int_{\mathcal{M}}L[\phi] + S_{\partial}[\phi].
\end{equation}
We may write the variation of the Lagrangian as $\delta (L[\phi]) = E[\phi]\cdot\delta\phi + \dd(\theta[\phi,\delta\phi])$. Using this, we see that the variation of the action takes the form
\begin{equation}
    \delta (S[\phi]) = \int_{\mathcal{M}} E[\phi]\cdot \delta\phi + \int_{\partial\mathcal{M}}\theta[\phi,\delta\phi] + \delta(S_\partial[\phi]).
\end{equation}
The boundary conditions are chosen such that the latter two terms cancel each other out; then the variational principle is well-defined, and we get equations of motion $E[\phi]=0$. Including now the deformation~\eqref{Equation: Peierls deformed action}, we see that the variation of the deformed action takes the form
\begin{equation}
    \delta (S_\lambda[\phi]) = \int_{\mathcal{M}} \qty(E[\phi] - \epsilon\lambda\fdv{A}{\phi}[\phi])\cdot \delta\phi,
\end{equation}
where we have already used the boundary conditions to set the boundary terms to zero. Thus, the deformed equations of motion are
\begin{equation}
    E[\phi] - \epsilon\lambda\fdv{A}{\phi}[\phi] = 0.
\end{equation}
The retarded and advanced solutions $\phi_r=\phi_0+\lambda\delta\phi_r$, $\phi_a=\phi_0+\lambda\delta\phi_a$ must both solve these deformed equations of motion.

At this point, we will make an assumption about the causal properties of the theory we are considering. Wherever the undeformed equations of motion hold, we will assume that knowledge of the field configuration $\phi|_{\mathcal{U}}$ in any subset $\mathcal{U}\subset\mathcal{M}$ of spacetime suffices to determine the field configuration $\phi|_{D(\mathcal{U})}$ in the domain of dependence $D(\mathcal{U})$ of $\mathcal{U}$,\footnote{The domain of dependence is defined by
\begin{equation}
    D(\mathcal{U}) = \{x\in\mathcal{M} \mid \text{every inextendible causal curve through $x$ intersects $\mathcal{U}$ at least once}\}.
\end{equation}
A causal curve is one whose tangent vector is nowhere spacelike.} up to gauge symmetry. This is a key property of any relativistic theory, and intuitively means that no signal can propagate faster than light. It holds even in relativistic theories with diffeomorphism invariance. In this case, the undeformed equations of motion hold everywhere outside the support of $A$. Thus, we can apply this assumption everywhere in $\mathcal{M}\setminus\operatorname{supp}(A)$.

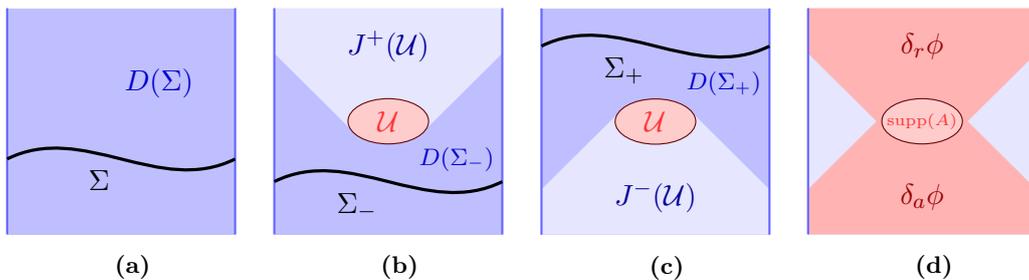
\begin{figure}
    \centering
    \begin{subfigure}{0.2\textwidth}
        \begin{tikzpicture}
            \fill[blue!25] (0,0) rectangle (3,3);
            \draw[thick,blue!60] (0,0) -- (0,3);
            \draw[thick,blue!60] (3,0) -- (3,3);
            \draw[very thick] (0,1) .. controls (1,1.5) and (2,0.5) .. (3,1) node[midway, below left] {$\Sigma$};
            \node[blue!80!black] at (2,2) {$D(\Sigma)$};
        \end{tikzpicture}
        \caption{}
    \end{subfigure}
    \begin{subfigure}{0.2\textwidth}
        \begin{tikzpicture}
            \fill[blue!25] (0,0) rectangle (3,3);
            \fill[blue!10] (1,1.4) -- (2,1.4) -- (3,2.4) -- (3,3) -- (0,3) -- (0,2.4) -- cycle;
            \draw[red!40!black,fill=red!20] (1.5,1.5) ellipse (0.53 and 0.3) node[red!80] {$\mathcal{U}$};
            \draw[thick,blue!60] (0,0) -- (0,3);
            \draw[thick,blue!60] (3,0) -- (3,3);
            \draw[very thick,shift={(0,-0.3)}] (0,1) .. controls (1,1.5) and (2,0.5) .. (3,1) node[midway, below left] {$\Sigma_-$};
            \node[blue!80!black] at (2.4,1) {\footnotesize$D(\Sigma_-)$};
            \node[blue!60!black] at (1.5,2.5) {$J^+(\mathcal{U})$};
        \end{tikzpicture}
        \caption{}
    \end{subfigure}
    \begin{subfigure}{0.2\textwidth}
        \begin{tikzpicture}
            \fill[blue!25] (0,0) rectangle (3,3);
            \fill[blue!10] (1,1.6) -- (2,1.6) -- (3,0.6) -- (3,0) -- (0,0) -- (0,0.6) -- cycle;
            \draw[red!40!black,fill=red!20] (1.5,1.5) ellipse (0.53 and 0.3) node[red!80] {$\mathcal{U}$};
            \draw[thick,blue!60] (0,0) -- (0,3);
            \draw[thick,blue!60] (3,0) -- (3,3);
            \draw[very thick,shift={(0,1.5)}] (0,1) .. controls (1,1.5) and (2,0.5) .. (3,1) node[midway, below left] {$\Sigma_+$};
            \node[blue!80!black] at (2.4,2) {\footnotesize$D(\Sigma_+)$};
            \node[blue!60!black] at (1.5,0.5) {$J^-(\mathcal{U})$};
        \end{tikzpicture}
        \caption{}
    \end{subfigure}
    \begin{subfigure}{0.2\textwidth}
        \begin{tikzpicture}
            \fill[blue!10] (0,0) rectangle (3,3);
            \fill[red!30] (1,1.4) -- (2,1.4) -- (3,2.4) -- (3,3) -- (0,3) -- (0,2.4) -- cycle;
            \fill[red!30] (1,1.6) -- (2,1.6) -- (3,0.6) -- (3,0) -- (0,0) -- (0,0.6) -- cycle;
            \draw[thick,blue!60] (0,0) -- (0,3);
            \draw[thick,blue!60] (3,0) -- (3,3);
            \draw[red!40!black,fill=red!20] (1.5,1.5) ellipse (0.53 and 0.3) node[red!80] {\tiny$\operatorname{supp}(A)$};
            \node[red!60!black] at (1.5,2.5) {$\delta_r\phi$};
            \node[red!60!black] at (1.5,0.5) {$\delta_a\phi$};
        \end{tikzpicture}
        \caption{}
    \end{subfigure}
    \caption{\mbox{\textbf{(a)}}\ The domain of dependence of a Cauchy surface is all of $\mathcal{M}$. \mbox{\textbf{(b)}}\ However, if we remove a subset $\mathcal{U}$, then the domain of dependence of a Cauchy surface $\Sigma_-$ to the past of $\mathcal{U}$ is $\mathcal{M}\setminus J^+(\mathcal{U})$. \mbox{\textbf{(c)}}\ Similarly, the domain of dependence of a Cauchy surface $\Sigma_+$ to the future of $\mathcal{U}$ is $\mathcal{M}\setminus J^-(\mathcal{U})$. \mbox{\textbf{(d)}} If $\mathcal{U}$ is the support of an observable $A$, then the retarded field variation $\delta\phi_r$ can be taken to vanish in $\mathcal{M}\setminus J^+(\mathcal{U})$, while the advanced field variation $\delta\phi_a$ can be taken to vanish in $\mathcal{M}\setminus J^-(\mathcal{U})$. Thus, $\delta_A\phi = \delta\phi_r - \delta\phi_a$ has support in $J^+(\mathcal{U})\cup J^-(\mathcal{U})$, i.e.\ within the domain of influence of the support of $A$.}
    \label{Figure: domain of dependence}
\end{figure}

Let us see what this assumption gets us. First, note that $\phi_r=\phi_0$ on a Cauchy surface $\Sigma_-$ sufficiently far in the past. By the definition of Cauchy surfaces, the domain of dependence of $\Sigma_-$ is the whole of spacetime $\mathcal{M}$. However, $\phi_r$ and $\phi_0$ obey the undeformed equations of motion only in $\mathcal{M}\setminus\operatorname{supp}(A)$, and $\Sigma_-$ is not a Cauchy surface for $\mathcal{M}\setminus\operatorname{supp}(A)$, because of the presence of causal curves that pass through the support of $A$. Thus, the domain of dependence of $\Sigma_-$ in $\mathcal{M}\setminus\operatorname{supp}(A)$ is strictly smaller than $\mathcal{M}\setminus\operatorname{supp}(A)$. Indeed (see Figure~\ref{Figure: domain of dependence}), it is given by $\mathcal{M}\setminus J^+(\operatorname{supp}(A))$, where $J^+$ denotes the causal future.\footnote{The causal future $J^+(\mathcal{U})$ of a set $\mathcal{U}$ is the subset of spacetime that can be reached by future-directed causal curves originating in $\mathcal{U}$.} Thus, by the causality assumption, $\phi_r=\phi_0$ at $\Sigma_-$ implies that $\phi_r=\phi_0$ in $\mathcal{M}\setminus J^+(\operatorname{supp}(A))$ (up to gauge symmetry). Therefore we can make a gauge choice such that $\delta\phi_r$ is non-vanishing only in $J^+(\operatorname{supp(A)})$. Similarly, $\phi_a=\phi_0$ on a Cauchy surface $\Sigma_+$ sufficiently far in the future, and the domain of dependence of this Cauchy surface in $\mathcal{M}\setminus\operatorname{supp}(A)$ is $\mathcal{M}\setminus J^-(\operatorname{supp}(A))$, where $J^-$ denotes the causal past.\footnote{The causal past $J^-(\mathcal{U})$ of a set $\mathcal{U}$ is the subset of spacetime that can be reached by past-directed causal curves originating in $\mathcal{U}$.} Therefore $\phi_a=\phi_0$ in $\mathcal{M}\setminus J^-(\operatorname{supp}(A))$ (up to gauge symmetry), so we can make a gauge choice such that $\delta\phi_a$ is non-vanishing only in $J^-(\operatorname{supp}(A))$. Overall, we can conclude that we can make a gauge choice such that $\delta_A\phi = \delta\phi_r-\delta\phi_a$ is non-vanishing only in $J^+(\operatorname{supp}(A))\cup J^-(\operatorname{supp}(A))$, a.k.a.\ the `domain of influence' of the support of $A$.

The Peierls bracket of $A$ and $B$ is
\begin{equation}
    \pb{A}{B}[\phi] = \delta_A B[\phi] = \int_{\mathcal{M}}\epsilon\,\fdv{B}{\phi}[\phi] \cdot \delta_A\phi[\phi].
\end{equation}
As we have discussed, $\fdv{B}{\phi}$ only has support in the support of $B$, while $\delta_A\phi$ can be chosen to only have support inside the domain of influence of the support of $A$ (since $B$ is assumed to be gauge-invariant, we are allowed to make this choice). Thus, if $A$ and $B$ are spacelike separated (meaning their supports are spacelike separated), the integrand above must vanish, and so we can conclude that $\pb{A}{B} = 0$. This is the definition of microcausality.

So much for the case of kinematically local observables in theories without diffeomorphism invariance. Let us now move on to the more relevant case at hand: relationally local observables in theories with diffeomorphism invariance. We will see that the argument goes very similarly. We just need to replace some field variations by relational field variations in certain key places.

An alternative to the following argument would be to map the entire variational principle to the local orientation space $\mathscr{O}$ of a parametrised frame. That is, we could write the action as a functional of the relational field configuration $\phi_R$ and the frame field, and then vary those directly to obtain some equations of motion that they would obey. We could then employ the causal structure on $\mathscr{O}$, defined as the pullback of the spacetime causal structure, as above. However, there are some potentially complicated questions about which variations of the relational field configuration and frame field are permissible, as well as issues concerning what happens when the frame is not surjective. Additionally, doing things in this way would restrict us to a single frame, whereas ideally we would like to know about the bracket of relationally local observables defined using distinct frames. For these reasons, we will do everything directly on spacetime.

Let $A:\mathcal{S}\to \RR$ be an observable constructed from a collection of gauge-invariant relationally local observables $A_i$, $i=1,2,\dots$. Using the chain rule with~\eqref{Equation: relationally local quantity} and~\eqref{Equation: relationally local map}, we may write the variation of $A$ as
\begin{equation}
    \delta(A[\phi]) = \int_{\mathcal{M}} A^{(0)}[\phi]\cdot(\delta\phi - \lie_{V_R[\phi,\delta\phi]}\phi) + A^{(1)}[\phi]\cdot\partial(\delta\phi - \lie_{V_R[\phi,\delta\phi]}\phi) + \dots
\end{equation}
For some observables $A^{(0)},A^{(1)},\dots$. We define the `relational support' of $A$  as
\begin{equation}
    \operatorname{supp}_R(A)[\phi] = \bigcup_k\,\operatorname{supp}(A^{(k)}[\phi]).
\end{equation}
It is the subset of spacetime over which $\delta(A[\phi])$ has a dependence on $\delta\phi - \lie_{V_R[\phi,\delta\phi]}\phi$. \jjvk{The relational support is the region of spacetime to which a relationally local observable is truly physically associated, after taking into account the gauge-fixing role played by the frame. For example, if we constructed a relationally local observable using a geodesic frame as described in Section~\ref{Section: geodesic example}, the relational support of that observable would be made up of endpoints of the geodesics that constitute the frame\pah{, rather than all the spacetime points along these geodesics. Note that $A[\phi]$ may also be smeared over either $\mathscr{O}$ or equivalently over $\mathcal{M}$, as discussed in Sec.~\ref{Section: general formalism / relational picture / single integrals smearing}}.}

\jjvk{For some further intuition on the relational support, consider a smeared relational observable of the kind in~\eqref{Equation: smeared relational observable spacetime}, which we repeat below for convenience:
    \begin{equation}
        O_{B,R;w}[\phi] = \int_{\mathcal{M}} w_R[\phi]\,1_{\mathcal{N}[\phi]}\, B[\phi], \text{ where } w_R[\phi] = (R[\phi])_* w.
    \end{equation}
    Here we are assuming $B$ is covariant and kinematically local.
    Under a field variation $\phi\to\phi+\delta\phi$, we have 
    \begin{equation}
        w_R[\phi]\,1_{\mathcal{N}[\phi]} \to w_R[\phi+\delta\phi]\, 1_{\mathcal{N}[\phi+\delta\phi]} = \qty(R[\phi+\delta\phi]\circ(R[\phi])^{-1})_*\big(w_R[\phi]\,1_{\mathcal{N}[\phi]}\big),
    \end{equation}
    so we can write
    \begin{nalign}
        O_{B,R;w}[\phi+\delta\phi] &= \int_{\mathcal{M}} \qty(R[\phi+\delta\phi]\circ(R[\phi])^{-1})_*\big(w_R[\phi]\,1_{\mathcal{N}[\phi]}\big)\, B[\phi+\delta\phi] \\
                                   &= \int_{\mathcal{M}} w_R[\phi]\,1_{\mathcal{N}[\phi]}\, \qty(R[\phi+\delta\phi]\circ(R[\phi])^{-1})^* B[\phi+\delta\phi] \\
                                   &= \int_{\mathcal{M}} w_R[\phi]\,1_{\mathcal{N}[\phi]}\, B\qty[\qty(R[\phi+\delta\phi]\circ(R[\phi])^{-1})^* (\phi+\delta\phi)]. \\
    \end{nalign}
    At linear order, 
    \begin{equation}
        \qty(R[\phi+\delta\phi]\circ(R[\phi])^{-1})^* (\phi+\delta\phi) = \phi+\delta\phi - \lie_{V_R[\phi,\delta\phi]}\phi,
    \end{equation}
    so we may use~\eqref{Equation: kinematically local quantity} to write
    \begin{equation}
        \delta(O_{B,R;w}[\phi]) = \int_{\mathcal{M}} w_{R[\phi]}\,1_{\mathcal{N}}[\phi]\, \Big( B'[\phi]\cdot(\delta\phi - \lie_{V_R[\phi,\delta\phi]}\phi) + B''[\phi] \cdot \partial (\delta\phi - \lie_{V_R[\phi,\delta\phi]}\phi) + \dots\Big).
    \end{equation}
    Thus, the relational support of the smeared relational observable $O_{B,R;w}$ is none other than the support of the spacetime smearing $w_R[\phi]$, or alternatively the image under the frame map $R[\phi]$ of the support of the local orientation space smearing $w$. Note that if the frame map $R[\phi]$ is continuous, then by choosing $w$ to have compact support, the relational support of $O_{B,R;w}$ will also be compact.
}

Let us assume that $A$ is a bulk observable, so $\operatorname{supp}_R(A)[\phi]$ doesn't touch the boundary $\partial\mathcal{M}$, and we can integrate by parts to write
\begin{equation}
    \delta(A[\phi]) = \int_{\mathcal{M}}\epsilon\, \frac{\delta_R A}{\delta_R\phi}[\phi]\cdot(\delta\phi - \lie_{V_R[\phi,\delta\phi]}\phi).
    \label{Equation: relational functional derivative}
\end{equation}
This defines $\frac{\delta_R A}{\delta_R\phi}$, which we call the `relational functional derivative' of $A$. Note that the relational support of $A$ coincides with the support of $\frac{\delta_R A}{\delta_R\phi}$.\footnote{\jjvk{In the case of the smeared relational observable, one may confirm that the relational functional derivative of $O_{B,R;w}$ is proportional to the ordinary functional derivative of $w_R[\phi]1_{\mathcal{N}[\phi]}B[\phi]$, provided that the pullback of $w_R[\phi]1_{\mathcal{N}[\phi]}$ to $\partial\mathcal{M}$ vanishes.}}

We make similar assumptions about $B$, i.e.\ that it is formed from a collection of gauge-invariant relationally local observables, and that its relational support doesn't touch the boundary. However, to keep things as general as possible, we will not assume that the frame used by $A$, which we have been calling $R$, is the same as the frame used by $B$, which we will call $\tilde{R}$. Then we can write
\begin{equation}
    \delta(B[\phi]) = \int_{\mathcal{M}}\epsilon\, \frac{\delta_{\tilde{R}} B}{\delta_{\tilde{R}}\phi}[\phi]\cdot(\delta\phi - \lie_{V_{\tilde{R}}[\phi,\delta\phi]}\phi)
    \label{Equation: relational B variation}
\end{equation}

Using~\eqref{Equation: relational functional derivative}, we find that the variation of the deformed action may be written
\begin{equation}
    \delta(S_\lambda[\phi]) = \int_{\mathcal{M}}\qty(E[\phi]\cdot\delta\phi - \epsilon\lambda\frac{\delta_R A}{\delta_R\phi}[\phi] \cdot (\delta\phi - \lie_{V_R[\phi,\delta\phi]}\phi)).
    \label{Equation: relational action deformation variation}
\end{equation}
The non-local dependence of $V_R[\phi,\delta\phi]$ on $\phi$ and $\delta\phi$ would appear to make the resulting deformation to the equations of motion similarly highly non-local and complicated. However, the general covariance of the theory means that this is in fact not the case.
\jjvk{%
    Indeed, by making a particular (field-dependent) gauge choice, we can write the equations of motion in a form in which this non-locality is eliminated.

    To see this, let $\tilde\phi[\phi] = (U[\phi])^*\phi$, where $U[\phi]\in H[\phi]$ is a small diffeomorphism that depends on $\phi$. In particular, let us choose $U[\phi]$ such that it agrees with $R[\phi]\circ (R[\phi_0])^{-1}$ in the relational support of $A$, and moreover that $U[\phi_0]$ is the identity everywhere in spacetime, where $\phi_0$ is the solution to the undeformed equations of motion. \pah{For $U[\phi]$ to be small it is important that $\operatorname{supp}_R(A)[\phi]$ does not intersect the boundary.} The field configuration $\tilde\phi$ \footnote{For notational simplicity, in the following we will not write out the dependence of $\tilde\phi$ on $\phi$ explicitly.} is physically equivalent to $\phi$, but it has been partially gauge-fixed such that the frame field $(R[\tilde\phi])^{-1}$ obeys
    \begin{multline}
        (R[\tilde\phi])^{-1} = (R[(U[\phi])^*\phi])^{-1} = \qty((U[\phi])^{-1}\circ R[\phi])^{-1} = (R[\phi])^{-1}\circ U[\phi] \\
        = (R[\phi])^{-1}\circ R[\phi] \circ (R[\phi_0])^{-1} = (R[\phi_0])^{-1},
    \end{multline}
    where the second line only holds in $\operatorname{supp}_R(A)$. Thus, the frame field $(R[\tilde\phi])^{-1}=(R[\phi_0])^{-1}$ is invariant \pah{under field variations} in the relational support of $A$. As a consequence, 
    \begin{equation}
        V_R[\tilde\phi,\delta\tilde\phi] = R[\tilde\phi+\delta\tilde\phi]\circ (R[\tilde\phi])^{-1}-\operatorname{Id}_{\mathcal{N}[\tilde\phi]} = 0 \text{ in } \operatorname{supp}_R(A).
        \label{Equation: V R vanishing relational support}
    \end{equation}
    (Note that $\tilde\phi+\delta\tilde\phi = (U[\phi+\delta\phi])^*(\phi+\delta\phi)$.)

    The deformed action is gauge-independent, so $S_\lambda[\phi] = S_\lambda[\tilde\phi]$. Thus, we can replace $\phi\to\tilde\phi$ everywhere on the right-hand side of~\eqref{Equation: relational action deformation variation}, yielding
    \begin{nalign}
        \delta(S_\lambda[\phi]) &= \int_{\mathcal{M}}\qty(E[\tilde\phi]\cdot\delta\tilde\phi - \epsilon\lambda\frac{\delta_R A}{\delta_R\phi}[\tilde\phi] \cdot (\delta\tilde\phi - \lie_{V_R[\tilde\phi,\delta\tilde\phi]}\tilde\phi)) \\
                                &= \int_{\mathcal{M}}\qty(E[\tilde\phi] - \epsilon\lambda \frac{\delta_R A}{\delta_R\phi}[\tilde\phi])\cdot \delta\tilde\phi,
    \end{nalign}
    where we used~\eqref{Equation: V R vanishing relational support} to reach the second line. Thus, the deformed equation of motion for $\tilde\phi$ is
    \begin{equation}
        E[\tilde\phi] - \epsilon\lambda\frac{\delta_R A}{\delta_R\phi}[\tilde\phi] = 0.
        \label{Equation: tilde phi deformed eom}
    \end{equation}
    This equation of motion is only deformed in the relational support of $A$. In particular, it does not have any contributions arising from the kinematical non-locality of the frame. It is worth reiterating that it is only satisfied when the gauge of $\tilde\phi$ obeys
    \begin{equation}
        (R[\tilde\phi])^{-1} = (R[\phi_0])^{-1} \text{ in } \operatorname{supp}_R(A).
        \label{Equation: tilde phi gauge}
    \end{equation}
    In a more general gauge, the equations of motion would take a kinematically non-local form. Note that~\eqref{Equation: tilde phi gauge} only partially fixes the gauge; the gauge of $\tilde\phi$ can be arbitrarily chosen outside $\operatorname{supp}_R(A)$. 

    The retarded and advanced solutions $\tilde\phi_r=\tilde\phi[\phi_r]$, $\tilde\phi_a=\tilde\phi[\phi_a]$ must both solve~\eqref{Equation: tilde phi deformed eom}. Away from the relational support of $A$,~\eqref{Equation: tilde phi deformed eom} reduces to the undeformed equations of motion $E[\phi] = 0$.
    Let us choose for $U[\phi]$ to be the identity sufficiently far in the past and to the future of $\operatorname{supp}_R(A)$, and define $\delta\tilde\phi_{r,a}$ by
    \begin{equation}
        \tilde\phi_{r,a} = \tilde\phi_0+\lambda\delta\pah{\tilde\phi}_{r,a}.
    \end{equation}
    Then, since $\tilde\phi_r=\phi_0$ on a Cauchy surface to the past, we must have that $\delta\tilde\phi_r$ vanishes outside $J^+(\operatorname{supp}_R(A))$, up to a choice of gauge.
    Similarly, since $\tilde\phi_a=\phi_0$ on a Cauchy surface to the future, we must have that $\delta\tilde\phi_a$ vanishes outside $J^-(\operatorname{supp}_R(A))$, up to a choice of gauge. Therefore, we can make a gauge choice such that $\delta_A\tilde\phi = \delta\tilde\phi_r - \delta\tilde\phi_a$ vanishes outside $J^+(\operatorname{supp}_R(A))\cup J^-(\operatorname{supp}_R(A))$, i.e.\ the domain of influence of $\operatorname{supp}_R(A)$.

    Finally, $\delta_A\phi=\delta\phi_r-\delta\phi_a$ is related to $\delta_A\tilde\phi$ by an infinitesimal small diffeomorphism of $\phi_0$, since
    \begin{nalign}
        \lambda\delta_A\tilde\phi +\order{\lambda^2} = \tilde\phi_r - \tilde\phi_a &= (U[\phi_r])^*\phi_r - (U[\phi_a])^*\phi_a\\
                                  & \begin{multlined}
                                      =\qty(U\qty[\phi_0+\lambda\delta\phi_r+ \order{\lambda^2} ])^*\qty(\phi_0+\lambda\delta\phi_r+ \order{\lambda^2} ) \\
                                      - \qty(U\qty[\phi_0+\lambda\delta\phi_a+ \order{\lambda^2} ])^*\qty(\phi_0+\lambda\delta\phi_a+ \order{\lambda^2} )
                                  \end{multlined}\\
                                  &= \lambda\qty(\delta_A\phi - \mathcal{L}_{W_r-W_a}\phi_0) + \order{\lambda^2},
    \end{nalign}
    where in the last line we are defining the $\order{1}$ vector fields $W_{r,a}$ via
    \begin{equation}
        U\qty[\phi_0+\lambda\delta\phi_{r,a}+\order{\lambda^2}] = \operatorname{Id} + \lambda W_{r,a} + \order{\lambda^2}.
    \end{equation}
    Since $U[\phi]$ is a small diffeomorphism, $W:=W_r-W_a$ is an infinitesimal small diffeomorphism, and taking the limit $\lambda\to 0$ we find
    \begin{equation}
        \delta_A\phi = \delta_A\tilde\phi + \lie_W\phi_0.
        \label{Equation delta A phi tilde phi small diffeo}
    \end{equation}
    Having obtained $\delta_A\phi$ for a particular background field configuration $\phi_0$, we can extend it to a function $\delta_A\phi = \delta_A\phi[\phi]$ of an arbitrary field configuration $\phi$ by repeating the above procedure.
}

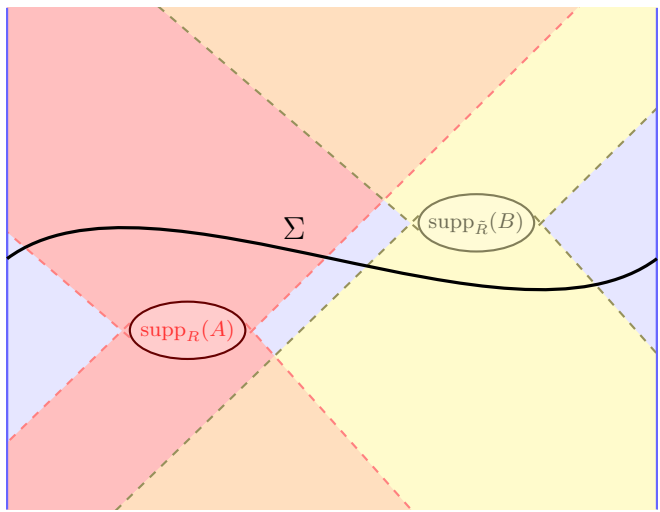
\begin{figure}
    \centering
    \begin{tikzpicture}[thick,scale=0.95]
        \fill[blue!10] (0,0) rectangle (9,7);
        \draw[thick,blue!60] (0,0) -- (0,7);
        \draw[thick,blue!60] (9,0) -- (9,7);
        \begin{scope}
            \clip (0,0) rectangle (9,7);
            \fill[red!25] (1.7,2.4) -- (-8.7,11.4) -- (12.3,11.4) -- (3.3,2.4);
            \fill[red!25] (1.7,2.6) -- (-8.7,-7.6) -- (12.3,-7.6) -- (3.3,2.6);
            \begin{scope}[shift={(4,1.5)}]
                \fill[yellow!25] (1.7,2.4) -- (-8.7,11.4) -- (12.3,11.4) -- (3.3,2.4);
                \fill[yellow!25] (1.7,2.6) -- (-8.7,-7.6) -- (12.3,-7.6) -- (3.3,2.6);
            \end{scope}
            \begin{scope}
                \clip (1.7,2.4) -- (-8.7,11.4) -- (12.3,11.4) -- (3.3,2.4);
                \begin{scope}[shift={(4,1.5)}]
                    \fill[orange!25] (1.7,2.4) -- (-8.7,11.4) -- (12.3,11.4) -- (3.3,2.4);
                \end{scope}
            \end{scope}
            \begin{scope}
                \clip (1.7,2.6) -- (-8.7,-7.6) -- (12.3,-7.6) -- (3.3,2.6);
                \begin{scope}[shift={(4,1.5)}]
                    \fill[orange!25] (1.7,2.6) -- (-8.7,-7.6) -- (12.3,-7.6) -- (3.3,2.6);
                \end{scope}
            \end{scope}
            \draw[red!50,dashed] (1.7,2.4) -- (-8.7,11.4) -- (12.3,11.4) -- (3.3,2.4);
            \draw[red!50,dashed] (1.7,2.6) -- (-8.7,-7.6) -- (12.3,-7.6) -- (3.3,2.6);
            \begin{scope}[shift={(4,1.5)}]
                \draw[yellow!50!black,dashed] (1.7,2.4) -- (-8.7,11.4) -- (12.3,11.4) -- (3.3,2.4);
                \draw[yellow!50!black,dashed] (1.7,2.6) -- (-8.7,-7.6) -- (12.3,-7.6) -- (3.3,2.6);
            \end{scope}
        \end{scope}
        \draw[thick,blue!60] (0,0) -- (0,7);
        \draw[thick,blue!60] (9,0) -- (9,7);
        \draw[red!40!black,fill=red!20] (2.5,2.5) ellipse (0.8 and 0.4) node[red!80, scale=0.9] {\footnotesize$\operatorname{supp}_R(A)$};
        \draw[yellow!40!black,fill=yellow!20] (6.5,4) ellipse (0.8 and 0.4) node[yellow!30!black, scale=0.9] {\footnotesize$\operatorname{supp}_{\tilde{R}}(B)$};
        \draw[very thick] (0,3.5) .. controls (2,5) and (7,2) .. (9,3.5) node[pos=0.45, above] {\large$\Sigma$};
    \end{tikzpicture}
    \caption{We can choose a Cauchy surface $\Sigma$ whose points are each contained within at most one of the domains of influence of $A$ and $B$, when they are spacelike separated.}
    \label{Figure: spacelike separated Cauchy surface}
\end{figure}

At this point, we can compute the Peierls bracket of $A$ and $B$ with the usual formula:
\begin{nalign}
    \pb{A}{B}[\phi] = \delta_A B[\phi] &= \int_{\mathcal{M}}\epsilon\frac{\delta_{\tilde{R}} B}{\delta_{\tilde{R}}\phi}[\phi] \cdot (\delta_A\phi - \lie_{V_{\tilde{R}}[\phi,\delta_A\phi]}\phi) \\
                                       &\jjvk{= \int_{\mathcal{M}}\epsilon\frac{\delta_{\tilde{R}} B}{\delta_{\tilde{R}}\phi}[\phi] \cdot (\delta_A\tilde\phi - \lie_{V_{\tilde{R}}[\phi,\delta_A\tilde\phi]}\phi),}
    \label{Equation: Peierls relational 1}
\end{nalign}
\jjvk{where the first line follows from~\eqref{Equation: relational B variation}, and the second line follows from~\eqref{Equation delta A phi tilde phi small diffeo} and the fact that the term in brackets is independent of the gauge of $\delta_A\phi$.}
Suppose $A$ and $B$ are spacelike separated, meaning their relational supports are spacelike separated. Then the support of $\delta_A\tilde\phi$ and the support of $\frac{\delta_{\tilde{R}} B}{\delta_{\tilde{R}}\phi}$ do not intersect, so we may write
\begin{equation}
    \pb{A}{B}[\phi] = - \int_{\mathcal{M}}\epsilon\frac{\delta_{\tilde{R}} B}{\delta_{\tilde{R}}\phi}[\phi] \cdot \lie_{V_{\tilde{R}}[\phi,\delta_A\jjvk{\tilde\phi}]}\phi.
    \label{Equation: Peierls relational 2}
\end{equation}
In general, $\lie_{V_{\tilde{R}}[\phi,\delta_A\tilde\phi]}\phi$ may not vanish in the support of $B$, so it may not be immediately clear that this integral vanishes.\footnote{\jjvk{We might try choosing a gauge in which $V_{\tilde{R}}[\phi,\delta_A\tilde\phi]=0$. However, it is not immediately clear that this can be done at the same time as requiring $\delta_A\tilde\phi$ to vanish outside of $\operatorname{supp}_{\tilde{R}}(B)$. Without this latter requirement, we cannot go from~\eqref{Equation: Peierls relational 1} to~\eqref{Equation: Peierls relational 2}.}} To see that it does, we can use the (equivalent) formula for the Peierls bracket in terms of the presymplectic form:
\begin{equation}
    \pb{A}{B}[\phi] = \Omega[\phi,\delta_A\phi,\delta_B\phi] \jjvk{= \Omega[\phi,\delta_A\tilde\phi,\delta_B\tilde\phi] = \int_\Sigma\omega[\phi,\delta_A\tilde\phi,\delta_B\tilde\phi]},
    \label{Equation: relational poisson bracket}
\end{equation}
\jjvk{where the second equality follows from the fact that small diffeomorphisms are degenerate directions of $\Omega$.}
Note that $\delta_A\tilde\phi,\delta_B\tilde\phi$ vanish outside the domains of influence of $\operatorname{supp}_R(A),\operatorname{supp}_{\tilde{R}}(B)$ respectively. If the relational supports of $A$ and $B$ are spacelike separated, then we can choose the Cauchy surface $\Sigma$ in such a way that
\begin{equation}
    \Sigma \cap \big(J^+(\operatorname{supp}_R(A))\cup J^-(\operatorname{supp}_R(A))\big) \cap \big(J^+(\operatorname{supp}_{\tilde{R}}(B))\cup J^-(\operatorname{supp}_{\tilde{R}}(B))\big) = \emptyset.
\end{equation}
This is illustrated in Figure~\ref{Figure: spacelike separated Cauchy surface}. Thus, on $\Sigma$, either $\delta_A\phi$ or $\delta_B\phi$ vanishes. Since \jjvk{$\omega[\phi,\delta_A\tilde\phi,\delta_B\tilde\phi]$} is locally formed from its arguments, and linear in \jjvk{$\delta_A\tilde\phi,\delta_B\tilde\phi$}, it must be the case that the integrand in~\eqref{Equation: relational poisson bracket} vanishes. Therefore, the bracket of $A$ and $B$ vanishes when they are spacelike separated, which is the definition of microcausality.

\jjvk{It should be noted that, conceptually speaking, the presymplectic form by itself is not sufficient to deduce microcausality. Indeed, the argument using the Peierls construction given above was essential for deducing the relevant causal properties of $\delta_A\phi$, $\delta_B\phi$. \pah{Specifically, it was essential for arguing that the \emph{relational} supports of the gauge-invariant observables $A$ and $B$ are key, rather than the kinematical supports of the non-invariant spacetime fields of which they are comprised}. (Of course, as we have already noted, the Peierls bracket and the Poisson bracket implied by the presymplectic form are equivalent.)}

Let us highlight three important facts about relational microcausality. First, in gravity the notion of `spacelike separated' is field-dependent. Indeed, it is the metric that determines the causal structure, and the metric is a dynamical entity. Thus, the observables $A$ and $B$ may be spacelike separated for some states, but not for others. The vanishing of their Poisson bracket is only guaranteed in states where they are spacelike separated. Second, this argument only works in the bulk \jjvk{(since for general frames it is not possible to find a small $U[\phi]$ that agrees with $R[\phi]\circ(R[\phi_0])^{-1}$ all the way up to the boundary)}, so if one of $A$ or $B$ has relational support containing the boundary, there is no guarantee that their bracket will vanish, even if they are spacelike separated. This is why this result is not in conflict with the brackets found in e.g.\ \cite{Harlow:2021dfp}. Finally, it should be noted that we can in principle construct relationally local observables in any kind of theory, even theories that are not gravitational. However, only in gravitational theories will (bulk) relational microcausality hold for all such observables. Indeed, the above argument relies heavily upon the fact that small diffeomorphisms are gauge symmetries, which is only true in gravitational theories, in order to substitute $\phi\to\tilde\phi$ in the deformed action $S_\lambda[\phi]$.

\jjvk{Our results go significantly beyond the brief arguments made for the relational microcausality \pah{of single-integral observables by Marolf} in~\cite{Marolf:2015jha}. Besides giving considerably more detail, what we have obtained works for any kind of dynamical reference frame \pah{(and fields to be described relative to that frame), including non-local ones such as the geodesic dressing frames which may touch the boundary,} whereas~\cite{Marolf:2015jha} only allowed for a frame provided by a set of kinematically local, fully covariant scalar fields.\pah{\footnote{\pah{In fact, it was stated in~\cite{Marolf:2015jha} that microcausality relations would no longer hold for non-local frames.}} For that reason}, the observables in~\cite{Marolf:2015jha} were not permitted to have any large diffeomorphism charges (e.g.\ Poincar\'e charges in asymptotically flat spacetimes), whereas the observables we consider can have such charges, as a consequence of the fact that in our case the frame $R[\phi]$ need not be covariant for all possible diffeomorphisms, but only gauge-covariant, i.e.\ covariant for small diffeomorphisms.} \pah{Our argument also explicitly works for dynamical frames that are neither surjective nor injective. Recall from Section~\ref{Section: general formalism / relational picture / single integrals smearing} that we can simply restrict the frame to a subregion of spacetime where the frame is invertible, while still obtaining a relational observable that is fully invariant under small diffeomorphisms. In particular, this relational observable can also be written as a (possibly gravitationally charged) single-integral observable with a covariant characteristic function vanishing outside that subregion.

Microcausality of dressed matter observables was also discussed in~\cite{Kabat}. We further note that Dittrich developed an (unpublished) microcausality argument~\cite{bianca} for relational observables describing a scalar field relative to locally constructed scalar reference fields in spacetimes without boundary. This argument was based on the canonical power-series construction in~\cite{Dittrich:2005kc} to which we relate in Section~\ref{Section: general formalism / covariant to canonical} below. The above can thus also be seen as a significant generalisation of that argument. 
}

It should be noted that the status of bulk microcausality in gravity has been much debated in the literature. One important contribution which drew the opposite conclusion to our own is~\cite{Donnelly:2015hta,Donnelly:2016rvo,Giddings:2015lla}. See also~\cite{Bodendorfer}. We will comment on this tension in the conclusion.

\subsubsection{Large diffeomorphisms}
\label{Section: large diffeos}

We would like to now describe how large diffeomorphisms affect relational observables. It is well-known that the charges that generate large diffeomorphisms are boundary integrals. Indeed, using~\eqref{Equation: omega noether}, we have
\begin{equation}
    \Omega[\phi,\delta\phi,\delta_V\phi] = \int_{\partial\Sigma}\big(\delta(Q_V[\phi]) - Q_{\delta V}[\phi] - \iota_V\theta[\phi,\delta\phi]\big).
    \label{Equation: large diffeo boundary}
\end{equation}
If $V$ is a vector field corresponding to an infinitesimal large diffeomorphism, this integral will not vanish, and for certain $V$ the right-hand side may be written as $\delta (H_V[\phi])$ for some charge $H_V[\phi]$ that may be purely written as a boundary integral~\cite{Iyer:1994ys,Lee:1990nz,Harlow:2019yfa}. Such field transformations are called integrable.

Poisson brackets with $H_V$ cause other observables to transform as if large diffeomorphisms were acting on the fields:
\begin{equation}\label{eq:largediffpbracket}
    \pb{A}{H_V}[\phi] = \dv{s}\left. A[\phi+s\lie_V\phi]\right|_{s=0}.
\end{equation}
As we will see in this section, even bulk relationally local observables can transform non-trivially under such active large diffeomorphisms. This is a manifestation of the fact that large diffeomorphisms correspond to global symmetries. It seems at first that this provides a counterexample to microcausality, since $\partial\Sigma$ is spacelike separated from many bulk points. However, this just reflects the fact that relational microcausality is only guaranteed \pah{to} hold between two bulk observables.

Recall that for a small diffeomorphism $f\in H[\phi]$, we have $R[f_*\phi] = f\circ R[\phi]$\jjvk{, so $\mathcal{N}[f_*\phi]=f(\mathcal{N}[\phi])$. Let us now generalise to the case of a more general (not necessarily small) $f\in G[\phi]$, in which case these equations may not hold. Assuming the frame is injective and smooth, we can write the transformation law of the frame field as
\begin{equation}
    (R[f_*\phi])^{-1} = C_f[\phi] \circ (R[\phi])^{-1} \circ f^{-1} \qq{in} \mathcal{N}[f_*\phi]\cap f(\mathcal{N}[\phi])
    \label{Equation: R-1 tfmn general f}
\end{equation}
for some diffeomorphism $C_f[\phi]$ of the local orientation space $\mathscr{O}$. 
Clearly $C_f[\phi]$ is the identity for small $f\in H[\phi]$.} 

For an example, let us consider again the geodesic dressing example considered in Section~\ref{Section: geodesic example}. Recall that $x_{\tau,z,W}[g]$ is the endpoint of a geodesic of length $\tau$ which starts at $z\in\partial\mathcal{M}$ with tangent vector $W-\hat{n}[g]$. For a large diffeomorphism $f$, we have
\begin{equation}
    x_{\tau,z,W}[f_*g] = f(x_{\tau,f^{-1}(z),X_{f^{-1}}(W)}[g]).
    \label{Equation: x large change}
\end{equation}
where $X_f$ is the transformation law of the bundle $\mathrm{T}\partial\mathcal{M}$, i.e.\ $X_f = \dd{f_\partial}$ where $f_\partial$ is the restriction of $f$ to $\partial\mathcal{M}$. To verify that this is true, let $\gamma_{\tau,z,W}[g]$ be the geodesic defining $x_{\tau,z,W}[g]$. Then, by the tensorial nature of the geodesic equation, $f(\gamma_{\tau,f^{-1}(z),f^*W}[g])$ is the geodesic according to the metric $f_*g$ of length $\tau$, starting from
\begin{equation}
    f(f^{-1}(z)) = z,
\end{equation}
with tangent vector
\begin{equation}
    X_f(X_{f^{-1}}(W) - \hat{n}[g]) = W - \hat{n}[f_*g].
\end{equation}
So, by definition,
\begin{equation}
    \gamma_{\tau,z,W}[f_*g] = f(\gamma_{\tau,f^{-1}(z),X_{f^{-1}}(W)}[g]),
\end{equation}
from which~\eqref{Equation: x large change} follows. Thus, large diffeomorphisms have a natural action on the geodesic reference frames. In particular, if we define
\begin{equation}
    C_f: (\tau,z,W)\mapsto (\tau,f(z),X_f(W)), \qquad f \in H,
\end{equation}
then~\eqref{Equation: R-1 tfmn general f} is obeyed.

\jjvk{Let us for simplicity assume now that the frame is surjective, so that $\mathcal{N}[\phi]=\mathcal{M}$, which implies \eqref{Equation: R-1 tfmn general f} holds everywhere in spacetime. Then we can invert the transformation law of the frame field to obtain
\begin{equation}
    R[f_*\phi] = f\circ R[\phi] \circ (C_f[\phi])^{-1}
    \label{Equation: R tfmn general f}
\end{equation}
The diffeomorphism $C_f[\phi]$ governs how relational observables transform under large diffeomorphisms. Indeed, we have
\begin{equation}\label{eq:relobs_largediff}
    O_{A,R}[f_*\phi] = (R[f_*\phi])^* A[f_*\phi] = (C_f[\phi])_* (R[\phi])^* f^* f_*  A[\phi] = (C_f[\phi])_* O_{A,R}[\phi].
\end{equation}}

A useful perspective on~\eqref{Equation: R tfmn general f} may be gained by viewing $R[\phi]:\mathscr{O}\to\mathcal{M}$ as an equivariant bundle whose transformation law is $C_f[\phi]$. This equivariant structure is mixed up with the field dependence of $R[\phi]$ in an interesting way. Indeed,~\eqref{Equation: R tfmn general f} replaces the usual equivariance property $\pi = f\circ \pi \circ (X_f)^{-1}$. The usual $X_{f\circ g} = X_f\circ X_g$ composition law is also replaced. Indeed, suppose $g\in G[\phi]$ and $f\in G[f_*\phi]$. Then $f\circ g\in G[\phi]$, and we may write
\begin{multline}
    f\circ g\circ R[\phi] \circ (C_{f\circ g}[\phi])^{-1} = R[f_*g_*\phi] = f\circ R[g_*\phi] \circ (C_f[g_*\phi])^{-1}\\
    = f \circ g \circ R[\phi] \circ (C_g[\phi])^{-1} \circ (C_f[g_*\phi])^{-1}.
\end{multline}
So the composition law becomes
\begin{equation}
    C_{f\circ g}[\phi] = C_f[g_*\phi]\circ C_g[\phi].
\end{equation}

With this point of view, the right-hand side of~\eqref{Equation: R tfmn general f} is none other than the pushforward of $R[\phi]$ through $f$. Thus, with this transformation law, the map $R[\phi]$ may be viewed as \emph{fully} covariant (i.e.\ covariant under \emph{all} spacetime diffeomorphisms). Similarly, the frame field obeys
\begin{equation}
    (R[f_*\phi])^{-1} = f_{*[\phi]} (R[\phi])^{-1},
\end{equation}
i.e.\ it is also \emph{fully} covariant (the notation $f_{*[\phi]}$ is supposed to take into account the fact that the transformation law and the pushforward are field dependent). Note that to take this point of view, we can no longer view the frame field as a scalar field on spacetime (though for small diffeomorphisms it behaves like one). Instead, it is a rather special type of field with transformation law $C_f[\phi]$, which can be quite unusual.

We may push $C_f[\phi]$ forward to a spacetime diffeomorphism
\begin{equation}
    f_R[\phi] = R[\phi]\circ C_f[\phi] \circ (R[\phi])^{-1}.
\end{equation}
Using~\eqref{Equation: R tfmn general f}, it can be seen that this is equivalent to
\begin{equation}
    f_R[\phi] = R[\phi]\circ (R[f_*\phi])^{-1} \circ f
\end{equation}
The diffeomorphism $f_R[\phi]$ \pah{need not necessarily lie in the same gauge equivalence class as $f$ (it may not even be large if $f$ is); however, it} \jjvk{does not depend on the gauge of $f$}. Indeed, suppose $f^1,f^2\in G[\phi]$ differ by a small diffeomorphism, so $f_*^1\phi\sim f_*^2\phi$, i.e.\ $f^2\circ (f^1)^{-1}\in H[f^1_*\phi]$. Then we have
\begin{multline}
    f^2_R[\phi] = R[\phi]\circ (R[f^2_*\phi])^{-1}\circ f^2 = R[\phi]\circ ((f^2\circ (f^1)^{-1})\circ R[f^1_*\phi])^{-1}\circ f^2 \\
    = R[\phi]\circ (R[f^1_*\phi])^{-1}\circ f^1\circ (f^2)^{-1}\circ f^2 = R[\phi]\circ (R[f^1_*\phi])^{-1}\circ f^1 = f^1_R[\phi].
\end{multline}
On the other hand, $f_R[\phi]$ is not typically a gauge-invariant function of $\phi$, as we shall see below.

More generally, we can allow $f=f[\phi]$ itself to depend on the field configuration. For each $\phi$, we must have $f[\phi]\in G[\phi]$. The action of this diffeomorphism yields the transformed relational observable $O_{A,R}^f$, which is defined by
\begin{equation}
    O_{A,R}^f[\phi] = O_{A,R}[(f[\phi])_*\phi] = (C_{f[\phi]}[\phi])_* O_{A,R}[\phi],
\end{equation}
and \jjvk{we set}
\begin{equation}
    f_R[\phi] = R[\phi]\circ C_{f[\phi]}[\phi]\circ (R[\phi])^{-1} = R[\phi]\circ(R[(f[\phi])_*\phi])^{-1}\circ f[\phi].
    \label{Equation f_R definition}
\end{equation}
For the transformation from $O_{A,R}$ to $O^f_{A,R}$ to be actually physically realisable, both $O_{A,R}$ and $O^f_{A,R}$ need to be gauge-invariant, which generically will only be true if $C_{f[\phi]}[\phi]$ is gauge-invariant, i.e.
\begin{equation}
    C_{f[g_*\phi]}[g_*\phi] = C_{f[\phi]}[\phi] \text{ for all } g\in H[\phi].
\end{equation}
In this case, for all $g\in H[\phi]$ we have
\begin{multline}
    f_R[g_*\phi] = R[g_*\phi] \circ C_{f[g_*\phi]}[g_*\phi] \circ (R[g_*\phi])^{-1}\\
    = g\circ R[\phi] \circ C_{f[\phi]}[\phi]\circ (R[\phi])^{-1}\circ g^{-1} = g \circ f_R[\phi] \circ g^{-1}.
\end{multline}
In other words, $f_R$ must be gauge-covariant.

Note that
\begin{equation}
    O_{A,R}^f[\phi] = (R[\phi])^* (f_R[\phi])_* A[\phi].
\end{equation}
We can observe that for the relational observable $O_{A,R}$, acting with the large diffeomorphism $f$ gives the same result as if we left the fields and the frame invariant, but changed $A$ to $A^f$, where
\begin{equation}
    A^f[\phi] = (f_R[\phi])_* A[\phi].
\end{equation}
Alternatively, we can leave the fields and $A$ invariant, but modify the frame from $R$ to $R^f$, where
\begin{equation}
    R^f[\phi] = (f_R[\phi])^{-1} \circ R[\phi] = R[\phi]\circ (C_{f[\phi]}[\phi])^{-1} = (f[\phi])^{-1} \circ R[(f[\phi])_*\phi].
\end{equation}
Thus, we get the identities
\begin{equation}
    O_{A,R}^f = O_{A^f,R} = O_{A,R^f}.
\end{equation}

For example, if $R$ is the geodesic dressing frame using a fixed boundary vector field $W_1$ described in Section~\ref{Section: geodesic example}, then $R^f$ is another geodesic frame using a different boundary vector field $(f[\phi])_*W_1$ (which could be field dependent, if $f$ is).

Suppose an observable $B$ is relationally local with respect to $R$, and let $B^f$ be defined by $B^f[\phi] = B[f[\phi]_*\phi]$. The effect of the large diffeomorphism $f[\phi]$ is to transform the observable from $B$ to $B^f$. But $B^f$ is in general no longer relationally local with respect to $R$, since there is no reason for
\begin{equation}
    (R[\phi])_*B^f[\phi] = (R[\phi])_*B[f[\phi]_*\phi]
\end{equation}
to be kinematically local (in general $f[\phi]$ can have a quite kinematically non-local dependence on $\phi$).
However, we do have
\begin{equation}
    (R^f[\phi])_*B^f[\phi] = (f[\phi])^*\qty(R\big[(f[\phi])_*\phi\big])_* B\big[(f[\phi])_*\phi\big].
\end{equation}
If $(R[\phi])_*B[\phi]$ is covariant (which is the case when $B=O_{A,R}$ for covariant $A$), then we have
\begin{equation}
    (R^f[\phi])_*B^f[\phi] = (f[\phi])^*(f[\phi])_*(R[\phi])_* B[\phi] = (R[\phi])_* B[\phi],
\end{equation}
which is kinematically local by the definition of $B$. Thus, for such $B$, the transformed observable $B^f$ is relationally local with respect to the transformed frame $R^f$. In other words, large diffeomorphisms can have a non-trivial effect on the meaning of relational locality. This corresponds to the change of frames $R\to R^f$, which can be understood as being implemented by the diffeomorphism $C_{f[\phi]}[\phi]$ acting on the local orientation space $\mathscr{O}$.

Let us now consider the case where the diffeomorphism $f[\phi]$ is close to the identity, so that it can be viewed as generated by a spacetime vector field $\xi[\phi]$. Using~\eqref{Equation: V_R definition} then allows us to write
\begin{equation}\label{eq:b1}
    R[(f[\phi])_*\phi]\circ (R[\phi])^{-1} = R[\phi+\lie_{\xi[\phi]}\phi]\circ (R[\phi])^{-1} = \operatorname{Id}_{\mathcal{N}[\phi]} + V_R[\phi,\lie_{\xi[\phi]}\phi]
\end{equation}
to leading order in $\xi[\phi]$. Comparing with~\eqref{Equation f_R definition}, we see that $f_R[\phi]$ is generated by the vector field
\begin{equation}
    \xi_R[\phi] = \xi[\phi] - V_R[\phi,\lie_{\xi[\phi]}\phi].
\end{equation}
\pah{While $\xi[\phi]$ is only unique up to small diffeomorphisms, note that $\xi_R[\phi]$ is unique, given that $f_R[\phi]$ only depends on the gauge equivalence class of large diffeomorphisms, as noted above.}
Since $f_R[\phi]$ is the pushforward of $C_{f[\phi]}[\phi]$ through $R[\phi]$, it is clear also that $C_{f[\phi]}[\phi]$ is generated by the local orientation space vector field
\begin{equation}\label{eq:b2}
    \tilde{\xi}_R[\phi] = (R[\phi])^* \xi_R[\phi].
\end{equation}
This is clearly a relational observable, formed using the frame and $\xi_R[\phi]$. However, in general it is not relationally local, since $\xi_R[\phi]$ in general is not kinematically local.

If a $\xi[\phi]$ is such that~\eqref{Equation: large diffeo boundary} may be written as $\delta(H_\xi[\phi])$ when $V=\xi[\phi]$, for some observable $H_\xi$, then $H_\xi$ generates the transformation $\phi\to\phi+\lie_\xi\phi$. If $O_{A,R}\pah{[\phi]=(R[\phi])^*A[\phi]}$ is a relational observable, then \pah{by \eqref{eq:relobs_largediff}} we have\footnote{\pah{This result also follows directly from \eqref{eq:largediffpbracket} by using that $O_{A,R}[\phi]=(R[\phi])^*A[\phi]$ and equations \eqref{eq:b1}--\eqref{eq:b2}.}}
\begin{equation}
    \pb{O_{A,R}}{H_\xi}[\phi] = \lie_{\tilde{\xi}_R[\phi]} O_{A,R}[\phi].
    \label{Equation: relational observable large diffeo}
\end{equation}
Thus, $H_\xi$ generates the orientation space diffeomorphism given by $\tilde{\xi}_R$.\footnote{It is interesting that the algebra of the observables $H_\xi$ is not the same as the Lie algebra of the corresponding vector fields $\tilde{\xi}_R$, essentially because of the field-dependence of the vector fields (for the algebra corresponding to field-independent vector fields on orientation space, see e.g.\ \cite{CEH}).} \pah{This provides a way to discuss time evolution in an invariant manner \jjvk{(see the next subsection)}.}

\jjvk{
    Note that throughout this discussion we have been assuming that the relational observable $O_{A,R}[\phi]=(R[\phi])_*A[\phi]$ has been formed using a fully covariant $A[\phi]$. However, more generally $A[\phi]$ need only be gauge-covariant for $O_{A,R}[\phi]$ to be gauge-invariant. In such a case, the above discussion must be modified, and~\eqref{Equation: relational observable large diffeo} need not hold. The most clear example of this is the case where $A[\phi]=(R[\phi])^{-1}$ is the frame field, which is indeed only gauge-covariant. Then $O_{A,R}[\phi] = (R[\phi])^*(R[\phi])^{-1}=(R[\phi])^{-1}\circ R[\phi]$ is a scalar field which maps $o\mapsto o$ for $o\in \mathscr{O}$. This field is independent of $\phi$ and so must have vanishing Poisson brackets with all other observables. On the other hand, $\lie_{\tilde{\xi}_R[\phi]} O_{A,R}[\phi] = -\tilde{\xi}_R[\phi]$.\footnote{Indeed, the frame field is a scalar, so it's pullback $O_{A,R}[\phi]$ to orientation space is also a scalar, and so transforms like
        \begin{equation}
            (C_{f[\phi]}[\phi])_*O_{A,R}[\phi] = O_{A,R}[\phi] \circ (C_{f[\phi]}[\phi])^{-1}.
        \end{equation}
        Now expanding $C_{f[\phi]}[\phi]=\operatorname{Id}+\tilde\xi_R[\phi]+\dots$ to linear order in $\tilde\xi_R[\phi]$, and using $O_{A,R}[\phi]=\operatorname{Id}$, we get the claimed relation.
    } Thus,~\eqref{Equation: relational observable large diffeo} is violated when $A$ is the frame field. As a consequence, we can understand the evolution generated by $H_\xi$ as a genuine transformation of relational observables $O_{A,R}$ (for covariant $A$) relative to the frame; the former change by the action of a certain diffeomorphism on $\mathscr{O}$, while the latter does not.\footnote{In a sense, $H_\xi$ may be understood as the generator of an `active diffeomorphism', because it transforms the relational fields $\phi_R$ according to the action of a diffeomorphism of $\mathscr{O}$. Then we may think of observables $O_{A,R}$ for which~\eqref{Equation: relational observable large diffeo} holds as `covariant' -- because the left-hand side is an active diffeomorphism, while the right-hand side is a passive diffeomorphism. For example, $O_{A,R}$ is covariant whenever $A$ is covariant, but when $A$ is the frame field $O_{A,R}$ is not covariant (indeed it does not change at all under active diffeomorphisms, and so is a background field on $\mathscr{O}$).}
}

\jjvk{
    Let us finally note a relationship between large diffeomorphisms and the non-locality of the frame. As we have tried to stress, the frame field $(R[\phi])^{-1}$ need not be a kinematically local object. However, if it is (for example as in the case of Brown-Kucha\v{r} dust, see Section~\ref{Section: more examples / dust}), then one may conclude that any \emph{bulk} relational observables formed using the frame will necessarily have trivial large diffeomorphism charges. This is essentially because the frame cannot `tell the difference' between large and small diffeomorphisms. More precisely, suppose that $x$ is some fixed point in the bulk $\mathcal{M}\setminus\partial\mathcal{M}$, and let $f\in G[\phi]$ be any on-shell diffeomorphism. Because $x$ is away from the boundary, we can always find a diffeomorphism $\tilde{f}$ which agrees with $f$ in a neighbourhood of $x$, but acts as the identity in a neighbourhood of the boundary, and which is therefore small. Then in a neighbourhood of $x$ we have
    \begin{equation}
        (R[f_*\phi])^{-1} = (R[\tilde{f}_*\phi])^{-1} = (R[\phi])^{-1}\circ\tilde{f} = (R[\phi])^{-1}\circ f,
    \end{equation}
    where the first equality follows from kinematical locality, and the second equality follows from gauge-covariance of the frame. This holds for any point $x$ in the bulk, so one may conclude that the frame field is in fact covariant for all large diffeomorphisms as well as small ones (when evaluated in the bulk). Thus, any relational observable $O_{A,R}[\phi] = (R[\phi])^* A[\phi]$ formed from this frame will be \emph{invariant} under large diffeomorphisms when evaluated at local orientations which map under $R[\phi]$ to points in the bulk (assuming $A[\phi]$ is fully covariant). Consequently, for a bulk relational observable to be non-trivially charged under large diffeomorphisms, the frame must be kinematically non-local. 

    In fact, this argument can be extended to include any frame which doesn't depend on the fields in a neighbourhood of the boundary. Such frames cannot tell the difference between large diffeomorphisms and small diffeomorphisms, meaning they must be fully covariant, and so only yield relational observables which are invariant under large diffeomorphisms. Thus, for a relational observable to have non-trivial large diffeomorphism charges, it must be formed from a frame which has a dependence on the boundary fields,\footnote{This may be related to a version of the `dressing theorem' of~\cite{Donnelly:2016rvo}.} for example like the geodesic frames discussed in Section~\ref{Section: geodesic example}.
}

\pah{
\subsubsection{Relational time evolution}}
\label{Section: relational time evolution}

Now consider the case where there is a special large diffeomorphism that corresponds to boundary time evolution, generated by a spacetime vector field $\zeta[\phi]$. This vector field is only uniquely determined up to small diffeomorphisms, but the observable $H_\zeta$ which generates the large diffeomorphism is unique. This is \pah{a} Hamiltonian of the theory. Since it is a boundary integral, it would na\"ively appear to generate no evolution in the bulk, which is one aspect of the so-called `problem of time'.

However, relational observables transform according to~\eqref{Equation: relational observable large diffeo}. Thus, $H_\zeta$ generates evolution with respect to the orientation space vector field $\tilde{\zeta}_R$. This evolution applies to \emph{all} relational observables \pah{that are charged under large diffeomorphisms}, including relationally local ones in the bulk. Thus, there is no `problem of time' for gravitationally charged relationally local bulk observables. \pah{We comment on the situation of relational observables with trivial large diffeomorphism charge shortly, encompassing also the case of an absent boundary.}

To make this especially explicit, we can construct a `clock function' in the bulk, which measures the time experienced by an observer with access to relationally local observables. We simply pick a gauge-covariant spacetime function $t[\phi]:\mathcal{N}[\phi]\to\RR$ satisfying
\begin{equation}
    \zeta_R[\phi](t[\phi]) = 1.
\end{equation}
One way to do this is to find a $\tilde{t}[\phi]:\mathscr{O}\to\RR$ obeying
\begin{equation}
    \tilde\zeta_R[\phi](\tilde{t}[\phi]) = 1.
\end{equation}
Since $\tilde\zeta_R$ is gauge-invariant, we can pick $\tilde{t}$ to be gauge-invariant\pah{, in which case it is a relational observable too. Thanks to \eqref{Equation: relational observable large diffeo}, the last equation entails (if $\tilde t$ is field-dependent)
\begin{equation}
\{\tilde t,H_\zeta\}[\phi]=1,
\end{equation}
and so the relational clock function $\tilde t[\phi]$ on the local orientation space is conjugate to the Hamiltonian $H_\zeta$.}
Then we simply set
\begin{equation}
    t[\phi] = (R[\phi])_* \tilde{t}[\phi] = \tilde{t}[\phi]\circ (R[\phi])^{-1}.
\end{equation}
This function $t[\phi]$ is the desired bulk clock function \pah{relative to which the bulk degrees of freedom evolve in a gauge-invariant manner. Since $\zeta_R[\phi]$ is uniquely defined for a given large diffeomorphism, so is $t[\phi]$. We can view it as a bulk-extended boundary time, given that the Hamiltonian $H_\zeta$ resides in the boundary algebra.

This time evolution does not apply to relational observables with trivial gravitational charge (as they commute with $H_\zeta$) and thus in particular does not apply to spacetimes without boundary. There are, however, ways to also define relational bulk time evolution for such cases and thereby resolve the `problem of time' in the classical setting more generally. The idea is essentially to invert the above logic. For simplicity, suppose that the frame $R[\phi]$ is injective and sufficiently differentiable so that the local orientation space $\mathscr{O}$ inherits a metric and causal structure from spacetime. The goal is then to find a (gauge-invariant) timelike orientation space vector field $\tilde\rho_R[\phi]$ such that not only $\mathcal{L}_{\tilde\rho_R}O_{A,R}[\phi]\neq0$ for gravitationally non-charged relational observables (there will exist many such vector fields), but also such that this transformation corresponds to a field-space vector field $\mathcal{V}_{\tilde\rho_R}$ that is both tangential to the space of solutions $\mathcal{S}$ and integrable. Since by \eqref{eq:phirvanish} gauge transformations act trivially on relational field configurations $\phi_R$, such a $\mathcal{V}_{\tilde\rho_R}$ must be a physical transformation (hence a symmetry) in at least some neighbourhoods of $\mathcal{S}$. In this case, we can find a non-vanishing gauge-invariant Hamiltonian via $\iota_{\mathcal{V}_{\tilde\rho_R}}\Omega=\delta H_{\tilde\rho_R}$  that generates this relational dynamics
\begin{equation}\label{eq:reltime}
    \{O_{A,R},H_{\tilde\rho_R}\}[\phi]=\mathcal{L}_{\tilde\rho_R}O_{A,R}[\phi].
\end{equation}
The relational observable $O_{A,R}$ then evolves in the orientation space clock function $\tilde t[\phi]$ obeying $\tilde\rho_R[\phi](\tilde t[\phi])=1$, which once more encodes in a gauge-invariant manner how $A[\phi]$ evolves relative to the dynamical spacetime bulk clock function $t[\phi]=\tilde t[\phi]\circ(R[\phi])^{-1}$.
We emphasize that $\tilde t$ defines the time ``in the perspective'' of the frame $R[\phi]$ and amounts to a choice of (possibly local) foliation in orientation space $\mathscr{O}$. Accordingly, it may (but need not necessarily) correspond to a choice of background frame in $\mathscr{O}$.

The above conditions on $\tilde\rho_R$ are quite restrictive \jjvk{and there is no guarantee that they can be satisfied in general}, but such vector fields do exist in many cases of interest, and we will see examples of them in Section~\ref{Section: more examples / pft}, where we discuss parametrised field field theory, and in Section~\ref{Section: more examples / dust}, where we investigate Brown-Kucha\v{r} dust models.\footnote{For a non-field-theoretic illustration in canonical language, see e.g.\ \cite{Hohn:2018iwn,Hohn:2018toe, Chataignier:2019kof}.} This bulk notion of relational dynamics applicable to non-charged observables is a covariant version of the one prominently appearing in the canonical literature on relational observables, e.g.\ see~\cite{Rovelli:1989jn,Rovelli:1990ph,rovelliQuantumGravity2004,Thiemann:2007pyv,Tambornino:2011vg,Dittrich:2005kc,Dittrich:2007jx}. This includes the notion of relational dynamics underlying the Page-Wootters formalism~\cite{Page:1983uc,Giovannetti:2015qha,Smith:2017pwx} which has been shown to be equivalent to the relational observables framework~\cite{Hoehn:2019fsy,Hoehn:2020epv}.

It should be noted that all the above relational time evolutions correspond to active diffeomorphisms on orientation space $\mathscr{O}$ as they act only on dynamical relational fields. In particular, the associated Hamiltonians do not act on the frame field as its corresponding relational observable $(R[\phi])^*(R[\phi])^{-1}=\operatorname{Id}_\mathscr{O}$ is the identity diffeomorphism on $\mathscr{O}$; it is the non-dynamical tautological observable answering the question ``what is the local configuration of the frame field at the event in spacetime where the frame field has local orientation $o\in\mathscr{O}$?''. This means that the frame does not evolve relative to itself. As $\operatorname{Id}_\mathscr{O}$ is not constant on $\mathscr{O}$, the frame may, however, transform under a passive diffeomorphism.

}

\subsection{Generalised and quasi-local dressings}
\label{Section: general formalism / further generalisations}

In this section, we will comment on some possible generalisations of the above formalism.

Up to now, we have been considering reference frames made up of local dressings, which take values as points in spacetime. However, we can use dressings which are instead valued in some auxiliary space $\mathcal{K}$ upon which $\operatorname{Diff}(\mathcal{M})$ acts. We then define a $\mathcal{K}$-dressing as an observable $k:\mathcal{S}\to\mathcal{K}$ with the gauge-covariance property
\begin{equation}
    k[f_*\phi] = \varphi_f \big(k[\phi]\big) \text{ for all } f\in H[\phi],
\end{equation}
where $\varphi_f$ denotes the action of $f\in\operatorname{Diff}(\mathcal{M})$ on $\mathcal{K}$.

A $\mathcal{K}$-dressing can be used to dress observables $A$ which take values as functions $A[\phi]:\mathcal{K}\to\mathcal{L}$, for some additional space $\mathcal{L}$. If $A$ has the property
\begin{equation}
    A[f_*\phi] = A[\phi]\circ \varphi_{f^{-1}} \text{ for all  } f\in H[\phi],
    \label{Equation: generalised dressable}
\end{equation}
then the dressed observable $O_{A,K}$ defined by
\begin{equation}
    O_{A,k}[\phi] = A[\phi](k[\phi])
\end{equation}
will be gauge-invariant, since
\begin{equation}
    O_{A,k}[f_*\phi] = A[f_*\phi](k[f_*\phi]) = A[\phi]\big(\phi_{f^{-1}}(\phi_f(k[\phi]))\big)
    = A[\phi](k[\phi]) = O_{A,k}[\phi] \text{ for all }f\in H[\phi].
\end{equation}

On the other hand, we could consider the case in which $\mathcal{L}$ itself also carries an action of $\operatorname{Diff}(\mathcal{M})$, which for convenience we will also denote $\varphi_f$, and $A$ doesn't obey~\eqref{Equation: generalised dressable} but instead satisfies
\begin{equation}
    A[f_*\phi] = \varphi_f\circ A[\phi] \circ \varphi_{f^{-1}} \text{ for all } f\in H[\phi].
    \label{Equation: generalised dressable 2}
\end{equation}
In this case, the dressed observable is not gauge-invariant, but satisfies
\begin{multline}
    O_{A,k}[f_*\phi] = A[f_*\phi](k[f_*\phi]) = \varphi_f \big(A[\phi]\big(\varphi_{f^{-1}}(\varphi_f(k[\phi]))\big)\\
        = \varphi_f A[\phi](k[\phi]) = \varphi_f( O_{A,k}[\phi]) \text{ for all } f\in H[\phi].
\end{multline}
Thus, $O_{A,k}$ is an $\mathcal{L}$-dressing, which we could now use to dress another observable $B$ taking values as a function on $\mathcal{L}$:
\begin{equation}
    O_{B,A,k}[\phi] = O_{B,O_{A,k}}[\phi] = B[\phi](O_{A,k}[\phi]) = B[\phi](A[\phi](k[\phi])).
\end{equation}
If $B$ behaves like the $A$ in~\eqref{Equation: generalised dressable}, then $O_{B,A,k}$ would be gauge-invariant. On the other hand, if $B$ behaves like the $A$ in~\eqref{Equation: generalised dressable 2}, then $O_{B,A,k}$ would be a dressing taking values in some new space, which we could then use to dress another observable, etc.

Let us define the universal $\mathcal{K}$-dressing space as the space of all possible $\mathcal{K}$-dressings:
\begin{equation}
    \mathscr{D}(\mathcal{K}) = \{k:\mathcal{S}\to\mathcal{K} \mid k[f_*\phi] = \varphi_f\big(k[\phi]\big)\}.
\end{equation}
This is a bundle over $\mathcal{K}$ with field-dependent projection $\pi[\phi]:\mathscr{D}(\mathcal{K})\to \mathcal{K}$ given by
\begin{equation}
    \pi[\phi](k) = k[\phi].
\end{equation}
A $\mathcal{K}$-frame $\mathscr{R}$ can then be defined as a subset of $\mathscr{D}(\mathcal{K})$, and a parametrised $\mathcal{K}$-frame can be understood as an injective map $R:\mathscr{O}\to\mathscr{D}(\mathcal{K})$. For each $\phi$ this gives a map $R[\phi]:\mathscr{O}\to\mathcal{K}$ via $R[\phi] = \pi[\phi]\circ R$.

If $\mathcal{K}$ and $\mathscr{O}$ have the appropriate structures (e.g.\ they are differential manifolds), and $R[\phi]$ is injective sufficiently smooth, then we can use $R$ to construct relational observables in basically the same way as with the case of local dressings. To be explicit, suppose $A$ is a local observable on $\mathcal{K}$. Thus, it is either local quantity valued, so $A:\mathcal{S}\to\mathcal{B}$ where $\mathcal{B}$ is a bundle over $\mathcal{K}$, or local map valued, so $A:\mathcal{S}\to\Gamma(\mathcal{B}_1,\mathcal{B}_2)$ where $\mathcal{B}_1,\mathcal{B}_2$ are bundles over $\mathcal{K}$. We say that $A$ is covariant if for all $f\in\operatorname{Diff}(\mathcal{M})$ we have
\begin{equation}
    A[f_*\phi] = X_{\varphi_f} (A[\phi]) \text{ for local quantities, } A[f_*\phi] = (\varphi_f)_* A[\phi] \text{ for local maps}.
\end{equation}
Here we are also assuming that the action of $f$ on $\mathcal{K}$ is sufficiently smooth that it is a diffeomorphism, i.e.\ $\varphi_f\in\operatorname{Diff}(\mathcal{K})$, so that the above is well-defined. Then the relational observable of $A$ relative to $R$ is
\begin{equation}
    O_{A,R}[\phi] = (X_{R[\phi]})^{-1}A[\phi] \text{ for local quantities, } O_{A,R}[\phi] = (R[\phi])^*A[\phi] \text{ for local maps}.
\end{equation}
If $A$ is covariant, then $O_{A,R}$ will be gauge-invariant.

Much of the previous discussion involving frames that use local dressings also applies to these generalised frames. Indeed, we get back the local dressing case by setting $\mathcal{K}=\mathcal{M}$, and $\varphi_f(x) = f(x)$ for $x\in\mathcal{M}$.

But we don't have to set $\mathcal{K}=\mathcal{M}$. We just need $\mathcal{K}$ to have a well-behaved action of $\operatorname{Diff}(\mathcal{M})$. For example, we could set $\mathcal{K}$ to be some bundle over $\mathcal{M}$. A point in $\mathcal{K}$ would then be associated to a point in $\mathcal{M}$, but it contains some extra local properties. For example, we could have $\mathcal{K}=\mathrm{T}\mathcal{M}$. A $\mathrm{T}\mathcal{M}$-dressing gives a point in $\mathcal{M}$ and a vector at that point. Actually, given any parametrised local reference frame $R:\mathscr{O}\to\mathscr{D}$ we can construct a large family of $\mathrm{T}\mathcal{M}$-dressings. The differential of $R[\phi]$ is a map
\begin{equation}
    \dd(R[\phi]):\mathrm{T}\mathscr{O} \to \mathrm{T}\mathcal{M}.
\end{equation}
For each $v\in\mathrm{T}\mathscr{O}$ the observable $W_v$ defined by
\begin{equation}
    W_v[\phi] = \dd(R[\phi])(v)
\end{equation}
is a $\mathrm{T}\mathcal{M}$ dressing, since
\begin{multline}
    W_v[f_*\phi] = \dd(R[f_*\phi])(v) = \dd(f\circ R[\phi])(v) = \big(\dd{f} \circ \dd(R[\phi])\big)(v)\\
    = \dd{f}\big(\dd(R[\phi])(v)\big) = \dd{f}(W_v[\phi]) = \varphi_f\big( W_v[\phi]\big) \text{ for all } f\in H[\phi],
\end{multline}
where we have identified the action $\varphi_f$ of $f\in \operatorname{Diff}(\mathcal{M})$ on $\mathrm{T}\mathcal{M}$ with the differential map $\dd{f}$. By varying over all $v\in\mathrm{T}\mathscr{O}$, we get an entire parametrised $\mathrm{T}\mathcal{M}$-frame $S$:
\begin{equation}
    S: \quad \mathrm{T}\mathscr{O}\to \mathscr{D}(\mathrm{T}\mathcal{M}), \quad v \mapsto W_v.
\end{equation}
The same general argument applies for other equivariant bundles $\mathcal{K}$ over $\mathcal{M}$ if we replace the differentials by the appropriate actions $X_f$ of $f\in\operatorname{Diff}(\mathcal{M})$.

Another interesting case is where $\mathcal{K}$ is the space of all spacetime subsets $\mathcal{U}\subset\mathcal{M}$. This space has a natural action of $\operatorname{Diff}(\mathcal{M})$ given by $\varphi_f(\mathcal{U}) = f(\mathcal{U})$. We could also consider $\mathcal{K}$ which consists of only a restricted class of subsets of $\mathcal{U}\subset\mathcal{M}$, so long as that class of subsets is closed under $\varphi_f$. For example, we could let $\mathcal{K}$ consist of all $k$-dimensional submanifolds of $\mathcal{M}$. We call $\mathcal{K}$-dressings, for $\mathcal{K}$ of this type, `quasilocal dressings', and we call the resulting $\mathcal{K}$-frames `quasilocal frames'. Later in the paper, we will give an example of a quasilocal frame involving minimal surfaces (see Section~\ref{Section: more examples / minimal surfaces}).

It should be noted that the image of any local reference frame is a quasilocal dressing. Indeed, a reference frame $\mathscr{R}$ is a subset of $\mathscr{D}$, and its image is given by
\begin{equation}
    \mathcal{N}[\phi] = \bigcup_{x\in\mathscr{R}}x[\phi].
\end{equation}
Thus, we have
\begin{multline}
    \mathcal{N}[f_*\phi] = \bigcup_{x\in\mathscr{R}}x[f_*\phi] = \bigcup_{x\in\mathscr{R}}f(x[\phi]) = f\qty(\bigcup_{x\in\mathscr{R}}x[\phi]) \\
    = f(\mathcal{N}[\phi]) = \varphi_f\big(\mathcal{N}[\phi]\big)\text{ for all }f\in H[\phi].
\end{multline}
Clearly, the image of a frame $\mathcal{N}[\phi]$ contains less information than the frame itself. That is, more than one frame $\mathscr{R}_1,\mathscr{R}_2$ can have the same image. It is interesting to ask if there are quasilocal dressings which are not the image of any frame. We will not address this question in the current work, instead leaving it for future exploration.

\subsection{Comparison with the canonical approach to relational observables}
\label{Section: general formalism / covariant to canonical}

Let us now give a discussion of how the relational observables we have constructed are related to the canonical ones described in~\cite{Dittrich:2005kc}. The arguments presented here are similar to those appearing in~\cite{Carrozza:2020bnj} \pah{for relational observables in non-gravitational gauge theories}.

The canonical formalism is based on the kinematical phase space, which may be constructed from the space of kinematical field configurations $\Gamma(\Phi)$ in the following way \pah{\cite{Lee:1990nz}}. First, one picks a Cauchy surface $\Sigma\subset\mathcal{M}$, which in general is field-dependent.\footnote{In gravitational theories $\Sigma=\Sigma[\phi]$ must be field-dependent to be a genuine Cauchy surface for all possible choices of the metric.} One then obtains a presymplectic form on $\Gamma(\Phi)$ by integrating the presymplectic current $\omega$ over this Cauchy surface:\footnote{Here we are assuming that boundary conditions have been dealt with by adding to $\omega$ an appropriate exact spacetime form.}
\begin{equation}
    \Omega^\Sigma[\phi,\delta_1\phi,\delta_2\phi] = \int_{\Sigma[\phi]}\omega[\phi,\delta_1\phi,\delta_2\phi].
\end{equation}
This presymplectic form has degenerate directions which induces an equivalence relation $\sim_\Sigma$ on $\Gamma(\Phi)$.\footnote{Essentially, two kinematical field configurations are equivalent if they can be connected by a curve which is everywhere tangent to a degenerate direction of $\Omega^\Sigma$.} The kinematical phase space on $\Sigma$ is then defined as the quotient space
\begin{equation}
    \mathcal{P}_{\text{kin.}}^\Sigma = \Gamma(\Phi)/\sim_\Sigma.
\end{equation}
Each element of $\mathcal{P}_{\text{kin}.}$ can be thought of as being labelled by the canonical initial data and conjugate momenta of the fields on $\Sigma$. The symplectic form $\Omega_{\text{kin.}}^\Sigma$ on this phase space is the unique 2-form obeying
\begin{equation}
    \big(\pi^\Sigma\big)^*\Omega_{\text{kin.}}^\Sigma = \Omega^\Sigma,
\end{equation}
where $\pi^\Sigma:\Gamma(\Phi)\to \mathcal{P}^\Sigma_{\text{kin.}}$ is the map from each kinematical field configuration to its equivalence class in the kinematical phase space. The kinematical Poisson bracket $\pb{\cdot}{\cdot}_\Sigma$ is the one associated with this symplectic form. Since this is a gauge theory, the kinematical state is subject to some constraints -- in particular physical states must lie on the constraint surface
\begin{equation}
    \mathcal{C} = \mathcal{S}/\sim_\Sigma\, = \pi^\Sigma(\mathcal{S}).
\end{equation}
The pullback of $\Omega_{\text{kin.}}^\Sigma$ to $\mathcal{S}$ is degenerate, with its degenerate directions corresponding to gauge transformations. If one reduces again by the degenerate directions of $\Omega_{\text{kin.}}^\Sigma$ on $\mathcal{S}$, it can be shown that one obtains the physical space of states $\mathcal{P}=\mathcal{S}/\sim$, which doesn't depend on the choice of $\Sigma$.

The full spacetime diffeomorphism group $\operatorname{Diff}(\mathcal{M})$ acts on $\Gamma(\Phi)$, but it isn't necessarily true that this gives a well-behaved action on $\mathcal{P}_{\text{kin.}}$. Ideally, we would define this action in the following way. Suppose $\phi^\Sigma\in \mathcal{P}^\Sigma_{\text{kin.}}$ is represented by the kinematical field configuration $\phi\in\Gamma(\Phi)$, so
\begin{equation}
    \phi^\Sigma = \pi^\Sigma[\phi].
\end{equation}
Under an active diffeomorphism $f\in\operatorname{Diff}(\mathcal{M})$, this transforms to
\begin{equation}
    Q_f[\phi^\Sigma] = \pi^\Sigma[f_*\phi],
\end{equation}
which we can use to define $Q_f:\mathcal{P}_{\text{kin.}}^\Sigma\to\mathcal{P}_{\text{kin.}}^\Sigma$. However, this definition only works if this equation also holds for any other representative $\tilde\phi$ of $\phi^\Sigma$, so
\begin{equation}
    Q_f[\phi^\Sigma] = \pi^\Sigma[f_*\tilde\phi].
\end{equation}
Thus, for $f$ to be well-defined we need
\begin{equation}
    \phi\mathop{\sim_\Sigma} \tilde\phi \implies f_*\phi\mathop{\sim_\Sigma} f_*\tilde\phi.
    \label{Equation: f acts kinematical phase space}
\end{equation}
Let us find a sufficient condition for this to be true. It suffices to consider $\phi$ and $\tilde\phi$ which are infinitesimally close, so let us write $\tilde\phi=\phi+\delta_1\phi$. The equivalence $\phi\sim\tilde\phi$ is then basically the statement that $\delta_1\phi$ is a degenerate direction of $\Omega^\Sigma$, i.e.
\begin{equation}
    \Omega^\Sigma[\phi,\delta_1\phi,\delta_2\phi] = \int_{\Sigma[\phi]}\omega[\phi,\delta_1\phi,\delta_2\phi] = 0 \text{ for all } \delta_2\phi.
    \label{Equation: OmegaSigma degenerate}
\end{equation}
Since $\omega$ is covariantly constructed from $\phi,\delta_1\phi,\delta_2\phi$, we can write
\begin{equation}
    \Omega^\Sigma[f_*\phi,f_*\delta_1\phi,f_*\delta_2\phi] = \int_{\Sigma[f_*\phi]}\omega[f_*\phi,f_*\delta_1\phi,f_*\delta_2\phi] = \int_{f^{-1}(\Sigma[f_*\phi])}\omega[\phi,\delta_1\phi,\delta_2\phi].
\end{equation}
If $f^{-1}(\Sigma[f_*\phi])= \Sigma[\phi]$, then by~\eqref{Equation: OmegaSigma degenerate} this must vanish for all $\delta_2\phi$, and so $f_*\phi\to f_*\phi+f_*\delta_1\phi=f_*\tilde\phi$ would have to be a degenerate direction of $\Omega^\Sigma$. If $f^{-1}(\Sigma[f_*\phi])= \Sigma[\phi]$ holds for all $\phi$, then this gives a sufficient condition for~\eqref{Equation: f acts kinematical phase space}. For convenience, we will assume that this is true for all diffeomorphisms $f$, i.e.\ that $\Sigma$ is covariant:
\begin{equation}
    \Sigma[f_*\phi] = f(\Sigma[\phi]) \text{ for all } f\in\operatorname{Diff}(\mathcal{M}).
    \label{Equation: covariant Sigma}
\end{equation}
Thus, $Q_f$ is well-defined for all $f\in\operatorname{Diff}(\mathcal{M})$. Moreover, it is clear that $Q_f$ gives a group action of $\operatorname{Diff}(\mathcal{M})$ on $\mathcal{P}_{\text{kin.}}^\Sigma$.\footnote{In more general situations, $\Sigma[f_*\phi]=f(\Sigma[\phi])$ will only hold for certain $f$, and $Q_f$ will only be well-defined for those $f$.}

Recall that for $\phi\in\mathcal{S}$ the subsets $G[\phi]\subset\operatorname{Diff}(\mathcal{M})$ is defined such that $f\in G[\phi]$ means $f_*\phi\in \mathcal{S}$. The analogous statement in the kinematical phase space is
\begin{equation}
    \phi^\Sigma\in\mathcal{C} \implies Q_f(\phi^\Sigma)\in\mathcal{C} \text{ for all } f\in G[\phi], \text{ where } \phi^\Sigma = \pi^\Sigma(\phi)
\end{equation}
This holds for all $\phi$ which represent $\phi^\Sigma$.

Consider an observable $A$ which takes the value $A[\phi]$ on the kinematical field configuration $\phi\in\Gamma(\Phi)$. Let us assume that on-shell this observable may be written purely in terms of the initial data and conjugate momenta on $\Sigma$. This means that
\begin{equation}
    A[\phi] = \tilde{A}[\pi^\Sigma[\phi]] \text{ for all } \phi \in \mathcal{S},
\end{equation}
where $\tilde{A}$ is some function on the constraint surface $\mathcal{C}$. Under an active diffeomorphism $\phi\to f_*\phi$, we have
\begin{equation}
    A[\phi] \to \tilde{A}[\pi^\Sigma[f_*\phi]] = \tilde{A}[Q_f[\pi^\Sigma[\phi]]] \text{ for all } \phi \in \mathcal{S}.
\end{equation}
If $A$ is a covariant local observable, then we know that $A[\phi]\to f_*(A[\phi])$. In combination this yields
\begin{equation}
    \tilde{A}[Q_f[\phi^\Sigma]] = f_*(\tilde{A}[\phi^\Sigma]) \text{ for all } \phi^\Sigma \in \mathcal{C}.
    \label{Equation: kinematical covariance}
\end{equation}

Suppose also that a parametrised frame $R:\mathscr{O}\to\mathscr{D}$ may be written on-shell purely in terms of the initial data and conjugate momenta on $\Sigma$, so
\begin{equation}
    R[\phi] = \tilde{R}[\pi^\Sigma[\phi]] \text{ for all } \phi\in\mathcal{S},
\end{equation}
for some $\tilde{R}[\phi]:\mathscr{O}\to\mathcal{M}$. For $f\in H[\phi]$, we have
\begin{equation}
    f\circ R[\phi] = R[f_*\phi] = \tilde{R}[\pi^\Sigma[f_*\phi]] = \tilde{R}[Q_f[\pi^\Sigma[\phi]]] \text{ for all } \phi\in\mathcal{S},
\end{equation}
so
\begin{equation}
    f\circ \tilde{R}[\phi^\Sigma] = \tilde{R}[Q_f[\phi^\Sigma]] \text{ for all } \phi^\Sigma\in\mathcal{C}.
\end{equation}

The relational observable
\begin{equation}
    O_{A,R}[\phi] = (R[\phi])^* A[\phi]
\end{equation}
formed from $A$ and $R$ may then on-shell be written in terms of the initial data and conjugate momenta via
\begin{equation}
    O_{A,R}[\phi] = \tilde{O}_{A,R}[\pi^\Sigma[\phi]],
\end{equation}
where
\begin{equation}
    \tilde{O}_{A,R}[\phi^\Sigma] = (\tilde{R}[\phi^\Sigma])^* \tilde{A}[\phi^\Sigma].
    \label{Equation: covariant and canonical O A R}
\end{equation}
Moreover, $\tilde{O}_{A,R}$ is by construction gauge-invariant on the constraint surface $\mathcal{C}$. Let us extend $\tilde{R}$ and $\tilde{A}$ arbitrarily to functions on all of $\mathcal{P}^\Sigma_{\text{kin.}}$. We may not be able to do this in such a way that $\tilde{O}_{A,R}$ is gauge-invariant away from $\mathcal{C}$.

Let us now fix a $\phi_0^\Sigma\in\mathcal{P}_{\text{kin.}}^\Sigma$, and decompose
\begin{equation}
    \tilde{R}[\phi^\Sigma] = \tilde{U}[\phi^\Sigma] \circ \tilde{R}_0, \text{ where } \tilde{R}_0 = \tilde{R}[\phi^\Sigma_0], \, \tilde{U}[\phi^\Sigma] = \tilde{R}[\phi^\Sigma]\circ (\tilde{R}[\phi^\Sigma_0])^{-1}.
\end{equation}
Here $\tilde{R}_0$ is a fixed diffeomorphism from $\mathscr{O}\to\mathcal{M}$, while $\tilde{U}_0[\phi^\Sigma]$ is a diffeomorphism from $\mathcal{M}$ to itself. We may then write
\begin{equation}
    \tilde{O}_{A,R}[\phi^\Sigma] = (\tilde{R}_0)_*(\tilde{U}[\phi^\Sigma])^* \tilde{A}[\phi^\Sigma].
\end{equation}
(Note that this step is unnecessary if $\mathscr{O}=\mathcal{M}$, in which case we can just set $\tilde{R}_0$ to the identity, and $\tilde{R}=\tilde{U}$.) Now, using the property~\eqref{Equation: kinematical covariance}, we have
\begin{equation}
    \tilde{O}_{A,R}[\phi^\Sigma] \approx (\tilde{R}_0)_*\tilde{A}\big[Q_{(\tilde{U}[\phi^\Sigma])^{-1}}[\phi^\Sigma]\big] = \tilde{A}_0\big[Q_{(\tilde{U}[\phi^\Sigma])^{-1}}[\phi^\Sigma]\big],
    \label{Equation: kinematical relational observable}
\end{equation}
where $\approx$ denotes equality on the constraint surface $\mathcal{C}$, a.k.a.\ `weak equality', and we are defining $\tilde{A}_0$ by $\tilde{A}_0[\phi^\Sigma] = (\tilde{R}_0)_*\tilde{A}[\phi^\Sigma]$.

At this point, we would like to write this relational observable in the form given in~\cite{Dittrich:2005kc}, i.e.\ as a power series of Poisson brackets. However, to do so we need the spacetime diffeomorphism $\tilde{U}[\phi^\Sigma]$ to be small, which is not always going to be true. However, let us now restrict to the case where $A$ is kinematically local (so $O_{A,R}$ is relationally local), and let us also restrict to evaluating $O_{A,R}[\phi]$ in the bulk, i.e.\ at local orientations in $\mathscr{O}$ which correspond to spacetime points away from $\partial\mathcal{M}$. Then~\eqref{Equation: kinematical relational observable} does not depend at all on the behaviour of $\tilde{U}[\phi^\Sigma]$ in a neighbourhood of $\partial\mathcal{M}$, so we can replace it by a different diffeomorphism $\tilde{U}'[\phi^\Sigma]$ which vanishes in a neighbourhood of $\partial\mathcal{M}$, but agrees with $\tilde{U}[\phi^\Sigma]$:
\begin{equation}
    \tilde{O}_{A,R}[\phi^\Sigma] \approx \tilde{A}_0\big[Q_{(\tilde{U}'[\phi^\Sigma])^{-1}}[\phi^\Sigma]\big].
\end{equation}
Note that such a $\tilde{U}'[\phi^\Sigma]$ is a small diffeomorphism. 

Let us define the map $F:\mathcal{P}^\Sigma_{\text{kin.}}\to\mathcal{P}^\Sigma_{\text{kin.}}$ by
\begin{equation}
    F[\phi^\Sigma] = Q_{(\tilde{U}'[\phi^\Sigma])^{-1}}[\phi^\Sigma].
\end{equation}
Since $\tilde{U}'[\phi^\Sigma]$ is a finite small diffeomorphism, $F$ is a finite gauge transformation. In order to decompose it using Poisson brackets, we need to be able to write it as the exponential of an infinitesimal gauge transformation, i.e.\ we need to be able to write $F = e^{\mathcal{X}}$ for some vector field $\mathcal{X}$ on $\mathcal{P}^\Sigma_{\text{kin.}}$. It is clear that this is not going to be possible in general. Indeed, suppose we can write the diffeomorphism $\tilde{U}'[\phi^\Sigma]$ as the exponential of some spacetime vector field $u[\phi^\Sigma]$, i.e.\footnote{Actually, this only need to be true up to a field-independent spacetime diffeomorphism. That is, we only need
\begin{equation}
    \tilde{U}'[\phi]=e^{u[\phi]} \circ \tilde{U}_0' ,
\end{equation}
where $\tilde{U}'_0$ is a field-independent spacetime diffeomorphism that \textit{cannot} be written as an exponentiated vector field. However, we can just absorb this factor of $\tilde{U}_0'$ into $\tilde{R}_0$ by redefining 
\begin{equation}
    \tilde{R}_0 \to \tilde{U}_0'\circ \tilde{R}_0, \qq{and} \tilde{U}'[\phi] \to \tilde{U}'[\phi]\circ (\tilde{U}_0')^{-1}.
\end{equation}
Then we will again be in the case~\eqref{Equation: small diffeo exponential}, and all the following arguments will hold.}
\begin{equation}
    \tilde{U}'[\phi^\Sigma] = e^{u[\phi^\Sigma]}.
    \label{Equation: small diffeo exponential}
\end{equation}
Then, if the map $f\mapsto Q_f$, $f\in\operatorname{Diff}(\mathcal{M})$ is sufficiently smooth, we will be able to write $F=e^{\mathcal{X}_u}$, for some vector field $\mathcal{X}_u$ that is determined by $u[\phi^\sigma]$. However, the smoothness of $f\mapsto Q_f$ is not in general guaranteed. Even if it is sufficiently smooth, it is known that there are elements of the diffeomorphism group that cannot be reached as in~\eqref{Equation: small diffeo exponential} by exponentiating a vector field~\cite{krieglConvenientSettingGlobal1997}.

Let us assume that these requirements are satisfied, so that we \emph{can} write $F = e^{\mathcal{X}_u}$. Then we may write
\begin{equation}
    \tilde{O}_{A,R} \approx e^{\mathcal{X}_u} \tilde{A}_0 = \sum_{n=0}^\infty \frac1{n!} (\mathcal{X}_u)^n(\tilde{A}_0).
    \label{Equation: tilde O A R X}
\end{equation}
Note that since $\tilde{U}'[\phi^\Sigma]$ vanishes in a neighbourhood of $\partial\mathcal{M}$, so too does $u[\phi^\Sigma]$. Thus, $u[\phi^\Sigma]$ corresponds to an infinitesimal small diffeomorphism, so $\mathcal{X}_u$ is an infinitesimal gauge transformation. Thus, when we contract the vector field $\mathcal{X}_u$ into the kinematical symplectic form $\Omega^\Sigma_{\text{phys.}}$, we must get a 1-form which vanishes when we pull it back to the constraint surface. Such a 1-form may in general be written
\begin{equation}
    \iota_{\mathcal{X}_u}\Omega^\Sigma_{\text{kin.}} = \alpha + \sum_i \beta_i \dd{\gamma_i},
\end{equation}
where $\alpha$ is a 1-form which vanishes at $\mathcal{C}$, and $\beta_i,\gamma_i$ are a set of functions satisfying $\gamma_i=0$ on $\mathcal{C}$. Then we have
\begin{equation}
\iota_{\mathcal{X}_u}\Omega^\Sigma_{\text{kin.}} \approx \sum_i\beta_i\dd{\gamma_i} = \sum_i\big(\dd(\beta_i\gamma_i) - \underbrace{\gamma_i}_{\approx 0}\dd{\beta_i}\big) \approx -\dd(C_u),
\end{equation}
where $C_u=-\sum_i\beta_i\gamma_i$. Thus, $C_u$ is the Hamiltonian generator of a transformation of $\phi_\Sigma$ that is equivalent to the gauge transformation $\mathcal{X}_u$ on the constraint surface $\mathcal{C}$ (but not necessarily away from $\mathcal{C}$). Note that $C_u=0$ on the constraint surface.

We can now replace every instance of $\mathcal{X}_u$ in~\eqref{Equation: tilde O A R X} with a Poisson bracket with $C_u$. In particular, defining the iterated Poisson bracket
\begin{equation}
    \pb{a}{b}^{n}_\Sigma = \underbrace{\{a,\{a,\{a,\dots,\{a,}_{n}b\}_\Sigma\dots\}_\Sigma\}_\Sigma\}_\Sigma,
\end{equation}
with the additional convention $\pb{a}{b}^{0}_\Sigma = b$, we may write
\begin{equation}
    \tilde{O}_{A,R} \approx \sum_{n=0}^{\infty} \frac{1}{n!} \{C_u,\tilde{A}_0\}^{n}_\Sigma.
\end{equation}
This is exactly the type of expression for canonical relational observables that was established in~\cite{Dittrich:2005kc}. Viewing $\tilde{O}_{A,R}$ as a function on $\mathcal{C}$, we may pull it back to a function on the space of solutions using $\pi^\sigma:\mathcal{S}\to\mathcal{C}$. By~\eqref{Equation: covariant and canonical O A R}, this function is exactly the relational observable $O_{A,R}$ on the space of solutions, so we have
\begin{equation}
    O_{A,R} = \sum_{n=0}^{\infty} \frac{1}{n!} \{C_u,\tilde{A}_0\}^{n}_\Sigma \circ \pi^\Sigma.
\end{equation}
Note that this equation holds only on the space of solutions, and not necessarily in the larger space $\Gamma(\Phi)$ of all kinematical field configurations. \pah{This is the gravity version of the} expression found in~\cite{Carrozza:2020bnj} \pah{for gauge theories}. Note that the map $\tilde{R}_0$ in $\tilde{A}_0[\phi^\Sigma] = (\tilde{R}_0)^*\tilde{A}[\phi^\Sigma]$ is independent of $\phi^\Sigma$, so we can factor it out of the Poisson brackets to write
\begin{equation}
    O_{A,R} = (\tilde{R}_0)^*\circ \sum_{n=0}^{\infty} \frac{1}{n!} \{C_u,\tilde{A}\}^{n}_\Sigma \circ \pi^\Sigma.
\end{equation}

This shows that our general formalism (under some restrictions on the reference frame) is compatible with the canonical approach. \pah{Together with Section~\ref{Section: general formalism / relational picture / single integrals smearing}, we have thus established equivalence of three \emph{a priori} distinct formulations of relational observables: the single integral one~\cite{DeWitt:1962cg,Marolf:1994wh,Giddings:2005id,Marolf:2015jha,Donnelly:2016rvo}, our covariant one, and the canonical power series formulation framework of~\cite{Dittrich:2004cb,Dittrich:2005kc,Dittrich:2006ee,Dittrich:2007jx} (the equivalence of the latter with the former two up to the mentioned restrictions).}  However, there are many advantages to using the covariant framework we have described instead of the canonical one. In addition to not imposing the restrictions on the frames that were required here, the covariant approach works for any type of spacetime, while the canonical setting requires a Cauchy surface, and so only works for globally hyperbolic spacetimes. Even if we are in a globally hyperbolic spacetime, it may still be practically difficult to construct an appropriate covariant Cauchy surface $\Sigma[\phi]$ obeying~\eqref{Equation: covariant Sigma}. Finally, in our opinion the covariant approach that we have described in this paper is much more conceptually accessible.

\section{More examples}
\label{Section: more examples}

Having described the abstract general formalism, our aim now is to discuss a few more examples of its application. These examples will not touch on every aspect of the formalism, but they will hopefully provide some intuition for a few of its ingredients, and demonstrate its versatility. 

First, we will discuss what are known as `parametrised field theories', which are a simple class of generally covariant field theories which can be constructed from ordinary field theories by introducing an `embedding field'~\cite{Isham:1984sb,Kuchar:1989bk,Kuchar:1989wz}. Then, we will consider the case of the Brown-Kucha\v{r} dust~~\cite{Brown:1994py}. As in the geodesic example discussed in Section~\ref{Section: geodesic example}, in both of these examples, we will construct reference frames by making a choice of dressings. That is, we find a way to parametrise spacetime via some gauge-covariant fields (the embedding fields in the case of parametrised field theory, and the dust scalars in the case of Brown-Kucha\v{r} dust). Each of these dressings are necessarily elements of the universal dressing space $\mathscr{D}$. We can then use these frames in conjunction with covariant observables to form relational observables. 

The third and final example involves minimal surfaces. Its main purpose is to demonstrate \pah{the generality of the formalism, and that it} can be applied in \pah{sufficiently flexible} ways \pah{that may be of particular} use in holography~\cite{MinimalSurfaces1,MinimalSurfaces2,MinimalSurfaces3,MinimalSurfaces4,MinimalSurfaces5}.

\subsection{Parametrised field theory}
\label{Section: more examples / pft}

Let us begin by considering the case of a parametrised field theory based on a free scalar field in flat spacetime. Such a theory describes the evolution of the scalar field with respect to an arbitrary set of dynamical coordinates. Introducing these dynamical coordinates \pah{as embedding fields} allows us to recast the usual action into a manifestly diffeomorphism invariant one. Although we will focus on the case of a scalar field, by introducing dynamical coordinates in this way one can construct diffeomorphism invariant parametrised field theories from almost any field theory on a fixed spacetime. As we will see, the dynamical coordinates provide perhaps the simplest kind of dynamical reference frame possible.

The original purpose of parametrised field theory was to study issues arising in the canonical approach to quantum gravity. The hope was that by using a simple theory made diffeomorphism invariant, one could understand the fundamental aspects of diffeomorphism invariance in quantum gravity~\cite{Isham:1984sb,Kuchar:1989bk,Kuchar:1989wz,Torre:1992rg}. It has also been studied in the covariant phase space setting, without boundary~\cite{Lee:1990nz,Torre:1992bg} and with boundary~\cite{Andrade:2010hx}.

In the following, we will describe the parametrised field theory based on a free scalar field, and show how it can be used to write down simple relational observables. We will also discuss frame reorientations.

We start by considering a $D$-dimensional flat spacetime equipped with some global inertial coordinates $X^\alpha$, with respect to which the components of the metric are $\eta_{\alpha \beta}=\operatorname{diag}(-1,1,\dots,1)$.\footnote{Actually, it is not strictly necessary for $X^\alpha$ to be global inertial coordinates -- indeed they could be any coordinates on spacetime, and the construction would still work.} The action for a free scalar field $\psi$ is then given by
\begin{equation}
    S[\psi] = -\frac12\int_{\mathscr{O}} \dd[D]{X} \eta^{\alpha \beta} \partial_\alpha \psi \partial_\beta \psi,
\end{equation}
where we denote this initial spacetime by $\mathscr{O}$ because it will eventually be understood as the local orientation space of a parametrised frame. This action is a functional of the scalar field $\psi$ only. To keep the presentation as simple and streamlined as possible, we will assume that the boundary conditions on $\psi$ are such that we can ignore any effects that occur at the boundary of $\mathscr{O}$ in the following. Such boundary effects would not change much about the presentation. Indeed, the previous action already has a well-defined variational principle for Dirichlet boundary conditions on $\psi$, while it does need some extra works for the Neumann and Robin boundary conditions (see~\cite{Andrade:2010hx}). The reader concerned about possible implications of boundary terms may, if they wish, view the following arguments as applying to compactified flat spacetimes only.

To fully understand the parametrised field theory action, it is convenient to introduce two copies of spacetime. We will keep using the symbol $\mathscr{O}$ for the initial spacetime, while we denote the other copy $\mathcal{M}$. The point of the parametrisation process is to enlarge the configuration space of the fields by the space of diffeomorphisms from the first copy of spacetime $\mathscr{O}$ to the second copy $\mathcal{M}$. It is quite natural that doing so requires us to `unfreeze' the global inertial coordinates $X^{\alpha}$, so let us consider another \textit{arbitrary} coordinate system $x^\mu$ on $\mathcal{M}$ such that $X^\alpha$ is understood as a function of the new coordinates $x^\mu$. In terms of the two copies of spacetime, it can naturally be understood as a diffeomorphism from $\mathcal{M}$ to $\mathscr{O}$
\begin{equation}
    X: \quad\mathcal{M} \to \mathscr{O}, \quad x \mapsto X(x).
\end{equation}
This diffeomorphism is sometimes known as an `embedding field'. \pah{It is an example of the reference frame field $(R[\psi])^{-1}$ in \eqref{eq:framefield}.}

We can now write the action as an integral over $\mathcal{M}$ by pulling everything back through this diffeomorphism. It becomes simply
\begin{equation}
    S[\psi] = -\frac12\int_{\mathcal{M}} \dd[D]{x} \sqrt{\abs{g}} g^{\mu\nu} \partial_\mu \bar{\psi} \partial_\nu \bar{\psi}.
\end{equation}
where $\bar{\psi} = X^*\psi = \psi\circ X$ and where $g=X^*\eta$ is the induced metric on $\mathcal{M}$ with components
\begin{equation}
    g_{\mu \nu} = \pdv{X^\alpha}{x^\mu} \pdv{X^\beta}{x^\nu} \eta_{\alpha\beta}.
\end{equation}

The action of the parametrised field theory is equivalent to the action of the free scalar field $\psi$, except we now view the function $X$ and the pulled back scalar $\bar\psi$ as the fundamental kinematical fields on which the action depends:
\begin{equation}
    S_{\text{PFT}}[X,\bar\psi] = -\frac12\int_{\mathcal{M}} \dd[D] x \sqrt{\abs{g}} g^{\mu\nu} \partial_\mu \bar{\psi} \partial_\nu \bar{\psi}.
\end{equation}
Note that while $X$ was involved in the construction of $\bar{\psi}$, we are at this stage considering them to be independent fields on $\mathcal{M}$. By construction, the parametrised field theory action is invariant under diffeomorphism on $\mathcal{M}$. In particular, under a diffeomorphism $f$ of $\mathcal{M}$, the fields transform as
\begin{align}
    \bar{\psi} &\rightarrow f_*\bar{\psi} = \bar\psi \circ f^{-1},\\
    \qquad X &\rightarrow X \circ f^{-1}, \label{Equation: X active diffeo}
\end{align}
which implies that the metric transforms as 
\begin{equation}
    g = X^*\eta \to (X\circ f^{-1})^*\eta = f_* X^*\eta = f_* g.
\end{equation}
Thus, both the scalar field $\bar\psi$ and metric $g$ transform by pushforward. Since the integrand in the action is covariantly constructed from them, the action will be invariant under diffeomorphisms.

The embedding field $X:\mathcal{M}\to\mathscr{O}$ is the frame field \pah{\eqref{eq:framefield}} of a reference frame, since it transforms like~\eqref{Equation: X active diffeo}. Let us construct this reference frame explicitly. For each $o\in O$, since $X$ is invertible we can define
\begin{equation}
    R(o)[\phi] = X^{-1}(o),
\end{equation}
where $\phi=(\bar\psi, X)$ denotes all the kinematical fields. The transformation law~\eqref{Equation: X active diffeo} implies that $R(o)$ is a local dressing:
\begin{equation}
    R(o)[f_*\phi] = (X\circ f^{-1})^{-1}(o) = f\big(X^{-1}(o)\big) = f\big(R(o)[\phi]\big).
\end{equation}
Thus, $R(o) \in \mathscr{D}$, and
\begin{equation}
    R:\quad \mathscr{O}\to\mathscr{D},\quad o\mapsto R(o)
\end{equation}
is a parametrised reference frame whose local orientation space is $\mathscr{O}$. Then $R[\phi]:\mathscr{O}\to\mathcal{M}$ is defined by $R[\phi]=\pi[\phi]\circ R$, where $\pi[\phi]:\mathscr{D}\to\mathcal{M}$ is the bundle projection from the universal dressing space to spacetime. In this case, we have
\begin{equation}
    R[\phi](o) = R(o)[\phi] = X^{-1}(o), \quad \text{so} \quad R[\phi] = X^{-1}.
\end{equation}
Thus, the frame field $(R[\phi])^{-1}=X$ is the embedding field, as claimed.

Relational observables using this frame are basically the same as the ordinary observables of the unparametrised scalar field theory. To see this let us consider the example of the relational observable formed from the scalar field $\bar\psi$ of the parametrised field theory (which is covariant). We have
\begin{equation}
    O_{\bar{\psi},R}[\phi] = (R[\phi])^*\bar\psi = X_*\bar\psi = \psi.
\end{equation}
Thus, $O_{\bar\psi,R}$ is just the original scalar field of the unparametrised theory. In this way, the reference frame $R$ essentially encodes the relationship between the unparametrised and parametrised theories.

The simple scenario of parametrised field theory provides a useful setting to gain some intuition about frame reorientations. Recall that to consider frame reorientations one must first pick out some degrees of freedom that one considers to be separated from those of the frame. This can be done using the structure of the kinematical or covariant phase space, as described previously. However, in this case there is a fairly obvious choice to make: we consider the embedding field $X$ to be made up of the frame degrees of freedom, while the parametrised scalar $\bar\psi$ is made up of the `other' degrees of freedom. A frame reorientation is a transformation of the fields that leaves the `other' degrees of freedom invariant, so in this case it is of the form
\begin{equation}
    X\to X', \quad \bar\psi \to \bar\psi.
    \label{Equation: PFT reorientation}
\end{equation}
However, of the transformations of this kind, only those which stay on-shell will be physically allowed.

To see which transformations stay on-shell, let us obtain the equations of motion by varying the action:
\begin{equation}
    \delta(S_{\text{PFT}}[X,\bar\psi]) = -\frac12\int_{\mathcal{M}}\dd[D]{x}\sqrt{\abs{g}} \Big(\big(\partial_\mu\bar\psi\partial_\nu\bar\psi-\frac12 g_{\mu\nu}g^{\rho\sigma}\partial_\rho\bar\psi\partial_\sigma\bar\psi\big)\delta(g^{\mu\nu}) + 2g^{\mu\nu}\partial_\mu\bar\psi\partial_\nu(\delta\bar\psi)\Big).
\end{equation}
Note that the only variation of the metric comes from a variation in $X$, which essentially means we can write 
\begin{equation}
    \delta (g^{\mu\nu}) = -\nabla^\mu \delta X^\nu - \nabla^\nu \delta X^\mu,
\end{equation}
where $\delta X$ is an arbitrary vector field on $\mathcal{M}$. Integrating by parts (and neglecting boundary terms), we find
\begin{equation}
    \delta(S_{\text{PFT}}[X,\bar\psi]) = -\frac12\int_{\mathcal{M}}\dd[D]{x}\sqrt{\abs{g}} \Big(2\nabla^\mu\big(\partial_\mu\bar\psi\partial_\nu\bar\psi-\frac12 g_{\mu\nu}g^{\rho\sigma}\partial_\rho\bar\psi\partial_\sigma\bar\psi\big)\delta X^\nu - 2g^{\mu\nu}\nabla_\mu\nabla_\nu\bar\psi\,\delta\bar\psi\Big).
\end{equation}
Thus, the equations of motion are
\begin{equation}
    g^{\mu\nu}\nabla_\mu\nabla_\nu \bar\psi = 0
    \label{Equation: bar psi eom}
\end{equation}
and
\begin{equation}
    \nabla^\mu\big(\partial_\mu\bar\psi\partial_\nu\bar\psi-\frac12 g_{\mu\nu}g^{\rho\sigma}\partial_\rho\bar\psi\partial_\sigma\bar\psi\big) = 0.
    \label{Equation: X eom}
\end{equation}
Note that the latter equation is just the conservation of the energy momentum tensor of $\bar\psi$. Using the product rule and carrying out some cancellations, it is actually equivalent to
\begin{equation}
    \partial_\rho\bar\psi\, g^{\mu\nu}\nabla_\mu\nabla_\nu\bar\psi = 0.
\end{equation}
Thus, if~\eqref{Equation: bar psi eom} holds, then so does~\eqref{Equation: X eom}. So in fact the only equation of motion we need to consider is~\eqref{Equation: bar psi eom}, which states that $\bar\psi$ obeys the massless wave equation with respect to the metric $g= X_*\eta$.

So let us suppose $\bar\psi$ and $X$ obey these equations of motion, and consider a transformation of the kind~\eqref{Equation: PFT reorientation}. Clearly, $\bar\psi$ must obey the massless wave equation on both the original metric $g=X_*\eta$ and the transformed metric $g' =X'_*\eta$. Generically, this requires $g=g'$, which means that 
\begin{equation}
    \eta = (X')^*X_* \eta = \big((X')^{-1}\circ X\big)_*\eta,
\end{equation}
i.e.\ $(X')^{-1}\circ X$ must be an isometry of $\eta$. Therefore, in this case on-shell frame reorientations are of the form
\begin{equation}
    X\to X' = X\circ f, \quad \bar\psi\to\bar\psi,
    \label{Equation: PFT on-shell reorientation}
\end{equation}
where $f\in\operatorname{Diff}(\mathscr{O})$ is a Poincar\'e transformation. If $X^\alpha$ is a set of global inertial coordinates, then $(X')^\alpha$ will also be a set of global inertial coordinates.

\jjvk{Note that the frame field $X$ is fully covariant. As a consequence (see Section~\ref{Section: large diffeos}), the relational observables formed using the frame will be invariant under all diffeomorphisms, including large ones. Despite this, the frame reorientations just obtained provide us with a notion of relational time evolution, consistent with Section~\ref{Section: relational time evolution}. Indeed, let $\tilde\rho = \pdv{X^0}$ be the vector field on $\mathscr{O}$ which generates evolution along the timelike coordinate $X^0$ in the set of global inertial coordinates $X^\alpha$, and let 
    \begin{equation}
        H_{\tilde\rho} = \frac12\int_{\Sigma_{\mathscr{O}}} \dd[D-1]{X} \big(\partial_0\psi\partial_0\psi + \delta^{ij}\partial_i\psi\partial_j\psi\big),
    \end{equation}
    where $\Sigma_{\mathscr{O}}$ is a constant $X^0$ surface in $\mathscr{O}$. This may be recognised as the Hamiltonian of the unparametrised scalar field $\psi$, but by substituting $\psi=X_*\bar\psi$ it may also be viewed as a Hamiltonian of the parametrised theory. Moreover, it may be confirmed that (up to gauge symmetry) it generates the transformation~\eqref{Equation: PFT on-shell reorientation}, with $f$ being time translation along $X_0$.
}

\pah{Lastly, note that in this simple example, the relational microcausality established in Section~\ref{Section: general formalism / relational phase space / microcausality} is equivalent to the standard notion of microcausality in the Minkowski spacetime $(\mathscr{O},\eta)$ of the original unparametrised field theory.}

\subsection{Brown-Kucha\v{r} dust and changes of frame}
\label{Section: more examples / dust}
Let us next consider how dust matter fields can be used to construct spacetime dressings. The dust model we will consider was first introduced by Brown and Kucha\v{r}~\cite{Brown:1994py}, who showed that it could be used to obtain a privileged set of spatial and temporal coordinates on spacetime. This is exactly what we need: the dust provides us with a dynamical coordinate system that transforms covariantly under small diffeomorphisms. Dust fields are often used to construct dynamical reference frames in this way in the canonical setting~\pah{\cite{Giesel:2007wi,Giesel:2020bht,Tambornino:2011vg,Husain:2011tk}}. Here, we show that they can similarly be used in the covariant setting.

At the end of this example, we will use the dust reference frame to explicitly describe an example of a change of frames, in particular showing how one can change from the geodesic frame discussed at the beginning of the paper to the dust one.

Let us consider a four-dimensional spacetime manifold $\mathcal{M}$, equipped with an arbitrary coordinate system $x^{\mu}$. Dust matter is then parametrised by the set of 8 scalars, $T$, $Z^{k}$, $M$ and $W_{k}$, where $k=1,2,3$. The dust action is given by
\begin{equation}
    S_{\text{dust}}[\phi] = -\frac{1}{2} \int_{\mathcal{M}} \dd[4]{x} \sqrt{\abs{g}} M ( g^{\mu \nu} U_\mu U_\nu + 1),
\end{equation}
where $\phi=(T,Z^{k},M,W_k,g)$, and the four-velocity field $U_\mu$ is defined by
\begin{equation}\label{eq:dustvel}
    U_\mu = -\partial_\mu T + \sum_{k=1}^3 W_k \partial_\mu Z^k .
\end{equation}
We will assume that $M=0$ in a neighbourhood of the boundary, which allows us to ignore boundary terms in the following.

In order to understand how the dust can be used to parametrise spacetime, we need to understand the physical significance of the dust parameters. To do so, let us look at the equations of motion and the dust energy-momentum tensor. The latter is given by
\begin{equation}
  T^{\text{dust}}_{\mu \nu} = -\frac{2}{\sqrt{\text{det}(g)}} \frac{\delta S_\text{dust}}{\delta g_{\mu \nu}} = M U_\mu U_\nu - \frac{M}{2} g_{\mu\nu} (g^{\rho\sigma}U_\rho U_\sigma +1) .
 \end{equation}
We can combine the energy momentum tensor with the equation of motion for the scalar field $M$. Indeed, this equation of motion tells us that $U$ is a unit timelike vector
\begin{equation}
  \frac{\delta S_\text{dust}}{\delta M} = g^{\mu\nu} U_\mu U_\nu + 1 = 0
  \label{Equation: U unit timelike}
\end{equation}
such that the second term of the energy momentum tensor vanishes on-shell. The remaining term is that of a perfect fluid with density $M$ and vanishing pressure.

The $T$ equation of motion is
\begin{equation}
    \nabla_\mu(M U^\mu) = 0,
    \label{Equation: conservation of mass}
\end{equation}
which is a conservation of mass equation. One of its consequences is that if $M=0$ at some point on a flow line generated by the vector field $U$, then $M=0$ everywhere along the flow line. So spacetime can be split into two regions. One region $\mathcal{N}[\phi]$ is defined by 
\begin{equation}
    \mathcal{N}[\phi] = \{x\in\mathcal{M}\mid M(x)\ne 0\},
\end{equation}
and is a congruence of flow lines of $U$. The complement region $\mathcal{M}\setminus\mathcal{N}[\phi]$ is defined by $M=0$.

Let us focus on the region $\mathcal{N}[\phi]$. The equations of motion of $W_k$ tells us that $Z^{k}$ are constant along the flow lines generated by $U$
\begin{equation}
  U^\mu\partial_\mu Z^{k} = 0.
  \label{eq:Zk_constant_U}
\end{equation}
Therefore, the three scalar fields $Z^{k}$ can be used to label each flow line. Combining~\eqref{eq:Zk_constant_U} with~\eqref{Equation: U unit timelike} yields
\begin{equation}
  U^\mu\partial_\mu T = 1,
  \label{eq:T_propertime_U}
\end{equation}
so $T$ measures proper time along each flow line. Finally, combining~\eqref{Equation: conservation of mass} with the conservation of the energy momentum tensor, we have (on-shell)
\begin{equation}
    0 = \nabla_\mu(MU^\mu U^\nu)  = M U^\mu\nabla_\mu U^\nu.
\end{equation}
Thus, the flow lines are geodesics.

Let $\mathcal{D}$ be the set of flow lines in $\mathcal{N}[\phi]$. This set is sometimes known as the `dust space'. Let $P:\mathcal{N}[\phi]\to\mathcal{D}$ be the map that takes each point $x\in\mathcal{N}[\phi]$ to the flow line containing it. Because the fields $Z^k$ are constant on the flow lines, they may be written
\begin{equation}
    Z^k = \tilde{Z}^k\circ P
\end{equation}
for some functions $\tilde{Z}^k:\mathcal{D}\to\RR$. We will assume that each flow line is uniquely determined by the values of the $Z^k$, i.e.\ that the function 
\begin{equation}
    \tilde{Z}:\quad \mathcal{D}\to\RR^3, \quad \tilde{Z}(l) = (\tilde{Z}^1(l),\tilde{Z}^2(l),\tilde{Z}^3(l))
\end{equation}
is injective. We will also assume that it is smooth. Furthermore, for simplicity, we will restrict consideration to field configurations such that the image $\mathscr{O}_\mathcal{D} = \tilde{Z}(\mathcal{D})\subset\RR^3$ is fixed. These assumptions imply that the inverse $\tilde{Z}^{-1}:\mathscr{O}_{\mathcal{D}}\to\mathcal{D}$ is well-defined and smooth.

By~\eqref{eq:T_propertime_U}, $T$ is a monotonic increasing function along each flow line. Thus, given a flow line $l\in\mathcal{D}$, and a time $t\in\RR$, there is a unique point $x\in l$ such that $T(x)=t$ (assuming \pah{$\mathcal{N}[\phi]$} is geodesically complete). Let us therefore define $\mathscr{O} = \RR\times\mathscr{O}_{\mathscr{D}}$ and 
\begin{equation}
    R_d[\phi]: \quad \mathscr{O}\to\mathcal{M},\quad (t,(z^1,z^2,z^3))\mapsto x\text{ such that } x\in \tilde{Z}^{-1}(z^1,z^2,z^3) \text{ and } T(x)=t.
\end{equation}
For each $o=(t,(z^1,z^2,z^3))\in\mathscr{O}$, the map $R_d(o):\phi\mapsto R_d[\phi](o)$ defines a local dressing. The map
\begin{equation}
    R_d:\quad \mathscr{O}\to\mathscr{D}, \quad o\mapsto R_d(o)
\end{equation}
is therefore a parametrised frame with local orientation space $\mathscr{O}$. Its image is $\mathcal{N}[\phi]$. The frame field can be seen to take the values given by $T,Z^k$:
\begin{equation}\label{eq:dustscalars}
    (R_d[\phi])^{-1}(x) = (T(x),(Z^1(x),Z^2(x),Z^3(x))).
\end{equation}
Given any covariant observable $A[\phi]$ \pah{(e.g.\ constructed from the metric)}, we can then write the relational observable
\begin{equation}
  O_{A,R_d} = (R_d[\phi])^* A[\phi].
\end{equation}

\begin{figure}
    \centering
    \begin{tikzpicture}[scale=1.5]
        \begin{scope}
            \fill[blue!10,shift={(0.2,-0.1)}] (1,0) -- (5.9,0) -- (8,1) -- (8,4) -- (4,4) -- (1,0);
            \fill[blue!10,shift={(1.2,-0.1)}] (0,0) -- (3,1) -- (3,4) -- (0,3) -- (0,0);
            \fill[red!10] (2.9,0.2) -- (3.1,0.5) -- (3.15,1.5) -- (3.3,2.1) -- (3.1,3.2) -- (6,3.2)-- (6.1,2.8) -- (6,1.5) -- (5.8,0.2);
            \fill[red!10] (6,3.2)-- (6.1,2.8) -- (6,1.5) -- (5.8,0.2) -- (7.4,1) -- (7.35,2) --(7.5,3.65);
            \fill[rounded corners=10pt,red!10] (2.9,0.2) -- (3.1,0.5) -- (3.15,1.5) -- (3.3,2.1) -- (3.1,3.2) -- (5,3.2) -- cycle;
            \fill[rounded corners=10pt,red!10] (5,1) -- (7.4,1) -- (7.35,2) -- (7.6,2.8) --(7.5,3.65);
            \draw[red!60,fill=red!10,dotted] (3.1,3.2) -- (6.2,3.2) -- (7.5,3.65) -- (4.5,3.65) -- cycle;
            \draw[red!60,dotted] (2.9,0.2) -- (5.8,0.2) -- (7.4,1) -- (4.2,1) -- cycle;
            \draw[blue!50,fill=purple!18,line width=0.8pt,rounded corners=10pt] (3,1.3)-- (6,1.3) -- (7.5,2.2) -- (4.5,2.2) -- cycle;
            \draw[red!60,rounded corners=10pt,dashed] (2.9,0.2) -- (3.1,0.5) -- (3.15,1.5) -- (3.3,2.1) -- (3.1,3.2);
            \draw[red!60,rounded corners=10pt,dashed] (5.8,0.2) -- (6,1.5) -- (6.1,2.8) -- (6.2,3.2);
            \draw[red!60,rounded corners=10pt,dashed] (7.4,1) -- (7.35,2) -- (7.6,2.8) --(7.5,3.65);
            \draw[red!60,rounded corners=10pt,dashed] (4.2,1) -- (4.35,1.5) -- (4.5,2) -- (4.55,3) -- (4.5,3.65);
            \node[purple] at (3.65,1.5) {$\Sigma$};
            \node[red!60!black] at (7.15,2.9) {\large$\mathcal{N}[\phi]$};
            \node[blue] at (2,1.65) {\large$M=0$};
        \end{scope}

        \foreach \a in {0,1}{
          \foreach \b in {1,2,3}{
            % \node[scale=0.8] at (2.6+\b*1 + \a*0.75, 0.4 + \a*0.40) {$\bullet$};
            % \node[scale=0.8] at (2.95+\b*1 + \a*0.65, 3.3 + \a*0.20) {$\bullet$};
            \draw[red!60!black,line width=2.6pt] (2.6+\b*1 + \a*0.75, 0.4 + \a*0.40) .. controls (2.95+\b*1 + \a*0.63, 1.65 + \a*0.20) .. (2.95+\b*1 + \a*0.65, 3.3 + \a*0.20);
            \draw[red!60,line width=1.4pt] (2.6+\b*1 + \a*0.75, 0.4 + \a*0.40) .. controls (2.95+\b*1 + \a*0.63, 1.65 + \a*0.20) .. (2.95+\b*1 + \a*0.65, 3.3 + \a*0.20);
            \fill (2.9+\b*1 + \a*0.64, 1.65 + \a*0.20) circle (0.04);
          }
        }
        \foreach \a in {0,1}{
          \foreach \b in {1,2,3}{
                    \draw[-{Latex},line width=0.9pt,blue] (2.9+\b*1 + \a*0.64, 1.65 + \a*0.20) --+ (0.1,0.7);
            }
        }
        \begin{scope}
            \clip (6,1) rectangle (8,2);
            \draw[blue!50,line width=0.8pt,rounded corners=10pt] (3,1.3)-- (6,1.3) -- (7.5,2.2) -- (4.5,2.2) -- cycle;
        \end{scope}
        \begin{scope}
            \clip (2,1) rectangle (7,1.6);
            \draw[blue!50,line width=0.8pt,rounded corners=10pt] (3,1.3)-- (6,1.3) -- (7.5,2.2) -- (4.5,2.2) -- cycle;
        \end{scope}
    \end{tikzpicture}
    \caption{The region described by the dust frame is $\mathcal{N}[\phi]$ (in pale red), while its complement (in pale blue) corresponds to the region of spacetime where $M$ vanishes. The partial Cauchy slice $\Sigma$ plays the role of the anchor for the dust frame. To each point of $\Sigma$ is associated a set of value $\tilde{Z}(l)$, denoting the flow line (in red), and the blue arrow corresponds to the velocity of each flow line at $\Sigma$.}
    \label{fig:dust_spacetime}
\end{figure}
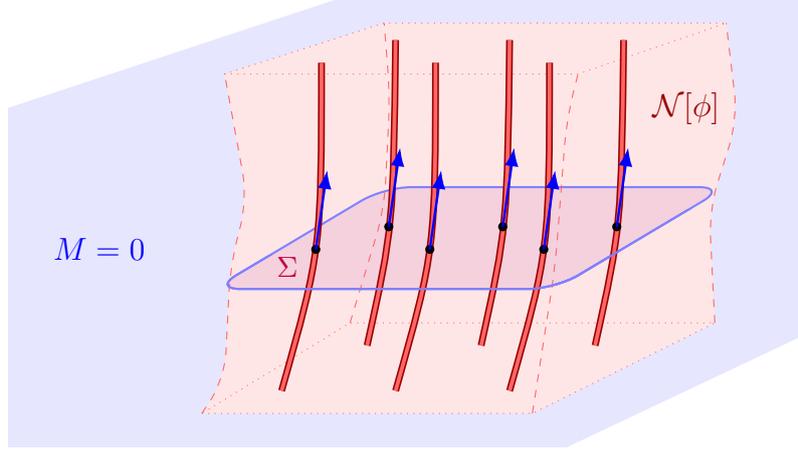

We can observe that the dust frame has very similar properties to the geodesic frame considered earlier in the paper. Both cases involve congruences of geodesics\pah{: the construction in Section~\ref{Section: geodesic example} invoked boundary-anchored timelike or spacelike geodesics, while here we use `Cauchy-slice-anchored' timelike geodesics}, \red{see Figure \ref{fig:dust_spacetime}}. 
However, it is interesting to note that the constructions happened in different orders. For the geodesic frame, we constructed some scalar fields by starting with a local orientation space made up of boundary points, boundary tangent vectors and a proper time variable. On the other hand, here we started with a set of eight bulk scalar fields, which led us to the local orientation space $\RR\times \mathscr{O}_{\mathscr{D}}$. \pah{Independently of these orders, the dust coordinate scalar fields in \eqref{eq:dustscalars} are the analogue of \eqref{eq:geoframe2} in the boundary-anchored construction.}

Let us now write down explicitly how a change of frame from the geodesic frame to the dust frame corresponds to a simple change of (dynamical) coordinates. Let us consider the case of an injective frame $R_g:\mathscr{O}\to\mathscr{D}$ constructed via boundary geodesics. Similarly, to what was described previously in Section~\ref{Section: geodesic example}, we fix a boundary tangent vector field $W$, and a boundary subregion $\mathcal{U}\subset\partial\mathcal{M}$, and take lengths in an interval $[0,\tau)$. The local orientation space of the parametrised frame is $\mathscr{O}_g = [0,\tau)\times \mathcal{U}$. As discussed previously \pah{around \eqref{eq:geoframe1}}, the frame field is given by
\begin{equation}
    (R_g[\phi])^{-1}: \quad x\mapsto (T_g(x), Z_g(x)),
\end{equation}
where $T(x)$ is the distance along a geodesic from $Z(x)$ to $x$ with tangent vector $\hat{n}-W(Z(x))$ at $Z(x)$. However, let us introduce a coordinate system $\phi:\mathcal{U}\to\RR^3$ for the boundary subregion $\mathcal{U}$. We can use this to reparametrise the frame, changing the local orientation space to $[0,\tau)\times \phi(\mathcal{U})$, and the frame field to
\begin{equation}
    (R_g[\phi])^{-1}: \quad x\mapsto (T_g(x),Z^1_g(x),Z^2_g(x),Z^3_g(x)),
\end{equation}
where $Z^k_g = \varphi^k\circ Z_g$. This is clearly a dynamical set of coordinates. The dust frame field is also a dynamical coordinate system
\begin{equation}
    (R_d[\phi])^{-1}: \quad x \mapsto (T(x),Z^1(x),Z^2(x),Z^3(x))
\end{equation}
where $Z^k$ and $T$ are the dust scalar fields. Let us write the geodesic frame coordinate system as $X_g^\mu$, $\mu=0,1,2,3$, and the dust frame coordinate system as $X^\alpha_d$, $\alpha=0,1,2,3$ where
\begin{nalign}
    X^0_g &= T_g, & X^0_d &= T, \\
    X^k_g &= Z^k_g, & X^k_d &= Z^k, & k=1,2,3.
\end{nalign}
Assuming that the images of the two frames have some overlap, changing from one frame to the other is done using the transition function $R_{g\to d}[\phi]:X_g\mapsto X_d$ between the two (dynamical) coordinate systems.

This transition function can also be used to relate the relational observables constructed using the two frames. For example, for the metric, the relational observables for the two frames are just the metric components in each coordinate system, which are related by
\begin{equation}
    g_{\mu\nu} = \pdv{X^\alpha_d}{X^\mu_g}\,\pdv{X^\beta_d}{X^\nu_g}\, g_{\alpha\beta}.
\end{equation}
 \pah{As noted around \eqref{eq:gframectrans}, for a \emph{fixed} field configuration, these are mere coordinate transformations, however, mapping gauge-invariant objects into one another; this illustrates the gauge-invariant, yet frame-dependent physics that our framework gives rise to (cf.\ the discussion on dynamical frame covariance in Section~\ref{sssec_dyncov}).

Next, let us turn to reorientations of the dust frame. To this end, it is convenient to go partially on-shell: as noted in~\cite{Brown:1994py}, one can solve \eqref{eq:Zk_constant_U} for the three scalars $W_k[T,Z^k,g]$ (the dust velocity in its own frame).   As in the case of parametrised field theory,  there is then a kinematical splitting between the dust frame scalars $(T,Z^k)$, on the one hand, and $(M,g)$, on the other (the `other' degrees of freedom).\footnote{Effectively, one now starts from a product kinematical field configuration space $\Gamma(\Phi)=\mathcal{Q}_{T,Z}\times\mathcal{Q}_{M,g}$.} From Section~\ref{sssec_reorient} it follows that a reorientation is given by a field dependent orientation space diffeomorphism $(R[\tilde\phi])^{-1}=F[\phi]\circ(R[\phi])^{-1}$, where $F[\phi] = (R[\tilde\phi])^{-1}\circ R[\phi]\in\rm{Diff}(\mathscr{O})$, leaving the `other' degrees of freedom untouched. That is,
\begin{equation}\label{eq:dustsym}
    (T,Z^k)\mapsto (T',Z'^k)=F[\phi](T,Z^k),\qquad (M,g)\mapsto(M,g).
\end{equation}
Since this relies on a kinematical splitting, the question is which such diffeomorphisms $F[\phi]$ are compatible with the space of solutions $\mathcal{S}$. Since the dust fields couple to gravity, it is clear that not all orientation space diffeomorphisms can be realized as symmetries. Indeed, as shown in~\cite{Brown:1994py}, only diffeomorphisms $F[\phi]\in\rm{Diff}(\mathscr{O})$ that preserve the dust four-velocity one-form \eqref{eq:dustvel} also preserve the space of solutions. These evidently leave $S_{\rm dust}[\phi]$ invariant and since \eqref{eq:dustsym} does not depend on $x\in\mathcal{M}$, these correspond to \emph{global} symmetries of the action that are generated by non-trivial charges. They furnish a non-trivial set, including the full set $\rm{Diff}(\mathscr{O}_\mathscr{D})$ of dust space diffeomorphisms (and more).

This brings us to the notion of relational bulk time evolution, as discussed in Section~\ref{Section: relational time evolution}. Arguably the simplest choice of timelike orientation space vector field is $\tilde\rho_R=\frac{\partial}{\partial t}$. It is clear that this choice corresponds to the evolution of $(M,g)$ relative to the proper time along the dust congruence, so we have as bulk clock function $t[\phi]=T$. This clock function is only unique relative to a choice of initial Cauchy slice, upon which \eqref{eq:T_propertime_U} generates a foliation of equal dust time surfaces in $\mathcal{M}$. The relational evolution will thus correspond to how $(M,g)$ evolve along with a choice of foliation in $\mathcal{M}$ (different such choices are related by a reorientation~\cite{Brown:1994py}), which translates to a choice of background foliation in $\mathscr{O}$. 
This relational evolution is gauge equivalent to a frame reorientation $T'=T+const$ that leaves all other variables in \eqref{eq:dustsym} invariant. It follows from~\cite{Brown:1994py} that this temporal frame reorientation is generated by the charge corresponding to mass conservation
\begin{equation}
    Q[\vartheta] =-\int_\Sigma d^3y\sqrt{\gamma} Mn_\mu U^\mu\vartheta
\end{equation}
with field-independent $\vartheta$, where $\gamma$ is the induced metric on a Cauchy slice $\Sigma$ and $n_\mu$ its future-pointing unit normal. The sought after Hamiltonian for \eqref{eq:reltime} is thus gauge equivalent to this charge $H_{\tilde\rho_R}\approx Q$. Concrete expressions for such a Hamiltonian have been derived in canonical language in~\cite{Giesel:2007wi} (see also~\cite{Brown:1994py,Tambornino:2011vg}). 

Our relational microcausality also applies to the relational observables relative to the dust frame and corresponds to a standard notion of microcausality on the deparametrised dust spacetime $\mathscr{O}$ (on which there is no diffeomorphism gauge symmetry).

 }

\subsection{Minimal surfaces}
\label{Section: more examples / minimal surfaces}

For our final brief example, we will define a class of quasilocal dressings (as described in Section~\ref{Section: general formalism / further generalisations}) based on \emph{minimal surfaces}. Such surfaces are of great interest in applications of quantum information theory to holography and quantum gravity, where they are key to understanding the relationship between local boundary degrees of freedom and local bulk degrees of freedom (see~\cite{MinimalSurfaces1,MinimalSurfaces2,MinimalSurfaces3,MinimalSurfaces4,MinimalSurfaces5} among many others). Here, we will mainly just focus on using them to demonstrate the versatility of the formalism. However, the relatively well understood quantum corrections to the holographic results, in combination with the ideas given below, may provide some insight into the potential quantum generalisation of the results in this paper.

Let $\mathcal{K}^{(d)}$ be the set consisting of all $d$-dimensional submanifolds of the $D$-dimensional spacetime manifold $\mathcal{M}$. This set clearly has an action of $\operatorname{Diff}(\mathcal{M})$, given by $\mathcal{U}\to f(\mathcal{U})$ for $f\in\operatorname{Diff}(\mathcal{M})$, $\mathcal{U}\in\mathcal{K}^{(d)}$.

Consider observables $B$ which for each $\phi$ are functions $B[\phi]$ on $\mathcal{K}^{(d)}$. Such observables are covariant if
\begin{equation}
    B[f_*\phi](\mathcal{U}) = B[\phi](f^{-1}(\mathcal{U})).
\end{equation}
Examples of covariant observables $B[\phi](\mathcal{U})$ include the spacetime volume of $\mathcal{U}$, the average value of a scalar field over $\mathcal{U}$, and also any topological invariants of $\mathcal{U}$. Although the last of these is gauge-invariant by definition, the former two are not. We would like a method for converting such gauge-dependent, covariant quantities into physical observables, so we need a dressing.

Let $\mathcal{A}\subset\partial\mathcal{M}$ be a $d$-dimensional submanifold of the boundary of $\mathcal{M}$, and let $\mathcal{K}_{\mathcal{A}}\subset\mathcal{K}^{(d)}$ be the set consisting of all $d$-dimensional submanifolds of $\mathcal{M}$ which are homologous to $\mathcal{A}$ (i.e.\ for all $\Upsilon\in\mathcal{K}_{\mathcal{A}}$, there is a $(d+1)$-dimensional submanifold $\Sigma_{\mathcal{A}}$ whose boundary consists of $\mathcal{A}$ and $\Upsilon$, oriented appropriately).

Let us now pick some covariant observable $B$ for which $B[\phi]$ is a function from $\mathcal{K}^{(d)}$ to $\RR$, and let
\begin{equation}
    \Upsilon_{B,\mathcal{A}}[\phi](\mathcal{A}) \text{ be the element of $\mathcal{K}_{\mathcal{A}}$ which minimises } B[\phi]:\mathcal{K}_{\mathcal{A}}\to\RR.
\end{equation}
We are implicitly assuming that $B$ and $\mathcal{A}$ are chosen such that this minimum exists and is unique.

Let us assume that the set of small diffeomorphisms $H[\phi]$ consists only of diffeomorphisms which leave the boundary invariant. Then $\Upsilon_{B,\mathcal{A}}$ is a quasilocal dressing. To see this, note that for any $\Upsilon\in\mathcal{K}_A$ not equal to $\Upsilon_{B,\mathcal{A}}[\phi]$ we have
\begin{equation}
    B[\phi]\big(\Upsilon_{B,\mathcal{A}}[\phi]\big) < B[\phi](\Upsilon),
\end{equation}
so for any $f\in H[\phi]$ we have
\begin{equation}
    B[\phi]\big(f^{-1}\big(f\big(\Upsilon_{B,\mathcal{A}}[\phi]\big)\big)\big) < B[\phi]\big(f^{-1}\big(f(\Upsilon)\big)\big).
\end{equation}
Using $B[\phi]\circ f^{-1} = B[f_*\phi]$ and writing $f(\Upsilon)=\Upsilon'$, we find
\begin{equation}
    B[f_*\phi]\big(f\big(\Upsilon_{B,\mathcal{A}}[\phi]\big)\big) < B[f_*\phi](\Upsilon').
\end{equation}
This holds for any $\Upsilon'\in f(\mathcal{K}_{\mathcal{A}})$ not equal to $f\big(\Upsilon_{B,\mathcal{A}}[\phi]\big)\in f(\mathcal{K}_{\mathcal{A}})$. Moreover, since $f$ is small it leaves the boundary invariant, so $f(\mathcal{K}_{\mathcal{A}})=\mathcal{K}_{\mathcal{A}}$. Thus, $f\big(\Upsilon_{B,\mathcal{A}}[\phi]\big)$ must minimise $B[f_*\phi]:\mathcal{K}_{\mathcal{A}}\to\RR$, i.e.
\begin{equation}
    \Upsilon_{B,\mathcal{A}}[f_*\phi] = f\big(\Upsilon_{B,\mathcal{A}}[\phi]\big) \text{ for all } f\in H[\phi].
\end{equation}
So $\Upsilon_{B,\mathcal{A}}[\phi]$ transforms like a dressing, as claimed.

We can use it to dress any covariant observable valued as a function on $\mathcal{K}^{(d)}$. In fact, we can go further, and construct an entire quasilocal frame of such minimal dressings. Such a frame can be parametrised by a space $\mathscr{O}^{(d)}$ of $d$-dimensional submanifolds of $\partial\mathcal{M}$. Then the frame is given by
\begin{equation}
    R_\Upsilon: \quad \mathscr{O}^{(d)}\to \mathscr{D}(\mathcal{K}^{(d)}), \quad \mathcal{A} \mapsto \Upsilon_{B,\mathcal{A}}.
\end{equation}
Such a frame will not be surjective, but it will be injective. We can use it to construct relational observables out of more complicated covariant quasilocal objects like tensors on $\mathcal{K}^{(d)}$.

In holography, when $d=D-2$ and $B$ is the area functional, the minimal surface $\Upsilon_{B,\mathcal{A}}[\phi]$ plays an important role in the concept of subregion duality. This roughly says that the boundary degrees of freedom in $\mathcal{A}$ are dual to the bulk degrees of freedom in a bulk subregion $\mathcal{W}_{\mathcal{A}}[\phi]$ known as the `entanglement wedge', which is the domain of dependence of a partial Cauchy surface bounded by $\mathcal{A}$ and $\Upsilon_{B,\mathcal{A}}[\phi]$. The map from $\Upsilon_{B,\mathcal{A}}[\phi]$ to $\mathcal{W}_{\mathcal{A}}[\phi]$ can itself be understood in terms of a field-dependent map $\mathcal{W}[\phi]:\mathcal{K}^{(D-2)} \to \mathcal{K}^{(D)}$. We will not go into an explicit derivation of the following, but since the domain of dependence of any region depends on the causal structure of spacetime, which itself depends covariantly on the metric, for all $f\in H[\phi]$ this map obeys
\begin{equation}
    \mathcal{W}[f_*\phi] = f\circ \mathcal{W}[\phi] \circ f^{-1}.
\end{equation}
In other words, it obeys~\eqref{Equation: generalised dressable 2}. Thus, if we dress $W$ with $\Upsilon_{B,\mathcal{A}}$, we get a $\mathcal{K}^{(D)}$-dressing, which is just the entanglement wedge:
\begin{equation}
    \mathcal{W}_{\mathcal{A}}[\phi] = \mathcal{W}[\phi]\big(\Upsilon_{B,\mathcal{A}}[\phi]\big).
\end{equation}
As before we can construct a parametrised quasilocal frame using this dressing:
\begin{equation}
    R_{\mathcal{W}}: \quad \mathscr{O}^{(D-2)}\to \mathscr{D}(\mathcal{K}^{(D)}), \quad \mathcal{A} \mapsto \mathcal{W}_{\mathcal{A}}.
\end{equation}

Now suppose we have a family of local dressings $x_\Upsilon$ where $\Upsilon\in\mathcal{K}^{(D)}$, such that
\begin{equation}
    x_\Upsilon[\phi] \text{ depends only on $\phi$ in $\Upsilon$}, \quad
    x_{f(\Upsilon)}[f_*\phi] = f(x_\Upsilon[\phi]), \qq{and}
    x_\Upsilon[\phi]\in \Upsilon.
\end{equation}
Thus, the dressing is formed from the fields only in $\Upsilon$ in a way that respects covariance, and moreover is valued as a point in $\Upsilon$. We can view this as a field-dependent map
\begin{equation}
    x[\phi] : \quad \mathcal{K}^{(D)}\to \mathcal{M}, \quad \mathcal{U} \mapsto x_\Upsilon[\phi]
\end{equation}
that obeys
\begin{equation}
    x[f_*\phi] = f\circ x[\phi] \circ f^{-1},
\end{equation}
i.e.\ \eqref{Equation: generalised dressable 2}. Thus, by composing $x$ with the quasilocal entanglement wedge dressing, we get a dressing
\begin{equation}
    x_{\mathcal{A}}[\phi] = x[\phi]\circ \mathcal{W}_{\mathcal{A}}[\phi],
\end{equation}
which depends on the fields only in the entanglement wedge, and is valued in the entanglement wedge. Such dressings may be useful in understanding relational locality in holographic theories that respect subregion duality.

Before ending this section, let us turn away from the holographic interpretation and instead describe a way in which one can use intersections of the quasilocal minimal surface dressings to construct local dressings. It is not clear that these dressings have any useful applications. However, despite their exotic nature, they can still be used in the general formalism we have described, and frames based on them obey all the properties we have discussed, such as relational microcausality.

Generically, the intersection of a codimension $m$ submanifold of $\mathcal{M}$ with a codimension $n$ submanifold of $\mathcal{M}$ will be a codimension $m+n$ submanifold of $\mathcal{M}$ (so long as $m+n\le D$). Suppose $D=pn$. If we have a collection of $p$ codimension $n$ submanifolds $\mathcal{U}_i$, their intersection will be a codimension $D$ submanifold of $\mathcal{M}$, i.e.\ a discrete set of points.

Let $\mathcal{A}_i\subset \mathcal{K}^{(D-n)}$, and let us assume that the intersection
\begin{equation}
    \bigcap_i \Upsilon_{B,\mathcal{A}_i}[\phi]
\end{equation}
consists of a single point $x_{\{\mathcal{A}_i\}}[\phi]\in\mathcal{M}$. Then this point is a local dressing, since
\begin{equation}
    \{x_{\{\mathcal{A}_i\}}[f_*\phi]\} = \bigcap_i\Upsilon_{B,\mathcal{A}_i}[f_*\phi] = \bigcap_i f\big(\Upsilon_{B,\mathcal{A}_i}[\phi]\big) = f\qty(\bigcap_i\Upsilon_{B,\mathcal{A}_i}[\phi]) = \qty{f\big(x_{\{\mathcal{A}_i\}}[\phi]\big)}.
\end{equation}
Moreover, if we let $\mathscr{O}=\big(\mathscr{O}^{(D-n)}\big)^p$, then
\begin{equation}
    R:\quad \mathscr{O}\to \mathscr{D},\quad (\mathcal{A}_1,\dots,\mathcal{A}_p) \mapsto x_{\{\mathcal{A}_i\}}
\end{equation}
is a frame.

\section{Conclusion}
\label{Section: conclusion}

In this paper, we have presented a general formalism for constructing and understanding dressed and relational observables in gravitational theories. These are based on dynamical reference frames, which we have formulated as sets of dressings, i.e.\ gauge-covariant space-time points. \pah{We refer to all these observables as relational observables as they describe the `other' degrees of freedom relative to the dynamical frame in a precise sense. The formalism gives rise to a completely general and gauge-invariant framework of dynamical frame covariance for generally covariant theories that resonates with recent efforts on establishing a `quantum frame covariance' ~\cite{delaHamette:2021oex,Hoehn:2021wet,Hoehn:2019fsy,Hoehn:2021flk,Castro-Ruiz:2021vnq,delaHamette:2020dyi,Giacomini:2017zju,Giacomini:2018gxh,Vanrietvelde:2018pgb,Vanrietvelde:2018dit,Hoehn:2020epv,Hohn:2018iwn,Castro-Ruiz:2019nnl,Ballesteros:2020lgl,Krumm:2020fws,delaHamette:2021iwx}.} This yields a version of general covariance which is more physical than the usual one. We have given several examples demonstrating our formalism, and described the way in which it generalises previous approaches.

At the same time, we have decisively addressed the issue of locality in generally covariant theories, arguing that it should be understood \emph{relationally}. In particular, we have defined the notion of relational locality, which is a dynamical frame-dependent version of locality. We have shown that this kind of locality satisfies all the requirements in the wishlist that we gave in the Introduction. It permits non-trivial gauge-invariant observables -- these are relational observables. It allows for non-trivial local dynamics, since \pah{gravitationally charged} relational observables transform under active large diffeomorphisms, including global time evolution, \pah{while observables with trivial gravitational charge are (under certain conditions) also subject to a relational bulk time evolution} (which doesn't require spacetime to have a boundary). They can be accessed by different mutually compatible reference frames in a way made precise by our formalism. And they respect microcausality -- at least in the bulk.

We have also shown how our covariant approach is equivalent (under certain conditions) to the canonical approach of~\cite{Dittrich:2004cb,Dittrich:2005kc,Dittrich:2006ee,Dittrich:2007jx}, and the single integral approach~\cite{DeWitt:1962cg,Marolf:1994wh,Giddings:2005id,Marolf:2015jha,Giddings2006,Donnelly:2016rvo}. Thus, all three formalisms are fundamentally treatments of the same physics.

An important classical step that we have left open is the understanding of the algebra of observables underlying our general formalism \pah{(for interesting steps in this direction in Jackiw-Teitelboim gravity, see \cite{Harlow:2021dfp})}. In particular, when the relational observables are based on a construction involving the boundary, it is expected that they at least encode information about the new physical degrees of freedom at the boundary that arise due to the gauge symmetry becoming physical. The study of this algebra and its relation with the usual boundary symmetry algebra is a mandatory step towards a better understanding of the physics arising at the boundary. 

\pah{

\subsection{Remarks on the perturbative non-locality arguments by Donnelly-Giddings}\label{ssec_DGcomparison}

In a series of works on gravitational locality, Donnelly and Giddings have reported a class of  gauge-invariant gravitationally dressed observables that are non-local and do \emph{not} obey microcausality~\cite{Donnelly:2015hta,Donnelly:2016rvo,Giddings:2015lla} (see also~\cite{Donnelly:2017jcd,Giddings:2018umg,Giddings:2018koz,Giddings:2019hjc}). It was further argued that such non-local properties of gravitational observables are generic, inhibit the definition of subsystems in terms of commuting subalgebras and thus call for a revision of the very notion of subsystems in gravitational contexts.

The fact that microcausality was not observed in those papers, but that it is in the present one, means that the type of locality proposed in those works cannot be compatible with the relational locality defined here. This is a puzzle whose solution we cannot yet claim to have. There are, however, a few comments we can make on the relation between the two notions of locality.

Most crucially, our approach is non-perturbative, whereas the one in~\cite{Donnelly:2015hta,Donnelly:2016rvo} is perturbative around Minkowski spacetime. More precisely, in the latter works, gravitational dressings of a scalar field $\psi$ are considered to linear order in $\kappa\propto\sqrt{G}$
\begin{equation}\label{eq:perturbrelobs}
    \Psi(x)=\psi(x^\mu+V^\mu(x))=\psi(x)+V^\mu(x)\partial_\mu\psi(x)+\order{V^2},
\end{equation}
where $x$ is a point in Minkowski spacetime and $V^\mu(x)$ is a vector dressing (of linear order in $\kappa$) that is a generally non-local functional of the metric perturbations (e.g.\ a gravitational Wilson line along a geodesic shot in from asymptotia). These observables thus have a non-local support in Minkowski spacetime and it is shown in~\cite{Donnelly:2015hta} that, e.g., $[\dot\Psi(x),\Psi(x')]$ can fail to vanish even when $x,x'$ are spacelike separated, where $[\cdot,\cdot]$ can be read as both a commutator or Poisson bracket. This can happen when the spacetime supports of the gravitational dressings $V^\mu(x),V^\mu(x')$ are causally related.

Note that the dressed observables in \eqref{eq:perturbrelobs} are nothing but linearised expansions of our relational observables $(R[\phi])^*\psi$ for the special case that the frame $R[\phi]=(U[\phi])^{-1}$ is a gauge-fixing diffeomorphism, generated by $V^\mu[\phi]$ as defined in \eqref{Equation: gauge-fixing diffeomorphism} and further discussed at the end of Section~\ref{subsection:relational_locality}. The local orientation space $\mathscr{O}$ of this frame is thus another copy of spacetime itself. Our above argument entails that the non-perturbative relational observables $(U[\phi])_*\psi$ obey microcausality, so how can it be that their perturbative counterparts $\Psi(x)$ do not?

One possible answer is as follows. While $(U[\phi])_*\psi$ is invariant under arbitrary small diffeomorphisms, $\Psi(x)$ is only invariant under linearised diffeomorphisms $x\mapsto x+\kappa\xi$. Indeed, choosing Minkowski spacetime as a background amounts to a partial gauge-fixing of the full small diffeomorphism group such that (to first order in $\kappa$) only linearised diffeomorphisms survive as residual gauge symmetry. In our relational microcausality argument it was crucial that we could impose a certain gauge-fixing to remove the  non-local dependence in the equations of motion, resulting in \eqref{Equation: tilde phi deformed eom}. In the perturbative setting, the residual gauge symmetry may not suffice to remove this non-locality from the equations of motion, in which case our argument would no longer work. This would suggest that the perturbative analysis does not in fact fully constrain the non-perturbative one, as claimed in~\cite{Donnelly:2015hta,Donnelly:2016rvo}. Indeed, the linearisation instabilities~\cite{fischer1973linearization,moncrief1975spacetime,Moncrief:1976un,Moncrief:1979bg} imply that not every perturbative solution can be extended to a non-perturbative one, essentially because the higher orders in the perturbation theory impose non-trivial constraints on the lower order gauge-invariant degrees of freedom.\footnote{It should be noted however, that these linearisation instabilties are most well understood in the case of spatially compact spacetimes. This is in contrast to the work of Donnelly and Giddings, which assumed the existence of an asymptotic boundary.} Thus, it is not obvious that every perturbatively gauge-invariant observable need necessarily correspond to a fully gauge-invariant one on the space of non-perturbative solutions.

Before discussing another possible answer, let us discuss a conceptual difference ensuing from the distinct gauge invariances: $(U[\phi])_*\psi$ is a non-trivial function on the local orientation space $\mathscr{O}$, but no longer has any dependence on fixed (i.e.\ field-independent) points $x$ in spacetime $\mathcal{M}$, in contrast to $\Psi(x)$. Hence, unlike with $\Psi(x)$, there is no sense in which we could refer to the relational observable $(U[\phi])_*\psi$ at some \emph{fixed} $x\in\mathcal{M}$ without introducing a gauge-fixing; we can only refer to $(U[\phi])_*\psi$ at a fixed point $o$ in orientation space $\mathscr{O}$ (which maps to the \emph{field-dependent} $x[\phi]=(U[\phi])^{-1}(o)\in\mathcal{M}$). For that reason, unlike in the perturbative case, there is no sense in which we could compute the Poisson bracket between the spacetime time derivative of $(U[\phi])_*\psi$ evaluated at \emph{fixed} $x\in\mathcal{M}$ and $(U[\phi])_*\psi$ evaluated at \emph{fixed} $x'\in\mathcal{M}$ without imposing a gauge-fixing. Instead, our non-perturbative microcausality argument implies, for instance, that $\left((U[\phi])_*\psi\right)(o)$ and  $\left((U[\phi])_*\psi\right)(o')$ Poisson commute at \emph{fixed} $o,o'\in\mathscr{O}$ if the corresponding \emph{field-dependent} relational support points $x=(U[\phi])^{-1}(o)$ and $x'=(U[\phi])^{-1}(o')$ in spacetime $\mathcal{M}$ are spacelike separated. 
 In particular, since $U[\phi]$ is a small diffeomorphism, we have that $o,o'$ are spacelike on orientation space $(\mathscr{O},(U[\phi])_*g)$ if the field-dependent $x,x'$ are spacelike in $(\mathcal{M},g)$. Now in the perturbative case, we have $o=(U[\phi])^{-1}(x)=x+V[\phi](x)+\order{V^2}$ and $o'=(U[\phi])^{-1}(x')=x'+V[\phi](x')+\order{V^2}$. In the construction of~\cite{Donnelly:2015hta,Donnelly:2016rvo,Bodendorfer} one treats $x,x'$ as fixed which here entails field-dependent $o,o'$. However, when attempting to import our relational microcausality argument to the perturbative case, we should invert these roles and expect commutation of, say, $\dot\Psi(x)=\dot\psi(o)$ and $\Psi(x')=\psi(o')$ when the \emph{fixed} $o,o'$ are spacelike separated in $\mathscr{O}$.\footnote{In the perturbative setting $o,o'$ need not necessarily be spacelike separated, even if $x,x'$ are because both relations are defined in terms of the \emph{same} flat Minkowski background metric and the dressing $V[\phi]$ may affect these relations. However, since $V[\phi]$ is linear in $\kappa$ this can only happen when $x,x'$ are infinitesimally close to being null separated. This could lead to an additional interesting difference between the relational and background microcausality properties. But it cannot explain the tension between our non-perturbative microcausality and its perturbative failure in~\cite{Donnelly:2015hta,Donnelly:2016rvo} as in the latter works $x,x'$ need not be infinitesimally close to null separated for $[\dot\Psi(x),\Psi(x')]\neq0$. }
 
A second possible answer concerns the methods. The construction in~\cite{Donnelly:2015hta} requires one to introduce an explicit gauge-invariance breaking term into the action in order to compute the brackets using gauge-dependent canonical commutation relations. This amounts to an \emph{a priori} gauge-fixing, i.e.\ \emph{before} constructing the brackets. By contrast, the covariant Peierls bracket is fully gauge-invariant by construction. We have exploited this to perform an \emph{a posteriori} partial gauge-fixing $\phi\to\tilde\phi$ in $S_\lambda[\phi]$ to simplify the evaluation of the bracket by removing the non-locality from the equations of motion. If we were to similarly introduce an \emph{a priori} gauge-fixing term here, the action would no longer be gauge-invariant, meaning we would not be able to do the \emph{a posteriori} substitution $\phi\to\tilde\phi$, and our argument would not work -- thus there may be some previously unnoticed problem with performing this kind of gauge-fixing before computing the bracket. A similar situation happens in the canonical setting \cite{Bodendorfer}, where an \emph{a priori} choice of gauge, the radial gauge, is made.

Finally, as an aside, we note that a distinction was drawn in~\cite{Donnelly:2016rvo} between the single-integral relational observables and the perturbatively dressed observables as in \eqref{eq:perturbrelobs}. It was further argued that single-integral observables only provide a state-dependent (i.e.\ classically configuration-dependent) and approximate notion of locality that in particular would not apply to the discussion of perturbations around empty Minkowski spacetime. While~\cite{Donnelly:2016rvo} was referring to single-integral observables with trivial gravitational charge, we have seen above that the dressed observables in \eqref{eq:perturbrelobs} are linearised expansions of our relational observables $(R[\phi])^*\psi$, which in turn, as seen in Section~\ref{Section: general formalism / relational picture / single integrals smearing}, are (possibly gravitationally charged) single-integral observables. In other words, the dressed observables used in~\cite{Donnelly:2016rvo} are perturbative expansions of single-integral relational observables, which are thus useful for studying perturbations of empty Minkowski spacetime.\footnote{Our relational formalism relies on gauge-covariant reference frames. Strictly speaking, we have only shown that the boundary-anchored frames that lead to gravitationally charged single-integral observables as in the geodesic dressing example of Section~\ref{Section: geodesic example} are gauge-covariant for finite boundaries, but we have not checked it for the asymptotically flat setting. However, it is not clear why this property should not carry over to the case of asymptotic boundaries.} Classically, the configuration dependence of the relational locality defined by the single-integral observables relative to a fixed frame is also not a problem as seen in this work: we can simply restrict a frame to a subregion of spacetime where it is bijective and cover spacetime with a relational atlas, i.e.\ a suitable collection of frames. If the relational atlas is not valid on the full space of solutions, we can invoke a field-space meta-atlas, as discussed in Section~\ref{sssec_relatlas}. In this manner, the relational notion of locality can encompass every classical configuration and all of spacetime. Classically, it is also clear that relational observables can define an exact notion of relational locality if one does not smear them non-trivially over orientation space $\mathscr{O}$ (in which case they do define an approximate locality).

We leave the resolution of the tension between the perturbative and non-perturbative observations on microcausality for future work. However, since a  non-perturbative analysis is more complete than a perturbative one, it is reasonable to conclude that bulk microcausality and locality are in fact compatible with gravity if defined in a relational -- and thus gauge-invariant -- manner. At least at the classical non-perturbative level, one can define the notion of bulk subsystems in terms of commuting subalgebras of relational observables. What \emph{is} different to the non-gravitational setting, however, is that, while subalgebras associated with spacelike separated bulk subregions commute, their elements need not commute with observables on the boundary (if there is one), even if the support of the latter is spacelike separated.\footnote{\jjvk{Note that this also happens in gauge field theories: asymptotically charged observables do not commute with the boundary charges that generate large gauge transformations.}}\textsuperscript{,}\footnote{\jjvk{This property was used to argue for a quantum error correction interpetation of holography in~\cite{QEC1}; for a connection to the current work, see the last paragraph of the conclusion.}} This happens when the bulk relational observables are gravitationally charged, in which case they transform non-trivially under large diffeomorphisms.
This is an observation that has a bearing on the discussion of boundary unitarity and holographic properties in quantum gravity (even without invoking the AdS/CFT duality)~\cite{Marolf:2008mf,Jacobson:2019gnm}.

%Next, in this paper we have mostly worked under the assumption of a finite boundary, while~\cite{Donnelly:2015hta,Donnelly:2016rvo} worked in the asymptotically flat setting; however it is not clear why the formalism in this paper could not extend to the case of asymptotic boundaries. %Finally, in~\cite{Donnelly:2015hta,Donnelly:2016rvo} brackets were computed using a canonical approach, in which a gauge-fixing term has to be introduced into the action; if we were to similarly introduce a gauge-fixing term here, the action would no longer be gauge-invariant, meaning we would not be able to do the substitution $\phi\to\tilde\phi$ in $S_\lambda[\phi]$, and our argument would not work -- thus there may be some previously unnoticed problem with performing this kind of gauge-fixing before computing the bracket.

}

\subsection{Quantum possibilities}

Before ending, we should address the quantum elephant in the room. Everything we have so far discussed has been entirely classical, but if our framework is to stand a chance of adequately describing the real world, at some point a quantum version of it will have to be put forward. Let us comment on a perspective provided by our results which may prove useful.

Previous approaches to constructing quantum versions of relational observables usually begin with a quantisation of the kinematical description, for example a canonical \pah{Dirac constraint} quantisation. \pah{In principle, t}his results in quantum versions of the frame field and the field to be observed. Both of these are gauge-dependent, but they can be combined to form a quantum relational observable, in a manner analogous to the classical construction we have described. \pah{While this is techically challenging in the field theory context, some progress has recently been made in the context of mechanical and cosmological models \cite{delaHamette:2021oex,Hoehn:2019fsy,Hoehn:2020epv,Chataignier:2019kof,Chataignier:2020fap,Chataignier:2020fys,Marolf:2009wp}.} The quantum relational observable is gauge-invariant, and so can be projected down to an operator acting on a physical Hilbert space. \pah{It is sometimes argued that there are issues with this approach} (see for example ~\cite{Giddings:2005id,Giddings2006,Donnelly:2016rvo,Marolf:2015jha}), with perhaps the main one stemming from the fact that the frame field is itself subject to \pah{wild} quantum fluctuations \pah{on the smallest scales, rendering it a useless relational locator for other fields}.\footnote{It is possible this is not really an issue in quantum gravity, where the local structure of spacetime is believed to break down at the Planck scale.} \pah{It is further argued that, t}o account for this, one must carry out some kind of smearing, \pah{yielding an approximate relational notion of locality,} but this procedure suffers from ambiguities.

\pah{
However, the situation might perhaps not be quite as grim. For example, one could make the same argument about reparametrisation-invariant mechanical or cosmological models, which can be viewed as field theories on a $(1+0)$-dimensional (space)time manifold. Also there the reference field (i.e.\ the clock) would appear to fluctuate. But it turns out that this is not a problem for relational observables, essentially because those fluctations are kinematical (the clock is by construction non-invariant). In particular, the clock degrees of freedom are \emph{redundant} on the manifestly gauge-invariant physical Hilbert space of states solving the constraints and so are their fluctuations. For that reason, they can be removed through a quantum symmetry reduction (essentially a gauge-fixing) procedure that maps into the `perspective' of the clock -- a unitarily equivalent quantum theory in which the clock no longer is treated as a dynamical degree of freedom \cite{Hohn:2018toe,Hoehn:2019fsy,Hoehn:2020epv}. As shown in \cite{Hoehn:2019fsy,Hoehn:2020epv,delaHamette:2021oex}, one example of such a quantum reduction is the Page-Wootters formalism \cite{Page:1983uc,Giovannetti:2015qha,Smith:2017pwx}, which describes the evolution of the `other' degrees of freedom relative to the clock. In simple cases, this quantum reduction is even unitarily equivalent to the quantization of the classically reduced (i.e.\ deparametrized) theory in which the clock is no longer treated as a dynamical variable \cite{Hohn:2018toe,Hohn:2018iwn,Hoehn:2019fsy}. But this is not in general true due to the generic inequivalence of Dirac and reduced quantization. Regardless, in these cases one can still construct relational observables relative to the kinematically fluctuating clock which encode a well-defined unitary time evolution (when the clock does not interact) and are equivalent to the quantum reduced theory in which no clock fluctuations occur \cite{Hoehn:2019fsy,Hoehn:2020epv,delaHamette:2021oex}. While quantum field theory is clearly a substantially more challenging setting, these observations suggest that possibly the kinematical fluctuations of reference fields relative to the spacetime background may not be such a fundamental obstacle after all and could similarly be `gauged away' through a reduction procedure (e.g., by extending the Page-Wootters formalsm to quantum field theory). \jjvk{This would be consistent with a key point of this paper: that the kinematical locality with respect to which the reference fields fluctuate is not physically meaningful.}

Alternatively, r}ather than quantising the kinematical description, it \pah{might} be more convenient to directly quantise the physical gauge-invariant observables that we have discussed. This would be a form of reduced quantisation. Doing so in an algebraic way would allow us to avoid canonical quantisation, and so preserve the covariant structure of the theory. \jjvk{It also avoids the issue of fluctuations in the frame field, since that field would never undergo direct quantisation.} It would be nice if we could apply the axioms of algebraic quantum field theory (AQFT)~\cite{Rejzner:2016hdj} to this covariant algebraic quantisation. Previously, this has not been possible because those axioms depend heavily on the existence of a physical notion of locality, which in gravity has not been precisely understood. The relational locality we have described fills this gap, and thus paves the way for an AQFT treatment of bulk gravitational physics, using a (relational version of) the standard AQFT axioms. 

\jjvk{The relational observables we have constructed all live in the local orientation spaces $\mathscr{O}$ of parametrised frames. Moreover, the local orientation spaces have a causal structure (the pullback of the spacetime causal structure). So it is reasonable to understand the quantisation of the algebra of observables of a given frame as yielding a QFT on the local orientation space. It is then natural to interpret this QFT using the language of quantum error correction~\cite{PreskillNotes,nielsen_chuang_2010}. In particular, the Hilbert space $\mathcal{H}_{\text{QFT}}^{\mathscr{O}}$ of the local orientation space QFT may be viewed as a code subspace of the gravitational Hilbert space $\mathcal{H}_{\text{gravity}}$. The code subspace consists of all states for which (in the classical limit), the frame map $R[\phi]$ is well-behaved. This represents a significant generalisation of the use of quantum error correction in holography~\cite{QEC1,QEC2,QEC3}, which works even in the non-holographic context. We will leave this fascinating avenue to forthcoming work.}

Finally, we would like to comment that the relational approach proposed in this paper should allow for a better understanding of the localisation of quantum information in gravity,\footnote{For an alternative perspective, see~\cite{Donnelly:2017jcd}.} to which our intuition from kinematical locality does not completely apply -- for example, an application of that intuition in the context of black hole evaporation leads to the famous firewall paradox~\cite{AMPS}. In holography, the local structure of the bulk is encoded in entanglement wedges (which may be thought of as a kind of bulk reference frame parametrised by boundary subregions, as described in Section~\ref{Section: more examples / minimal surfaces}) but we have learned in recent years that the quantum corrected entanglement wedge~\cite{MinimalSurfaces4,MinimalSurfaces5}\footnote{Incidentally, the results of these papers may provide some clues on how to quantise more general dynamical frames.}, and hence the local structure of the quantum bulk, can deviate significantly from the classical version in the vicinity of evaporating black holes~\cite{Islands1,Islands2,Islands3,Islands4,Islands5,Islands6}. In essence, quantum information is shared between the outside of the black hole (in Hawking radiation), and the inside of the black hole (in the now well-known `islands'). This result is believed to extend beyond holography. By understanding which dynamical reference frames can be quantised, and how they change when they are quantised, we should be able to better characterise how and when \pah{\emph{relationally}} local quantum information is shared in this way, in a more general setting.

\section*{Acknowledgements}
Thank you to Sylvain Carrozza, Bianca Dittrich, Stefan Eccles and Julian de Vuyst for helpful discussions. CG was supported by the Alexander von Humboldt Foundation. This work was supported in part by funding from Okinawa Institute of Science and Technology Graduate University.

\appendix

\section{Small diffeomorphisms in evolving spacetimes}
\label{Appendix: small diffeos}

Let us consider the usual dynamical set up involving evolution between two spacelike surfaces $\Sigma_1$ and $\Sigma_2$, whose boundaries $\partial\Sigma_1,\partial\Sigma_2$ are connected by a timelike surface $\mathcal{I}$. In such a case, $\partial\mathcal{M} = \Sigma_1\cup\Sigma_2\cup\mathcal{I}$. The boundary conditions on $\mathcal{I}$ are a part of the specification of the dynamics of the theory itself, whereas the boundary conditions on $\Sigma_1,\Sigma_2$ specify the states on these surfaces. Thus, within any given theory, the boundary conditions at $\mathcal{I}$ must be obeyed at all times, while those on $\Sigma_1,\Sigma_2$ can be violated if we wish to consider a change of initial or final state.

\begin{figure}[H]
    \centering
    \begin{tikzpicture}[scale=1.5,line width=0.8pt]
        \fill[blue!10] (0,0) .. controls (0.2,1) and (0.2,2) .. (0,3) -- (3,3) .. controls (2.8,2) and (2.8,1) .. (3,0) -- (0,0);
        \fill[blue!25] plot [smooth cycle, tension=0.8] coordinates {(0,0) (0.8,0.2) (1.8,0.6) (3,0) (2,-0.3) (0.9,-0.6)};
        \fill[blue!25,shift={(0,3)}] plot [smooth cycle, tension=0.8] coordinates {(0,0) (0.8,0.2) (1.8,0.6) (3,0) (2,-0.3) (0.9,-0.6)};
        \draw[blue!80] (0,0) .. controls (0.2,1) and (0.2,2) .. (0,3);
        \draw[blue!80] (3,0) .. controls (2.8,1) and (2.8,2) .. (3,3);
        \draw[blue!50,line width=0.4pt] (2,-0.3) .. controls (1.95,0.7) and (1.95,1.7) .. (2,2.7);
        \draw[blue!50,line width=0.4pt] (0.9,-0.6) .. controls (0.95,0.4) and (0.95,1.4) .. (0.9,2.4);
        \draw[dashed] plot [smooth cycle, tension=0.8] coordinates {(0,0) (0.8,0.2) (1.8,0.6) (3,0) (2,-0.3) (0.9,-0.6)};
        \begin{scope}
            \clip (0,0) rectangle (3,-0.7);
            \draw plot [smooth cycle, tension=0.8] coordinates {(0,0) (0.8,0.2) (1.8,0.6) (3,0) (2,-0.3) (0.9,-0.6)};
        \end{scope}
        \draw[shift={(0,3)}] plot [smooth cycle, tension=0.8] coordinates {(0,0) (0.8,0.2) (1.8,0.6) (3,0) (2,-0.3) (0.9,-0.6)};

        \node[blue!80!black] at (1.45,1.3) {\Large$\mathcal{M}$};
        \node at (1.45,0) {\Large$\Sigma_1$};
        \node at (1.45,3) {\Large$\Sigma_2$};
        \draw[decorate,decoration={brace,amplitude=10pt}] (-0.1,0.1) -- (-0.1,2.9);
        \node[left] at (-0.3,1.5) {\Large$\mathcal{I}$};
    \end{tikzpicture}
    \label{Figure: Sigma I}
    \caption{We consider a spacetime $\mathcal{M}$ whose boundary $\partial\mathcal{M}$ consists of an initial spacelike surface $\Sigma_1$, a final spacelike surface $\Sigma_2$, and a timelike surface $\mathcal{I}$.}
\end{figure}
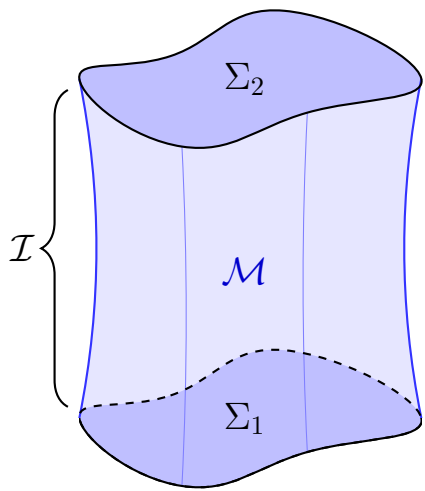

What we say here will apply to any theory with this setup, but for concreteness let us consider general relativity coupled to a scalar, as in Section~\ref{Section: geodesic example}, and use Dirichlet boundary conditions.\footnote{Note that generally $\partial\mathcal{M}$ will not be smooth at the `corners' $\Sigma_{1,2}\cap \mathcal{I}$. To account for this one technically should include Hayward terms in the action~\cite{Hayward}, but in the interest of not overcomplicating the discussion we shall omit these. In their presence the variational principle is still consistent with the Dirichlet boundary conditions we are using.} Let us consider active diffeomorphisms $f:\mathcal{M}\to\mathcal{M}$ on the fields
\begin{equation}
    g_{ab}\to f_* g_{ab}, \qquad \psi \to f_*\psi,
\end{equation}
and restrict to the group $G$ of those which preserve the boundary conditions at $\mathcal{I}$. Any diffeomorphism $f:\mathcal{M}\to\mathcal{M}$ must map $\partial\mathcal{M}$ to itself, so let $f_\partial:\partial\mathcal{M}\to\partial\mathcal{M}$ be the restriction of $f$ to $\partial\mathcal{M}$.\footnote{For simplicity, we are only considering finite boundaries $\partial\mathcal{M}$. In the case of asymptotic boundaries, one usually takes $\mathcal{M}$ to be a subset of some larger manifold $\overline{\mathcal{M}}$, and one can consider the action on the dynamical fields of the group of diffeomorphisms of $\overline{\mathcal{M}}$ -- such diffeomorphisms do not need to map $\partial\mathcal{M}$ to itself. Eventually one takes the limit as $\mathcal{M}$ grows to span all of $\overline{\mathcal{M}}$, but even in this limit diffeomorphisms which do not preserve $\partial\mathcal{M}$ can play a role in the theory.} We have
\begin{equation}
    f\in G \iff (f_\partial)_*h_{ab}|_{\mathcal{I}} = h_{ab}|_{\mathcal{I}}, \,(f_\partial)_*\psi|_{\mathcal{I}} = \psi|_{\mathcal{I}}.
    \label{Equation: on-shell diffeos}
\end{equation}
In other words $f_\partial$ must be an isometry of the induced metric on $\mathcal{I}$, and it must leave the value of $\psi$ on $\mathcal{I}$ unchanged. If a diffeomorphism satisfies these conditions, then it is allowed in this theory.

In particular, diffeomorphisms are allowed to change the induced metric and value of $\psi$ on the initial and final surfaces $\Sigma_1,\Sigma_2$. As we have already said, this should be interpreted as a change of initial and final state. However, for a class of diffeomorphisms, even though the induced metric and $\psi$ may change on $\Sigma_1,\Sigma_2$, the physical initial and final states may not -- such diffeomorphisms are gauge transformations, and are called `small diffeomorphisms'. To determine which diffeomorphisms are small, we can use one of the fundamental features of gauge transformations -- that they can be applied independently to the initial and final state. Thus, a small diffeomorphism of the state on $\Sigma_1$ is one that can be performed without altering the state on $\Sigma_2$, and vice versa. In our case, any diffeomorphism $f$ for which $f_\partial$ is the identity acting on $\mathcal{I}$ will be small, because there is then nothing that relates the actions of $f_\partial$ at $\Sigma_1$ and $\Sigma_2$.

Diffeomorphisms which are not small are called `large diffeomorphisms', and are not gauge symmetries, but rather genuine symmetries of the theory, because they involve a deterministic relationship between the change in the initial state and the change in the final state. In our case, such diffeomorphisms must act non-trivially at $\mathcal{I}$, while still obeying~\eqref{Equation: on-shell diffeos}. In fact, if we assume that $\partial\mathcal{M}$ is connected, and $\partial\Sigma_1,\partial\Sigma_2$ are not empty, then any such diffeomorphism continuously connected to the identity will be large, because any Killing vector field of the induced metric on $\mathcal{I}$ is uniquely determined by its value and gradient at any point. Thus, the action of $f_\partial$ near $\partial\Sigma_1\subset \mathcal{I}$ is determined by the action of $f_\partial$ near $\partial\Sigma_2\subset \mathcal{I}$, and vice versa.

Note that, in this scenario, $H$ is always a normal subgroup of $G$. To see this, note that any small diffeomorphism $f\in H$ may be written as $f=f_1\circ f_2$, where $f_1,f_2$ act trivially at $\Sigma_2,\Sigma_1$ respectively. Then, for any $h\in G$, at $\Sigma_2$ we have $h\circ f_1\circ h^{-1}=h\circ h^{-1}=\operatorname{Id}$, and at $\Sigma_1$ we have $h\circ f_2\circ h^{-1}= h\circ h^{-1}=\operatorname{Id}$. Thus, $h\circ f_1\circ h^{-1},h\circ f_2\circ h^{-1}$ act trivially at $\Sigma_2,\Sigma_1$ respectively, so they are small. Therefore, $h\circ f\circ h^{-1} = h\circ f_1\circ h^{-1}\circ h\circ f_2\circ h^{-1}$ is also small. Thus, $H$ is a normal subgroup of $G$ as claimed. The upshot of this is that we can form the quotient group $G/H$, which is a group of genuine physical symmetries for the theory.

\section{Covariant phase space diffeomorphism charges}
\label{Appendix: noether}

Let $L[\phi]$ be the Lagrangian density top form of a generally covariant theory. Under a field variation we may write at linear order
\begin{equation}
    \delta(L[\phi]) = E[\phi]\cdot\delta\phi + \dd(\theta[\phi,\delta\phi]).
\end{equation}
The equations of motion are $E[\phi]=0$, and $\theta$ is the presymplectic potential density. Given two field variations $\delta_1\phi,\delta_2\phi$ we can define the presymplectic current
\begin{equation}
    \omega[\phi,\delta_1\phi,\delta_2\phi] = \delta_1\big(\theta[\phi,\delta_2\phi]\big) - \delta_2\big(\theta[\phi,\delta_1\phi]\big) - \theta[\phi,\delta_{12}\phi],
\end{equation}
where $\delta_{12}\phi$ is the field space Lie bracket of $\delta_1\phi$ and $\delta_2\phi$. Let us assume we are always on-shell so that we always have $E=0$.

Consider the field variation $\lie_W\phi$ for some spacetime vector field $W$. We have
\begin{equation}
    \dd(\iota_W L[\phi]) = \lie_W(L[\phi]) = \dd(\theta[\phi,\lie_W\phi]),
\end{equation}
so $\theta[\phi,\lie_W\phi]-\iota_W L[\phi]$ is a closed form for all $W$. The results of~\cite{WaldClosedForms} imply that it is actually an exact form for all $W$, so we can write
\begin{equation}
    \theta[\phi,\lie_W\phi] = \iota_W L[\phi] + \dd(Q_W[\phi])
\end{equation}
for some form $Q_W[\phi]$ that is linear in $W$, called the `Noether charge density'.

We also have
\begin{equation}
    \lie_W\theta[\phi,\delta\phi] = \iota_W\underbrace{\dd(\theta[\phi,\delta\phi])}_{=\delta(L[\phi])} + \dd(\iota_W\theta[\phi,\delta\phi]),
\end{equation}
and the field space Lie bracket of $\delta\phi$ with $\lie_W\phi$ is $\lie_{\delta W}\phi$, since
\begin{equation}
    \delta(\lie_W\phi) - \lie_W(\delta\phi) = \lie_{\delta W}\phi.
\end{equation}
Here we are allowing in general for $W$ to depend on $\phi$.

Combining the above, we find
\begin{nalign}
    \omega[\phi,\delta\phi,\lie_W\phi] &= \delta\big(\theta[\phi,\lie_W\phi]\big) - \lie_W\big(\theta[\phi,\delta\phi]\big) - \theta[\phi,\lie_{\delta W}\phi]\\
                                       &= \delta\big(\iota_W L[\phi] + \dd(Q_W[\phi])\big) - \dd(\iota_W\theta[\phi,\delta\phi]) - \iota_W\delta (L[\phi]) - \iota_{\delta W} L[\phi] - \dd(Q_{\delta W}[\phi])\\
                                       &= \dd(\delta\big(Q_W[\phi]\big) - Q_{\delta W}[\phi] - \iota_W\theta[\phi,\delta\phi]),
\end{nalign}
where we have used $\delta(\iota_W L[\phi]) = \iota_{\delta W} L[\phi] + \iota_W \delta(L[\phi])$.

If we set $\delta\phi = \lie_V\phi$ for some other vector field $V$ in the above, we get
\begin{nalign}
    \omega[\phi,\lie_V\phi,\lie_W\phi] &= \dd(\lie_V\big(Q_W[\phi]\big) - Q_{\lie_V W}[\phi] - \iota_W\theta[\phi,\lie_V\phi]) \\
                                       &= \dd(\iota_V\dd(Q_W[\phi]) - Q_{[V,W]}[\phi] - \iota_W\big(\iota_VL[\phi] + \dd(Q_V[\phi])\big))\\
                                       &= \dd(\iota_V\dd(Q_W[\phi])-\iota_W\dd(Q_V[\phi]) - Q_{[V,W]}[\phi] + \iota_V\iota_WL[\phi]).
\end{nalign}

\section{Glossary}
\label{Appendix: Glossary}
\begin{description}
    \item[Local quantity] An element of a bundle $\mathcal{B}$ over a space $\mathcal{A}$. We say that the quantity is local on $\mathcal{A}$.
    \item[Bundle projection] The map $\pi:\mathcal{B}\to\mathcal{A}$ which gives the point in $\mathcal{A}$ to which a local quantity in $\mathcal{B}$ is associated.
    \item[Transformation law] The map $X_f$ giving the action of $f\in\operatorname{Diff}(\mathcal{A})$ on $\mathcal{B}$, where $\mathcal{B}$ is an equivariant bundle over $\mathcal{A}$. We will almost always assume that the bundles we use are equivariant.
    \item[Field] A section of a bundle $\pi:\mathcal{B}\to\mathcal{A}$, i.e.\ a map $s:\mathcal{A}\to\mathcal{B}$ satisfying $\operatorname{Id}_{\mathcal{A}}=\pi\circ s$. The space of all such sections is denoted $\Gamma(\mathcal{B})$.
    \item[Local map] A map $\sigma:\mathcal{B}_1\to\mathcal{B}_2$ between two bundles $\pi_1:\mathcal{B}_1\to\mathcal{A}$, $\pi_2:\mathcal{B}_2\to\mathcal{A}$ satisfying $\pi_1=\pi_2\circ\sigma$. The space of all such maps is denoted $\Gamma(\mathcal{B}_1,\mathcal{B}_2)$.
    \item[Pushforward] The action of $f\in\operatorname{Diff}(\mathcal{A})$ on fields and local maps given by
        \begin{equation*}
            f_*s = X_f\circ f^{-1}, \qquad f_*\sigma = X^2_f\circ \sigma\circ X^1_{f^{-1}}
        \end{equation*}
        respectively. If the definitions of transformation laws $X_f$ extends appropriately to $f:\mathcal{A}_1\to\mathcal{A}_2$ for $\mathcal{A}_1\ne\mathcal{A}_2$, then so does the pushforward.
    \item[Pullback] The pushforward of the inverse, $f^*=(f^{-1})_*$.
    \item[Kinematical field bundle] A bundle $\Phi$ over spacetime $\mathcal{M}$, whose elements are the possible local values taken by the kinematical fields.
    \item[Kinematical field configuration] A section $\phi\in\Gamma(\phi)$ of the kinematical field bundle.
    \item[Space of solutions] The space $\mathcal{S}\subset\Gamma(\Phi)$ of kinematical field configurations that are allowed within a given theory (thus, $\mathcal{S}$ consists of kinematical field configurations which are sufficiently smooth, which obey the equations of motion, and which obey the boundary conditions). We call kinematical field configurations in $\mathcal{S}$ \textbf{on-shell}.
    \item[Observable] Any object with a dependence on the kinematical field configuration. For an observable $A$, we denote its value for any fixed  $\phi\in\mathcal{S}$ as $A[\phi]$.
    \item[Local observable] An observable $A$ such that for each $\phi\in\mathcal{S}$, $A[\phi]$ is a local quantity or local map. If $A[\phi]$ is local on spacetime, then we call $A$ a \textbf{local spacetime observable}.
    \item[Active diffeomorphism] The pushforward action of $\operatorname{Diff}(\mathcal{M})$ on $\Gamma(\Phi)$.
    \item[Passive diffeomorphism] The action of $\operatorname{Diff}(\mathcal{A})$ on local quantities on $\mathcal{A}$.
    \item[Covariance] The property of a local spacetime observable which says that it transforms in the same way under active and passive diffeomorphisms.
    \item[On-shell diffeomorphism] An element of the set $G[\phi]\subset\operatorname{Diff}(\mathcal{M})$, where $\phi\in\mathcal{S}$, consisting of spacetime diffeomorphisms which stay on-shell, i.e.\ if $f\in G[\phi]$ then $f_*\phi\in\mathcal{S}$.
    \item[Kinematical locality] The property of a local spacetime observable $A$ which says that $A[\phi]$ only depends on $\phi$ and its derivatives at the spacetime point to which $A[\phi]$ is local.
    \item[Small diffeomorphism] An on-shell diffeomorphism which does not change the physical state, i.e.\ which is a gauge transformation. Such diffeomorphisms form a subset $H[\phi]\subset G[\phi]$.
    \item[Large diffeomorphism] An on-shell diffeomorphism which is not small.
    \item[Gauge-covariance] The property of a local spacetime observable which says that it transforms in the same way under active and passive \emph{small} diffeomorphisms.
    \item[Local dressing] A gauge-covariant spacetime point.
    \item[Universal dressing space] The set $\mathscr{D}$ containing all possible local dressings. For each $\phi\in\mathcal{S}$, this is a bundle over spacetime, where the projection map $\pi[\phi]:\mathscr{D}\to\mathcal{M}$ is given by evaluating each local dressing.
    \item[Dynamical reference frame] A subset $\mathscr{R}\subset\mathscr{D}$ of the universal dressing space. Often we just call this a \textbf{frame}.
    \item[Image of a frame] The region of spacetime $\mathcal{N}[\phi]=\pi[\phi](\mathscr{R})$.
    \item[Surjective frame] A frame whose image is all of spacetime.
    \item[Injective frame] A frame for which the restriction of $\pi[\phi]$ to $\mathscr{R}$ is injective.
    \item[Bijective frame] A frame which is both injective and surjective.
    \item[Parametrisation of a frame] A bijection $R$ from some space of parameters $\mathscr{O}$ to a frame $\mathscr{R}$. Each element of $\mathscr{O}$ is called a \textbf{local orientation}.
    \item[Parametrised frame] An injective map $R:\mathscr{O}\to\mathscr{D}$. Such a map is both a frame and a parametrisation at the same time: the frame $\mathscr{R}$ is the image of $R$, and the parametrisation is the restriction of $R$ to the bijective map $\mathscr{O}\to\mathscr{R}$.
    \item[Frame field] An observable which for each $\phi\in\mathcal{S}$ is a section $s[\phi]$ of $R[\phi]=\pi[\phi]\circ R:\mathcal{O}\to\mathcal{N}[\phi]$. When the frame is injective, there is only one frame field $s[\phi]=(R[\phi])^{-1}$.
    \item[Global orientation] A possible configuration of a frame field, i.e.\ $s[\phi]$ for a fixed $\phi$. The space of all global orientations is denoted $\mathcal{O}$.
    \item[Relational observable] An observable $O_{A,R}$, where $A$ is covariant and $R$ is a parametrised frame defined by $O_{A,R}[\phi]=(R[\phi])^*A[\phi]$ if $A[\phi]$ is a local map, or $O_{A,R}[\phi] = X_{(R[\phi])^{-1}}(A[\phi])$ if $A[\phi]$ is a local quantity.
    \item[Relational locality] The property of a local observable $A$ on a local orientation space $\mathscr{O}$ of a parametrised frame $R:\mathscr{O}\to\mathscr{D}$ which says that $(R[\phi])_*A[\phi]$ is kinematically local if $A[\phi]$ is a local map, or $X_{R[\phi]}(A[\phi])$ is kinematically local if $A[\phi]$ is a local quantity.
    \item[Change of frames] A procedure for transforming between the perspectives of two frames. If the frames are parametrised, this involves a (usually field-dependent) map between their local orientation spaces.
    \item[Frame reorientation] A change in the kinematical field configuration which causes the global orientation of a frame to change, but which leaves some set of `other' fields fixed.
    \item[Relational atlas] A collection of injective reference frames whose images cover spacetime.
    \item[Relational field variation] The field variation defined by $\delta_R\phi = \delta\phi - \lie_{V_R[\phi,\delta\phi]}$, where $\delta\phi$ is any field variation, and $V_R[\phi,\delta\phi]$ is the spacetime vector field generating the diffeomorphism $R[\phi+\delta\phi]\circ (R[\phi])^{-1}$ for some (parametrised) frame $R$.
    \item[Relational field configuration] The pullback of the kinematical field configuration through a frame, i.e.\ $\phi_R = (R[\phi])^*\phi$.
    \item[Relational functional derivative] Given a (parametrised) frame, and an observable $A$ constructed from a set of relationally local observables, the object $\frac{\delta_R A}{\delta_R\phi}$ appearing in the expansion $\delta (A[\phi]) = \int_{\mathcal{M}}\frac{\delta_R A}{\delta_R\phi}\cdot\delta_R\phi$, where $\epsilon$ is the volume form on spacetime $\mathcal{M}$.
    \item[Relational support] The support of the relational functional derivative of an observable.
\end{description}

\printbibliography
\end{document}